# SAT.STFR.FRQ (UWA) DETAIL DESIGN REPORT (MID)

| | |
|---|---|
| **Document Number** | SKA-TEL-SADT-0000390 |
| **Revision** | 2.0 |
| **Author** | Sascha Schediwy, David Gozzard |
| **Date** | 2018-07-24 |
| **Status** | RELEASE |

| Name | Designation | Affiliation | Signature |
|---|---|---|---|
| *Owned by:* | | | |
| Sascha Schediwy | SADT.SAT.STFR.FRQ (UWA) Lead Designer | University of Western Australia | *signed* |
| *Approved by:* | | | |
| Jill Hammond | SADT Project Manager | University of Manchester | *Jill Hammond (Jul 24, 2018)* |
| *Released by:* | | | |
| Rob Gabrielczyk | SADT System Engineer | AEON Engineering Ltd. | *Robert Gabrielczyk (Jul 24, 2018)* |

# DOCUMENT HISTORY

| Revision | Date Of Issue | Engineering Change Number | Comments |
|---|---|---|---|
| 0.1 | 2017-08-01 | - | First release. |
| 0.2 | 2017-12-14 | - | Second release; post down select. |
| 0.3 | 2018-02-20 | - | Following Technical Author review. |
| 1.0 | 2018-02-20 | - | MH: Signature preparation, major release for SADT CDR delivery to the SKAO. |
| 1.1 | 2018-05-22 |  | Edit to include SKA IP disclaimer for public release |
| 1.2 | 2018-06-15 | ECP-180006 | With edits to complete OARs following CDR |
| 1.3 | 2018-07-18 |  | MH: Amendments to revision numbers referenced for AD43, RD2. Addition of revision number for RD59. Correction to document number referenced for AD56. Addition of reference to OAR cover ECP-180006 |
| 2.0 | 2018-07-18 |  | MH: Signature preparation, major release for SADT CDR delivery to the SKAO in final July submission. |

# REVIEW HISTORY

| Revision | Date Of Review | Reviewer | Comments |
|---|---|---|---|
|  |  |  |  |

# DOCUMENT SOFTWARE

| | Package | Version | Filename |
|---|---|---|---|
| **Wordprocessor** | Word | 2013 | SKA-TEL-SADT-0000390-SADT.STFR.FRQ-DDD-01-SAT.STFR.FRQ (UWA) Detail Design Report (MID) - post DS_rev8.docx |
| **Block diagrams** | PowerPoint | 2013 |  |
| **Other** | SolidEdge | ST4 |  |
|  | DesignSpark | 8.0 |  |
|  | Excel | 2013 |  |
|  | Java | SE 8 |  |

# ORGANISATION DETAILS

| Name | University of Manchester |
|---|---|
| Registered Address | Jodrell Bank Centre for Astrophysics<br>Alan Turing Building<br>The University of Manchester Oxford Road<br>Manchester, UK<br>M13 9PL |
| Fax. | +44 (0)161 275 4247 |
| Website | www.manchester.ac.uk |

## INTELLECTUAL PROPERTY DISCLAIMER

The design work for this phase synchronisation system was performed by the UWA as a member of the SaDT consortium contributing to the SKA project and is subject to the SKA IP policy.

# TABLE OF CONTENTS

# LIST OF FIGURES

# LIST OF TABLES

# 1 EXECUTIVE SUMMARY

The Square Kilometre Array (SKA) project [RD1] is an international effort to build the world's most sensitive radio telescope operating in the 50 MHz to 14 GHz frequency range. Construction of the SKA is divided into phases, with the first phase (SKA1) accounting for the first 10% of the telescope's receiving capacity. During SKA1, a Low-Frequency Aperture Array (LFAA) comprising over a hundred thousand individual dipole antenna elements will be constructed in Western Australia (SKA1-LOW), while an array of 197 parabolic-receptor antennas, incorporating the 64 receptors of MeerKAT, will be constructed in South Africa (SKA1-MID).

Radio telescope arrays, such as the SKA, require phase-coherent reference signals to be transmitted to each antenna site in the array. In the case of the SKA, these reference signals are generated at a central site and transmitted to the antenna sites via fibre-optic cables up to 175 km in length [RD2]. Environmental perturbations affect the optical path length of the fibre and act to degrade the phase stability of the reference signals received at the antennas, which has the ultimate effect of reducing the fidelity and dynamic range of the data [RD3]. Given the combination of long fibre distances and relatively high frequencies of the transmitted reference signals, the SKA needs to employ actively-stabilised frequency transfer technologies to suppress the fibre-optic link noise [RD4] in order to maintain phase-coherence across the array.

Since 2011, researchers at the University of Western Australia (UWA) have led the development of an 'SKA phase synchronisation system' designed specifically to meet the scientific needs and technical challenges of the SKA telescope. This system [RD5] is based on the transmission of actively stabilised phase-coherent reference signals generated at the Central Processing Facility (CPF), and then transmitted via separate optical fibre links to each antenna site. The frequency transfer technique at the core of the SKA phase synchronisation system is an evolution of the Atacama Large Millimeter Array (ALMA) distributed 'photonic Local Oscillator (LO) system' [RD6], incorporating key advances made by the international frequency metrology community over the last decade [RD7], [RD8] and [RD9], as well as novel innovations developed by UWA researchers [RD10] and [RD11].

Two variants of the SKA phase synchronisation system have been designed. Each one has been optimised specifically for its respective telescope:

- For SKA1-MID, the required microwave (MW) shift is generated using a Dual-parallel Mach-Zehnder Modulator (DPM), biased to generate Single-sideband Suppressed-carrier (SSB-SC) modulation [RD10].
- For SKA1-LOW, the Radio Frequency (RF) shift is generated using a simpler Acousto-optic Modulator (AOM) [RD11].

This results in two systems that easily meet the SKA functional performance requirements, as demonstrated by laboratory testing [RD10], [RD11], [RD12] and [RD13], overhead fibre field trials [RD14], [RD15] and [RD16], and astronomical verification [RD17], [RD18] and [RD19], yet maximise robustness and maintainability while keep complexity and costs to a minimum. Following an extensive technical down-select process, in October 2017 a MW-frequency variant from the UWA was selected by the Square Kilometre Array Office (SKAO) to be the phase synchronisation system for SKA1-MID.

The key innovation of the SKA phase synchronisation system was finding a way to use AOMs as servo-loop actuators for RF and MW-frequency transfer [RD10] and [RD11]. The large servo bandwidth and infinite feedback range of these servo-loop AOMs ensures that the stabilisation system servo-loops never require integrator resets. The SKA phase synchronisation system also utilises AOMs to generate static frequency shifts at the antenna sites to mitigate against unwanted reflections that are inevitably present on real-world links. Reflection mitigation is absolutely essential for the SKA phase synchronisation system, as there is no way to guarantee that all links will remain completely free of reflections over the lifetime of the project.

The SKA phase synchronisation system has the servo-loop electronics and the vast majority of all other optical and electronic components located at the CPF, greatly simplifying maintenance. A single high-quality Frequency Synthesiser (FS), tied to the SKA master clock, is used to generate phase coherent reference signals, and these are distributed to the Transmitter Modules (TMs) which are then used to transmit the optical signals across each fibre link. The TMs incorporate the servo-loop AOMs, and these are able to add an independent and unique RF-scale frequency offset – in the optical domain – to the common transmission frequency for each link. This avoids any possibility of common frequencies at each antenna site to ensure any stray RF emissions will not be coherent if picked up by the receivers.

The Receiver Modules (RMs) for the SKA phase synchronisation system have a very small form-factor and contain only a minimum number of simple optical and analogue electronic components, making them extremely robust to external environmental perturbation. In addition, they are designed to be capable of being mounted directly on the SKA1-MID antenna indexer alongside the receiver. Currently, the SADT interface with DISH is in the antenna pedestal, and the DISH Consortium are required to build a second frequency transfer system to transmit the reference signals up the cable wraps to the indexer. After a successful down-select, the DISH consortium and SKA Office (SKAO) have agreed to an Engineering Change Proposal (ECP) to correct this inefficiency.

A small form-factor, industry standard, Oven-controlled Crystal Oscillator (OCXO) is incorporated into the RM to provide phase coherence at timescales shorter than the light round-trip time of the fibre link. The OCXO is tied to the incoming reference signals using a simple, encapsulated Phase Locked Loop (PLL) based on the proven design implemented by the Australian SKA Pathfinder (ASKAP). This is particularly important, as it has been shown that using multiple MW-FSs can easily lead to a significant loss of coherence, even if the transmission frequency is being successfully stabilised [RD18], [RD19] and [RD20].

The SKA phase synchronisation system is designed in such a way as to also stabilise the non-common optical fibre paths in the TMs. This effectively stabilises the TMs at the same time as the fibre link, making the equipment in the CPF extremely robust to external environmental changes. The optical phase sensing allows for the use of Faraday Mirrors (FMs) to give maximum detected signal at the servo photodetector without requiring any initial polarisation alignment, or any ongoing polarisation control or polarisation scrambling.

The SKA phase synchronisation system has been extensively tested:

- Using standard metrology techniques in a laboratory setting [RD12] and [RD13], with signals transmitted over metropolitan fibre links and fibre spools under all required conditions;
- On 186 km of overhead fibre at the South African SKA site [RD14], [RD15] and [RD16];
- Using astronomical verification with the Australian Telescope Compact Array (ATCA) for SKA1-MID [RD18], [RD19] and [RD20], and the ASKAP for SKA1-LOW [RD19].

This has demonstrated that the SKA phase synchronisation system is fully compliant with all SKA requirements, as well as demonstrating functionality of critical practical factors that are not captured by these requirements.

Furthermore, UWA researchers in partnership with MeerKAT and University of Manchester (UoM) engineers, have developed the detailed designs into a set of mass manufacture archetypes, effectively getting a head-start at addressing manufacturing issues that may be encountered by contractors during the construction phase. The first set of mass manufacture archetypes for SKA1-LOW were completed in Q2, 2016 [RD21]; and for SKA1-MID in Q1, 2017 [RD22]. All aspects of the mass manufacture design are openly available and are provided with sufficient detail so that any firm with expertise in optical and electronic assembly can to reproduce these systems with minimal domain expert input. An optical technology consultancy firm was employed to provide an independent review of the labour costs associated with assembly and testing (see Appendix 8.9.3).

All sub-elements of the SKA phase synchronisation system have been designed to be hot-swappable, enabling simple installation and easy maintainability (especially as the vast majority are located at the CPF. The system is designed so that during commissioning, only one free parameter needs to be optimised per link.

Prior to the technology down-select process, the detailed design presented in a previous version of this document had been critically assessed by the following independent domain experts:

- Gijs Schoonderbeek from the ASTRON Netherlands Institute for Radio Astronomy;
- Larry D'Addario from the Jet Propulsion Laboratory;
- Johan Burger from SKA South Africa.

The input from this review was used to update this document. As part of the SADT Consortium-led technology down-select process, this document was reviewed further by the following independent domain experts:

- William Shillue from the National Radio Astronomy Observatory;
- Miho Fujieda from the National Institute of Information and Communications Technology;
- Sven-Christian Ebenhag from the Swedish Research Institute (RISE).

Again, this document was updated taking into account the feedback provided.

These reviews have built confidence in the detailed design and ensured that the SKA phase synchronisation system is the best possible phase synchronisation solution for the SKA telescope.

## 2 INTRODUCTORY SECTIONS

### 2.1 Purpose

This document provides the detailed design of the SKA phase synchronisation system for the SKA1-MID telescope developed by UWA. The design is used to derive a cost model for deploying the frequency dissemination equipment on both telescopes.

The detailed design report concerns the following aspects:

- The solution description which includes:

    - Technical background;
    - Solution overview;
    - Solution design justification;
    - Details of hardware and software;
    - Safety and security;
    - Integration;
    - Interoperability;
    - Costing.

- Evaluation which includes:

    - Testing and verification of the design;
    - Construction of a mass manufacture archetype;
    - Independent assessment of the solution.

- Conclusion, and recommendations for further development and procurement.
- Statement of Compliance indicating the system compliance with each of the SKA requirements.

The detailed design of the SKA phase synchronisation system reflects the current baseline (Rev 2) of the SKA Programme [AD1] and Level 1 requirements as per Revision 10 [AD2] at the time of writing.

### 2.2 Scope

This report describes the full extent of the SKA phase synchronisation system within the SKA's Signal and Data Transport (SADT) consortium's Synchronisation and Timing (SAT) network. The system receives an electronic reference signal from the SAT clocks at the SKA1-MID CPF, and transfers the full stability of the reference signal across the SAT network to each receptor pedestal. At the receptor pedestal, an electronic copy of the reference signal is provided to the downstream receiver electronics. The SKA phase synchronisation system is controlled and monitored using the LMC with the required Local Infrastructure (LINFRA) provided by SADT.

### 2.3 Intended Audience

This design report is to be used within the SADT Consortium, by the SKAO, and other design consortia within SKA. A draft of this report formed part of the Body of Evidence for the SADT Consortium down-select process of the SKA phase synchronisation system. This report will form part of the documentation for the SADT Consortium Critical Design Review (CDR).

The design work for this phase synchronisation system was performed by the UWA as a member of the SaDT consortium contributing to the SKA project and is subject to the SKA IP policy.

## 2.4 Applicable and Reference Documents

### 2.4.1 Applicable Documents

The documents in Table 1 are applicable to this document.

| [AD1] | P. Dewdney, SKA1 System Baseline (v2) Description. SKA Organisation, 2015. SKA-TEL-SKO-0000308 (rev. 01) |
|---|---|
| [AD2] | W. Turner. SKA Phase 1 System (Level 1) Requirements Specification. SKA Organisation, 2017. SKA-TEL-SKO-0000008 (rev. 11) |
| [AD3] | A. Wilkinson and M. Pearson. STFR Frequency Dissemination System Down-Select Methodology. SKA Signal and Data Transport Consortium, 2017. SKA-TEL-SADT-0000524 (rev. 3.0): p. 31. |

**Table 1 Applicable documents**

### 2.4.2 Reference Documents

The documents in Table 2 are referenced in this document.

| [RD1] | P. Dewdney, SKA1 System Baseline V2 Description. SKA Organisation, 2015. SKA-TEL-SKO-0000308 (rev. 1): p. 58 |
|---|---|
| [RD2] | R. Oberland, NWA Model SKA1-MID. SKA Signal and Data Transport Consortium, 2017. SKA-TEL-SADT-0000523 (rev. 6.0): p. 14 |
| [RD3] | J.F. Cliche and B. Shillue, Precision timing control for radioastronomy: maintaining femtosecond synchronization in the Atacama Large Millimeter Array. Control Systems, IEEE, 2006. 26(1): p. 19-26 |
| [RD4] | K. Grainge, et al., Square Kilometre Array: The radio telescope of the XXI century. Astronomy Reports, 2017. 61(4): p. 288-296 (see Appendix 8.2.1 of this document) |
| [RD5] | S.W. Schediwy, et al., Phase Synchronization For The Mid-Frequency Square Kilometre Array. In prep. for Publications of the Astronomical Society of Australia (2018) (see Appendix 8.2.3 of this document) |
| [RD6] | B. Shillue, S. AlBanna, and L. D'Addario. Transmission of low phase noise, low phase drift millimeter-wavelength references by a stabilized fiber distribution system. in Microwave Photonics, 2004. MWP'04. 2004 IEEE International Topical Meeting on. 2004 |
| [RD7] | S.M. Foreman, et al., Remote transfer of ultrastable frequency references via fiber networks. Review of Scientific Instruments, 2007. 78(2): p. 021101-25 |
| [RD8] | O. Lopez, et al., High-resolution microwave frequency dissemination on an 86-km urban optical link. Applied Physics B: Lasers and Optics, 2010. 98(4): p. 723-727 |
| [RD9] | K. Predehl, et al., A 920-Kilometer Optical Fiber Link for Frequency Metrology at the 19th Decimal Place. Science, 2012. 336(6080): p. 441-444 |
| [RD10] | S.W. Schediwy, et al., Stabilized microwave-frequency transfer using optical phase sensing and actuation. Optics Letters, 2017. 42(9): p. 1648-1651 (see Appendix 8.2.6 of this document) |
| [RD11] | D.G. Gozzard, et al., Simple Stabilized Radio-Frequency Transfer with Optical Phase Actuation. Photonics Technology Letters, IEEE, 2018. 30(3): p.258 (see Appendix 8.2.5 of this document) |
| [RD12] | S.W. Schediwy and D.G. Gozzard, Pre-PDR Laboratory Verification of UWA's SKA Synchronisation System. SKA Signal and Data Transport Consortium, 2014. SKA-TEL-SADT-0000616: p. 16 (see Appendix 8.3.1 of this document) |
| [RD13] | S. Schediwy and D. Gozzard, Pre-CDR Laboratory Verification of UWA's SKA Synchronisation System. SKA Signal and Data Transport Consortium, 2017. SKA-TEL-SADT-0000620: p. 13 (see Appendix 8.3.2 of this document) |
| [RD14] | S.W. Schediwy and D.G. Gozzard, UWA South African SKA Site Long-Haul Overhead Fibre Field Trial Report. SKA Signal and Data Transport Consortium, 2015. SKA-TEL-SADT-0000109: p. 20 (see Appendix 8.3.3 of this document) |

**Table 2 Reference documents**

# 3 ACRONYMS AND DEFINITION OF TERMS

## 3.1 Acronyms

| | |
|---|---|
| AAA | Authentication, Authorization, and Accounting |
| ADC | Analogue-to-digital Converter |
| ALMA | Atacama Large Millimeter Array |
| AOM | Acousto-optic Modulator |
| ASKAP | Australian SKA Pathfinder |
| ATCA | Australia Telescope Compact Array |
| BOM | Bill of Materials |
| CCD | Command and Control Device |
| CDR | Critical Design Review |
| CIA | Confidentiality, Integrity, and Availability |
| CIN | Configuration Identification Number |
| COTS | Commercial Off-The-Shelf |
| CPF | Central Processing Facility |
| DDS | Direct Digital Synthesiser |
| DfM | Design for Manufacturability |
| DPM | Dual-parallel Mach-Zehnder Modulator |
| ECP | Engineering Change Proposal |
| EICD | External Interface Control Document |
| e-MERLIN | Enhanced Multi Element Radio Linked Interferometry Network |
| EDFA | Erbium-Doped Fibre Amplifier |
| EMI | Electromagnetic Interference |
| FDP | Field-deployable Prototype |
| FIT | Failures In Time |
| FM | Faraday Mirror |
| FP | Fibre Patch |
| FS | Frequency Synthesiser |
| GUI | Graphical User Interface |
| H-maser | Hydrogen-maser |
| ICRAR | International Centre for Radio Astronomy Research |
| IICD | Internal Interface Control Document |
| LFAA | Low-Frequency Aperture Array |
| LINFRA | Local Infrastructure |
| LO | Local Oscillator |
| LRU | Line Replaceable Unit |
| MeerKAT | The South African precursor array being built on site in the Karoo |
| MI | Michaelson Interferometer |
| MS | Microwave Shift |
| MW | Microwave |
| MRO | Murchison Radioastronomy Observatory |
| MTBF | Mean Time Before Failure |
| MZI | Mach-Zehnder Interferometer |

| NPL | National Physical Laboratory |
|---|---|
| NSDN | Non-Science Data Network |
| NTFN | National Time and Frequency Network |
| OA | Optical Amplifier |
| OCXO | Oven-controlled Crystal Oscillator |
| OS | Optical Source |
| OTDR | Optical Time-domain Reflectometry |
| PBS | Product Breakdown Structure |
| PCB | Printed Circuit Board |
| PDR | Preliminary Design Review |
| PIC | Peripheral Interface Controller |
| PLL | Phase Locked Loop |
| RAMS | Reliability, Availability, Maintainability, and Safety |
| RD | Rack Distribution |
| RF | Radio Frequency |
| RFI | Radio Frequency Interference |
| RISE | Swedish Research Institute |
| RM | Receiver Module |
| RPF | Remote Processing Facility |
| SADT | Signal and Data Transport |
| SAT | Synchronisation and Timing |
| SCFO | Sample Clock Frequency Offset |
| s.d. | standard deviation |
| SG | Signal Generator |
| SKA | Square Kilometre Array |
| SKA1 | Square Kilometre Array phase 1 |
| SKA1-LOW | Low-frequency aperture array of SKA1 |
| SKA1-MID | Mid-frequency array of receptors of SKA1 |
| SKAO | Square Kilometre Array Office |
| SR | Sub Rack |
| SSB-SC | Single-sideband Suppressed-carrier |
| STFR | System for the Time and Frequency Reference |
| TBD | To be determined |
| THU | Tsinghua University |
| TM | Transmitter Module |
| TM | Telescope Manager |
| UoM | University of Manchester |
| UWA | University of Western Australia |
| VCO | Voltage-controlled Oscillator |
| WR | White Rabbit |
| WSOI | Wider System-of-interest |

## 3.2 Definition of Terms

**Analogue-to-digital converter (ADC)**
Key device within the LFAA and DISH receiver used to sample an incoming analogue waveform, using the STFR reference signals as its clock.

**Atacama Large Millimeter Array (ALMA)**
Radio telescope array operating between 31 and 1,000 GHz situated in the high mountains of Chile.

**Acousto-optic modulator (AOM)**
Optoelectronic device that can be used to apply a fixed or variable Radio Frequency (RF) shift onto a transmitted optical signal.

**Australian SKA Pathfinder (ASKAP)**
Australian SKA pathfinder radio telescope array.

**Australia Telescope Compact Array (ATCA)**
Australia's current workhorse radio telescope array.

**CLOCK**
The SADT work-package that includes the SKA Hydrogen-maser (H-Maser) ensemble and timescale.

**Command and Control Device (CCD)**
Unit integrated within the Sub Rack (SR) of the SKA phase synchronisation system for controlling and recording signals on the attached Transmitter Modules (TMs).

**Critical Design Review (CDR)**
Second-stage review conducted by the Square Kilometre Array Office (SKAO) on SKA Consortia design elements.

**Configuration Identification Number (CIN)**
Unique identification number assigned to each SADT Line Replaceable Unit (LRU).

**Commercial off-the-shelf (COTS)**
Equipment that can be purchased in their entirety as a single entity from a commercial supplier.

**Central Processing Facility (CPF)**
Environmentally controlled structure which houses (amongst other things) all the FRQ (UWA) elements other than the Receiver Modules s) and Optical Amplifier.

**Direct Digital Synthesiser (DDS)**
4-channel Printed Circuit Board (PCB) surface mount chip located on each TM, which takes a RF reference signal from the Signal Generator (SG), to produce the servo-loop Local Oscillator (LO) signal, as well as other ancillary signals required by the TM.

**DISH**
SKA Consortium responsible for producing the main physical elements of the SKA1-MID telescope.

**Dual-parallel Mach-Zehnder Modulator (DPM)**
Optoelectronic device that can be used to apply a series of complex modulation states (up to MW frequencies) onto a transmitted optical signal.

**Engineering Change Proposal (ECP)**
Formal process for varying elements of the SKA telescope design.

**External Interface Control Document (EICD)**
Document defining an interface between an SADT sub-element and a sub-element belonging to another SKA consortia.

**Enhanced Multi Element Radio Linked Interferometry Network (e-MERLIN)**

Radio telescope array, situated in the UK and linked by optical fibre, with similar maximum baselines as the SKA1-MID (but comprising only seven receptors).

**Electromagnetic Interference (EMI)**
Self-generated unwanted RF or MW-frequency signals.

**Faraday Mirror (FM)**
Optical device that reflects incoming light with a 90° turn of polarisation.

**Fibre Patch (FP) lead**
An STFR FRQ (UWA) hardware element: a short length of optical fibre cable.

**FRQ**
SADT work-package which incorporates the frequency dissemination technology at the core of the SKA phase synchronisation system.

**FRQ (UWA)**
The SKA phase synchronisation system from UWA.

**Frequency Synthesiser (FS)**
An STFR FRQ (UWA) hardware element: A high-quality MW-frequency signal source, referenced to the SKA clock enable, which is used to provide a static signal to the Microwave Shift (MS) LRU.

**Hydrogen-maser (H-maser)**
A high-precision frequency reference.

**International Centre for Radio Astronomy Research (ICRAR)**
World-class astronomy institute joint venture between the University of Australia and Curtin University.

**Internal Interface Control Document (IICD)**
Document defining an interface between two SADT sub-elements.

**Low-Frequency Aperture Array (LFAA)**
SKA Consortium responsible for producing the main physical elements of the SKA1-LOW telescope.

**LMC**
SADT local management and control for the SAT network.

**Local Oscillator (LO)**
The RF signal within each TM that is the reference against which the incoming optical link RF signal is compared to produce the servo-loop error signal.

**Line Replaceable Unit (LRU)**
Single reproducible hardware building block of the SADT work package elements; can be Commercial Off-The-Shelf (COTS) or bespoke.

**Meer Karoo Array Telescope (MeerKAT)**
South African SKA pathfinder radio telescope array.

**Michelson Interferometer (MI)**
Optical interferometer, where one arm comprises the entire fibre link, and the other short arm contained within the TM, provides the optical reference signals that for the link stabilisation servo loop.

**Microwave Shift (MS)**
An FRQ (UWA) hardware element: Applies an electronic signal from the Microwave (MW) Synthesiser to the optical signal from the Optical Source (OS) to produce two optical signals separated by the MW-frequency reference signal on two separate optical fibres.

**Microwave (MW)**
For the purposes of this document; frequencies between 1 MHz and 1 GHz.

**Murchison Radioastronomy Observatory (MRO)**
This is a designated radio quiet zone located near Boolardy station in Western Australia that currently host two main instruments; the Murchison Widefield Array and the Australian Square Kilometre Array Pathfinder. It is also the Australian site for the Square Kilometre Array.

**Mach-Zehnder Interferometer (MZI)**
Fiberised two-arm optical interferometer localised at the Central Processing Facility (CPF), used to produce the MW (SKA1-MID) or radio (SKA-LOW) frequency separation between the two transmitted optical signals.

**National Physical Laboratory (NPL)**
The United Kingdom's national measurement institute. Design authority of the SAT.clock.

**National Time and Frequency Network (NTFN)**
Australian ARC-funded project with the aim to develop continental-scale time and frequency transfer technology.

**Optical Amplifier (OA)**
An FRQ (UWA) hardware element: Bi-directional erbium-doped fibre amplifier, used to amplify optical signals on the longest SKA1-MID links.

**Oven-controlled Crystal Oscillator (OCXO)**
Ultra-low noise, thermally stabilised oscillator located with the RM to provide coherence for the SKA phase synchronisation system at timescales shorter than the light round trip time of the fibre link.

**Optical Source (OS)**
An FRQ (UWA) hardware element. This is a single high-coherence 1552.52 nm laser that is the maser source for all optical signals.

**Optical time-domain Reflectometry (OTDR)**
Technique for identifying the locations of large unwanted reflections in optical fibre links, as well as their total optical loss.

**Product Breakdown Structure (PBS)**
Hierarchical structure that identifies all LRUs within the SADT Consortium using unique Configuration Identification Numbers.

**Printed Circuit Board (PCB)**
Mechanical support for electronic components, conductive tracks, pads and other features etched from copper sheets on a non-conductive substrate.

**Preliminary Design Review (PDR)**
First-stage review conducted by the SKAO on SKA Consortia design elements.

**Peripheral Interface Controller (PIC)**
A family of microcontrollers made by the company Microchip Technology.

**Phase Locked Loop (PLL)**
A simple electronic circuit that includes a voltage-driven oscillator which constantly adjusts to match the frequency of an input signal.

**Rack Distribution (RD)**
An FRQ (UWA) hardware element, used to distribute RF, optical- and MW- frequency signals from their sources to the SRs.

**Radio Frequency (RF)**
For the purposes of this document; frequencies between 1 and 100 GHz (as per the common definition within radio engineering).

**Radio Frequency Interference (RFI)**
Unwanted RF or MW-frequency signals caused by external sources.

**Receiver Module (RM)**

An FRQ (UWA) hardware element. It is a hot-swappable, small form-factor enclosed module which houses fiberised optics and electronics, including a clean-up Oven-controlled Crystal Oscillator (OCXO) in a Phase Locked Loop (PLL) to convert the stabilised signal received across the fibre link, into an electronic signal which is passed to the LFAA or DISH.

**Remote Processing Facility (RPF)**

Environmentally controlled structure which houses (amongst other things) the FRQ (UWA) RMs and LFAA analogue-to-digital receivers.

**Signal and Data Transport (SADT)**

The signal and data transport element of phase 1 of the SKA telescope.

**SADT Consortium**

Collection of organisations, including UWA, led by the University of Manchester (UoM) to design the SADT element of the SKA Phase 1 telescope.

**Synchronisation and Timing (SAT)**

SADT network used to disseminating references signals and absolute time.

**standard deviation (s.d.)**

In statistics, the standard deviation is a measure that is used to quantify the amount of variation or dispersion of a set of data values.

**Signal Generator (SG)**

An FRQ (UWA) hardware element. It is the source of RF signals used to reference the Direct Digital Synthesiser (DDS) chips in each TM.

**SKA-LOW**

The low frequency SKA telescope located in Australia.

**SKA-MID**

The mid frequency SKA telescope located in South Africa.

**Square Kilometre Array Office (SKAO)**

The organisation that is the design authority for the SKA telescope and is the client of the SADT Consortium.

**Sub Rack (SR)**

An FRQ (UWA) hardware element. It consists of a 3U, 19" rack mount enclosure that houses 16 hot-swappable TMs; as well as a common power supply, the command and control module, and internal optical-, RF-, and MW-distribution units.

**Single-sideband Suppressed-carrier (SSB-SC)**

A modulation type where only one sideband is generated and all other optical signals, including the carrier, are suppressed.

**System for the Time and Frequency Reference (STFR)**

SADT sub-element that comprises the SKA phase synchronisation system and the SKA system for disseminating absolute time.

**Tsinghua University (THU)**

Design authority of the alterative SKA phase synchronisation system in the SADT technology down-select.

**Transmitter Module (TM)**

An FRQ (UWA) hardware element. It is a hot-swappable PCB, mounted within the SR, that includes fiberised optics and the servo-loop electronics for stabilising the transmitted reference signals to be transmitted across the fibre link.

**University of Manchester (UoM)**

Lead institute of the SADT Consortium.

**UTC**
SADT work-package responsible for disseminating absolute time.

**University of Western Australia (UWA)**
Design authority for the SKA phase synchronisation system, the detailed design of which is described in this document.

**Voltage-controlled Oscillator (VCO)**
Frequency source, the output frequency of which can be controlled given a range of input voltages.

# 4 SOLUTION DESCRIPTION

## 4.1 Technical Background

### 4.1.1 SKA Telescope Phase Synchronisation

Phase synchronisation systems have been successfully used on other radio telescope arrays including the Enhanced Multi Element Radio Linked Interferometry Network (e-MERLIN) in the UK [RD27] and [RD28] and the ALMA [RD6] and [RD29].

This report describes the detailed design of the SKA-MID variant of the SKA phase synchronisation system based on actively stabilised MW-frequency transfer via optical fibre [RD10]. A separate report describes the detailed design of the SKA-LOW variant [RD30].

Figure 1 shows the SKA1-MID network layout as illustrated in [RD2].

**Figure 1 SKA1-MID network layout**

### 4.1.2 Actively Stabilised Frequency Transfer via Optical Fibre

In both SKA telescopes, the electronic master reference signals originate from an ensemble of three H-Masers, situated in the CPF for each telescope. For SKA1-MID, the reference signals are transmitted from the South African CPF over optical links to the 197 antenna pedestals, including 64 retrofitted MeerKAT antennas, as shown in Figure 1 .

The reference signals are converted into the optical domain and then transmitted over the optical fibre link where they acquire phase noise due to environmental disturbances on the link. This degrades the phase-stability and thus the coherence of the reference signals, which has the ultimate effect of reducing the fidelity and dynamic range of the data.

The SKA telescope will employ actively stabilised frequency transfer technology to suppress these environmental disturbances and ensure phase coherence across the array. A simplified diagram of a generalised stabilised frequency transfer system is shown in Figure 2.

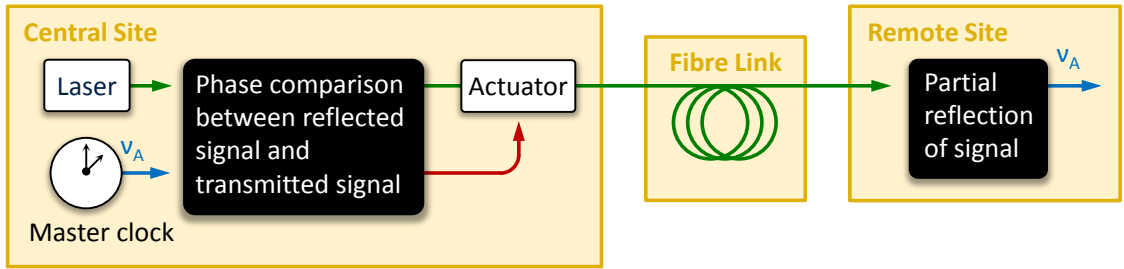

**Figure 2 Simplified schematic of a generalised stabilised frequency transfer system**

1. The reference signals are transmitted from the central site across the fibre link to the remote site.
2. At the remote site, a partial reflection of the transmitted reference signal (having made a round-trip through the link) is compared against a copy of the original transmitted reference signal.
3. The environmental disturbances on the link are encoded as the phase difference between these two signals. This comparison is used to generate a correction signal, which is then used to control an actuator that modifies the outgoing signal in such a way as to supress the effect of the environmental disturbances.
4. At the remote site, the now stabilised reference signals are converted into the electronic domain, and supplied to the telescope's Analogue-to-digital Converters (ADCs).

### 4.1.3  Designing to SKA Requirements

The SKA phase synchronisation system must meet a series of performance and operational requirements to allow the SKA to operate at maximum capacity. These are described in Table 3.

| Requirement | Reference | Description |
|---|---|---|
| Functional performance | §7.1 | Concerns the coherence loss, phase drift, and jitter requirements that must be met for optimal operation of the telescope. |
| Normal operating conditions | §7.2 | Concerns the climatic condition the system will be subjected to, conformance with the design and layout of the telescope, and other practical considerations such as RFI and system monitoring. The full list of relevant requirements is given in §7. |
| Key additional requirements | §7.3 | |

**Table 3 SKA phase synchronisation system performance and operational requirements**

## 4.2  Solution Overview

### 4.2.1  Solution Development

Since 2011, researchers at the University of Western Australia (UWA), have led the development of an SKA phase synchronisation system designed specifically to meet the scientific needs and technical challenges of the SKA telescope. A concept document (Appendix 8.1.1) and paper [RD31] were developed as part of the Australian National Time and Frequency Network (NTFN) project This led to a collaboration with UoM and the UK's National Physical Laboratory (NPL) from mid-2012. The collaboration was formalised through the funding of a UWA collaboration grant in 2013.

The UWA-UoM-NPL proto-consortium produced a document in mid-2013 (Appendix 8.1.2) that compared the pros and cons of four classes of phase synchronisation solutions for the SKA. These four techniques where then down-selected by the UWA-UoM-NPL proto-consortium to the one that would evolve to become the SKA phase synchronisation system described in this report.

With the formation of the UoM-led Signal and Data Transport (SADT) consortium in December 2013, several other members, including Tsinghua University (THU), joined the consortium. Researchers from THU proposed an alternative phase synchronisation system for the SKA [RD32] based on a 'phase-conjugation' technique first demonstrated in 1988 [RD33], and then further developed for radio astronomy by NTFN colleagues from UWA [RD34].

An SADT Consortium-coordinated technology down-select process selected the UWA solution for the SKA1-MID telescope, and the THU solution for the SKA1-LOW telescope.

### 4.2.2 Solution Description Summary

The SKA phase synchronisation system [RD5] is based on the transmission of actively stabilised phase-coherent reference signals generated at the CPF, and then transmitted via separate optical fibre links to each antenna site. The star-shaped network topology of such a phase synchronisation system, conveniently matches the fibre topology of the SKA's data network, which transmits the astronomical data from the antenna sites to the CPF. The frequency transfer technique at the core of the SKA phase synchronisation system is an evolution of ALMA's distributed 'photonic LO system' [RD6], incorporating key advances made by the international frequency metrology community over the last decade [RD7], [RD8] and [RD9], as well as novel innovations developed by UWA researchers [RD10] and [RD11].

Just as with the ALMA system, the SKA phase synchronisation system transmits the reference signals as a sinusoidal optical modulation encoded as the difference between two optical-frequency signals. Given the lower operating frequency of the SKA compared to ALMA, the two optical-frequency signals can be generated using a single laser and applying a frequency shift in one arm of a Mach-Zehnder Interferometer (MZI); thereby avoiding the differential phase noise from offset-locking two independent lasers as per the ALMA system.

Two variants of the SKA phase synchronisation system were designed, each one optimised specifically for each SKA telescope:

- For SKA1-MID, the required microwave (MW) shift is generated using a Dual-parallel Mach-Zehnder Modulator (DPM) biased to generate Single-sideband Suppressed-carrier (SSB-SC) modulation [RD10].
- For SKA1-LOW, the RF shift is generated using a simpler Acousto-optic Modulator (AOM) [RD11].

This results in two systems that easily meet the SKA coherence requirements, as demonstrated by:

- Laboratory testing [RD10], [RD11], [RD12] and [RD13].
- Overhead fibre field trials [RD14], [RD15] and [RD16].
- Astronomical verification [RD17], [RD18] and [RD19].

Complexity and costs are kept to a minimum.

The basis of the frequency transfer technique at the core of the SKA phase synchronisation system is that the optical signal from one laser is split into two arms of an MZI. In one arm, either a static MW-frequency shift (SKA1-MID) or static RF shift (SKA1-LOW) is applied to the optical signal. When the two arms are recombined at the output of the MZI, this results in two optical signals on a single fibre with either RF or MW-frequency separation. This signal is transmitted over the optical fibre link to the remote telescope sites where it is translated into the electronic domain by a photodetector. Part of the signal is reflected back to the transmitter site, where it is mixed with a copy of the transmitted optical signals, and the frequency modulation is extracted with a local photodetector. In the case of the SKA1-MID system, this electronic signal is then mixed with a copy of the MW signal, to produce an error signal that has encoded the fluctuations of the link. Applying this to the drive signal of an AOM closes the servo loop and effectively cancels the link noise for the remote site.

The colours for signal types used in all schematic diagrams throughout the document are shown in Table 4.

| Colour | Signal |
|---|---|
| Green | Optical frequencies |
| Black | Power/DC/audio frequencies |
| Blue | Microwave (MW) frequencies |
| Orange | Absolute time signal |
| Red | Radio frequencies |
| Pink | Ethernet/serial |

**Table 4 Figure signal colour codes**

### 4.2.3 Analytical Derivation of Solution

This section provides the detailed analytical derivation of the MW-frequency transfer technique at the core of the SKA phase synchronisation system for SKA1-MID. Further information on this technique is published in [RD10].

As shown in Figure 3, an optical signal with frequency $\nu_L$, generated by a laser located at the CPF, is injected into two arms of an MZI. A DPM is located in Arm 1 of the MZI, and an MW frequency of $\nu_{DP}$ is applied to the DPM electronic inputs. The phase of the electronic inputs, and the DPM voltage biases, are tuned to generate SSB-SC modulation, thereby producing a static MW-frequency shift $\nu_{DP}$ on the optical signal. Arm 2 of the MZI contains an AOM, which adds the servo AOM RF shift $\nu_{A\text{-}srv}$ and the servo actuation signal $\Delta\nu_{A\text{-}srv}$ to the optical signal. In addition, the optical signals in each of the two arms of the MZI pick up undesirable non-common phase noise $\Delta\phi_{MZI,i}$ (where *i* is the index representing the optical signals in Arm 1 and Arm 2 of the MZI).

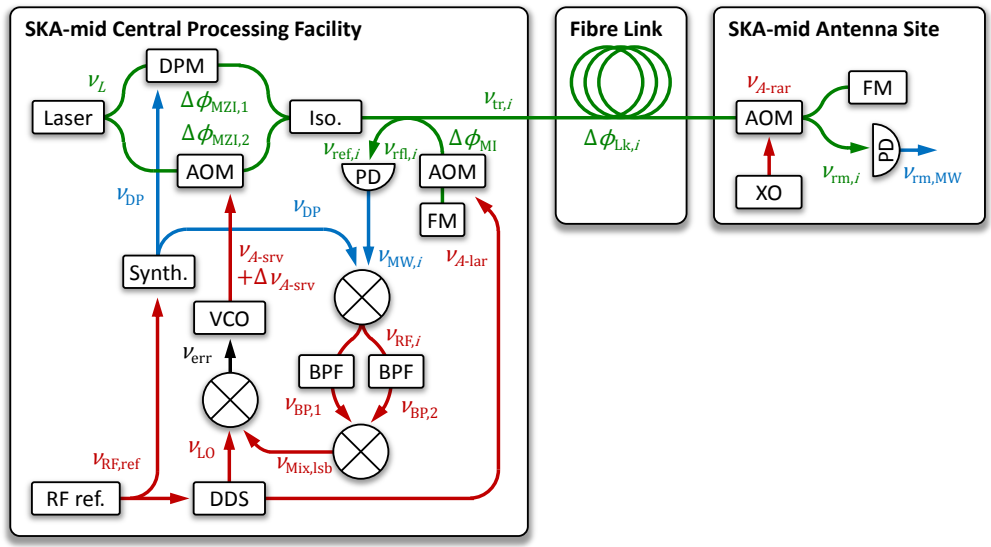

**Figure 3 Simplified schematic of the SKA1-MID stabilised MW-frequency transfer technique**

Just as is the case in standard stabilised optical transfer techniques [RD35], the optical signals then enter a Michelson Interferometer (MI) via an optical isolator (to prevent reflections returning to the laser). The fibre link is incorporated into the long arm of the MI, with the short arm providing the physical reference for the optical phase sensing. The optical reference signals $\nu_{ref,i}$ at the photodetector are:

$$\nu_{ref,1} = \nu_L + \nu_{DP} + \tfrac{1}{2\pi}\left(\Delta\dot{\phi}_{MZI,1} + 2\Delta\dot{\phi}_{MI}\right)$$

(I)

and:

$$\nu_{ref,2} = \nu_L + (1+\Delta)\nu_{A\text{-}srv} + \tfrac{1}{2\pi}\left(\Delta\dot\phi_{MZI,2} + 2\Delta\dot\phi_{MI}\right)$$

(II)

where $\Delta\phi_{MI}$ is the undesirable phase noise picked up by the optical signals passing through the MI reference arm.

A 'local anti-reflection' AOM that applies a static frequency offset of $\nu_{A\text{-}lar}$ can be incorporated into either the reference arm (as shown in Figure 3 ) or the long arm the MI. This allows the servo electronics to distinguish $\nu_{ref,i}$ from unwanted reflections on the link.

The anti-reflection AOMs are not critical for the technique, but are useful for practical implementation on fibre links that may contain unwanted reflections.

With the local anti-reflection AOM placed in the long arm, the optical signals transmitted across the fibre link $\nu_{tr,i}$ are:

$$\nu_{tr,1} = \nu_L + \nu_{DP} + \nu_{A\text{-}lar} + \tfrac{1}{2\pi}\Delta\dot\phi_{MZI,1}$$

(III)

and:

$$\nu_{tr,2} = \nu_L + (1+\Delta)\nu_{A\text{-}srv} + \nu_{A\text{-}lar} + \tfrac{1}{2\pi}\Delta\dot\phi_{MZI,2}$$

(IV)

As these signals are transmitted across the fibre link, they pick-up phase noise $\Delta\phi_{Lk,i}$ from random optical path length changes in the link that are unique to their specific transmitted frequency.

At the antenna site, the two optical signals pass through a remote anti-reflection AOM with a static frequency of $\nu_{A\text{-}rar}$ to produce the following two remote signals $\nu_{rm,i}$:

$$\nu_{rm,1} = \nu_L + \nu_{DP} + \nu_{A\text{-}lar} + \nu_{A\text{-}rar} + \tfrac{1}{2\pi}\left(\Delta\dot\phi_{MZI,1} + \Delta\dot\phi_{Lk,1}\right)$$

(V)

and:

$$\nu_{rm,2} = \nu_L + (1+\Delta)\nu_{A\text{-}srv} + \nu_{A\text{-}lar} + \nu_{A\text{-}rar} + \tfrac{1}{2\pi}\left(\Delta\dot\phi_{MZI,2} + \Delta\dot\phi_{Lk,2}\right)$$

(VI)

At the remote site, the signal is split into two fibre paths, with one set of signals going to a photodetector and the other to a Faraday Mirror (FM). At the photodetector the beat between the two optical signals $\nu_{rm,1}$ and $\nu_{rm,2}$ is recovered as the antenna site remote MW-frequency electronic signal:

$$\nu_{rm,MW} = (1+\Delta)\nu_{A\text{-}srv} - \nu_{DP} + \tfrac{1}{2\pi}\left(\Delta\dot\phi_{MZI,2} - \Delta\dot\phi_{MZI,1} + \Delta\dot\phi_{Lk,2} - \Delta\dot\phi_{Lk,1}\right)$$

(VII)

On the other fibre path, a FM reflects the two optical signals back to the CPF across the fibre link, with each signal receiving additional optical shifts $\nu_{A\text{-}rar}$ and $\nu_{A\text{-}lar}$ when passing through the remote and local AOMs a second time. In addition, the two optical signals pick up another copy of $\Delta\phi_{Lk,i}$ from the fibre link. The two reflected optical signals $\nu_{rfl,i}$ then strike the servo photodetector to produce:

$$\nu_{rfl,1} = \nu_L + \nu_{DP} + 2(\nu_{A\text{-}lar} + \nu_{A\text{-}rar}) + \tfrac{1}{2\pi}\left(\Delta\dot\phi_{MZI,1} + 2\Delta\dot\phi_{Lk,1}\right)$$

(VIII)

and:

$$\nu_{rfl,2} = \nu_L + (1+\Delta)\nu_{A\text{-}srv} + 2(\nu_{A\text{-}lar} + \nu_{A\text{-}rar}) + \frac{1}{2\pi}\left(\Delta\dot{\phi}_{MZI,2} + 2\Delta\dot{\phi}_{Lk,2}\right)$$

(IX)

The mixing of the two reference frequencies $\nu_{ref,i}$ and the two reflected optical frequencies $\nu_{rfl,i}$ results in six primary electronic mixing products $\nu_{MW,j}$ (where $j$ is the index 1 to 6). Of those, the two MW-frequency signal mixing products that are crucial for this technique are:

$$\nu_{MW,1} = \nu_{ref,1} - \nu_{rfl,2}$$
$$= \nu_{DP} - (1+\Delta)\nu_{A\text{-}srv} - 2(\nu_{A\text{-}lar} + \nu_{A\text{-}rar}) + \frac{1}{2\pi}(\Delta\dot{\phi}_{MZI,1} - \Delta\dot{\phi}_{MZI,2}) + \frac{1}{\pi}(\Delta\dot{\phi}_{MI} - \Delta\dot{\phi}_{Lk,2})$$

(X)

and:

$$\nu_{MW,2} = \nu_{ref,2} - \nu_{rfl,1}$$
$$= (1+\Delta)\nu_{A\text{-}srv} - \nu_{DP} - 2(\nu_{A\text{-}lar} + \nu_{A\text{-}rar}) + \frac{1}{2\pi}(\Delta\dot{\phi}_{MZI,2} - \Delta\dot{\phi}_{MZI,1}) + \frac{1}{\pi}(\Delta\dot{\phi}_{MI} - \Delta\dot{\phi}_{Lk,1})$$

(XI)

Given an appropriate selection of AOM frequencies, the other four mixing products (and all intermodulations) occur at different frequencies.

Once in the electronic domain, the MW-frequency signals $\nu_{MW,j}$ can be mixed with a copy of $\nu_{DP}$, to produce the following two critical electronic RF signals:

$$\nu_{RF,1} = -(1+\Delta)\nu_{A\text{-}srv} - 2(\nu_{A\text{-}lar} + \nu_{A\text{-}rar}) + \frac{1}{2\pi}(\Delta\dot{\phi}_{MZI,1} - \Delta\dot{\phi}_{MZI,2}) + \frac{1}{\pi}(\Delta\dot{\phi}_{MI} - \Delta\dot{\phi}_{Lk,2})$$

(XII)

and:

$$\nu_{RF,2} = (1+\Delta)\nu_{A\text{-}srv} - 2(\nu_{A\text{-}lar} + \nu_{A\text{-}rar}) + \frac{1}{2\pi}(\Delta\dot{\phi}_{MZI,2} - \Delta\dot{\phi}_{MZI,1}) + \frac{1}{\pi}(\Delta\dot{\phi}_{MI} - \Delta\dot{\phi}_{Lk,1})$$

(XIII)

The electronic signal path is then split, with each path containing a band-pass filter that is centred on one of the above frequencies. These filters reject the opposing signal, as well as the other unwanted mixing products ($\nu_{MW,3}$ to $\nu_{MW,6}$) and any frequency intermodulations. The signals $\nu_{RF,1}$ and $\nu_{RF,2}$ are then mixed to produce the lower-sideband:

$$\nu_{Mix,\text{lsb}} = \nu_{RF,2} - \nu_{RF,1} = 2\left((1+\Delta)\nu_{A\text{-}srv} + \frac{1}{2\pi}(\Delta\dot{\phi}_{MZI,2} - \Delta\dot{\phi}_{MZI,1} + \Delta\dot{\phi}_{Lk,2} - \Delta\dot{\phi}_{Lk,1})\right)$$

(XIV)

A low-pass filter is used to reject the upper-sideband and other products. Finally, $\nu_{Mix,\text{lsb}}$ is mixed with the servo LO $\nu_{LO}$ (set to $2\nu_{A\text{-}srv}$) to produce the servo error signal:

$$\nu_{err} = 2\left(\Delta\nu_{A\text{-}srv} + \frac{1}{2\pi}(\Delta\dot{\phi}_{MZI,2} - \Delta\dot{\phi}_{MZI,1} + \Delta\dot{\phi}_{Lk,2} - \Delta\dot{\phi}_{Lk,1})\right)$$

(XV)

The servo error signal is applied to a Voltage-controlled Oscillator (VCO) with a nominal frequency $\nu_{A\text{-}srv}$. The VCO output goes to the servo AOM, thereby closing the servo loop. When the servo is engaged, $\nu_{err}$ is driven to zero so:

$$\Delta \nu_{A\text{-}srv} = -\frac{1}{2\pi}(\Delta\dot\phi_{MZI,2} - \Delta\dot\phi_{MZI,1} + \Delta\dot\phi_{Lk,2} - \Delta\dot\phi_{Lk,1})$$

(XVI)

Substituting this into Equation 7 shows that the undesirable non-common phase noise picked up in the fibre link, as well as the phase noise from the MZI in the CPF, is cancelled out (within the light round-trip bandwidth and other practical gain limitations). This gives:

$$\nu^*_{rm,\text{MW}} = \nu_{A\text{-}srv} - \nu_{DP}$$

(XVII)

where $\nu^*_{rm,\text{MW}}$ is the MW remote signal at the antenna site the with the servo engaged.

This derivation demonstrates analytically that the MW-frequency transfer technique is effective at transferring the stability of a reference signal from the CPF to the antenna site (within the light round-trip bandwidth and other practical gain limitations). The design schematic in Figure 3 however, shows only a point-to-point link. §4.2.4 describes how this MW-frequency transfer technique is engineered to create a practical realisation of the SKA phase synchronisation system that is able to provide reference signals to all 197 SKA1-MID antenna sites.

### 4.2.4 Practical Realisation of Solution

This section describes how the frequency transfer technique described in §4.2.3, is able to be engineered into a practical realisation of the SKA phase synchronisation system for SKA1-MID.

The SKA phase synchronisation system is an element within the SADT Consortium's Synchronisation and Timing (SAT) network. The full SADT element name of the SKA phase synchronisation system is SADT.SAT.STFR.FRQ (UWA), where 'STFR' is an acronym of 'System for the Time and Frequency Reference' and 'FRQ' is an abbreviation of 'Frequency'. The other elements within the SAT network are:

- SAT.CLOCK, the SKA's H-Maser clock ensemble and timescale;
- SAT.UTC, the system for disseminating absolute time;
- SAT.LMC, the local monitor and control system for the SAT network.

The output of the FRQ (UWA) element is passed to the DISH Consortium. A simplified schematic of the SKA1-MID network architecture, showing the interrelationships between the various network elements and DISH, is shown in Figure 4.

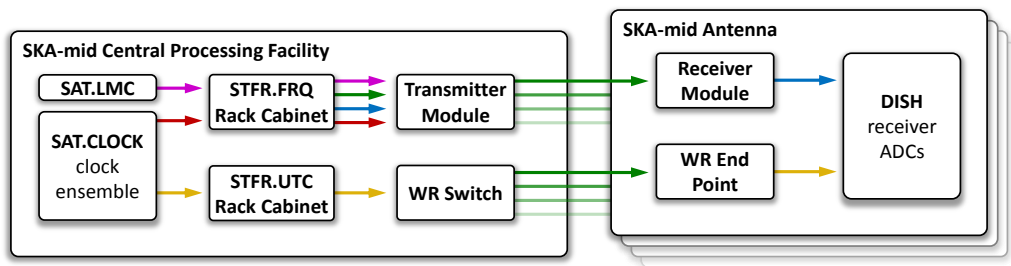

**Figure 4 Simplified schematic of the SKA1-MID SAT network**

The SKA phase synchronisation system for SKA1-MID, is made up of eleven hot-swappable, LRUs as defined in the SADT Product Breakdown Structure (PBS). These, along with their unique Configuration Identification Number (CIN) are given below:

- Rack cabinet (341-022900)
- OS (341-022400)

- FS (341-022500)
- MS (341-022600)
- SG (341-023100)
- Rack Distribution (RD) (341-022800)
- SR (341-022700)
- TM (341-022100)
- Fibre Patch (FP) lead (341-023200)
- Optical Amplifier (OA) (341-022200)
- RM (341-022300)

The combination of these LRUs comprise the entire FRQ system. The TM (341-022100) and the RM (341-022300) were already shown in Figure 4. Figure 5 shows the simplified schematic of the hardware housed in the FRQ rack cabinet.

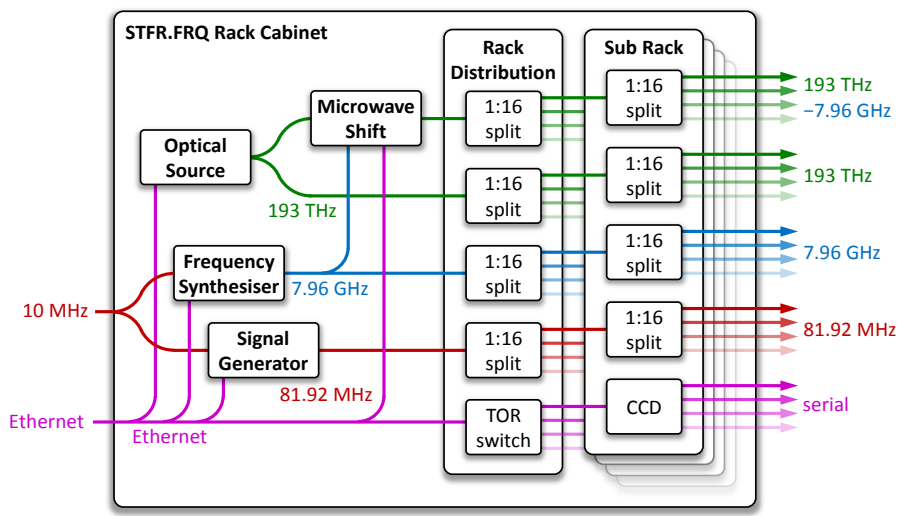

**Figure 5 Simplified schematic of the SKA1-MID FRQ rack cabinet**

The SKA1-MID FRQ rack cabinet incorporates:

- OS (341-022400)
- FS (341-022500)
- MS (341-022600)
- SG (341-023100)
- RD (341-022800)
- SR (341-022700).

**NOTE 1:** The rack cabinet (341-022900) is a collection of rack cabinet accessories including equipment, heat management and cable management. LINFRA is responsible for supplying the actual rack cabinet hardware.

The FRQ rack cabinet takes input from the SAT clocks and the LMC, and produces a series of outputs that are passed to the TM. The TM and RM simplified schematics are shown in Figure 6.

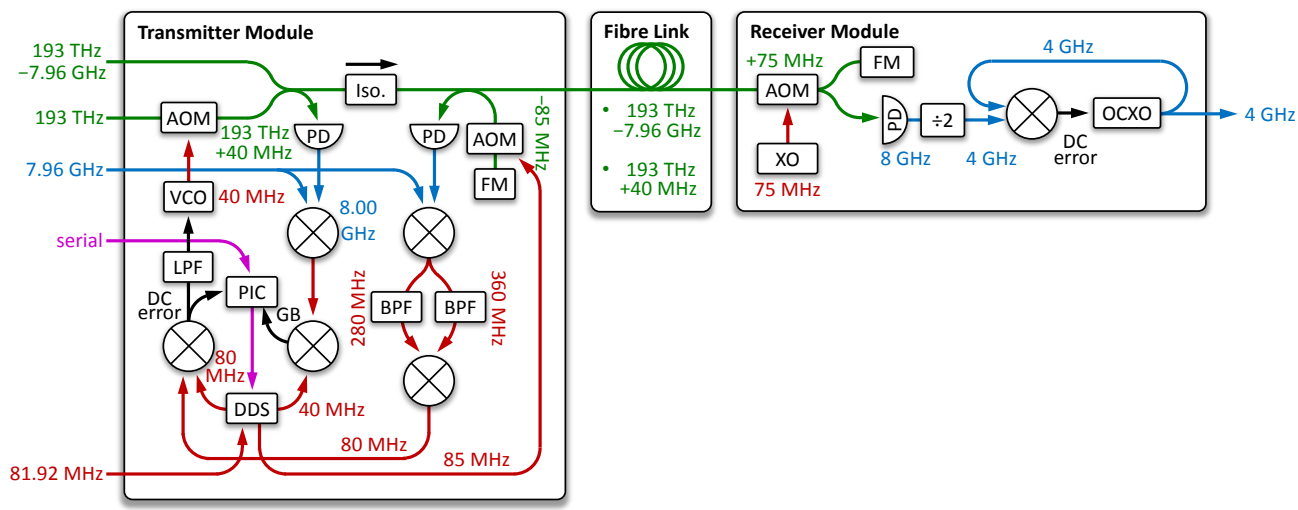

**Figure 6 Simplified schematic of the SKA1-MID TM, fibre link, and RM**

The TM transmits the reference signal, via a short FP lead (341-023200) (not shown) to the input of the fibre link, and then to the RM. The TM contains the servo-loop electronics and all monitor and control hardware. The RM includes a clean-up OCXO in a simple PLL. The output is passed to DISH.

**NOTE 2:** The OA (341-022200) has not been shown in the above figures. It is only required on the longest 14 of the total 197 SKA1-MID fibre links.

The detailed design of the SKA phase synchronisation system hardware for SKA1-MID is provided in greater detail in §4.4.

## 4.3 Solution Design Justification

The design justifications are supported by verification and measurement, however, this section only includes a summary of the key results from a limited sub-set of measurements. The full set of campaigns conducted to demonstrate that the SKA phase synchronisation system is fully compliant with the SKA requirements is described in §5.

The SKA phase synchronisation system has been extensively tested using:

- Standard metrology techniques in a laboratory setting [RD12] and [RD13], with signals transmitted over metropolitan fibre links under all required conductions.

- On overhead fibre up to 186 km in length at the South African SKA site [RD14], [RD15] and [RD16].

- Astronomical verification with the Australian Telescope Compact Array (ATCA) for SKA1-MID [RD18], [RD19] and [RD20], and the Australian SKA Pathfinder (ASKAP) for SKA1-LOW [RD19].

This has demonstrated that the SKA phase synchronisation system is fully compliant with all SKA requirements, as well as demonstrated functionality of critical practical factors that are not captured by the requirements. The full list of all relevant requirements is shown in a series of tables in §7.

The principal requirements of the SKA phase synchronisation system dictate that the distributed reference signals to be stable enough, over all relevant timescales, to ensure a sufficiently low coherence loss of the telescope array. Therefore, measurements of the frequency transfer stability of the SKA phase synchronisation system form the core data set against which these functional performance requirements (listed in §4.3.2) are evaluated.

Before describing these design justifications, two critical aspects relating to the measurement methods and equipment used to estimate frequency stability, are highlighted in §4.3.1.

### 4.3.1 Measurement Methods and Equipment

There are two aspects of measurement methods and equipment:

- Impact of the measurement device on the estimate of phase coherence (see §4.3.1.1);
- Impact of transmission frequency on absolute frequency stability (see §4.3.1.2).

#### 4.3.1.1 Impact of the Measurement Device on the Estimate of Phase Coherence

This relates to the primary equipment traditionally used in laboratory testing by the global metrology community to estimate frequency stability, the frequency counter. A frequency counter records the value of frequency as a function of time. From these data, a range of statistical algorithms (for example, Allan deviation) can be used to calculate estimates of frequency stability [RD36]. Different models of frequency counters (and related similar equipment) however, can produce data sets with various biases; and the aforementioned statistical algorithms must be selected according to the bias of the dataset in order to ensure accurate estimates.

Figure 7 shows two estimates of fractional frequency stability of the SKA phase synchronisation system derived from two measurements made concurrently using an Agilent 53132A frequency counter (Λ-weighted measurements with dead-time, dark blue circles) and a Microsemi 5125A phase noise test set (Π-weighted, dead-time free, light blue triangles).

The stability measurement using the Agilent counter follows a power-law close to $\tau^{-1/2}$ while the Microsemi measurement is closer to $\tau^{-1}$. The reason for this discrepancy is that measurements made with dead-time (in the manner performed at UWA) are known to bias the fractional frequency stability from $\tau^{-1}$ to $\tau^{-1/2}$ [RD37]. It is common practice in frequency metrology to use dead-time counters such as the Agilent 53132A (although they do not produce Allan deviation measurements) to produce fractional frequency stability estimates as long as the measurement setup is clearly stated.

For fractional frequency stability measurements made with dead-time, both white phase noise and white frequency noise follow the power-laws of $\tau^{-1/2}$ [RD37]. In the same way that Allan deviation measurements cannot distinguish between white phase noise and flicker phase noise (both having a slope of $\tau^{-1}$), measurements made using a counter with dead-time cannot distinguish between white phase and white frequency noise.

The dead-time free Microsemi measurements (which directly output Allan deviation values) show the noise processes to be predominantly white phase noise. For the UWA system, Microsemi (dead-time free) measurements must be used to accurately calculate the coherence loss instead of Agilent frequency counter measurements; however, measurements made with other frequency counters are still totally valid for comparative purposes. More detail on this point is provided in [RD26].

It should also be noted that the functional performance of the SKA phase synchronisation system was also independently verified using direct measurements with existing astronomical radio interferometers, including the Australia Telescope Compact Array (ATCA) [RD18], [RD19] and [RD20] and the ASKAP [RD19].

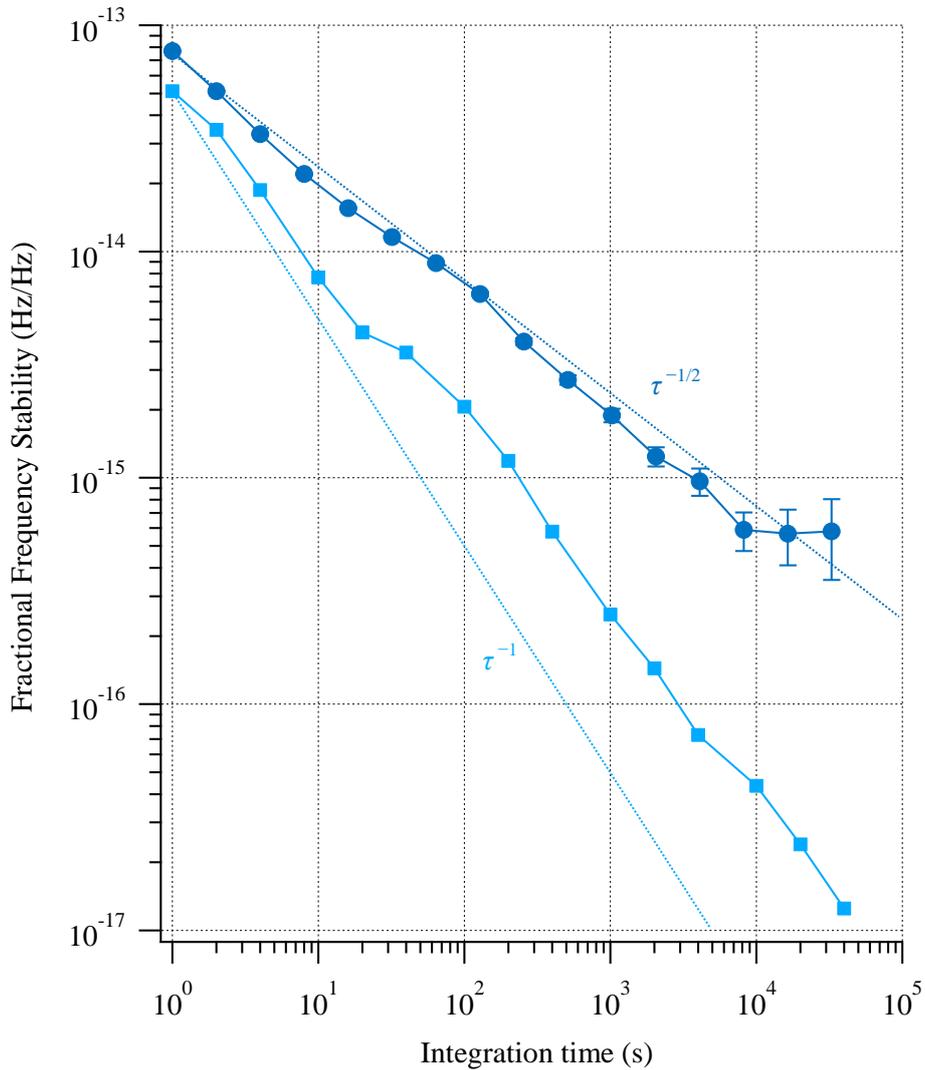

**Figure 7 Fractional frequency stability of the SKA phase synchronisation system (from [RD26])**

#### 4.3.1.2   Impact of Transmission Frequency on Absolute Frequency Stability

The second aspect of measurement methods and equipment relates to the selection of frequencies for the transmitted reference signals. Due to practical constraints, including the availability of electronic and optical components, as well as the interface requirements of the radio telescopes used in field trials, it was not always possible to transmit the exact reference signals frequency called for in the SKA phase synchronisation system design. However, as outlined below, the absolute frequency transfer stability of the SKA phase synchronisation system is largely insensitive to the frequency of the transmitted reference signal [RD16].

Figure 8 shows the absolute frequency stability of the 20 MHz transfer over 144 km of ground fibre taken from [RD16] extrapolated to the predicted performance of a 166 km link (orange). This is compared to 160 MHz transfer over 166 km of metropolitan fibre link (red) using the modulated photonic signal transfer system from [RD11] on which the 20 MHz transfer system is based. The stability is also compared to 8,000 MHz transfer over the 166 km link (blue) using a related design from [RD10].

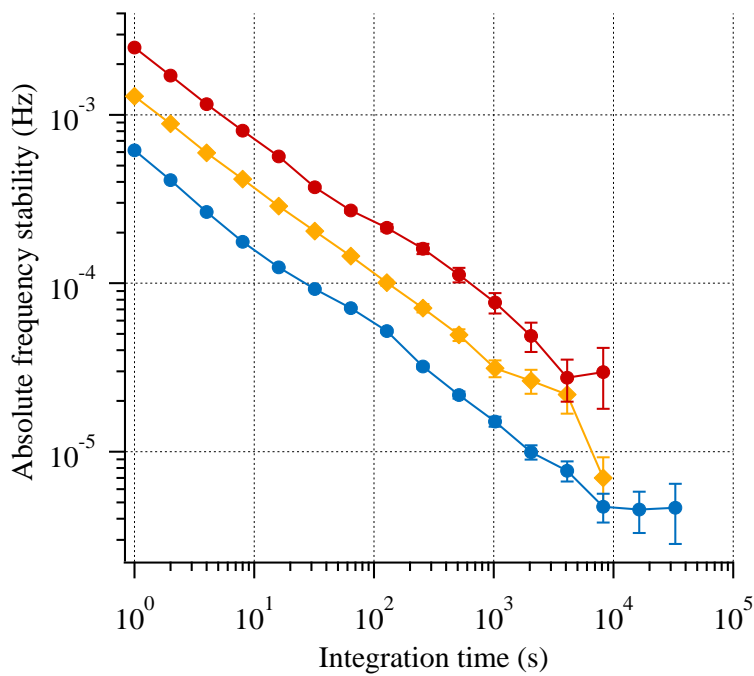

Figure 8 Absolute frequency stability of transmissions over 166 km of buried fibre (from [RD16])

Even though the highest and lowest frequencies differ by a factor of 400, the absolute frequency transfer stabilities are within less than a factor of four of each other. The remaining differences in the absolute stability levels shown by the three traces can be attributed to the fact that the experimental setups were not identical. Although based on the same design, the 20 MHz and 160 MHz transmission systems used different models of photodetectors, filters, mixers, and other electronics, and transmitted over a differently configured link. The design of the 8,000 MHz stabilised transfer system, although related to the design transmitting the lower frequencies, utilised substantially different electronics to accommodate the higher frequencies involved. For the case of the SKA phase synchronisation system at the transmitted stability levels reported here, the independence of the absolute frequency stability from the transmitted frequency value means that the fractional frequency stability increases with increasing transmission frequency. More detail on this point is provided in [RD16].

4.3.1.3 Noise Floor and Calibration process of standard test equipment

Finally, a note on the fractional frequency resolution (this has been referred to be others as the 'noise floor,' 'sensitivity,' or 'measurement repeatability') of the standard test equipment such as the Agilent 53132A frequency counter or Microsemi 5125a phase noise test set: it is important to note that the maximum permissible input frequency of these devices is typically far below our SKA transmission frequency. Therefore, we always use a self-referenced microwave-to-radio frequency down-mix procedure when using these devices (see for example, Figure 12 and Figure 13 of [RD13]). As the mixing product maintains the stability of the input, we effectively get higher fractional frequency resolution compared to using the device in its standard configuration. We take this into account (i.e. divide by the mixing ratio) when calculating the reported fractional frequency stability (ADEV).

4.3.2 **Functional Performance Requirements**

The key innovation of the SKA phase synchronisation system was finding a way to use AOMs as servo-loop actuators for modulated frequency transfer [RD10] and [RD11]. Previously, these devices had only been used as actuators for optical frequency transfer [RD38]. AOMs have large servo bandwidths as well an infinite feedback range (meaning the servo loop will never require an integrator reset, a common issue with practical implementation of most other techniques).

All 'standard' modulated frequency transfer techniques [RD8], including ALMA's [RD39], require group-delay actuation to compensate the physical length changes of the fibre link. For practical deployments, this usually involves a combination of fibre stretcher (medium actuation speed and very limited range) in series with at least one thermal spool (slow actuation speed and physically bulky). For the number of fibres, link distances, and servo loop actuation ranges encountered by the SKA, especially for SKA1-MID overhead fibre which required actuation ranges hundreds of times greater than comparable-length underground fibre [RD14] and [RD15], standard modulated frequency transfer is totally impractical. Modulated 'phase conjugation' techniques [RD33] have been demonstrated over longer distances [RD34] than standard modulated frequency transfer, and this method was proposed by THU researchers for the SKA for SKA1-MID.

The SKA phase synchronisation system also utilises AOMs to generate static frequency shifts at the antenna sites to mitigate against unwanted reflections that are inevitably present on real-world links. This mitigation strategy cannot be implemented with modulated frequency transfer techniques, as the carrier and sidebands would be shifted by same frequency. Therefore, these modulated transfer techniques require the returned signal to be rebroadcast at either a different modulation frequency, optical wavelength, or on a different fibre core, to avoid frequency overlap from unwanted reflections on the link. These reflection mitigating methods can bring about additional complications, including those resulting from optical polarization and chromatic dispersion, which in turn requires further complexity. However, reflection mitigation is essential for the SKA phase synchronisation system, as there is no way to guarantee that all links will remain completely free of reflections over the lifetime of the project.

These advantages ensure that the SKA phase synchronisation system can easily meet the functional performance requirements in practical realisation of the system while achieving maximum robustness, minimum cost, simple installation, and easy maintainability. §4.3.2.1 through §4.3.2.3 describe the solution design justifications with respect to:

- Coherence requirements (§4.3.2.1);
- Phase drift requirement (§4.3.2.2);
- Jitter requirement (§4.3.2.3).

#### 4.3.2.1 Coherence Requirements

The coherence requirements for SKA1-MID, as defined in [AD3], are:

*SAT.STFR.FRQ shall distribute a frequency reference with no more than 1.9% maximum coherence loss, within a maximum integration period of 1 second, over an operating frequency range of between 350MHz and 13.8GHz (SADT.SAT.STFR.FRQ_REQ-2268); and*

*SAT.STFR.FRQ shall distribute a reference frequency with no more than 1.9% coherence loss for intervals of 1 minute, over an operating frequency range of between 350MHz and 13.8GHz (SADT.SAT.STFR.FRQ_REQ-2692).*

These requirements are driven by the need for coherence over the correlator integration time (approximately one second) and the time for in-beam calibration (one minute), and arise from random deviations in phase. They are expressed in terms of the coherence loss caused by the phase difference between the two frequency signals delivered at two receptors, or in other words, per baseline. The coherence requirements are interpreted as a requirement to be met on the baseline with the worst-case stability, and not as an overall (averaged) coherence loss over the array [RD40]. Measurements are taken on a single frequency transfer link, but then adjusted to provide the coherence between a base-line formed by two antennas referenced by signals transmitted over two independent optical links.

The dominant factor affecting the value of coherence loss realisable with the SKA phase synchronisation system is the choice of frequency of the transmitted reference signals. Therefore, the system is designed to transmit the highest MW frequency practicable given the constraints of its constituent key hardware elements. The reason for this, as described in §4.3.1.2, is that the absolute frequency transfer stability of the SKA phase synchronisation system is largely insensitive to the frequency of the transmitted reference signal [RD16]. This means (all other things being equal) transmitting higher frequency reference signals results in greater fractional frequency stabilities, and therefore lower coherence loss values. The moderating factor is that the DISH interface specifies a 4 GHz[1] interface (evidently a convenient value from which to synthesise the various frequencies required by the five planned DISH receivers).

Investigations of readily available MW-frequency electronic and optoelectronic components led to the conclusion that there seems to be two major relevant technology breaks:

- One at around 10 GHz (8-12 GHz);

- One at around 20 GHz (18-26 GHz).

As it was desired to keep the electronics complexity of the RM to a minimum, only divide-by-2*N* frequency conversion schemes were considered. This led to the investigation of transfer frequencies of 4 GHz, 8 GHz, and 16 GHz. Several preferred components were found to be limited to operation below 10 GHz (most notably the preferred MW-frequency photodetector). As it was desired to minimise the risk of not meeting the SKA1-MID coherence requirements, 8 GHz transfer over 4 GHz transfer was selected (see further discussion in §6.3.3).

Table 5 shows a summary of evaluated coherence loss using a variety of analyses and input data sets ranging from laboratory demonstration to astronomical verification.

| Analysis # | Analysis Method | Measuring Instrument | Key Relevant Measurement Parameters | Evaluated Coherence Loss at 1 sec | Excess over Requirement | Evaluated Coherence loss at 1 min | Excess over Requirement |
|---|---|---|---|---|---|---|---|
| Analysis 1 | Laboratory demonstration; ALMA method [RD41], [RD42] and [RD43] | Microsemi 5125A Test Set | 166 km urban conduit (scaled to √2×175 km) | $4.4 \times 10^{-6}$ (0.00044%) | 4,320 | $1.6 \times 10^{-5}$ (0.0016%) | 1,190× |
| Analysis 2 | Laboratory demonstration; Thompson, Moran, and Swenson method [RD44] | Microsemi 5125A Test Set | 166 km urban conduit (scaled to √2×175 km) | $7.4 \times 10^{-6}$ (0.00074%) | 2,560 | $8.0 \times 10^{-5}$ (0.0080%) | 239× |
| Analysis 3 | Laboratory demonstration; Thompson, Moran, and Swenson method [RD44] | Agilent 53132A frequency counter | 166 km urban conduit (scaled to √2×175 km) | $1.7 \times 10^{-5}$ (0.0017%) | 1,120 | $1.1 \times 10^{-4}$ (0.011%) | 173× |
| Analysis 4 | Laboratory demonstration; SADT method [RD40] | Agilent 53132A frequency counter | 166 km urban conduit | $5.0 \times 10^{-6}$ * (0.00050%) | 3,800 | $2.3 \times 10^{-4}$ (0.023%) | 83× |
| Analysis 5 | Astronomical verification; SADT method [RD40] | ATCA astronomical interferometer | 77 km rural buried | $2.5 \times 10^{-7}$ * (0.000025%) | 76,000 | $9.0 \times 10^{-6}$ (0.00090%) | 2,110× |

*Extrapolated down to 1s.

**Table 5 Compliance against SKA1-MID coherence requirements**

The coherence loss values demonstrate that regardless of which analysis is used to estimate the coherence loss, the SKA phase synchronisation system is more than three orders-of-magnitude below the SKA1-MID coherence loss for one second requirement, and two orders-of-magnitude below the SKA1-MID coherence loss for one minute requirement, for the most challenging conditions to be faced for SKA1-MID (highest operating frequency and over the longest fibre link lengths that will be encountered).

---

[1] See §4.3.4.1.2 regarding a proposal for changing the interface to 3.96 GHz + Δ$F·k(A_n)$.

This outcome justifies the functional performance of the SKA phase synchronisation system. While the system does not achieve the same raw fractional frequency transfer performance of some other published MW-frequency transfer techniques [RD8] and [RD45], the technique is able to optimise other key design parameters (maximum robustness, minimum cost, simple installation, easy maintainability), while exceeding the functional performance needs of the SKA with ample margin. It also justifies the design choice to develop a separate less-complex RF transfer system variant [RD11] of the SKA phase synchronisation system specifically designed to be optimised for SKA1-LOW [RD5].

The detailed explanation of the experimental methods for Analysis 1 and Analysis 2 are given in:

- S.W. Schediwy, D.R. Gozzard, S. Stobie, J.A. Malan, and K. Grainge. *Stabilized microwave-frequency transfer using optical phase sensing and actuation*. Optics Letters **42** (2017) 1648.
  Appendix 8.2.6, Reference [RD10];

- §4 of: S.W. Schediwy and D.G. Gozzard, *Pre-CDR Laboratory Verification of UWA's SKA Synchronisation System*. SADT Report **620** (2017) 1-79, Appendix 8.3.2, Reference [RD13].

**NOTE 3:**   The *Optics Letters* journal paper presents the data taken with the Agilent 53132A frequency counter, as this paper was targeted at a frequency metrology audience. This is also the dark blue trace in Figure 7 . As mentioned, the measurements were also concurrently recorded using the Microsemi 5125A Test Set (light blue trace from Figure 7 ). These data, and the coherence loss analysis for Analysis 1 and Analysis 2 was presented in:

- D.R. Gozzard, S.W. Schediwy, R. Dodson, et al. *Astronomical verification of a stabilized frequency reference transfer system for the Square Kilometre Array*. The Astronomical Journal **154** (2017) 1, Appendix 8.2.9 [RD17];

- §3.2 of: D.G. Gozzard and S.W. Schediwy, *SKA1-mid Astronomical Verification*. SADT Report **524** (2016) 1-29, Appendix 8.3.5, [RD18].

The raw frequency counter data (which produced the dark blue trace in Figure 7 ) was submitted (along with several other similar data in other configurations) to the SADT Consortium for processing using the refined SADT technique described in detail in [RD40]. In addition, three raw phase solutions data sets (from a total of dozens of nearly identical data sets) from the astronomical verification field trials using ATCA were also submitted.

The phase coherence analysis of the raw frequency counter data and one representative ATCA phase solution data set, and therefore the coherence loss results for Analysis 3 and Analysis 4 are given in §4 of S.W. Schediwy and D.G. Gozzard, *Pre-CDR Laboratory Verification of UWA's SKA Synchronisation System*. SADT Report **620** (2017) 1-79, Appendix 8.3.2 [RD13].

All these results are also summarised in [RD5].

The experimental parameters outlined form the baseline for all other related measurements described in this report. Any variation from these parameters that was required to be made in order to test against the full range of normal operating conditions discussed in §4.3.3 and key additional requirements discussed in §4.3.4, are explicitly stated where relevant. Further details of experimental parameters for each of the 14 test campaigns are summarised in §5.

### 4.3.2.2 Phase Drift Requirement

The phase drift requirement is [AD3]:

*SAT.STFR.FRQ shall distribute a reference frequency to a performance allowing a maximum of 1 radian phase drift for intervals up to 10 minutes, over an operating frequency range of between*

*350 MHz and 13.8 GHz (SADT.SAT.STFR.FRQ_REQ-2693).*

This requirement is driven by the need for the phase on a calibrator, measured each side of an observation of up to 10 minutes, not to change by more than a radian during the observation. This is to ensure no wrap-round ambiguity in the phase solution. This is a requirement on phase drift, which includes both the systematic and random phase fluctuations [RD40].

The key parameter affecting phase drift is the accuracy of the frequency transfer technology. That is, if the frequency delivered is offset from the nominal value, then this will result in a linear phase drift; if that offset changes with time then so will the phase drift. The SKA phase synchronisation system is based on frequency transfer technology that is inherently phase-accurate. By essentially being a PLL, it is the phase delivered at the antenna site that is stabilised (through actuation of the transmitted frequency). Other design choices, such as only having one master FS, ensures that the phase accuracy is maintained. It has been shown that using multiple MW-FSs can easily lead to a significant loss of coherence, even if the transmission frequency is successfully stabilised [RD18], [RD19] and [RD20]. This is because for a synthesiser to maintain its output frequency relative to its reference, it must change the phase of its output as its internal temperature changes in response to variations in ambient temperature. Table 6 shows a summary of evaluated phase drifts ranging from laboratory demonstration to astronomical verification.

| Analysis # | Analysis Method | Measuring Instrument | Key Relevant Measurement Parameters | Evaluated Phase Drift at 10 mins | Excess over Requirement |
|---|---|---|---|---|---|
| Analysis 1 | Laboratory demonstration; direct phase drift measurement | Agilent 34401A Multimeter | 166 km urban conduit | 14 µrad (1 s.d.) | 71,400× |
| Analysis 2 | Astronomical verification; drift of interferometer phase solution | ATCA astronomical interferometer | 77 km rural buried | 0.08 rad (1 s.d.) | 12.5× |

**Table 6 Compliance against SKA1-MID Phase Drift requirements**

The phase drift results determined from the direct phase drift measurement in Analysis 1, shows that the SKA phase synchronisation system is almost five orders-of-magnitude below the SKA1-MID phase drift requirement (REQ-2693). The higher drift value from the astronomical verification analysis is put down to non-common elements of the ATCA interferometer, as outlined in [RD17] and [RD18]. The 10 minute phase drift is dependent on both systematic and random phase fluctuations, however, as it is a long-timescale process, it should be dominated by the systematic effects of the transmitter and receiver units [RD46]. This means it should not vary significantly with the increase in link length from 77 km or 166 km, to 175 km.

The detailed explanation of the experimental methods and results for Analysis 1 is given in §4 of: S.W. Schediwy and D.G. Gozzard, *Pre-CDR Laboratory Verification of UWA's SKA Synchronisation System*. SADT Report **620** (2017) 1-79, Appendix 8.3.2 [RD13].

**NOTE 4:** In this report, the figures are of frequency counter data showing only residual noise because any effects of temperature change are removed by the stabilisation system; exactly what the stabilisation system is designed to do. The residual noise limit in this case is likely the amplitude noise-to-phase noise coupling in the servo system's RF mixer. If any physically-attributable trend were evident in these data, the system could (and should) be improved to address it and remove it.

The results are also summarised in [RD5].

The detailed explanation of the experimental methods and results for Analysis 2 is given in:

- D.R. Gozzard, S.W. Schediwy, R. Dodson, et al. *Astronomical verification of a stabilized frequency reference transfer system for the Square Kilometre Array*. The Astronomical Journal **154** (2017) 1, Appendix 8.2.9 [RD17];

- D.G. Gozzard and S.W. Schediwy, *SKA1-mid Astronomical Verification*. SADT Report **524** (2016) 1-29,

Appendix 8.3.5 [RD18].

#### 4.3.2.3 Jitter Requirement

The jitter requirement is [AD3]:

*Jitter shall be equal to or less than 74 femtoseconds for Mid as defined by External Interface Control Document (EICD) 300-0000000-026_02_SADT-DSH_ICD.*

Any frequency transfer technique is ultimately limited in the bandwidth of frequencies of noise which can suppressed by the light round-trip time of the fibre link. This is because random noise processes that occur at time-scales faster than the light round-trip time will have de-cohered between the time of the light going one way and then returning. This means the servo loop cannot sense this noise, and therefore cannot correct for it.

To overcome this limitation, the SKA phase synchronisation system incorporates an industry-standard small form-factor OCXO into the RM to provide phase coherence at timescales shorter than the light round-trip time of the fibre link. The OCXO is tied to the incoming reference signals using a simple encapsulated PLL. The circuit design and mechanical enclosure is based off the proven design used by ASKAP.

The jitter for the SKA1-MID is calculated from the phase noise as given in the datasheet of the preferred supplier's OCXO, between 10 Hz and the maximum offset frequency (as per SADT guidelines given in §6.3 of [RD40]). Table 7 shows a summary of the evaluated jitter, determined from the phase noise data of the SKA1-MID OCXO.

| Analysis # | Analysis Method | Evaluated Jitter | Excess over Requirement |
|---|---|---|---|
| Analysis 1 | SADT-DSH_ICD; 74 fs [RD47] | 28.3 fs | 2.6× |

**Table 7 Compliance against SKA1-MID jitter requirement**

Analysis 1 shows that the jitter of the preferred supplier's OCXO exceeds the requirement value specified in the DISH ICD by a factor of 2.6.

The phase noise data and jitter calculation are described in §4 of: S.W. Schediwy and D.G. Gozzard, *Pre-CDR Laboratory Verification of UWA's SKA Synchronisation System*. SADT Report **620** (2017) 1-79, Appendix 8.3.2 [RD13].

These results are also summarised in [RD5].

There is confidence that the supplier's datasheet phase noise is not only accurate, but can be realised in practice. This is because the timing jitter of proxy OCXO used in an early prototype system (described in §3 of [RD13]) was also evaluated. For this OCXO, the phase noise was measured using two different instruments and experimental methods, and then the timing jitter was evaluated from these two data sets (as well as the specification phase noise) using these two different calculators. All analyses were with within 4 fs of each other.

Furthermore, an evaluation conducted by the SADT Consortium [RD40], demonstrates that the 74 fs value listed is unnecessarily low given the SKA1-MID digitisation effective number of bits specified in the SKA Level 1 requirement (SKA1-SYS_REQ-2190). The SADT Consortium evaluation found that the most stringent receiver case actually only calls for a jitter value less than 288 fs. Furthermore, analysis by the SKAO of the jitter budget is higher still, listing the most stringent value at 350 fs.

The OCXO in the current SKA1-MID detailed design uses an industry standard Europack form-factor. If the requirement is relaxed in the future[2] the OCXO could simply be replaced with a lower quality device,

---

[2] Or eliminated completely if the SADT interface is kept within the receptor pedestal, and the DISH consortium will continue to have to provide a separate clean-up OCXO at the indexer next to the ADC. See §5.3.1.2 for further discussion

thereby significantly lowering the cost of the proposed design.

In addition, the cross-over frequency of the PLL (that is, the boundary frequency between where the frequency transfer system is dominant compared with the clean-up OCXO) can be set to be optimised given the phase noise of the clean-up OCXO and the average (or worst) residual phase noise of the frequency transfer system (see [RD13]). While the phase noise of the oscillator is known, the residual phase noise of the frequency transfer system can depend on the particular noise characteristics of the fibre link infrastructure, which will not be completely known until after installation. It is likely, however, that the optimal cross-over frequency will very likely be between 1 Hz and 1 kHz (and likely to be between 10 Hz or 100 Hz).

The current SKA architecture has the FRQ-DISH interface in the pedestal, while the ADCs are in the indexer, thereby requiring a secondary frequency transfer system (with its own clean-up OCXO). Clearly, the logical conclusion is to move the FRQ RM to the indexer and combine these two systems. In November 2017, following the outcome of the technology down-select, key representatives from SADT and the SKAO (taking input from the DISH consortium) had a meeting to plan for an ECP to relax the jitter requirement and change the interface location to the antenna indexer. Relaxing the jitter specification will drastically increase the range of available suitable compact oscillators to be used in the case of the RM moving to the indexer. Further work is planned and funded to verify the functionality of the UWA RM on the indexer of the SKA Precursor telescope MeerKAT (see §6.2.2).

It is believed that the following redefinition of the SADT-DISH interface is the most sensible way forward:

- SADT is responsible for the provision of a stabilised frequency reference signal to the receptor (either pedestal or indexer) with a coherence budget of 1.9% (or TBD) for timescales slower than the clean-up OCXO cross-over frequency (perhaps 10 Hz or 100 Hz, TBD).

- DISH is responsible for the clean-up OCXO (as it should really be closely tied to the ADC) and therefore the coherence budget (1.9%, TBD) at timescales faster than the clean-up OCXO cross-over frequency.

### 4.3.3 Normal Operating Conditions

The normal operating condition requirements described in this section are as follows:

- Ambient temperature and humidity (see §4.3.3.1);
- Wind Speed (see §4.3.3.2);
- Seismic resilience (see §4.3.3.3);
- Telescope configuration (see §4.3.3.4);
- Overhead fibre (see §4.3.3.5).

As show in §4.3.2, the functional performance of the SKA phase synchronisation system has been proven using a number different measurement techniques. These techniques include:

- Allan deviation derived from a Microsemi 5125A phase noise test set;
- Fractional frequency computed from data logged with an Agilent 53132A frequency counter;
- Calculation of the coherence loss directly from phase noise data under astronomical verification conditions;
- Direct measurements of the phase drift using an Agilent 34410A digital multimeter;
- Phase drift analysis from astronomical verification data;

on this.

- Measurements of the short time-scale phase noise using a Microsemi 5125A phase noise test and a Rohde & Schwarz FSWP.

These comparisons have provided confidence that any of these measurement techniques may be used to accurately evaluate the performance of the SKA phase synchronisation system.

The full range of normal operating conditions could not be evaluated using all of the measurement techniques outlined above. For a range of laboratory demonstration and field trials, the instruments that produce true Allan deviation or phase noise measurements were not always available, and so the much smaller and cheaper Agilent frequency counter was often selected as the most suitable instrument to make the measurements of the system performance.

### 4.3.3.1 Ambient Temperature and Humidity Requirements

Ambient temperature and humidity requirements exist for the following:

- Within the receptor pedestal (see §4.3.3.1.1);
- Within the repeater shelters (see §4.3.3.1.2);
- Within the CPF (see §4.3.3.1.3);
- Operating over the fibre link (see §4.3.3.1.4).

#### 4.3.3.1.1 Within the Receptor Pedestal

The ambient temperature and humidity requirements for the components of SKA1-MID that are located within the Electromagnetic Interference (EMI) shielded cabinet (316-010000) within the receptor pedestal, as defined in [AD3], are:

*SAT.STFR.FRQ components sited within the "EMI Shielded Cabinet" (316-010000) shall withstand, and under normal operating conditions operate within STFR.FRQ Functional Performance requirements in an ambient temperature between -5°C and +50°C, where the rate of change of temperature is a maximum of ±3°C every 10 minutes (SADT.SAT.STFR.FRQ_REQ-305-075).*

*SAT.STFR.FRQ components sited within the "EMI Shielded Cabinet" (316-010000) shall withstand, and under normal operating conditions operate within STFR.FRQ Functional Performance requirements in a non-condensing relative humidity environment between 40% and 60% (SADT.SAT.STFR.FRQ_REQ-305-076).*

As outlined in §4.2.4, the elements of the SKA phase synchronisation system that are present at this location are:

- RM (341-022300)
- OA (341-022200).

The RM and OA are located inside the SKA1-MID receptor pedestal, inside a double-skinned EMI shielded cabinet alongside a significant range of other powered equipment [RD2]. The extreme temperature range and temperate gradients are evidently envisaged in the case of a power-up, following an extended down-time period. It is during these periods that it is expect the largest potential magnitude of impact could occur. The largest magnitude of any effect will occur during greatest rate of change of temperature. Non-condensing humidity cannot affect the system as the photons are kept confined inside the glass medium of the optical fibre at all times, and electrons are kept inside metallic transmission lines at all times.

The RM and OA located in the EMI shielded cabinet are designed to be robust against the required temperature and humidity range. The OA is a COTS unit designed specifically to operate in harsh, remote locations. In addition, the Erbium-Doped Fibre Amplifiers (EDFAs) are in-loop within the stabilisation servo, so even if temperature changes were to cause an impact, any such changes are suppressed by the stabilisation system. The RM is designed to use only a minimum number of electronic components, and these are housed inside a small form-factor metallic screening enclosure. All electronic and optical components in the RM have been specified to function properly in excess of this temperature and humidity range. The design of the RM also makes its functional performance insensitive to the environmental changes.

The performance of the system has been verified by subjecting the RM and OA to temperature and humidity changes in excess of the SKA requirements. Table 8 shows a summary of evaluated coherence loss tested using a laboratory demonstration against the ambient temperature and humidity requirements for equipment located at the receptor pedestal.

| Analysis # | Analysis Method | Measuring Instrument | Key Relevant Measurement Parameters | Evaluated Coherence Loss at 1 sec | Excess over Requirement | Evaluated Coherence loss at 1 min | Excess over Requirement |
|---|---|---|---|---|---|---|---|
| Analysis 1 | Laboratory demonstration; Thompson, Moran, and Swenson method [RD44] | Agilent 53132A frequency counter | RM; temp. range −8°C to +51°C temp. gradient +3°C/10 mins, −13°C/10 mins; relative humidity 9% to 99%; 2 m fibre patch | $9.4 \times 10^{-7}$ (0.000094%) | 20,200 | $6.2 \times 10^{-6}$ (0.00062%) | 3,060× |
| Analysis 2 | Laboratory demonstration; Thompson, Moran, and Swenson method [RD44] | Agilent 53132A frequency counter | OA; temp. range −8°C to +57°C temp. gradient +12°C/10 mins, −18°C/10 mins; relative humidity 9% to 99%; 2 m fibre patch | $1.9 \times 10^{-6}$ (0.00019%) | 10,000 | $1.2 \times 10^{-5}$ (0.0012%) | 1,580× |

**Table 8 Compliance against SKA1-MID ambient temperature and humidity requirements receptor pedestal**

The system continued to perform to specification, exceeding the 1 s and 60 s coherence loss requirements by four orders-of-magnitude and three orders-of-magnitude respectively.

The detailed explanation of the experimental methods and results for Analysis 1 and Analysis 2 is given in §6 of: S.W. Schediwy and D.G. Gozzard, *Pre-CDR Laboratory Verification of UWA's SKA Synchronisation System*. SADT Report **620** (2017) 1-79, Appendix 8.3.2 [RD13].

Regarding the impact on phase drift, the raw frequency offset time traces in [[RD18]] show absolute no indication of phase drift with changes in temperature or humidity.

In addition, a proxy SKA1-LOW OCXO (same company, same product range, different output frequency) has been tested. The specification sheet indicates operation between −40 to +85°C. This is the same specification as the SKA1-MID OCXO selected for the detailed design for mass manufacture of the SKA phase synchronisation system.

#### 4.3.3.1.2 Within the Repeater Shelters

The ambient temperature and humidity requirements for the components of SKA1-MID that are located within the inner repeater shelter" (340-052000) or outer repeater shelter" (340-53000), as defined in [AD3], are:

*SAT.STFR.FRQ components sited within the "Inner_Repeater_Shelter" (340-052000) or "Outer_Repeater_Shelter" (340-53000) shall withstand, and under normal operating conditions operate within STFR.FRQ Functional Performance requirements in an ambient temperature between*

*+18°C and +26°C (SADT.SAT.STFR.FRQ_REQ-305-077).*

*SAT.STFR.FRQ components sited within the "Inner_Repeater_Shelter" (340-052000) or "Outer_Repeater_Shelter" (340-53000) shall withstand, and under normal operating conditions operate within STFR.FRQ Functional Performance requirements in a non-condensing, relative humidity environment between 40% and 60% (SADT.SAT.STFR.FRQ_REQ-305-078).*

As outlined in §4.2.4, the only element of the SKA phase synchronisation system that is present at this location is the OA (341-022200).

The OA is a COTS unit designed specifically to operate in harsh, remote locations. In addition, the EDFAs are in-loop within the stabilisation servo, so even if temperature changes were to cause an impact, any such changes are suppressed by the stabilisation system. The performance of the system has been verified by subjecting the OA to temperature and humidity changes in excess of the SKA requirements. Table 9 shows a summary of evaluated coherence loss tested using a laboratory demonstration against the ambient temperature and humidity requirements for equipment located at the repeater shelter.

| Analysis # | Analysis Method | Measuring Instrument | Key Relevant Measurement Parameters | Evaluated Coherence Loss at 1 sec | Excess over Requirement | Evaluated Coherence loss at 1 min | Excess over Requirement |
|---|---|---|---|---|---|---|---|
| Analysis 1 | Laboratory demonstration; Thompson, Moran, and Swenson method [RD44] | Agilent 53132A frequency counter | temp. range −8°C to +57°C temp. gradient +12°C/10 mins; −18°C/10 mins; relative humidity 9% to 99%; 2 m fibre patch | $1.9 \times 10^{-6}$ (0.00019%) | 10,000 | $1.2 \times 10^{-5}$ (0.0012%) | 1,580× |

**Table 9 Compliance against SKA1-MID ambient temperature and humidity requirements repeater shelters**

The OA was subjected to a temperature range of −8°C to +51°C with temperature gradients of +3°C/10 mins and −13°C/10 mins. The system continued to perform to specification, exceeding the 1 s and 60 s coherence loss requirements by four orders-of-magnitude and three orders-of-magnitude respectively.

The detailed explanation of the experimental methods and results for Analysis 1 is given in §6 of: S.W. Schediwy and D.G. Gozzard, *Pre-CDR Laboratory Verification of UWA's SKA Synchronisation System*. SADT Report **620** (2017) 1-79, Appendix 8.3.2 [RD13].

Regarding the impact on phase drift, the raw frequency offset time traces in [RD13] show absolute no indication of phase drift with changes in temperature or humidity.

The equipment located at the CPF has no impact on the timing jitter because for the SKA phase synchronisation system the short-term phase noise is entirely determined by the clean-up OCXO in the RM.

### 4.3.3.1.3 Within the CPF

The ambient temperature and humidity requirements for the components of SKA1-MID that are located within the CPF, as defined in [AD3], are:

*SAT.STFR.FRQ components sited within the CPF shall withstand, and under normal operating conditions operate within specification, a fluctuating thermal environment between +18°C and +26°C (SADT.SAT.STFR.FRQ_REQ-305-079).*

*SAT.STFR.FRQ components sited within the CPF shall withstand, and operate within, a fluctuating non-condensing, relative humidity environment between 40% and 60% (SADT.SAT.STFR.FRQ_REQ-305-080).*

As outlined in §4.2.4, the elements of the SKA phase synchronisation system that are present at this

location are:

- Rack cabinet (341-022900)
- OS (341-022400)
- FS (341-022500)
- MS (341-022600)
- SG (341-023100)
- RD (341-022800)
- SR (341-022700)
- TM (341-022100)
- FP lead (341-023200).

The above listed equipment will be located inside the SKA1-MID CPF [RD2]. The moderate temperature range reflect that this equipment will be housed inside a large environmentally controlled facility.

Nonetheless, from previous experience of other frequency transfer systems (see §3 of [RD19]), the conditions (particularly acoustic-frequency vibrations) even in these type of facilities can potentially impact the equipment. For this reason, the SKA phase synchronisation system is designed in such a way as to also stabilise the optical wavelength-scale perturbations of the non-common optical fibre paths across the CPF equipment (see §4.2.3), by as much as 120 dB (see §4 of [RD19]). This makes the equipment in the CPF extremely robust to external environmental changes.

The performance of the system has been verified by subjecting complete transmitter system to temperature and humidity changes in excess of the SKA requirements. Non-condensing humidity cannot affect the system as the photons are kept confined inside the glass medium of the optical fibre at all times, and electrons are kept inside metallic transmission lines at all times. The tests and the performance of the system are described in detail in [RD13] The test showed the system to be fully compliant with the requirements. Table 10 shows a summary of evaluated coherence loss tested using a laboratory demonstration against the ambient temperature and humidity requirements for equipment located at the CPF.

| Analysis # | Analysis Method | Measuring Instrument | Key Relevant Measurement Parameters | Evaluated Coherence Loss at 1 sec | Excess over Requirement | Evaluated Coherence loss at 1 min | Excess over Requirement |
|---|---|---|---|---|---|---|---|
| Analysis 1 | Laboratory demonstration; Thompson, Moran, and Swenson method [RD44] | Agilent 53132A frequency counter | temp. range +16°C to +30°C; temp. gradient ≤0.6°C/10 mins; relative humidity 9% to 99%; 2 m fibre patch | $2.3 \times 10^{-6}$ (0.00023%) | 8,260 | $1.5 \times 10^{-5}$ (0.0015%) | 1,270× |

**Table 10 Compliance against SKA1-MID ambient temperature and humidity requirements CPF**

The complete transmitter system was subjected to a temperature range of +16°C to +30°C with temperature gradients of 0.6°C/10 mins. The system continued to perform to specification, exceeding the 1 s and 60 s coherence loss requirements by three orders of magnitude.

The detailed explanation of the experimental methods and results for Analysis 1 is given in §6 of: S.W. Schediwy and D.G. Gozzard, *Pre-CDR Laboratory Verification of UWA's SKA Synchronisation System*. SADT Report **620** (2017) 1-79, Appendix 8.3.2 [RD13].

Regarding the impact on phase drift, the raw frequency offset time traces in [RD13] show absolute no indication of phase drift with changes in temperature or humidity.

The equipment located at the CPF has no impact on the timing jitter because for the SKA phase synchronisation system the short-term phase noise is entirely determined by the clean-up OCXO in the RM.

#### 4.3.3.1.4 Operating over the Fibre Link

The ambient temperature and humidity requirements for the components of SKA1-MID that operate over the fibre link, as defined in [AD3], are:

*SAT.STFR.FRQ equipment and fibre located in non-weather protected locations shall be sufficiently environmentally protected to survive, and perform to specification for all ambient temperatures of between -5°C and +50°C (SADT.SAT.STFR.FRQ_REQ-2798).*

*SAT.STFR.FRQ equipment and fibre located in non-weather protected locations shall be sufficiently environmentally protected to survive, and perform to specification for rates of change of ambient temperature of up to ±3°C every 10 minutes (SADT.SAT.STFR.FRQ_REQ-2798).*

The original plan was for the fibre cables carrying the synchronisation signals to be buried to minimise their exposure to the environment. Not surprisingly, overhead cables swing in the breeze a lot more than buried cables, causing extremely large phase noise perturbation on the transmission [RD14] and [RD15]. Burying the cables comes at huge cost, with a trenched route for the three spiral arms of SKA-mid, the telescope to be built in South Africa, expected to cost €16.6M more than the overhead option.

It was not clear however, as to whether all frequency transfer techniques could operate robustly and with sufficient stability over these links. Therefore, UWA initiated a plan (assisted by SKA SA colleagues) to characterise the difference between the overhead fibre and buried fibre [RD14] and [RD15] and then went on to demonstrate the SKA phase synchronisation system operating up to 186 km of overhead fibre while subjected to extremes of temperature and humidity. Not only did the system operate robustly, but with a performance indistinguishable between overhead and underground fibre links [RD14], [RD15] and [RD16]. This result led the SKAO to adopt overhead distribution (saving the project €16.6M), and prompted THU to follow-up with similar tests. This justifies the use of AOMs as servo-loop actuators in this design, as AOMs have large servo bandwidths as well an infinite feedback range.

To verify the performance of the SKA phase synchronisation system operating in non-weather protected locations, it was tested over 154 km and 186 km of aerial suspended fibre at the South African SKA site. Table 11 shows a summary of evaluated coherence loss tested against the ambient temperature and humidity requirements for the system operating over the fibre link.

| Analysis # | Analysis Method | Measuring Instrument | Key Relevant Measurement Parameters | Evaluated Coherence Loss at 1 sec | Excess over Requirement | Evaluated Coherence loss at 1 min | Excess over Requirement |
|---|---|---|---|---|---|---|---|
| Analysis 1 | South African SKA site overhead fibre field trial; Thompson, Moran, and Swenson method [RD44] | Agilent 53132A frequency counter | 154 km SKA SA overhead fibre temp. range 0.9°C to 14.8°C; temp. gradient ±1.2°C/10 mins; rel. humidity 21% to 97% | $4.20\times10^{-5}$ (0.00420%) | 452 | $2.16\times10^{-4}$ (0.0216%) | 88.0× |
| Analysis 2 | South African SKA site overhead fibre field trial; Thompson, Moran, and Swenson method [RD44] | Agilent 53132A frequency counter | 186 km SKA SA overhead fibre temp. range 7.1°C to 10.1°C; temp. gradient ±0.8°C/10 mins; rel. humidity 21% to 97% | $1.41\times10^{-4}$ (0.0141%) | 134 | $7.26\times10^{-4}$ (0.0726%) | 26.2× |

**Table 11 Compliance against SKA1-MID temperature and humidity requirements for the fibre link**

The weather conditions during the test, exposed the transmission cable to a range of relative humidities greater than those demanded by the SKA requirements. The fibre however, was subjected to roughly half the temperature range and temperature change gradient required. Nonetheless, environmental testing of a limited range of environmental parameters over the full length of the actual overhead fibre cable planned to be used in SKA1-MID, is demonstrably preferable over limited length of bare fibre over the full range of environmental parameters. This is because testing 150 km+ of fibre cable within an environmental chamber is not possible. Even testing 150 km of bare fibre is inconceivable as the spools would occupy a much greater volume that any commercially-available environmental chamber that exists. Also, more importantly, the tight packing of glass, plastic, and air on a spool would provide high levels of insulation to the inner layers, making fast temperature changes across all 150 km unlikely to be practicably achievable. Shorter lengths, perhaps 25 km or 50 km of spooled bare fibre, are practicable but would result in a lower magnitude of tested effect as demonstrated with test parameters used:

- Regarding the temperature gradient: Using 25 km of spooled fibre, it is better to show a ±1.2°C/10min test on 154 km (185°C·km) of the exact same fibre cable that will be used on SKA1-MID, than ±3°C/10min over 25 km (75°C·km) of bare fibre spool; the magnitude of any effects would be 2.5 times greater with the reported test.

- Regarding the temperature range: Using 25 km of spooled fibre, it is better to show 0.9°C to 14.8°C on 154 km (2141°C·km) of the exact same fibre cable that will be used on SKA1-MID, than −5°C to 50°C on 25 km (1375°C·km) of bare fibre spool; the magnitude of any effects would be 1.5 times greater with the reported test.

Under these conditions, the coherence losses, even over the longest link, exceed the SKA 1 s and 60 s requirements by around two orders-of-magnitude and one order-of-magnitude respectively.

Given that the frequency transfer data from these tests, exposed to roughly half of the temperature parameters of the normal operating conditions show absolutely no residual relationship with temperature [RD14], [RD15] and [RD16], it can be concluded that the system will continue to function two and three orders-of-magnitude below the phase coherence requirement over the temperature parameters of the full normal operating conditions.

The detailed explanation of the experimental methods and results for Analysis 1 and Analysis 2 are given in:

- D.R. Gozzard, S.W. Schediwy, B. Wallace, et al. *Stabilized Modulated Photonic Signal Transfer Over 186 km of Aerial Fiber*. Submitted to Transactions on Ultrasonics, Ferroelectrics and Frequency Control (2017), Appendix 8.2.8 [RD16].

- §2 of: S.W. Schediwy and D.G. Gozzard, *UWA South African SKA Site Long-Haul Overhead Fibre Field Trial Report*. SADT Report **109** (2015) 1-20, Appendix 8.3.3 [RD14].

Regarding the impact on phase drift, the raw frequency offset time traces in [RD14] show absolute no indication of phase drift with changes in temperature or humidity.

The equipment located at the CPF has no impact on the timing jitter because for the SKA phase synchronisation system, the short-term phase noise is entirely determined by the clean-up OCXO in the RM.

**NOTE 5:** The overhead fibre tests described above were conducted with a prototype of the SKA phase synchronisation system that transmitted an RF of 20 MHz, rather than the designed 8 GHz MW-frequency transmission for the SKA1-MID system described in this report. While a number of ways that the SKA phase synchronisation system functions nearly identically across this broad range of frequencies has been analytically and empirically demonstrated, a risk remains that a previously unknown and unexpected factor breaks this relationship. If this were the case, it could be possible that the coherence loss could be adversely impacted. In order to retire this risk, UWA and SKA SA have arranged joint follow-up South African overhead fibre field trials that will test 6.8 GHz MW-frequency transfer. Rather than only testing overhead fibre frequency transfer however, the plan calls to integrate the SKA phase synchronisation system with the SKA1-MID precursor telescope, MeerKAT. The RM will be installed in the MeerKAT L-band receiver enclosure mounted on the receptor indexer. The reason for 6.8 GHz transfer is that L-band receiver clock is 1.72 GHz, and so 4 × 1.72 GHz produces the closest frequency to the nominal SKA1-MID 8 GHz transmission frequency. More information about this plan is provided in §6.2.

#### 4.3.3.2 Wind Speed Requirement

The wind speed requirement for SKA1-MID, as defined in [AD3], is:

*SAT.STFR.FRQ equipment and fibre located in non-weather protected locations shall be sufficiently environmentally protected to survive and perform to specification under normal SKA telescope operating wind conditions up to wind speeds of 40 km/hr (SADT.SAT.STFR.FRQ_REQ-2798).*

Following the passing of ECP 160013, the majority of the fibre links in the three spiral arms of SKA1-MID will be deployed on overhead fibre. This means that the SKA phase synchronisation system has to be able to operate effectively on optical links up to 175 km where the majority of this distance is deployed on overhead infrastructure and will therefore be exposed to the wind. As was reported in [RD14] and [RD15], overhead links required actuation ranges hundreds of times greater than comparable-length unground fibre, because the action of the wind loading on the fibre spans. In addition, the SKA already must deal with significantly longer maximum transmission distances than previous radio telescope arrays, such as ALMA [RD29].

This justifies the use of AOMs as servo-loop actuators in this design, as AOMs have large servo bandwidths as well an infinite feedback range (meaning the servo loop will never require an integrator reset, a common issue with practical implementation of most other techniques).

Table 12 shows a summary of compliance against the SKA1-MID wind speed requirements.

| Analysis # | Analysis Method | Measuring Instrument | Key Relevant Measurement Parameters | Evaluated Coherence Loss at 1 sec | Excess over Requirement | Evaluated Coherence loss at 1 min | Excess over Requirement |
|---|---|---|---|---|---|---|---|
| 1 | South African SKA Site Overhead Fibre Field Trial; Thompson, Moran, and Swenson method [RD44] | Agilent 53132A frequency counter | 32 km overhead fibre; wind speed ≤ 60 km/hr | $3.44 \times 10^{-6}$ (0.000344%) | 5,520 | $1.77 \times 10^{-5}$ (0.00177%) | 1,070× |
| 2 | South African SKA Site Overhead Fibre Field Trial; Thompson, Moran, and Swenson method [RD44] | Agilent 53132A frequency counter | 154 km overhead fibre; wind speed ≤50 km/hr | $4.20 \times 10^{-5}$ (0.00420%) | 452 | $2.16 \times 10^{-4}$ (0.0216%) | 88.0× |
| 3 | South African SKA Site Overhead Fibre Field Trial; Thompson, Moran, and Swenson method [RD44] | Agilent 53132A frequency counter | 186 km overhead fibre; wind speed ≤50 km/hr | $1.41 \times 10^{-4}$ (0.0141%) | 134 | $7.26 \times 10^{-4}$ (0.0726%) | 26.2× |

**Table 12 Compliance against SKA1-MID wind speed requirements**

Under these conditions, the coherence losses, even over the longest link, exceed the SKA 1 s and 60 s requirements by around two orders-of-magnitude and one order-of-magnitude respectively.

The detailed explanation of the experimental methods and results for Analysis 1, Analysis 2, and Analysis 3 is given in:

- D.R. Gozzard, S.W. Schediwy, B. Wallace, et al. *Stabilized Modulated Photonic Signal Transfer Over 186 km of Aerial Fiber*. Submitted to Transactions on Ultrasonics, Ferroelectrics and Frequency Control (2017), Appendix 8.2.8 [RD16].

- §2 of: S.W. Schediwy and D.G. Gozzard, *UWA South African SKA Site Long-Haul Overhead Fibre Field Trial Report*. SADT Report **109** (2015) 1-20, Appendix 8.3.3 [RD14].

The stabilised signal transfer system was able to operate continuously throughout these weather conditions. No cycle slips were recorded for the 24 hours of the 153.6 km link measurements or the 16 hours of the 186.2 km link measurements coinciding with the periods of higher wind speeds. Also, no cycle slips were recorded for 48 hours of the 153.6 km link measurement coinciding with the period of larger thermal gradients. No obvious correlation between phase fluctuation magnitude and wind speed is evident.

Regarding the impact on phase drift, the raw frequency offset time traces in [RD14] show absolute no indication of phase drift with changes in temperature or humidity.

The equipment located at the CPF has no impact on the timing jitter because for the SKA phase synchronisation system the short-term phase noise is entirely determined by the clean-up OCXO in the RM.

As noted in §4.3.3.1.4, but stated here again for completeness, the overhead fibre tests described were conducted with a prototype of the SKA phase synchronisation system that transmitted an RF of 20 MHz, rather than the designed 8 GHz MW-frequency transmission for the SKA1-MID system described in this report. While a number of ways that the SKA phase synchronisation system functions nearly identically across this broad range of frequencies has been analytically and empirically demonstrated, a risk remains that a previously unknown and unexpected factor breaks this relationship. If this were the case, it could be possible that the coherence loss could be adversely impacted.

In order to retire this risk, UWA and SKA SA have arranged joint follow-up South African overhead fibre field trial that will test 6.8 GHz MW-frequency transfer. But rather than only testing overhead fibre frequency transfer, the plan calls to integrate the SKA phase synchronisation system with the SKA1-MID precursor telescope, MeerKAT. The RM will be installed in the MeerKAT L-band receiver enclosure mounted on the receptor indexer. The reason for 6.8 GHz transfer is that L-band receiver clock is 1.72 GHz, and so 4 × 1.72 GHz produces the closest frequency to the nominal SKA1-MID 8 GHz transmission frequency. More information about this plan is provided in §6.2.

### 4.3.3.3 Seismic Resilience Requirement

The seismic resilience requirement for SKA1-MID, as defined in [AD3], is:

*SAT.STFR.FRQ components shall be fully operational subsequent to seismic events resulting in a maximum instantaneous peak ground acceleration of 1 m/s$^2$. Note: Seismic events include underground collapses in addition to earthquakes (SADT.SAT.STFR.FRQ_REQ-2798).*

The SKA phase synchronisation system is designed in such a way as to also stabilise the optical wavelength-scale perturbations of the non-common optical fibre paths in the TM. This effectively stabilises the TM as well as the fibre link making the equipment in the CPF extremely robust to external environmental changes. The RM for the SKA phase synchronisation system has very small form-factor and contains only a minimum number of simple optical and analogue electronic components, making it extremely robust to extremal environmental perturbation.

Table 13 shows a summary of compliance against the SKA1-MID seismic resilience requirement.

| Analysis # | Analysis Method | Measuring Instrument | Key Relevant Measurement Parameters | Evaluated Acceleration | Equipment Functional | Excess over Requirement |
|---|---|---|---|---|---|---|
| 1 | Field Trial; car trip | 3-Axis Accelerometer | 10 ms integration | 4 m/s$^2$ <br> 3 m/s$^2$ <br> 10 m/s$^2$ | Yes | 4× <br> 3× <br> 10× |
| 2 | Rotating equipment | NA | We are located on Earth's surface | 9.81 m/s$^2$ | Yes | 9.81× |

**Table 13 Compliance against SKA1-MID seismic resilience requirement**

All equipment comprising the SKA phase synchronisation system was shown to exceed the SKA1-MID seismic resilience requirement by a factor of 10. The detailed explanation of the experimental methods and results for Analysis 1 and Analysis 2 are given in Section 8 of: S.W. Schediwy and D.G. Gozzard, *Pre-CDR*

*Laboratory Verification of UWA's SKA Synchronisation System*. SADT Report **620** (2017) 1-79 Appendix 8.3.2, [RD18].

In addition, all equipment comprising the SKA phase synchronisation system was driven:

- 1,600 km by road (including over 550 km of unsealed roads) from Perth to the Murchison Radioastronomy Observatory (MRO) and back for astronomical verification with ASKAP [RD19];

- At least 7,500 kilometres from Perth to the Paul Wilde Observatory and back for astronomical verification with ATCA [RD18], [RD19] and [RD20].

Also, all equipment was located on Earth's surface where it experienced an average acceleration of 9.81 m/s at all times. It is highly likely that all equipment was rotated through several orientations in this field without adverse effect.

### 4.3.3.4 Telescope Configuration Requirement

The telescope configuration requirement for SKA1-MID, as defined in [AD3], is:

*For SKA1_MID, SAT.STFR.FRQ shall provide the MID Reference Frequency "Disseminated Reference Frequency Signal" to 133 SKA receptors and 64 MeerKAT receptors as defined by SKA-TEL-INSA-0000537 SKA1_Mid Configuration Coordinates (SADT.SAT.STRF.FRQ_REQ-2712).*

The SKA1-MID telescope comprises a total of 197 remote antenna sites, separated from the CPF by up to 175 km [RD2] of optical fibre link. As introduced in §4.2.4, and described in greater detail in §4.4.1, at the CPF, the SKA phase synchronisation system consists of 13 SRs (plus one hot-swappable spare SR), each containing 16 TMs. This system is therefore able to service all 197 optical fibre links, with 11 hot-swappable spare TMs (plus an additional 16 TMs in the spare SR).

The 197 actively-used TMs then send phase-coherent reference signals via separate optical fibre links to each of the 197 antenna sites. The star-shaped network topology of such a phase synchronisation system, conveniently matches the fibre topology of the SKA's data network, which transmits the astronomical data from the antenna sites to the CPF.

As already demonstrated in §4.3.2, the SKA phase synchronisation system can deliver the reference signals to even the furthest antenna sites while meeting all the functional performance requirements. This includes exceeding the SKA1-MID coherence loss for one second requirement and one minute requirement by more than three and two orders-of-magnitude respectively and exceeding the phase drift requirement by over five orders-of-magnitude. The jitter requirement is not dependent on the telescope configuration as the short-term phase noise is entirely determined by the clean-up OCXO in the RM.

### 4.3.3.5 Overhead Fibre Requirement

The overhead fibre requirement for SKA1-MID, as defined in [AD3], is:

*ECP 160013 recommending overhead fibre be used between the CPF and a location close to each receptor for SKA1 Mid has been approved. This means that the above coherence, phase drift and jitter requirements must be met over at least the longest overhead fibre span as defined by SKA-TEL-INSA-0000537 SKA1_Mid Configuration Coordinates (ECP 160013).*

Following the passing of ECP 160013, the majority of the fibre links in the three spiral arms of SKA1-MID will be deployed on overhead fibre. This means that the SKA phase synchronisation system has to be able to operate effectively on optical links up to 175 km where the majority of this distance is deployed on overhead infrastructure and will therefore be exposed the wind. As was reported in [RD14] and [RD15], overhead links required actuation ranges hundreds of times greater than comparable-length unground fibre, because the action of the wind loading on the fibre spans. In addition, the SKA already must deal with significantly longer maximum transmission distances than previous radio telescope arrays, such as ALMA [RD29]. This justifies the use of AOMs as servo-loop actuators in this design, as AOMs have large servo bandwidths as well an infinite feedback range and rapid response.

To verify the performance of the system on the overhead fibre, the SKA phase synchronisation system was tested over 32 km, 154 km, and 186 km of aerial suspended fibre at the South African SKA site. Table 14 shows a summary of evaluated coherence loss tested for various lengths of continuous operation.

Under these conditions, the coherence losses, even over the longest link, exceed the SKA 1 s and 60 s requirements by around two orders-of-magnitude and one order-of-magnitude respectively. In addition, the data above represents 88 hours of combined continuous operation without single cycle slip.

The detailed explanation of the experimental methods and results for Analysis 1 is given in:

- D.R. Gozzard, S.W. Schediwy, B. Wallace, et al. Stabilized Modulated Photonic Signal Transfer Over 186 km of Aerial Fiber. Submitted to Transactions on Ultrasonics, Ferroelectrics and Frequency Control (2017), Appendix 8.2.8 [RD16].

- §2 of: S.W. Schediwy and D.G. Gozzard, *UWA South African SKA Site Long-Haul Overhead Fibre Field Trial Report*. SADT Report **109** (2015) 1-20, Appendix 8.3.3 [RD14].

| Analysis # | Analysis Method | Measuring Instrument | Key Relevant Measurement Parameters | Evaluated Coherence Loss at 1 sec | Excess over Requirement | Evaluated Coherence loss at 1 min | Excess over Requirement |
|---|---|---|---|---|---|---|---|
| 1 | South African SKA site overhead fibre field trial; Thompson, Moran, and Swenson method [RD44] | Agilent 53132A frequency counter | 154 km overhead fibre; 24 hours continuous operation | $4.88 \times 10^{-5}$ (0.00488%) | 396 | $2.51 \times 10^{-4}$ (0.0251%) | 75.7× |
| 2 | South African SKA site overhead fibre field trial; Thompson, Moran, and Swenson method [RD44] | Agilent 53132A frequency counter | 154 km overhead fibre; 48 hours continuous operation | $4.20 \times 10^{-5}$ (0.00420%) | 452 | $2.16 \times 10^{-4}$ (0.0216%) | 88.0× |
| 3 | South African SKA site overhead fibre field trial; Thompson, Moran, and Swenson method [RD44] | Agilent 53132A frequency counter | 186 km overhead fibre; 16 hours continuous operation | $1.41 \times 10^{-4}$ (0.0141%) | 134 | $7.26 \times 10^{-4}$ (0.0726) | 26.2× |

**Table 14 Compliance against SKA1-MID overhead fibre requirement**

Additional information on the characteristics of overhead fibre and impact for frequency transfer are outlined in D.R. Gozzard, S.W. Schediwy, B. Wallace, et al. *Characterization of optical frequency transfer over 154 km of aerial fiber*. Optics Letters **42** (2017) 2197.

Regarding the impact on phase drift, the raw frequency offset time traces in [RD14] show absolute no indication of phase drift with changes in temperature or humidity.

The equipment located at the CPF has no impact on the timing jitter because for the SKA phase synchronisation system, the short-term phase noise is entirely determined by the clean-up OCXO in the RM.

As noted §4.3.3.1.4, but stated here again for completeness, the overhead fibre tests described were conducted with a prototype of the SKA phase synchronisation system that transmitted an RF of 20 MHz, rather than the designed 8 GHz MW-frequency transmission for the SKA1-MID system described in this report. While a number of ways that the SKA phase synchronisation system functions nearly identically across this broad range of frequencies have been analytically and empirically demonstrated, a risk remains that a previously unknown and unexpected factor breaks this relationship. If this were the case, it could be possible that the coherence loss could be adversely impacted.

In order to retire this risk, UWA and SKA SA have arranged a joint follow-up South African overhead fibre field trial that will test MW-frequency transfer. A part of €14.4k in funding has been secured to conduct this test in 2018. More information about this plan is provided in §6.2.

### 4.3.4 Key Additional Requirements

The key additional requirements described in this section are as follows:

- Offset frequency (see §4.3.4.1);
- Monitoring (see §4.3.4.2);
- RFI (see §4.3.4.3);
- Space (see §4.3.4.4).

#### 4.3.4.1 Offset Frequency Requirements

Offset frequency requirements exist for the following:

- Current SKA offset frequency architecture (see §4.3.4.1.1);
- Offset frequency architecture proposed by TALON (see §4.3.4.1.2).

##### 4.3.4.1.1 Current SKA Offset Frequency Architecture

The offset frequency requirements for SKA1-MID, as defined in [AD3], are:

*It shall be possible to set the sample-clock frequency for each receptor for each band to the nominal sample rate plus or minus N times a frequency offset, where N is an integer, from zero up to half the number of receptors (SKA1-SYS_REQ-3385).*

*There shall be an identical integer number of samples between the time-stamped one second marks from each receptor (SKA1-SYS_REQ-3386).*

*The frequency offset shall be such that there is at least 10 kHz of frequency difference to prevent unwanted cross-correlator output (SKA1-SYS_REQ-3387).*

*The maximum of the frequency offset shall be 1% (TBC) of the science bandwidth (SKA1-SYS_REQ-3388).*

In the SKA phase synchronisation system design, a 10 MHz reference frequency signal is provided by the SAT clocks. As shown in Figure 9, this is used to reference two high-quality, commercial grade FSs:

- An MW-frequency FS (341-022500);
- An RF SG (341-023100).

The 7.96 GHz output of the FS is split, with part going to the MS (341-022600).

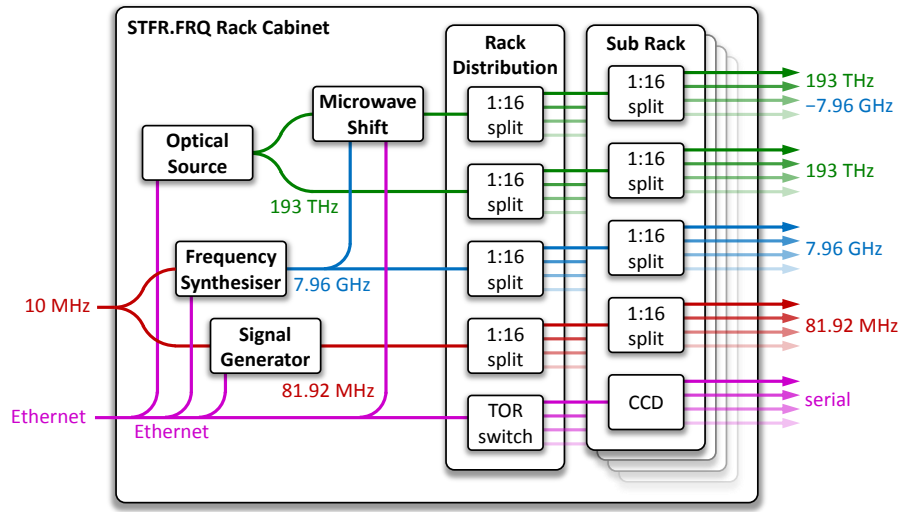

**Figure 9 Simplified schematic of the SKA1-MID FRQ rack cabinet (copy of Figure 5)**

In addition, a 193 THz optical-frequency signal is produced by a high-coherence laser within the OS (341-022400). The process is as follows:

1. The output of the OS is split, with a part also going to the MS. A DPM inside the MS is tuned SSB-SC modulation. This effectively produces a static frequency shift on the optical signal, resulting in a new optical signal with a frequency of 193 THz – 7.96 GHz.
2. Both optical signals, and a copy of the MW signal, are passed to the RD (341-022800) which, in combination with an SR (341-022700), effectively delivers these signals to every TM (341-022100) of the system.
3. The SG is used to produce an 81.92 MHz RF signal, which is likewise passed via the RD and SR, to every TM.

At the TM (see Figure 10):

1. The 81.92 MHz RF signal is passed to a DDS chip – for this detailed design, this is currently an *Analog Devices* AD9959 (four-channel, 500 MSPS, 10-bit), but any number of similar other devices will produce the same results.
2. The 81.92 MHz reference is internal multiplied by 4× to produce a system clock of 327.68 MHz.
3. The DDS is used to output a nominal 80 MHs signal to act as the LO for the frequency stabilisation system servo loop.
4. The error signal on that loop drives a VCO with a nominal frequency of 40 MHz, and this feeds to the servo AOM[3].
5. The AOM produces a nominal 40 MHz positive frequency shift on the passing 193 THz optical signal to produce a new optical signal at a frequency of 193 THz + 40 MHz.
6. At the RM, which is located at the SKA1-MID Antenna site, a photodetector recovers the frequency difference between the two optical frequencies, which in this case is −7.96 GHz + 40 MHz = 8.00 GHz.

---

[3] The explanation for the choice of +40 MHz for the servo AOM is as follows: The preferred supplier's AOMs are only available with low drive-power frequency at shifts of around only ±40 MHz and ±80 MHz (low power is an important condition for meeting the SKA power requirement as discussed in §4.3.6). The ±80 MHz frequency shifts are better suited as the anti-reflection AOMs, as this produces a greater frequency separation between the two RF servo signals (see *SKA Excel Frequency Calculator* in Appendix 8.8.2), making the RF signals easier to filter. As the servo AOM frequency could not also be around ±80 MHz due to interference, this only left ±40 MHz. +40 MHz was chosen over −40 MHz, as a model for testing already existed. This frequency therefore, also determines the −7.96 GHz frequency of the MS of the DPM, as the system calls for a transmission frequency of 8.00 GHz (as per §4.3.2.1).

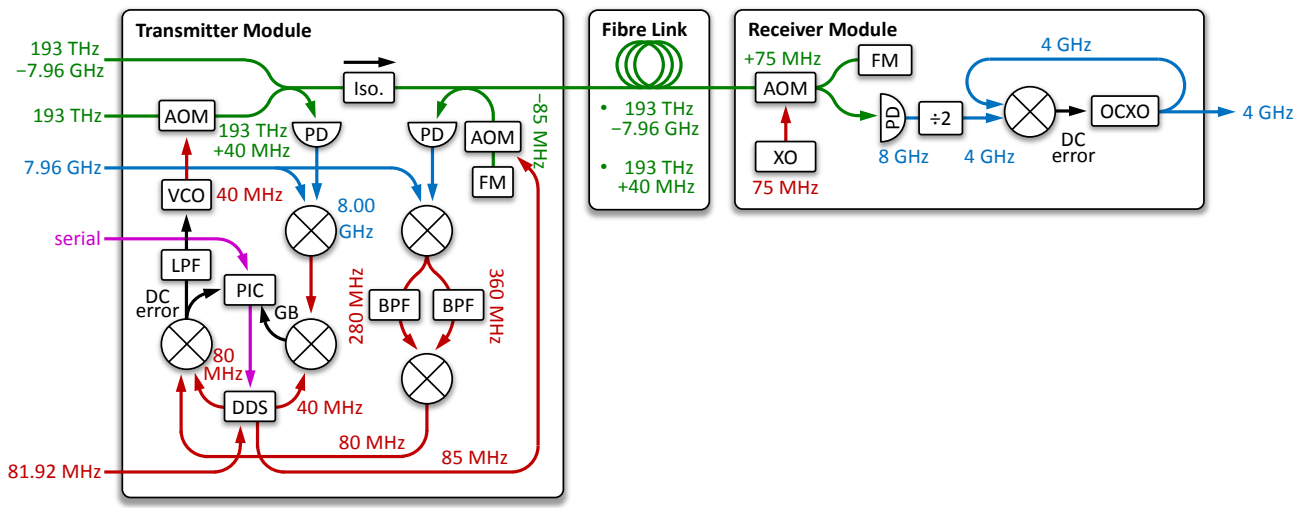

**Figure 10 Simplified schematic of the SKA1-MID TM (copy of Figure 6)**

Therefore, any frequency offset in the LO signal from the DDS will map directly onto the 8.00 GHz signal delivered to the RM (see §4.2.3 for the analytical derivation of this relationship). Further, due to the factor-of-two frequency divider at RM, any frequency offset in the LO signal from the DDS will produce half such offset on the 4.00 GHz signal. As each TM has its own DDS, independent and unique frequencies can be generated to each SKA1-MID antenna and avoid common frequencies at each antenna as per requirement SADT.SAT.STFR.FRQ_REQ-3383 (specified as a separate requirement in §4.3.4.3).

The required minimum frequency offset to prevent unwanted cross-correlator output is 10 kHz (SKA1-SYS_REQ-3387) and it must be possible to produce ±*N* number of independent offsets where *N* is an integer, from zero up to half the number of receptors, so *N* ~ ±100 (SKA1-SYS_REQ-3385). In addition, all frequencies must comprise integer number of samples between the time-stamped one second marks from each receptor (SKA1-SYS_REQ-3386). This implies that the assumed typical AOM offset range will ±2 MHz, with the 40 MHz AOM frequencies ranging of 38 MHz to 42 MHz.

To achieve this range, the DDS must produce an 80 MHz signal with offsets of ±2*N*× 10 kHz, over a range of 78 MHz to 82 MHz, and that each of these frequencies must be produced 'exactly'. A DDS will only produce output signals that 'exactly' track the input reference at some deterministic frequency values. With a system clock of the aforementioned 327.68 MHz however, the DDS can meet all of these requirements. This can be simulated using the DDS calculator at this link: http://www.analog.com/designtools/en/simdds/. In addition, this has been confirmed experimentally, as the SKA1-MID prototype of the SKA phase synchronisation system uses this model of DDS, and is accurate to less than 20 µHz [RD10] (this measurement is limited by the sensitivity of the frequency counter).

The maximum frequency offset for the SKA phase synchronisation system is limited by the optical bandwidth of the AOMs. As shown in Figure 11, for the preferred manufacture's 40 MHz AOMs, the 3 dB optical bandwidth is approximately ±4.5 MHz, so the assumed ±2 MHz offset can be achieved with negligible impact on the optical link power, and therefore the operation of the system.

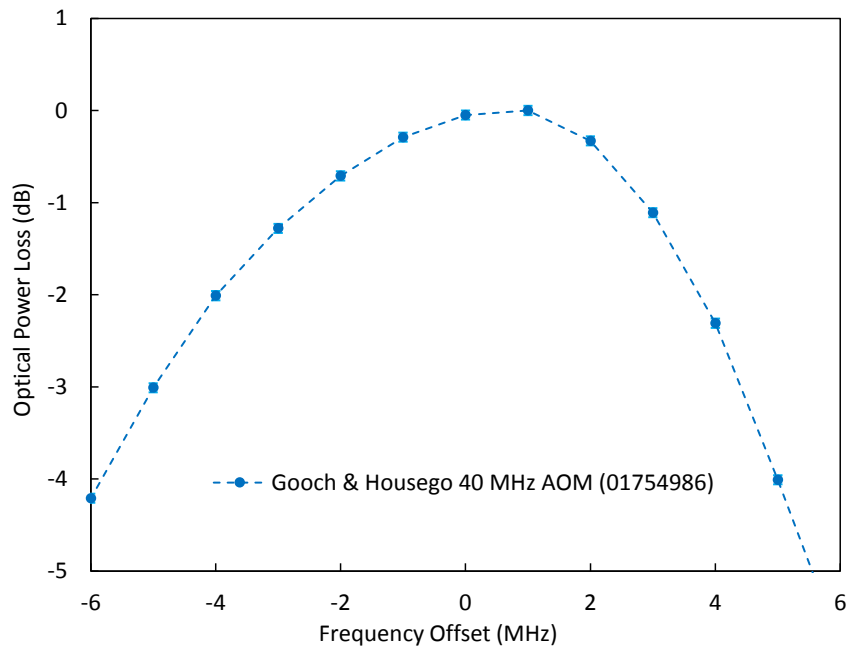

**Figure 11 Measured optical bandwidth of the Gooch & Housego AOMs**

ECP 160042 also specifies that the maximum frequency offset needs to be ±10 MHz (SKA1-SYS_REQ-3388), presumably in readiness for the larger number of antennas in SKA Phase 2. One way that this larger range could be achieved is by simply changing the design to use the preferred manufacture's 80 MHz AOMs, which have 3 dB optical bandwidths of ±10.5 MHz.

#### 4.3.4.1.2 Offset Frequency Architecture Proposed by TALON

A new Sample Clock Frequency Offset (SCFO) scheme has been proposed by TALON [RD48] in order to assist with the mitigation of strong out-of-band Radio Frequency Interference (RFI) and digitiser artefacts. The SKAO has requested an assessment as to whether this proposed offset scheme can be accommodated by the SKA phase synchronisation system. The analysis here indicates that the proposed offset scheme can indeed be accommodated, however, it is strongly recommended that the analysis is reviewed by someone with domain expertise on DDSs.

The offset scheme requires that reference clock $F_R(A_n)$ (denoted in this document by $\nu_{rm,MW}$ be in the form:

$$F_R(A_n) = 3.96 \times 10^9 + k(A_n) \cdot \Delta F$$

(XVIII)

where the offset-step $\Delta F$ has a preferred value of 1.8 kHz, and the offset index $k(A_n)$ is a positive integer between 1 and 2,222.

Therefore, $F_{R-low}$ is 3,960.000,0 MHz, and $F_{R-high}$ is 2,222 * 1.8 kHz = 3.999,6 MHz (note, this is not 4 MHz) higher in frequency than $F_{R-low}$ at 3,963.999,6 MHz (with a nominated mid-point at $F_{R-mid}$ at 3,961.999,8 MHz). The SKA phase synchronisation system must be able to produce these reference clock signals, and all signals at 1.8 kHz-spaced intermediate frequencies, with nominally ideal accuracy.

The SKA phase synchronisation system currently is designed to transmit 2* $\nu_{rm,MW}$ (although direct transmission of $\nu_{rm,MW}$ is being considered to improve simplicity, see §6.3.3), with $\nu_{rm,MW}$ being formed through the addition of the static DPM MW-frequency shift $\nu_{DP}$ and the active RF shift from the servo AOM $\nu_{A-srv}$. (The AOM has nominal centred frequency of 40 MHz, with optical transmission dropping as the frequency is offset further from the centre as shown in Figure 11.)

The frequency $v_{A-srv}$ is set by the servo loop's LO frequency $v_{LO}$, with $v_{LO}$ = 2* $v_{A-srv}$, due to the architecture of the servo design which double-passes the AOM, with $v_{LO}$ currently being provided by an *Analog Devices* AD9959 four channel DDS. The DDS must therefore be able to produce $v_{LO}$ = 2* $v_{A-srv}$ ± $\frac{1}{2}k(A_n) \cdot \Delta F$, with accurate frequencies at 2* 1.8 kHz intervals.

This can be achieved by setting the DDS internal reference frequency to 235.9296 MHz. This value is factorised by $2^{20} * 5^2 * 3^2$, thereby resulting in exact integer hertz every 225 Hz ($5^2 * 3^2$ = 225). As 225 Hz is a factor of 1.8 kHz, this allows the requisite $v_{LO}$ values to be generated exactly.

At the nominated mid-point $F_{R-mid}$, the reference clock frequency is 3,961.999,8 MHz. Therefore, $v_{LO-mid}$ is 44,445 * 1.8 kHz = 80.001,0 MHz, resulting in $v_{A-srv-mid}$ of 40.000,5 MHz. This determines that $v_{DP}$ must be set to 7,883.9991 MHz in order to transmit 7,923.999,6 MHz, and therefore deliver a reference clock frequency of 3,961.999,8 MHz. With $v_{DP}$ set, all other $v_{LO}$ can be determined. The following two examples at the frequency extremes are provided below.

- For $F_{R-low}$ the transmitted frequency is 2* 3,960.000,0 MHz = 7,920.000,0 MHz. With $v_{DP}$ fixed as indicted above, this requires $v_{A-srv-low}$ to be set to 36.000,9 MHz (an offset of 3.999,1 MHz below the nominal AOM frequency), resulting in $v_{LO-low}$ of 2* 36.000,9 MHz = 72.001,8 MHz. This values is generated by an integer of 40,001 times 1.8 kHz.

- For $F_{R-high}$ the transmitted frequency is 2* 3,963.999,6 MHz = 7,927.999,2 MHz. This requires $v_{A-srv-high}$ to be set to 44.000,1 MHz (an offset of 4.000,1 MHz above the nominal AOM frequency), resulting in $v_{LO-high}$ of 2* 44.000,1 MHz = 88.000,2 MHz. This values is generated by an integer of 48,889 times 1.8 kHz.

The inaccuracy of these two example values was experimentally confirmed to be less than 20 µHz (limited by the sensitivity of the frequency counter).

The DDS internal reference frequency to 235.9296 MHz, can be provided by an external reference of 58.982,4 MHz (multiplied internally by a factor of four).

#### 4.3.4.2 Monitoring Requirement

The monitoring requirement for SKA1-MID, as defined in [AD3], is:

*At least the following STFR component parameters shall be monitored: the Lock signal (indicating that the STFR system is functioning correctly); the Control voltage (giving an indication of how much control is still available to keep the STFR locked); and the Phase measurement (showing the corrections which have been applied to the frequency to compensate for the effects of changes in the fibre connecting the transmit and receive units of the STFR.FRQ system) (SADT.SAT.STFR.FRQ_REQ-2280).*

The SKA telescope is monitored and controlled by the Telescope Monitor system, and this system is interfaced via all other subsystems of the SKA telescope via a monitor and control system 'local' to each SKA Consortium work package. For the SKA phase synchronisation system, this SAT.LMC, and its interface is with the SR (341-022700). The SAT.LMC interface with the SR is with the Command and Control Device (CCD), which is a subsystem of the SR element. SAT.LMC interfaces with the CCD via a single Ethernet connector on the SR as shown in Figure 9. The primary CCD microcontroller then communicates with individual Peripheral Interface Controller (PIC) microcontroller chips located on to each of the sixteen TMs within the SR as shown in Figure 10. This communication is conducted using the RS485 serial transmission standard via the SR's backplane. The CCD is responsible for monitoring all the required STFR component parameters mentioned above.

As shown in Figure 10, a copy of the servo-loop error signal is fed into the PIC microcontroller chip. This is the basis of the required 'Lock signal'. As shown in §4.2.3, the servo-loop error signal is always driven to zero volts when the loop is locked. After low-pass filtering any high frequency leakage through the mixer, the locked error signal coming directly from the mixer typically has a residual plus-and-minus spread around zero volts of a few millivolts. With the servo-loop disengaged, the error signal will swing between the full range of the mixer output (typically a few hundred millivolts). The PIC microcontroller is programmed to report a positive Lock signal if the RMS voltage is below a specified threshold voltage (say 50 millivolts), and a negative Lock signal if the RMS is above this value.

As the SKA phase synchronisation system use AOMs as the servo-loop actuators, a 'Control voltage' does not exist. The AOMs suppresses the phase fluctuations of the reference signal transmitted on the fibre link, by actuating in the frequency of the reference signal. As phase is the integral of frequency, the combination of VCO and AOM effectively provide the servo-loop's integrator. This also explains the servo loop has an infinite feedback range and will never requires an integrator reset[4]. This part of the requirement is therefore not applicable.

The 'Phase measurement' (sometime referred to as 'glass box') is conducted using an independent out-of-loop system incorporated into the TM[5]. As shown in Figure 10, a photodiode is placed immediately after the output of a MZI. In the case of SKA1-MID:

1. The received electronic signal is mixed with a copy of the MW-signal sent to the DPM, producing an RF signal.
2. This signal is mixed with a reference signal provided by the DDS, using an IQ-mixer to produce two DC voltage signals (generated with a 90° phase offset).
3. The DC voltage signals are recorded by a PIC microcontroller chip.

Using those two inputs, the microcontroller can then continuously determine the accumulated phase change applied by the SKA phase synchronisation system without any phase ambiguity.

A mock-up implementation of both the Lock signal and Phase measurement monitoring were demonstrated to the SADT Consortium during the face-to-face meeting in Perth in 2016. Figure 12 is a photograph of the SADT laboratory at UWA showing the mock-up monitoring set-up. A prototype SR of the SKA phase synchronisation system can be seen on the optical table in the foreground, with the oscilloscope located on the shelf above the optical table shown recording one of the monitoring signals.

---

[4] The servo-loop is limited by the speed by which the fibre link is changing; that is the maximum magnitude of the Doppler shift that the AOM can apply to the reference signal. As the bandwidth of the AOM is several MHz (Figure 11), this is many of orders-of-magnitude greater than the most extreme situations encountered.

[5] Using an independent out-of-loop system, provides a more accurate measurement of applied phase changes, compared to simply monitoring the servo-loop Control voltage as is done with other stabilisation system. This is because the transfer function of an actuator (relationship between Control voltage and applied correction) is subject to change with time and environmental parameters.

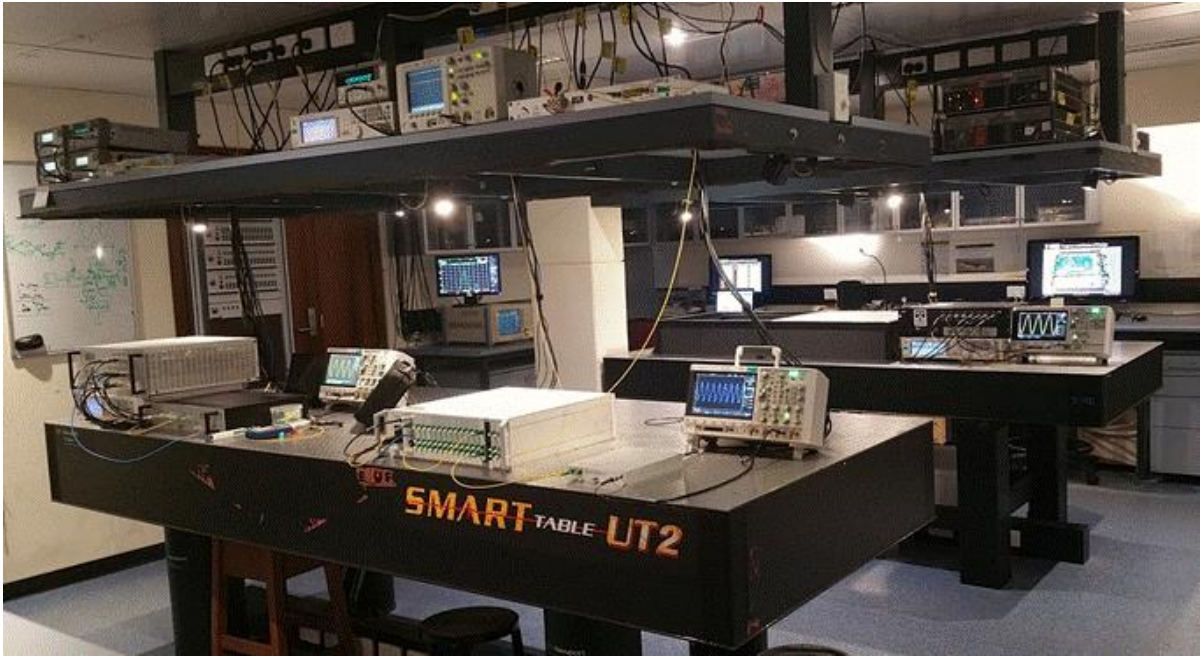

**Figure 12 Photograph of the SADT laboratory at UWA showing the mock-up monitoring set-up**

#### 4.3.4.3 RFI Requirements

The RFI requirements for SKA1-MID, as defined in [AD3], are:

*SAT.STFR.FRQ components emitting electromagnetic radiation within frequency intervals for broad and narrow band cases shall be within the SKA RFI/EMI Threshold Levels as defined in SKA-TEL-SKO-0000202-AG-RFI-ST-01 (SADT.SAT.STFR.FRQ_REQ-2462).*

*SAT.STFR.FRQ shall provide a "Disseminated Reference Frequency Signal" to each Receptor (MID), which shall be independently and uniquely settable for each Receptor. RFI emissions due to the dissemination of common frequency signals into the array shall not interact detrimentally with the Telescope's Receivers (SADT.SAT.STFR.FRQ_REQ-3383).*

The purpose of these requirements is to ensure that no equipment which is part of the SKA produces RFI at a level high enough to interact detrimentally with the operation and function of the telescope. The SKA phase synchronisation system has several design elements which add to ensure low self-generated RFI emission.

In regard to the 'electromagnetic radiation requirement' (SADT.SAT.STFR.FRQ_REQ-2462), arguably the most critical hardware element of the system in terms of RFI, is the RM. It is to be located within the SKA1-MID receptor pedestal at only a nominal 10 m separation from the receptor receivers. The system is designed to use only a minimum number of electronic components at the RM, and these are housed inside a small form-factor metallic screening enclosure which is based on the proven design used on MeerKAT. The enclosure design has a receded lid with many fasting points, shielded DC power feed-throughs, and FC optical connectors (the connector smallest non-metallic cross-section).

The RM also keeps the number of different electronics signals to a minimum; although it does have to include a clean-up OCXO in a PLL operating at a different frequency. This minimise any potential impact, this device is in its own encapsulated enclosure within the RM.

The TM PCB uses coplanar waveguides for transmission of its RF signals on a four layer PCB (so two layers of ground shielding to the adjoining TMs to limit cross talk); and coplanar waveguides inside a separate shielded enclosure for the MW-frequency signals (based on the proven-EMC MeerKAT designs). In addition, the SR mechanical enclosure can be upgraded with EMC baffles if required (these were not in place in the prototype EMC testing).

All these design choices, plus many more not described here, combine to provide a system that easily surpasses the SKA emitted electromagnetic radiation requirement SADT.SAT.STFR.FRQ_REQ-2462. Table 15 shows a summary of compliance against SKA1-MID RFI requirements for a number of key frequencies relevant for each location.

| Analysis # | Analysis Method | Measuring Instrument | Key Relevant Measurement Parameters | Relevant Frequency | Broad(Narrow)-Band Fieldstrength Threshold Mask | Maximum Measured Fieldstrength | Excess over Broad(Narrow) - Requirement |
|---|---|---|---|---|---|---|---|
| 1 | Field trail; Independent EMC Test; | Rohde % Schwarz ESU26; ETS-Lindgern semi-anechoic chamber; 30 MHz to 25.5 GHz. | *CPF Hardware*: OS FS MS SG RD SR TM | 80 MHz | 97(114) dBμV/m | 29 dBμV/m | 68(85) dB |
| | | | | 280 MHz | 110(116) dBμV/m | 29 dBμV/m | 81(87) dB |
| | | | | 360 MHz | 78(82) dBμV/m | 27 dBμV/m | 51(55) dB |
| | | | | 7.96 GHz | 108(95) dBμV/m | 81 dBμV/m | 27(14) dB |
| 2 | Field trail; Independent EMC Test; | Rohde % Schwarz ESU26; ETS-Lindgern semi-anechoic chamber; 30 MHz to 25.5 GHz. | *Receptor Hardware*: RM OA | 80 MHz | 97(114) dBμV/m | 25 dBμV/m* | 72(89) dB |
| | | | | 7.96 GHz | 108(95) dBμV/m | 58 dBμV/m | 50(37) dB |

*below measurement noise floor

**Table 15 Compliance against SKA1-MID RFI requirements**

As shown in Table 15, the SKA phase synchronisation system exceeds, for all but one signal, the broad- and narrow-band SKA1-MID RFI requirements by more than five and four orders-of-magnitude respectively. For the 7.96 GHz signal emanating from the CPF hardware, the excess over requirement is 27 dB and 14 dB for the broad- and narrow-band case respectively.

It is known that this 7.96 GHz signal originates from the high-power amplifier MS (the FS was tested independently and showed that the signal is below the measurement noise floor). This amplifier produces an output power of +23.8 dBm to drive both inputs of the DPM. The EMC field trials for the SKA1-MID transmitter equipment was performed with a 'Field Deployable Prototype'. This does not take advantage of any of the aforementioned EMC shielding design principles that are incorporated into the design for mass manufacture which forms the basis for the design presented in this document. Most critically:

- The amplifier does not have any de-coupling on its power lines;
- It is not enclosed in a shielded enclosure;
- The transmission line between amplifier and DPM is not optimised for low emission MW-frequency transmission.

With implementation of easy best-practice design it is expected that the measured field strength will be significantly reduced for even greater excess over requirement.

The detailed explanation of the experimental methods and results for Analysis 1 and Analysis 2 is given in §10 of: S.W. Schediwy and D.G. Gozzard, *Pre-CDR Laboratory Verification of UWA's SKA Synchronisation System*. SADT Report **620** (2017) 1-79, Appendix 8.3.2 [RD13].

Regarding the 'common frequency signals' requirement (SADT.SAT.STFR.FRQ_REQ-3383), the SKA phase synchronisation system meets this requirement due to design. A single high-quality FS, tied to the SKA master clock, is used to generate the reference signals, and these are disseminated via the TMs across each fibre link to each receptor. Critically, each TM incorporates a frequency shifting AOM. This can add an independently and uniquely settable offset frequency to the nominal MW-frequency reference signals. §4.3.4.1 describes the technical details of how this offset frequency is generated.

As this frequency shift is applied in the optical domain, it means that there is ever only a single, coherent MW-frequency electronic signal present at the CPF, yet the reference signals arrive at each receptor with independent frequencies. This avoids any possibility of common frequencies at each antenna site to ensure any stray RF emissions will not be coherent if picked up by the receivers.

The details of this architecture are described in §4.2.4 and in [RD5].

Some further notes on the measurement process:

- The tests were conducted up to 25.5 GHz, with all measurement apparatus calibrated. However, the geometric 'antenna factor' part of the calibration was not available for frequencies above 18 GHz. At worst, this would result in the reporting slightly inaccurate magnitudes of any observed signals above 18 GHz. No signals however, were detected above 8 GHz, as the highest frequency present in the UWA system is 8 GHz. The second and third order harmonic at 16 GHz and 24 GHz were measured (via direct sampling of the coaxial cable) to be greater than 40 dB and 60 dB down respectively.

- The EMC data presented are the raw values recorded by the calibrated measurement system, however, as the anechoic chamber could not accommodate the specified 3m measurement distance, the threshold masks were adjusted for distance (and measurement bandwidth) in accordance with standard formulae (and also presented in SKA-TEL-SKO-0000202-AG-RFI-ST-01). The measurements were conducted at 1m with the threshold masks adjusted for 3m. As both distances are well within the near-field, then, if anything, the adjustment is actually overly aggressive. In addition, the resolution bandwidth had to be adjusted as it cannot be set as a percentage of the measurement frequency (a higher resolution value was chosen to ensure narrow-band emission was not missed).

- Immunity from external radian was not explicitly tested as this was not part of the down-select criteria. While no formal immunity test have been conducted, every time the equipment was used, immunity tests were effectively being conducted. The laboratory and neighbouring rooms are filed with numerous pieces of low frequency and high frequency RFI emitting equipment and no unwanted effects were ever experienced (confirmed with antennas set-up in the laboratory). In addition, the equipment inside correlator rooms at ATCA and ASKAP have also been successfully deployed.

#### 4.3.4.4 Space Requirement

The space requirement for SKA1-MID, as defined in [AD3], is:

*The Candidate's solution shall meet the following maximum space requirements as defined by the space allocated to SAT.STFR.FRQ by the SADT NWA model SKA-TEL-SADT-0000523-MOD_NWAModelMid Revision 3.0. (SADT NWA Model SKA1-mid Rev. 03)*

The correlator of the SKA telescope is effectively a world-class supercomputer, and this machine takes up a large amount of space. On the other hand, the CPF that houses the correlator, must be constructed alongside the SKA's antennas, in a very remote location, so the significant construction costs constrain to keep the facility as small as possible. For similar reasons, the SKA receptor pedestals that house the majority of the receptor's equipment, as well as any intermediate shelters, are made as small as possible to keep costs down. For these reasons, space at these three locations is at a premium, and so all SKA equipment located there must be made as small as possible.

The SKA phase synchronisation system is designed to have a particularly small footprint in the receptor pedestal; the mass manufacture design is 243×185×35 mm (see §4.4.2.11). This is primarily because the relevant equipment, the RM, was always intended to be mounted in the receptor indexer alongside the receiver ADCs; but also as a consequence of the design choice to have as much of the system located within the CPF (as this results in much lower CAPEX installation costs and OPEX maintenance costs). The equipment within the RM consists only of:

- an anti-reflection AOM;
- a fiberised mirror;
- a photodetector;
- a small form-factor clean-up OCXO;
- a single PCB.

For the remaining equipment at the CPF, the SKA phase synchronisation system is able to keep volume down through the novel use of very compact AOMs as a servo-loop actuator for MW-frequency transfer systems [RD10]. Standard stabilised MW-frequency transfer techniques, for example [RD8], require group-delay actuation to compensate for the physical length changes of the fibre link. For practical deployments over long links, this usually involves implementing a combination of fibre stretcher (medium actuation speed and very limited range) in a series with a thermal spool (slow actuation speed and physically bulky). In addition, as the TMs are present at this location in the highest quantity, the design focuses on minimising their footprint by incorporating all optical, electronic, and mechanical elements onto a single 'Eurocard' form-factor PCB (see §4.4.1.8).

Finally, at the intermediate shelter sites, the system does not require a bulk and potentially complex optical-electronic-optical regeneration system, but can instead simply utilise a compact OA to boost optical signal strength. The compliance against the SKA1-MID space requirement is summarised in Table 16.

| Analysis # | Analysis Method | Key Relevant Measurement Parameters | Space Requirement | Evaluated Space | Excess over Requirement |
|---|---|---|---|---|---|
| 1 | Construction of mass manufacture archetypes; detailed 3D physical models computer | *CPF hardware*: Rack cabinet OS FS MS SG RD SR TM | 62U | 62U | 1× |
| 2 | Construction of mass manufacture archetypes; detailed 3D physical models computer | *Receptor hardware*: RM | 1U | 2U | 2× |
|  |  | *Receptor hardware*:* RM OA | 2U |  | 1× |
| 3 | Construction of mass manufacture archetypes; detailed 3D physical models computer | *Shelter hardware:* OA | 4U | 4U | 1× |

*OAs are required only in the pedestals of the three longest SKA1-MID links (Ant 004, Ant 008, and Ant 133) and in the pedestals of Ant 122 and Ant 129; see §4.4.1.10 for details on OA locations.

**Table 16 Compliance against SKA1-MID space requirement**

As outlined in §4.2.4, the elements of the SKA phase synchronisation system that are present at the SKA1-MID CPF are:

- RC (341-022900)
- OS (341-022400)

- FS (341-022500)
- MS (341-022600)
- SG (341-023100)
- RD (341-022800)
- SR (341-022700)
- TM (341-022100)
- FP lead (341-023300)

**NOTE 6:** The rack cabinet is a collection of rack cabinet accessories that include equipment for heat management and cable management. 16 TMs are mounted inside every SR.

On the 'CPF (KAPB) inc MeerKAT' tab of the SADT NWA Model SKA1-mid Rev. 03 [RD2], the FRQ (UWA) equipment is shown to occupy 62 rack units (U). Table 17 shows the rack unit breakdown.

| Equipment | Rack unit size (U) |
|---|---|
| SR | 42 |
| RC | 8 |
| OS | 2 |
| FS | 2 |
| MS | 2 |
| SG | 2 |
| RD | 4 |

**Table 17 FRQ (UWA) equipment rack occupancy details**

This is exactly the situation for the SKA phase synchronisation system. Figure 13 shows a render taken from the 3D physical model of the SKA1-MID equipment located at the CPF.

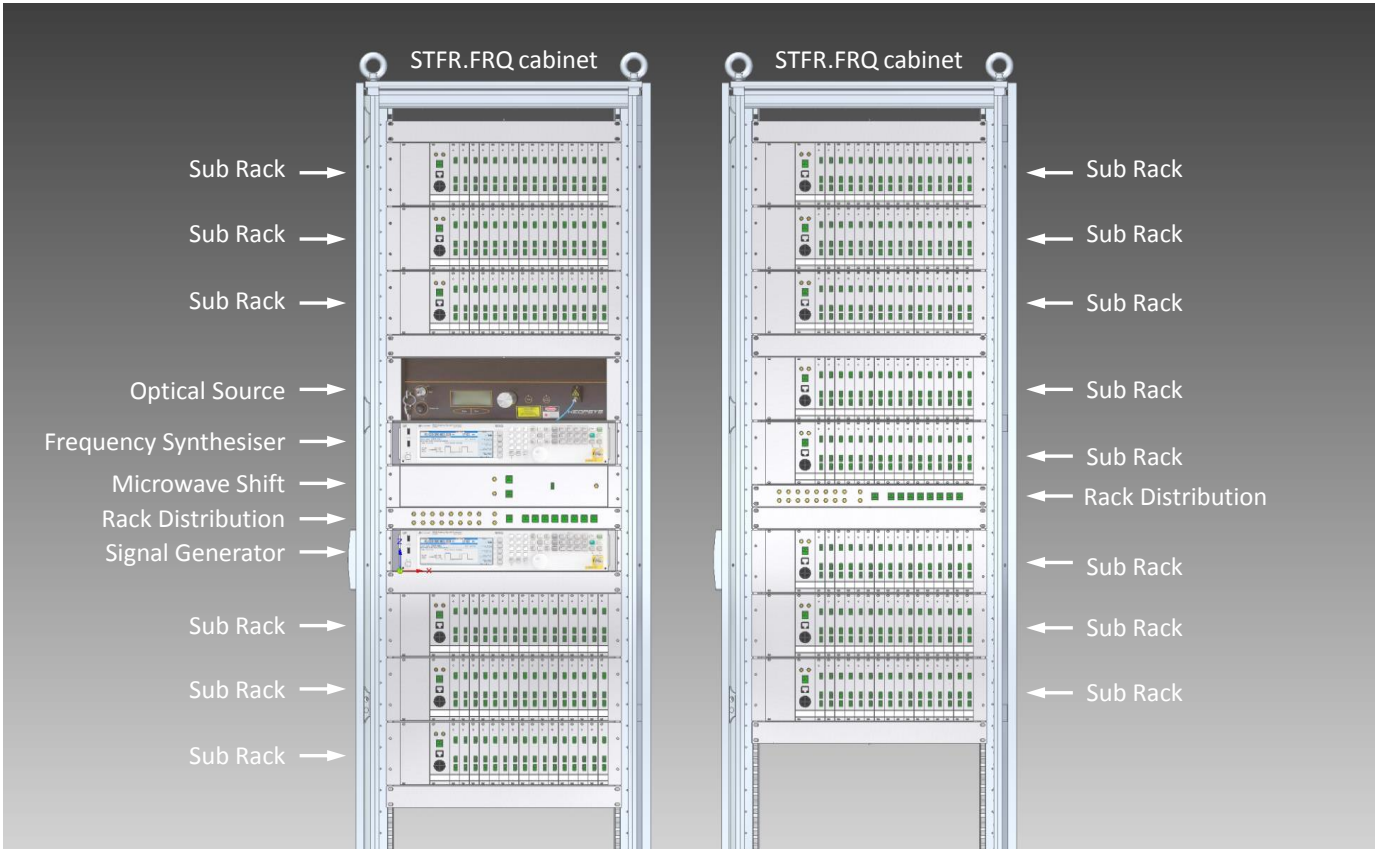

**Figure 13 SKA1-MID equipment located at the CPF**

As outlined in §4.2.4, the elements of the SKA phase synchronisation system that are present at the SKA1-MID pedestal are:

- RM (341-022300)
- OA (341-022200)

On the 'Pedestal(Outer)' and 'Pedestal(Inner)' tabs of the SADT NWA Model SKA1-mid Rev. 03 [RD49], the FRQ (UWA) RM is shown to occupy 1 rack units (U) in each of the 133 SKA1-MID locations. The maximum allocation of 2U is for the THU phase synchronisation system.

**NOTE 7:** As per the Level 1 Requirements Rev 10 plus ECPs, the 64 receptors of MeerKAT will be integrated into SKA1-MID; however, the SADT NWA Model SKA1-mid Rev. 03 has not yet been updated to include the MeerKAT pedestals (currently only placeholder information is available in the 'Pedestal(MeerKAT)' tab. Nonetheless, the design of the SKA phase synchronisation system must cater for all 197 SKA1-MID antennas. This is exactly the situation for the SKA phase synchronisation system. Figure 14 shows a photograph of a prototype SKA1-MID RM.

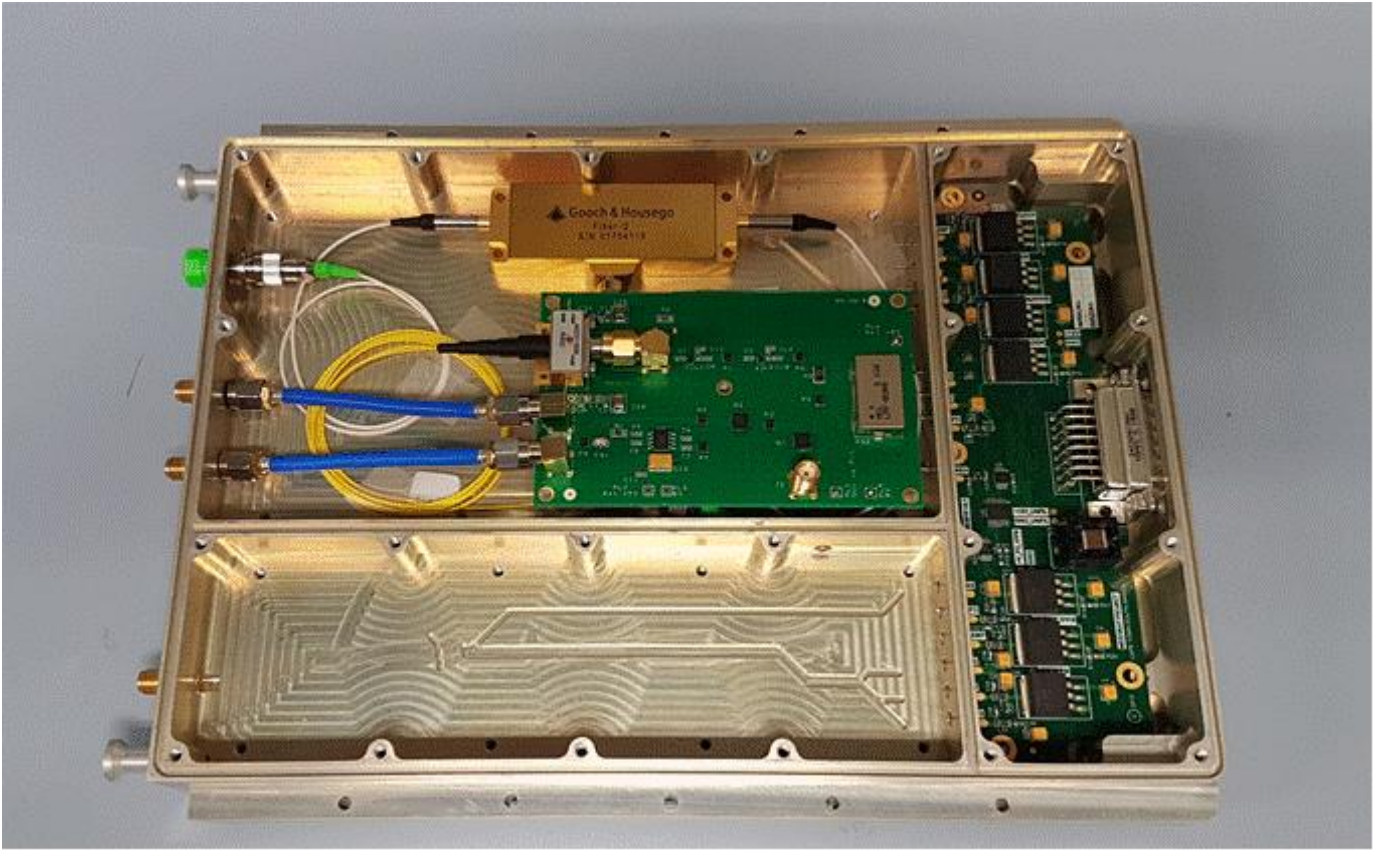

Figure 14 SKA1-MID equipment located at the receptor antenna

Figure 14 shows the key optical and electronic components as a mock-up inside a MeerKAT 'time and frequency enclosure'. On MeerKAT, these enclosures are mounted inside the MeerKAT receiver enclosure, which is itself then mounted on the receptor indexer. Figure 14 demonstrates that not only can the RM easily fit within the allotted 1U for mounting in the receptor pedestal, but it is even compatible with mounting directly on the indexer next to the ADCs of the receivers[6]. More information about this option is provided in §6.2.2, and also [RD22].

The antennas on the longest links require the use of the UWA design 1U bidirectional OA, as shown in Figure 15. Four are located within each of the three outer shelters, and an additional five in the five furthers outer pedestals. This is the exact situation shown in the 'Shelter(Outer)' and 'Pedestal(Outer)' tabs of the SADT NWA Model SKA1-mid Rev. 03 [RD2].

---

[6] This would negate the need for the DISH Consortium to build a second frequency transfer system to transmit the reference signals up the cable wraps to the indexer. After down-select, the DISH consortium has agreed to an ECP to correct this inefficiency.

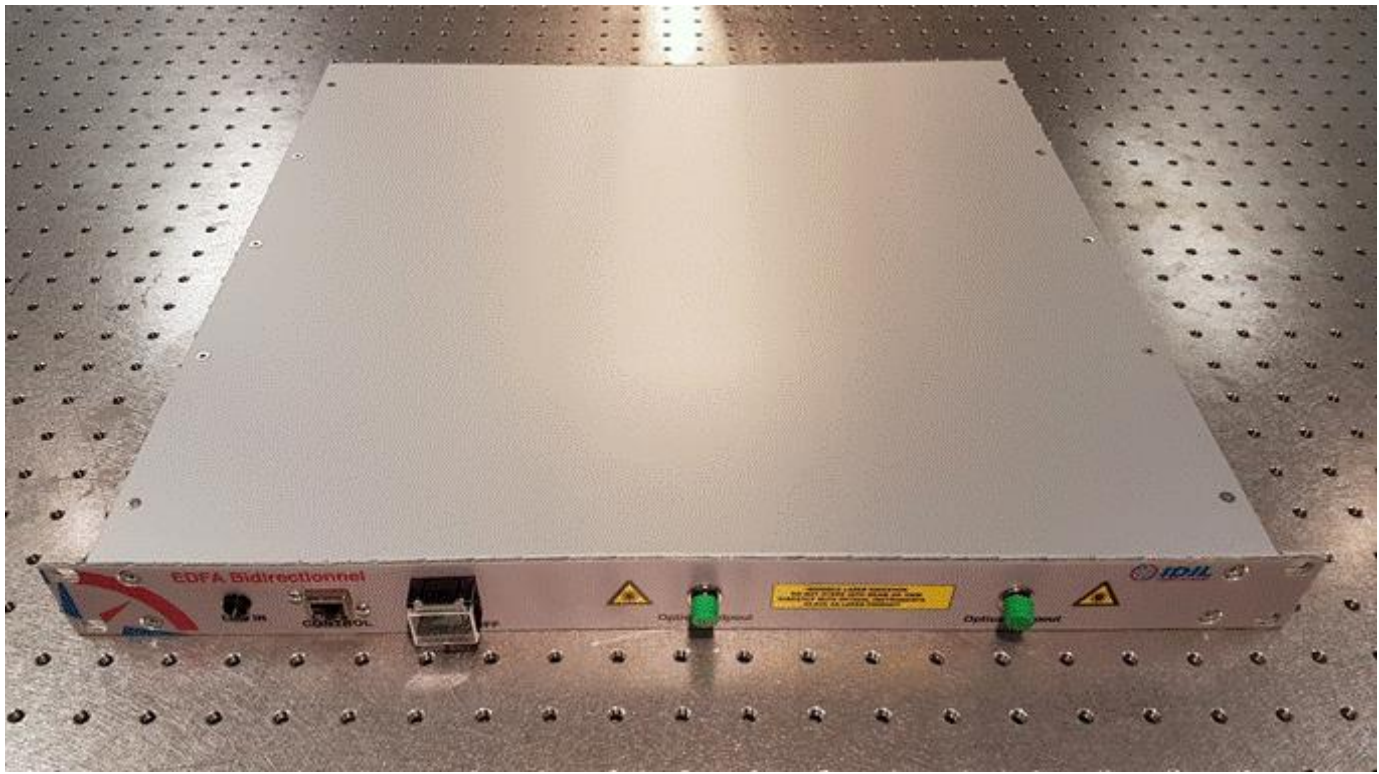

**Figure 15 SKA1-MID equipment located at the shelter (and some pedestals)**

Renders and/or photographs of the individual hardware elements for both the CPF and Remote Processing Facility (RPF) locations are provided in the relevant sub-sections of §4.4.1, and further details are provided in §4.4.2.

The *Solid Edge* Detailed mechanical design files are included as a zipped file pack in Appendix 8.5.1.

Further information of the design process is provided in [RD22], which is included in Appendix 8.3.7.

**NOTE 8:** This is a recursive requirement, as this requirement was based on the estimate of the required space values previously provided by the design teams.

### 4.3.5 Availability Requirements

The space requirement for SKA1-MID, as defined in [AD3], is:

*SAT.STFR.FRQ (end-to-end system excluding fibre) shall have 99.9% "Inherent Availability" (SADT.SAT.STFR.FRQ_REQ-3245).*

In order to achieve its scientific objectives, the SKA telescope must have a high overall duty cycle. As many of the telescope sub-systems, including the SKA phase synchronisation system, are critical for operation, each subsystem must have an even higher inherent availability. The SKA phase synchronisation system has been budgeted with needing to meet an inherent availability 99.9%.

The SADT Consortium contracted the independent firm Frazer-Nash Consultancy, to conduct a Reliability, Availability, Maintainability, and Safety (RAMS) analysis covering each of the SADT sub-elements, including the SKA phase synchronisation system. Table 18 shows the summary table of SKA1-MID failure modes severity classification taken from the Frazer-Nash Consultancy input worksheet.

These data was analysed within the SADT Consortium using an algorithm supplied by SKAO to compute an overall inherent availability. For the SKA phase synchronisation system, the overall inherent availability was determined to be 99.828% [RD50].

As the SKA phase synchronisation system requires three AOMs per link, the AOMs are the overall dominant cost component. As demonstrated in the UWA Detailed Cost Model (§4.10) however, the dominant cost component and large quantify in use in the system, the AOMs represent the biggest risk in terms of reliability, availability, and maintainability. However, the Frazer-Nash Consultancy's RAMS analysis concludes that the AOM risk criticality is 'tolerable'.

Furthermore, the following mean time before failure (MTBF) information has been provided by the preferred AOM supplier Gooch & Housego:

- A total of 14,163 AOM devices have been shipped to customers since the product was released in 2009;
- During this time, there has been only 1 in-service failure where the problem was attributed to faulty materials or manufacturing;
- This excludes "out of box" failures discovered prior to the device being put into service;
- It also excludes failures due to improper handling (leading to fibre damage) or customer error such as excessive optical or RF power

Assuming all AOMs are immediately put into service and used 24/7 then we have a failures in time (FIT) rate of 2.8 and MTBF of 359M hours.

Of course, the above calculation will overestimate MTBF as it is not know the actual hours each AOM has been used, nor do we know how many older devices may now be retired from service; however, even if a much lower average usage time is assumed (e.g. 1% of the time), the calculated MTBF would still be greater than 3M hours (greater than 300 years).

All aspects of the SKA phase synchronisation system detailed design is:

- Engineered using best practice principles;
- Developed in partnership with UoM, CSIRO, SKA SA;
- Based on the proven designs from previous interferometer telescope array (ALMA, ASKAP, ATCA) and from the international metrology community;
- Uses only industry standard high-volume commercial components.

All indications, based on testing and use of the various prototypes, and more importantly, mass manufacture archetypes over the last three-and-a-half years, is that the SKA phase synchronisation system will prove to have an inherent availability over the project lifetime significantly exceeding the stated requirement.

| Level | Class | Consequence to person or environment | Count |
|---|---|---|---|
| Catastrophic | I | - Total system loss;<br>- High precision time data loss;<br>- Low precision time data loss;<br>- Science data loss;<br>- A failure that results in death. | 0 |
| Critical | II | - Major equipment failure;<br>- Major system damage;<br>- Loss of major science programming;<br>- Loss of significant time period of science programming;<br>- Serious injury. | 0 |
| Major | III | - Partial loss of science programme;<br>- Failure causing minor system damage resulting in degraded output, operation or availability;<br>- Failure resulting in minor injury requiring first aid. | 9 |
| Minor | IV | - A failure not serious enough to cause injury;<br>- A failure not serious enough to cause system damage, but will result in unscheduled maintenance or repair. | 15 |
| Not Credible | N/A | Failure mode not credible during normal operation. | 0 |

**Table 18 SKA1-MID failure modes severity classification**

### 4.3.6 Power Requirements

The RFI requirements for SKA1-MID, as defined in [AD3], are:

*SAT.STFR.FRQ components, sited within the receptor shielded cabinet, shall not exceed nominal power consumption of 50 Watts (SADT.SAT.STFR.FRQ_REQ-305-113).*

*The total power consumption of combined SAT.STFR.FRQ components located in the CPF (MID) shall be no more than 2.3 kWatts (SADT.SAT.STFR.FRQ_REQ-305-163).*

*The total power consumption of combined SAT.STFR.FRQ components located in any single instance of the "Inner_Repeater_Shelter" (340-052000) or "Outer_Repeater_Shelter" (340-053000) shall be no more than 160 Watts (SADT.SAT.STFR.FRQ_REQ-305-164).*

The cost of electrical power is one of the primary total cost of ownership limitations for the SKA telescope. The need for world-class power-hungry supercomputing processing being required at very remote locations leads the need to limit the power use of all other SKA subsystems.

The SKA phase synchronisation system for SKA1-MID is designed using a combination LRU building blocks (see §4.2.4). The two blocks that are replicated with the highest quantity in the design are the TM and RM. These are built using a bespoke PCB using a surface-mount component design. On these PCBs, the most power-hungry component is the high-power amplifiers used to drive the two AOMs in the TM and one AOM in the RM. Low drive-power AOMs were specifically selected which allowed the TM power to be kept to within 6 W. Other building blocks are requested in much lower quantities, and so the power use per unit is less critical. The compliance against SKA1-MID Power Requirements is summarised in Table 19.

| Analysis # | Analysis Method | Measuring Instrument | Key Relevant Measurement Parameters | LRU | Evaluated Power Use | Power Requirement | Excess over Requirement |
|---|---|---|---|---|---|---|---|
| 1 | Laboratory demonstration; equipment spec. sheets | Emona EL-302RD Power supply; | Location; Receptor shielded cabinet | Total | 5 W | 50 W | 10.0× |
|  |  |  |  | RM | 5 W |  |  |
| 2 | Laboratory demonstration; equipment spec. sheets | Emona EL-302RD Power supply; | Location; CPF | Total | 2,230 W | 2.3 kW | 1.0× |
|  |  |  |  | Rack cabinet | 8× 24 W |  |  |
|  |  |  |  | OS | 80 W |  |  |
|  |  |  |  | FS | 250 W |  |  |
|  |  |  |  | MS | 80 W |  |  |
|  |  |  |  | SG | 250 W |  |  |
|  |  |  |  | RD | 2× 0 W |  |  |
|  |  |  |  | SR | 13× 10 W* |  |  |
|  |  |  |  | TM | 208× 6 W |  |  |
| 3 | Laboratory demonstration; equipment spec. sheets | Emona EL-302RD Power supply; | Location; Repeater Shelter | Total | 50 W | 160 W | 3.2× |
|  |  |  |  | OA | 50 W |  |  |

*Note, apart from supplying a small number of active components, the primary task of the SR power supply is to provide the power to the 16 TMs. Therefore, the power listed in this cell only includes the power of the components specific to the SR, and the power supply inefficiency of providing power to the TMs. Only 13 of the 14 installed SRs are powered at any one time, hence a total of 208 powered Transmitter Units.

**Table 19 Compliance against SKA1-MID power requirements**

**NOTE 9:** This is recursive requirement, as this requirement was based on the estimate of the power consumption values previously provided by the design teams.

### 4.3.7 Other Key System Parameters

The other key system parameters described in this section are as follows:

- Reflection mitigation (see §4.3.7.1);
- MS stability (see §4.3.7.2);
- Ability to deliver the reference signals directly to the receptor indexer (see §4.3.7.3);
- Miscellaneous key system parameters (see §4.3.7.4).

#### 4.3.7.1 Reflection Mitigation

The SKA phase synchronisation system also utilises AOMs to generate static frequency shifts at the antenna sites to mitigate against unwanted reflections that are inevitably present on real-world links. This mitigation strategy cannot be implemented with modulated frequency transfer techniques, as the carrier and sidebands would be shifted by same frequency. Therefore, these modulated transfer techniques require the returned signal to be rebroadcast at either a different modulation frequency, optical wavelength, or fibre core, to avoid frequency overlap from unwanted reflections on the link. These reflection mitigating methods then can bring about additional complications, including those resulting from optical polarization and chromatic dispersion, which in turn requires further complexity. Reflection mitigation however, is absolutely essential for the SKA phase synchronisation system, as there is no way to guarantee that all links will remain completely free of reflections (even if they are free of reflections at the start of operations) over the lifetime of the project.

The SKA phase synchronisation system has been successfully deployed on UWA-Pawsey 31 km fibre link which contains large optical reflections typical of a link that uses optical connectors. The Optical Time-domain Reflectometry (OTDR) trace for one 15.5 km half of the loop-back is shown in Figure 16. Several large reflections are evident in the trace.

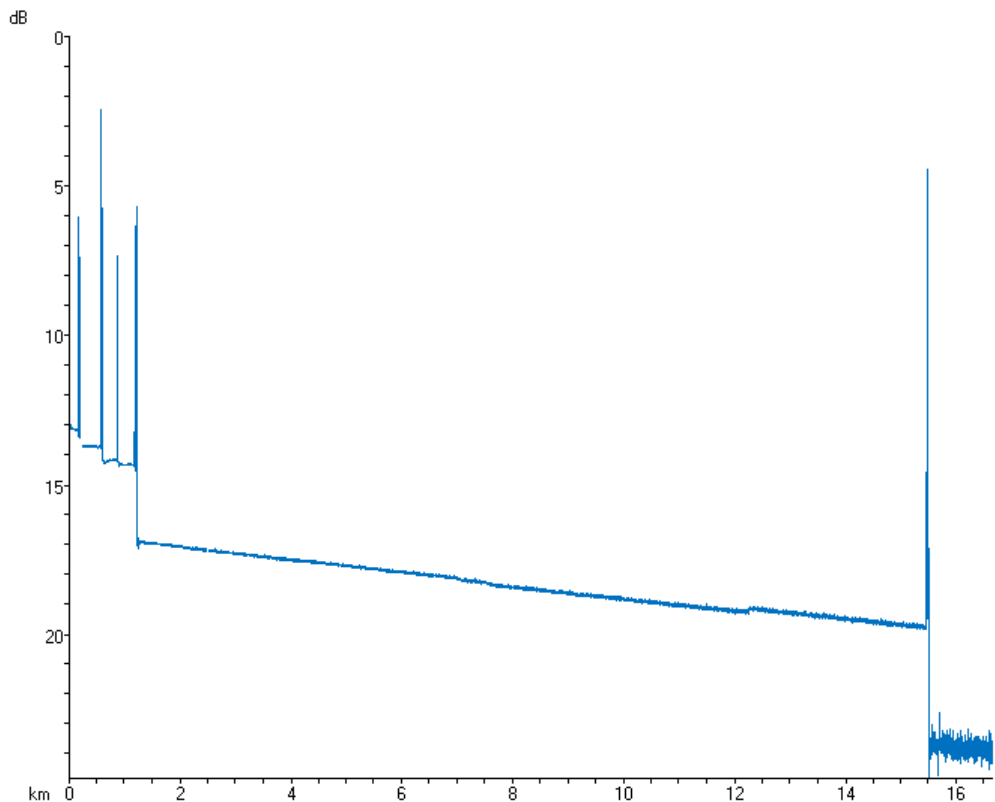

**Figure 16 Optical time reflectometry trace of the UWA-Pawsey 31 km active fibre link (from [RD12])**

#### 4.3.7.2 Microwave Shift Stability

The key component of the MS element of the SKA phase synchronisation system is a DPM. The device incorporates a photonic chip with two nested MZIs inside a master MZI. Each of the three MZIs contains an electro-optic modulator capable of MW-frequency modulation as well as tuning the relative optical path length difference at the optical wavelength scale. The DPM uses external tuning voltages to controls these biases, with the MS requiring the DPM to be tuned to generate SSB-SC modulation. Differential thermal gradients across the DPM can alter the parameters required to achieve the optimal modulation state. In the SKA phase synchronisation system, the opposite sideband (the suppressed sideband) must be suppressed by at least 10 dB (preferably 20 dB) compared to the primary sideband to ensure no technical complication with the frequency transfer technique[7]. Under ideal conditions, the opposite sideband can be supressed by around 80 dB. Figure 17 shows a photograph of the DPM mounted inside a small machined aluminium block inside the MS prototype optics enclosure.

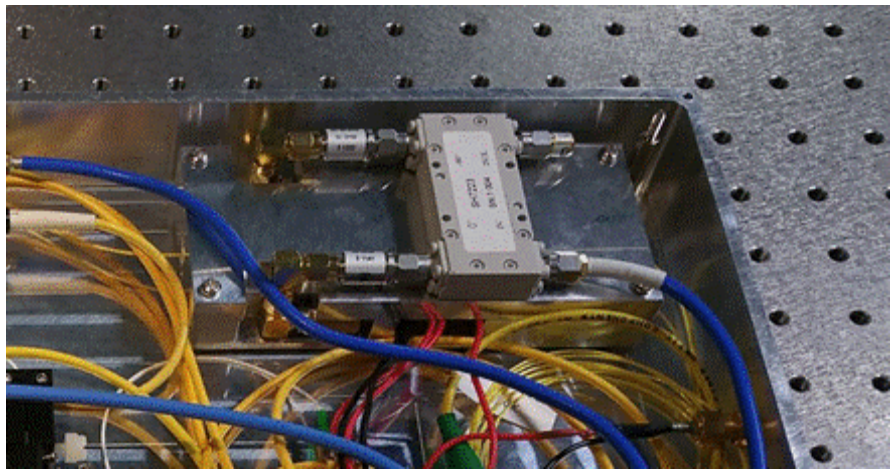

**Figure 17 Photograph of the DPM and machined aluminium block**

The drift of these biases are due to complex time dependent relationships between the applied bias voltage and the optical phase shift on the DPM. The approach is to include a slow servo loop, which is all internal to the MS, to maintain the ideal modulation state of the DPM. The detailed schematic of this setup is shown in Figure 18.

---

[7] There is no first-order issue with having all three signals present concurrently (as can be verified using the SKA Excel Frequency Calculator) and it has been experimentally demonstrated that the transfer stability is not impacted. Having the other sideband and carrier not sufficiently suppressed however, could cause some second-order technical issues. For example, moving power into other optical-frequency signals lowers the power in required optical signal, and those other signals could saturate key components (such as the photodetector) and also produce more intermodulations frequencies.

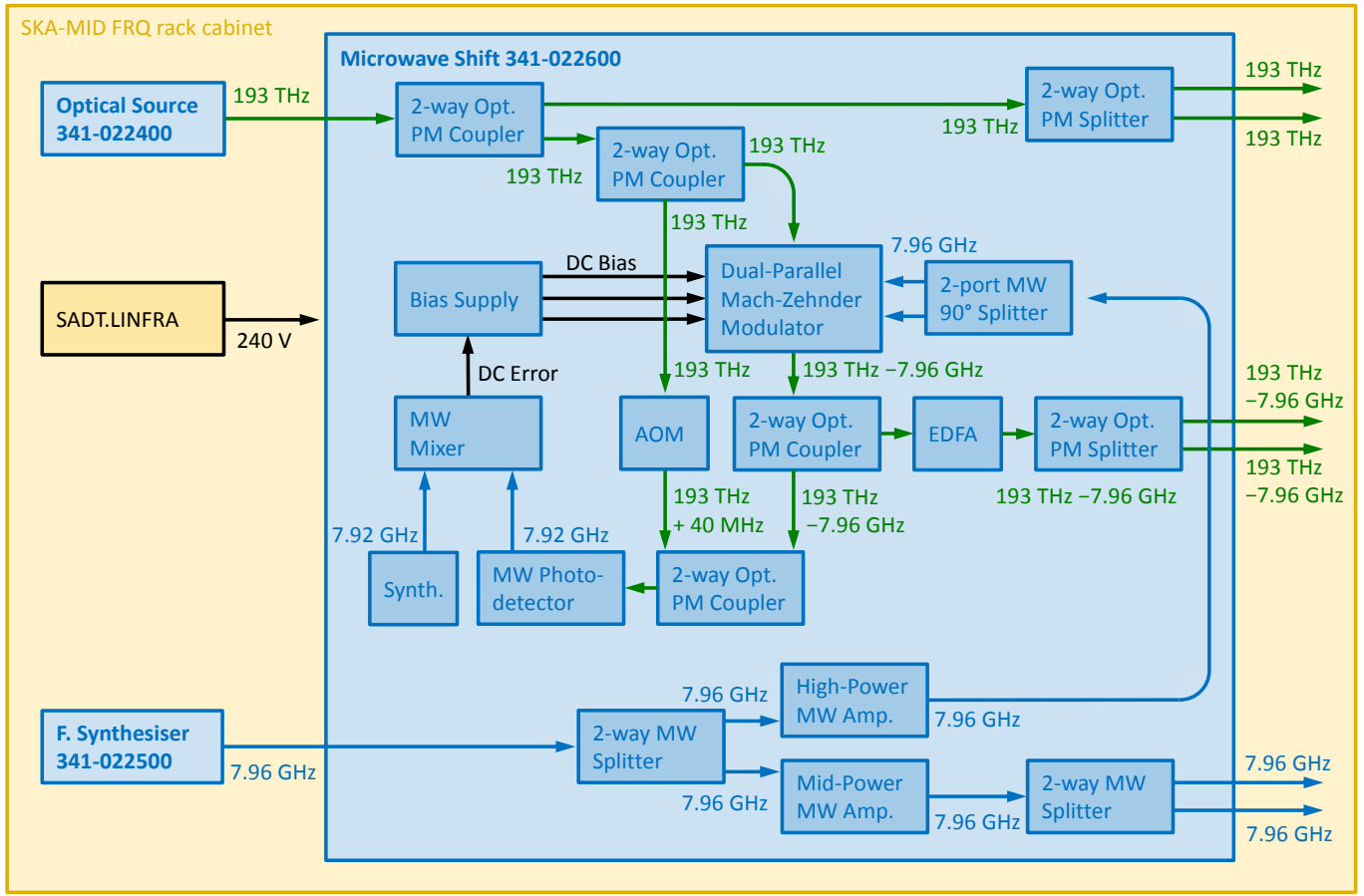

**Figure 18 SKA1-MID MS (341-022600) simplified schematic layout**

Without such a system, the modulation state of the DPM can only be maintained within the MS specification for a period of over a day. Figure 19 shows a plot of the suppression of opposite sideband of the DPM as a function of elapsed time after the bias controls are left to float, for the with and without this aluminium block insulation.

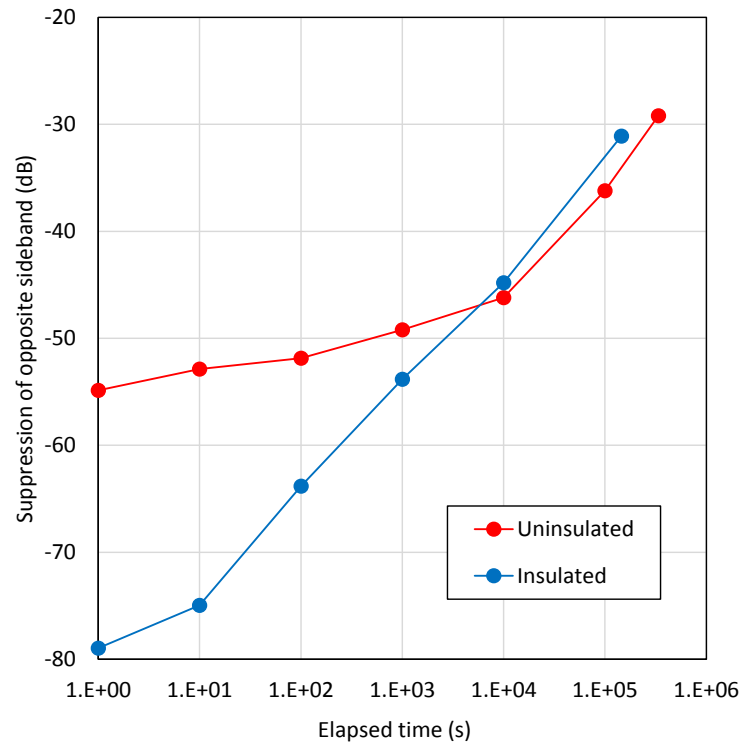

**Figure 19 DPM sideband suppression as a function of elapsed time after the bias controls are left to float**

### 4.3.7.3 Ability to Deliver the Reference Signals Directly to the Receptor Indexer

The RM for the SKA phase synchronisation system is capable of being mounted directly on the SKA1-MID antenna indexer alongside the receiver. Just as is done with most telescopes, including MeerKAT, the reference signals are delivered to the location of the receiver ADCs. Currently, the SADT interface with DISH is in the antenna pedestal, and the DISH Consortium are required to build a second frequency transfer system to transmit the reference signals up the cable wraps to the indexer where the receiver ADCs are located. After CDR, the DISH consortium and SKAO will enact an ECP to address this costly inefficacy. This is not possible with implementations of the SKA phase synchronisation system that require bulky hardware located at the antenna site (or hardware that cannot deliver the much more stringent EMC qualification required for the indexer location).

### 4.3.7.4 Miscellaneous Key System Parameters

The SKA phase synchronisation system use of optical phase sensing allows for the use of FMs to give maximum detected signal at the servo photodetector, as is done with stabilised optical transfer. This removes the need for any initial polarisation alignment, or any ongoing polarisation control or polarisation scrambling. The technique can be deployed on standard fibre links and does not require specialty fibre in the fibre link (such as dispersion compensation or polarisation maintaining fibre).

The MW signal being transmitted on the fibre link arises from only two optical signals, not three as is the case for standard intensity modulation commonly used in RF or MW transfer. Using only two optical signals ensures that that the maximum signal power is available at the antenna site regardless of link length.

Simple and cheap bi-directional OAs are deployed to extend the range of transmission, and, therefore, potentially complex electronic signal re-generation systems are not required. This also eliminates the potential for a remotely located single point of failure affecting multiple end stations if one re-generation systems are used in a branching system to supply multiple end points. The SKA phase synchronisation system therefore requires only a single laser, reducing system complexity.

## 4.4 Hardware

This section is split into the following sub-sections:

- Detailed Design Overview (see §4.4.1);
- Mechanical Detailed Design (see §4.4.2);
- Optical Detailed Design (see §4.4.3);
- Electronic Detailed Design (see §4.4.4).

### 4.4.1 Detailed Design Overview

The SKA1-MID equipment is split into the following eleven LRUs:

- Rack cabinet (341-022900)
- OS (341-022400)
- FS (341-022500)
- MS (341-022600)
- SG (341-023100)
- RD (341-022800)
- SR (341-022700)
- TM (341-022100)
- FP lead (341-023200)
- OA (341-022200)
- RM (341-022300).

The interdependencies of the FRQ LRUs are shown in Figure 20.

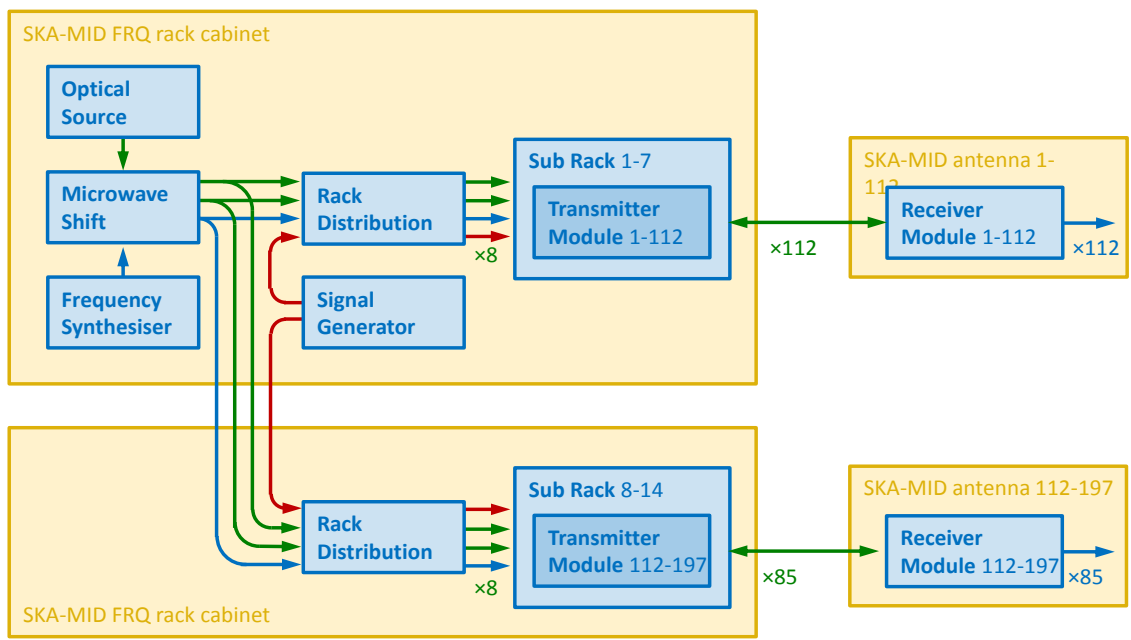

**Figure 20 SKA1-MID overview simplified schematic layout**

**NOTE 10:** The OA is required only on long links; using the current SAT.LINFRA design, this is for a total of 12

links for SKA1-MID. See Figure 44 for more information on the OA network distribution. The FP lead joins the TM to the start of the fibre link. The visual representation of the FRQ LRUs that are located in the SKA1-MID rack cabinet are shown in Figure 21.

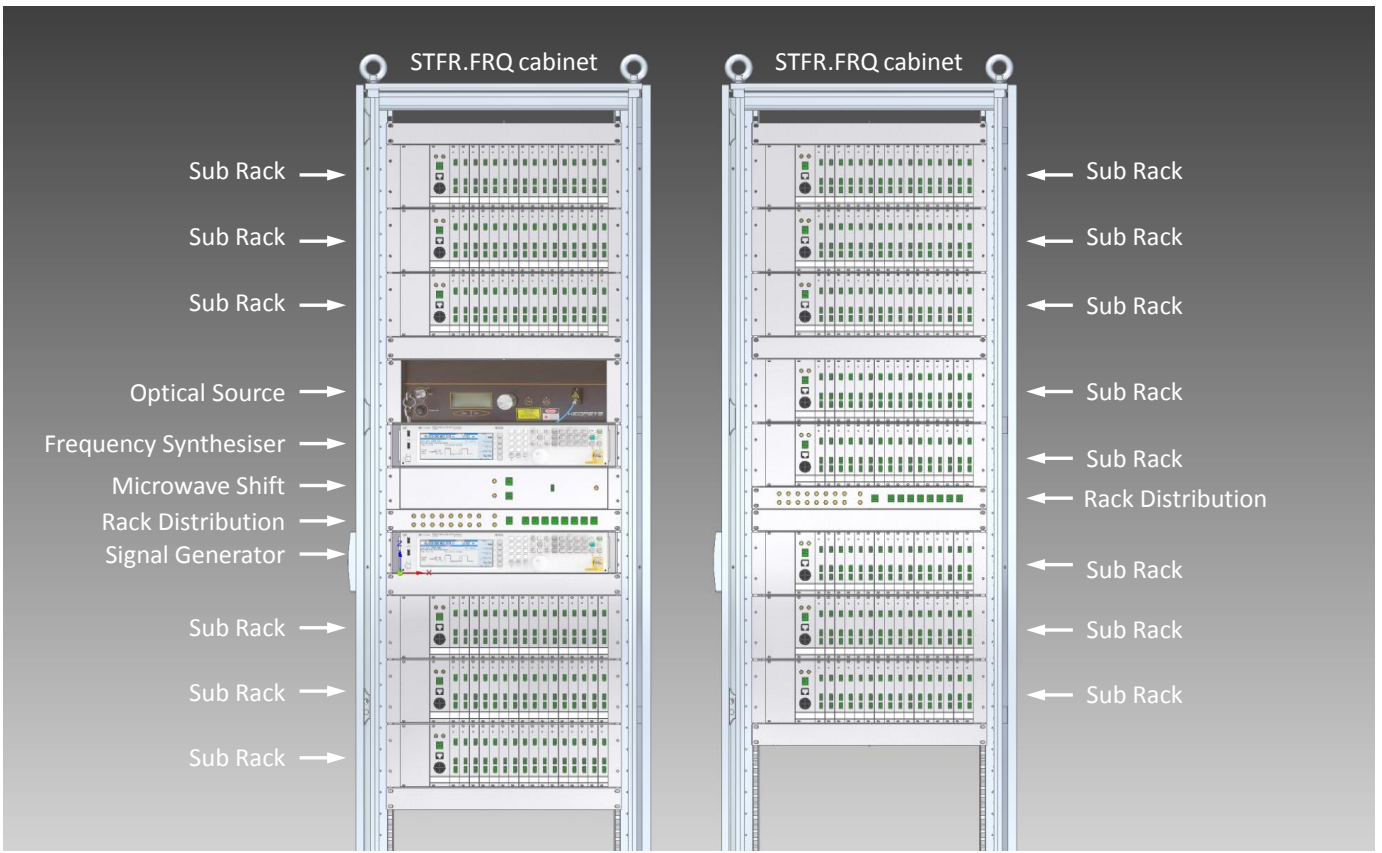

**Figure 21 SKA1-MID overview 3D render**

### 4.4.1.1  Rack Cabinet (341-022900)

This LRU is a collection of rack cabinet accessories including equipment for heat management and cable management, and the labour to install this equipment. A photograph of the rack cabinet (341-022900) is shown in Figure 22.

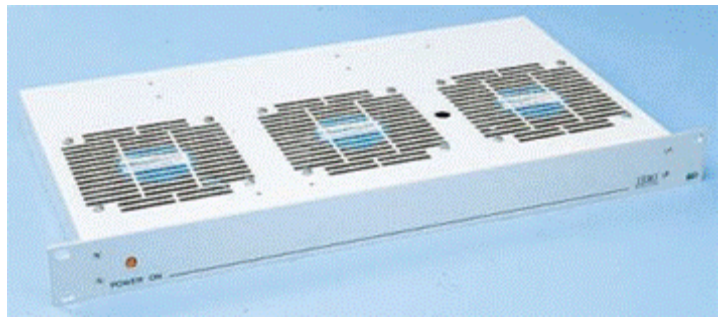

**Figure 22 Photograph of the SKA1-MID rack cabinet (341-022900)**

### 4.4.1.2 Optical Source (341-022400)

The simplified schematics layout for the OS (341-022400) is shown in Figure 23.

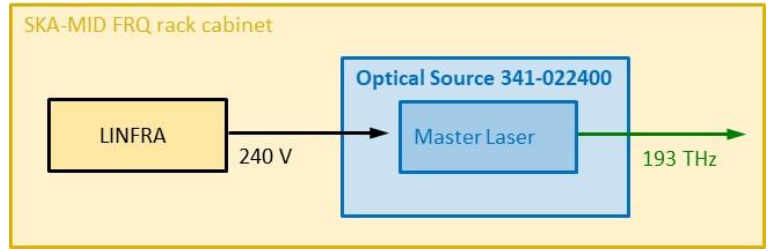

**Figure 23 SKA1-MID OS (341-022400) simplified schematic layout**

A photograph of the OS (341-022400) is shown in Figure 24.

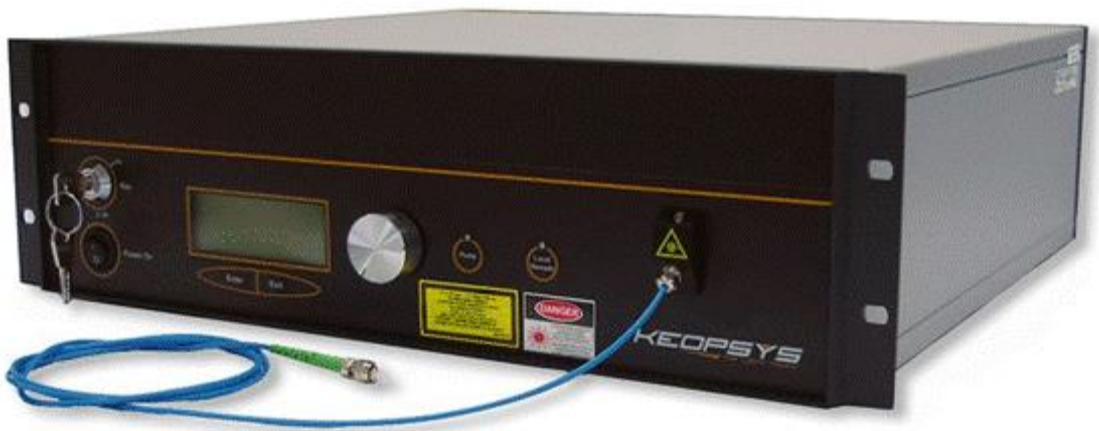

**Figure 24 Photograph of the SKA1-MID OS (341-022400)**

### 4.4.1.3 Frequency Source (341-022500)

The simplified schematics layout for the FS (341-022500) is shown in Figure 25.

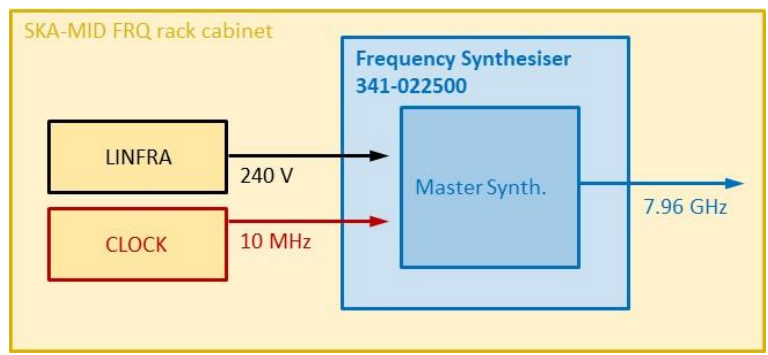

**Figure 25 SKA1-MID FS (341-022500) simplified schematic layout**

Note: the phase stability of the 7.96 MHz signal produced by FS is not because the DPM in MS is inside the servo control loop. That is, any drift in FS, and therefore the DPM, is corrected by the TM servo. Having FS referenced to CLOCK is not critical.

A photograph of the FS (341-022500) is shown in Figure 26.

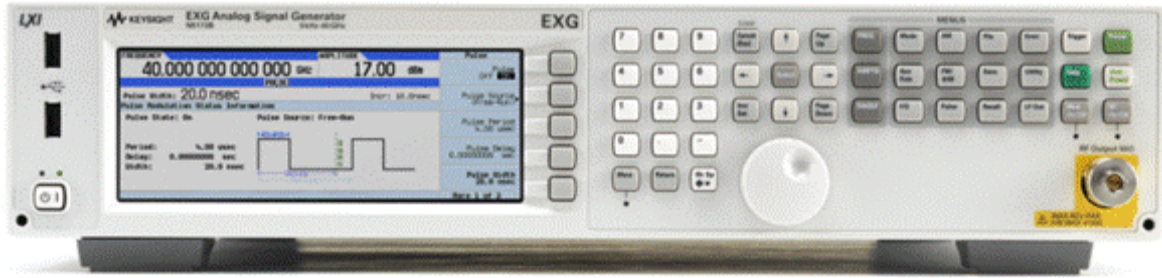

**Figure 26 Photograph of the SKA1-MID FS (341-022500)**

### 4.4.1.4 Microwave Shift (341-022600)

The simplified schematics layout for the MS (341-022600) is shown in Figure 27.

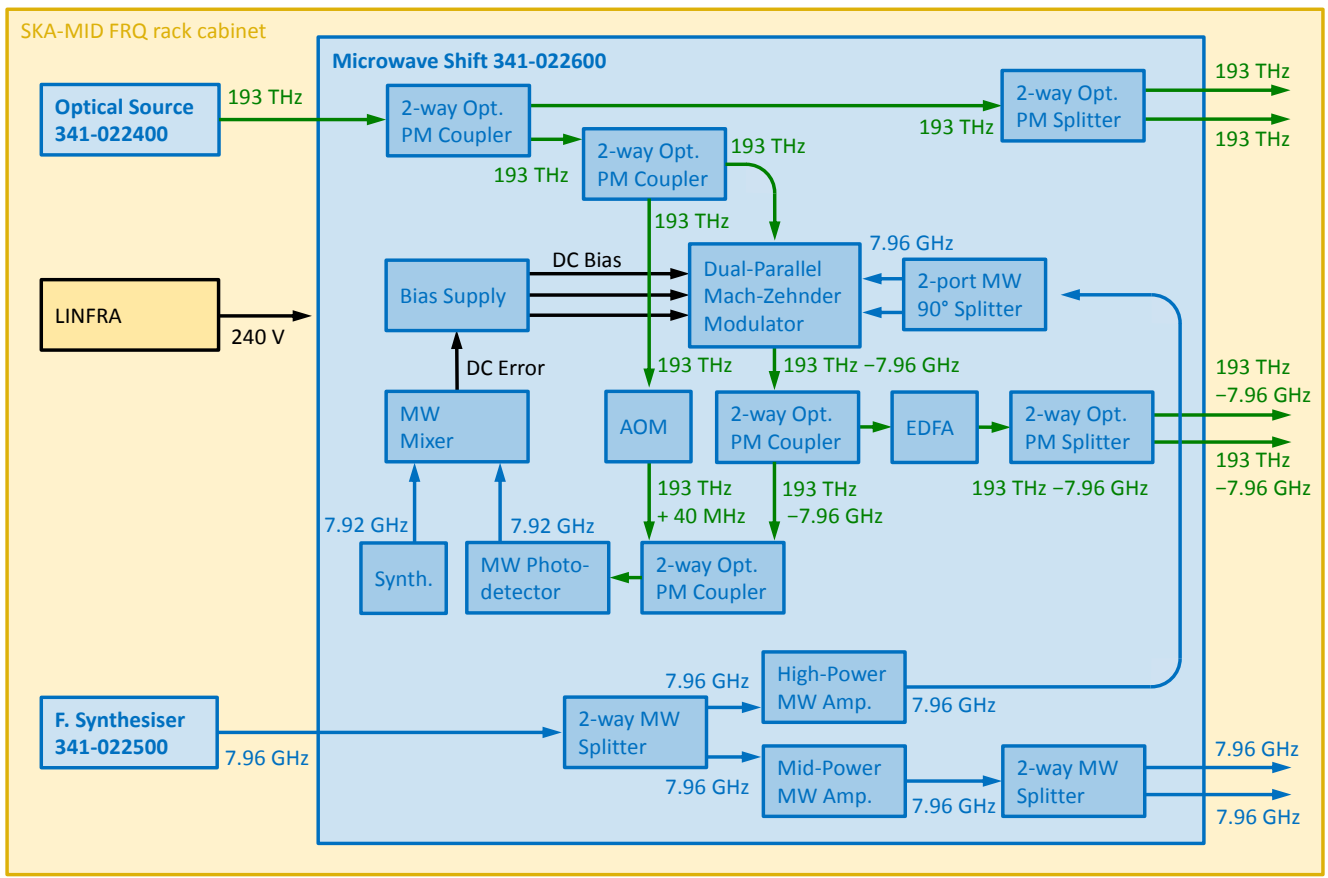

**Figure 27 SKA1-MID MS (341-022600) simplified schematic layout**

A photograph collage of the key elements of the MS (341-022600) is shown in Figure 28.

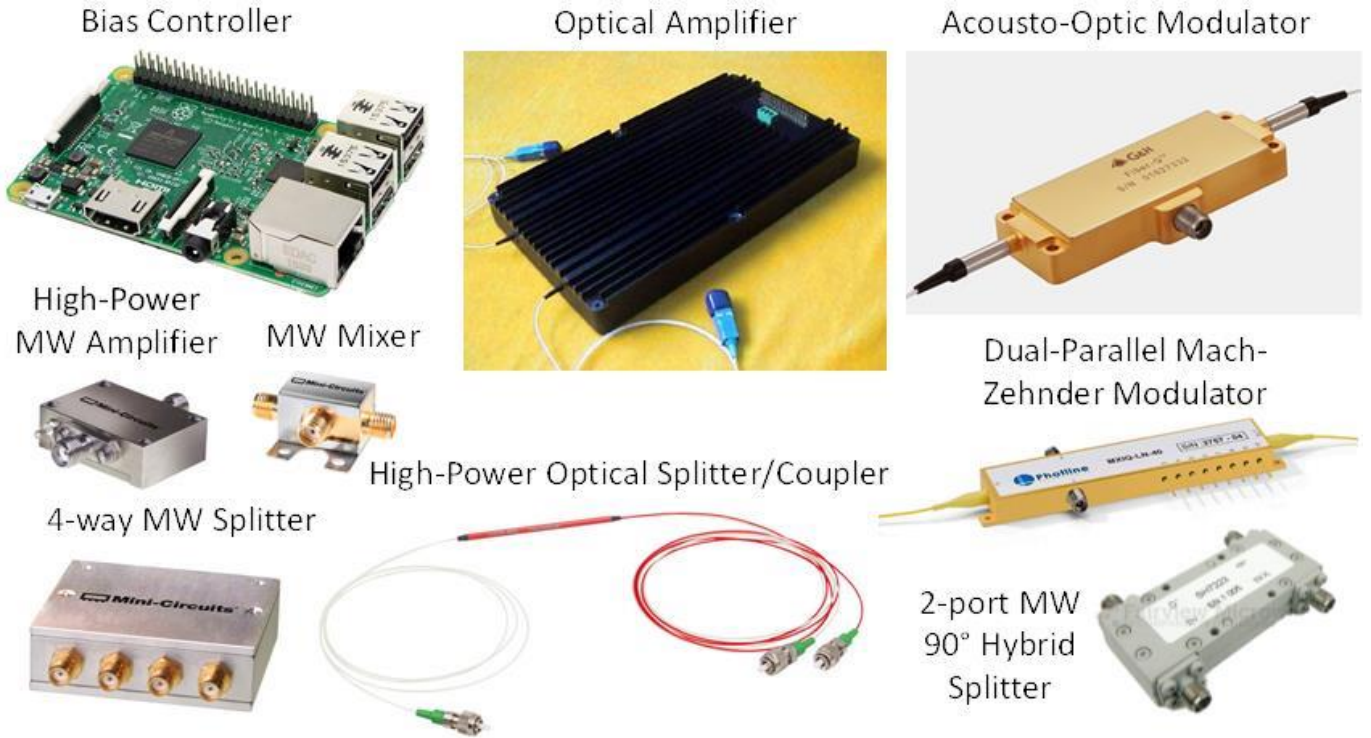

**Figure 28 SKA1-MID MS (341-022600) key elements**

The 3D render (external) for the MS (341-022600) is shown in Figure 29.

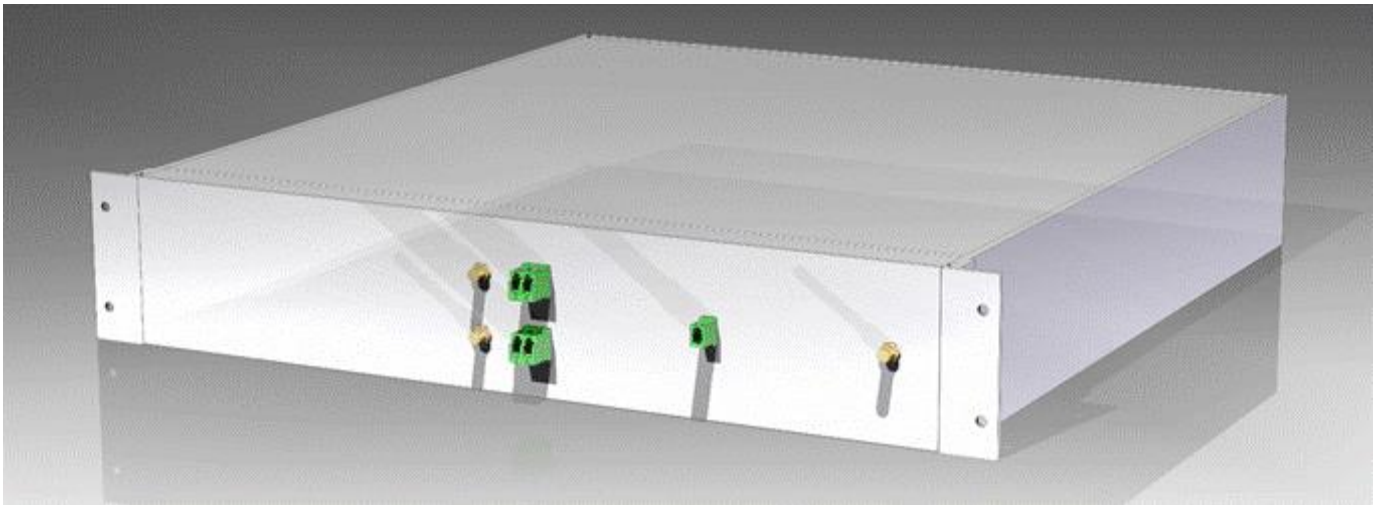

**Figure 29 SKA1-MID MS (341-022600) 3D render (external)**

### 4.4.1.5 Signal Generator (341-023100)

The simplified schematics layout for the SG (341-023100) is shown in Figure 30.

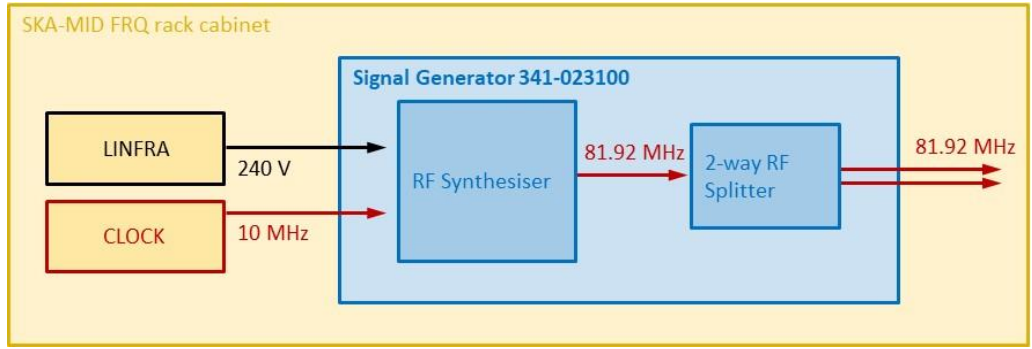

Figure 30 SKA1-MID SG (341-023100) simplified schematic layout

A photograph of the SG (341-023100) is shown in Figure 31.

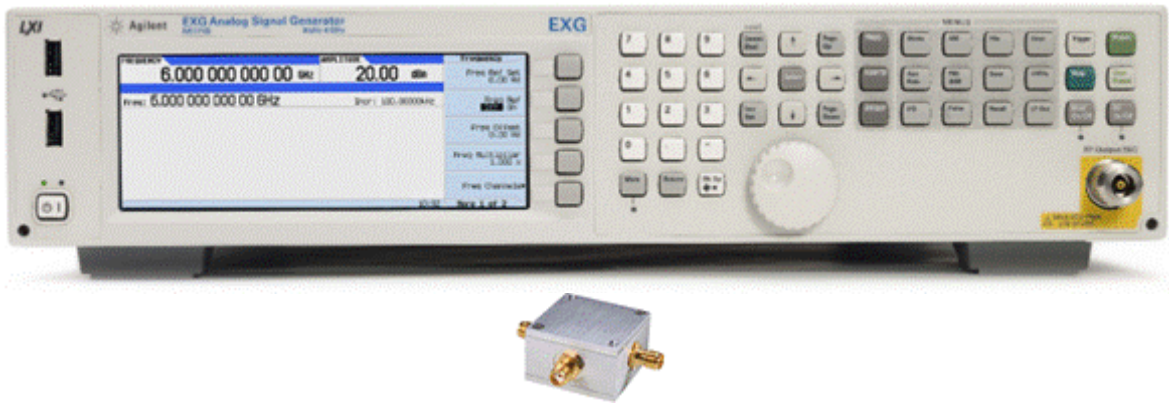

Figure 31 SKA1-MID SG (341-023100) photograph collage

### 4.4.1.6 Rack Distribution (341-022800)

The simplified schematics layout (internal) for the RD (341-022800) is shown in Figure 32.

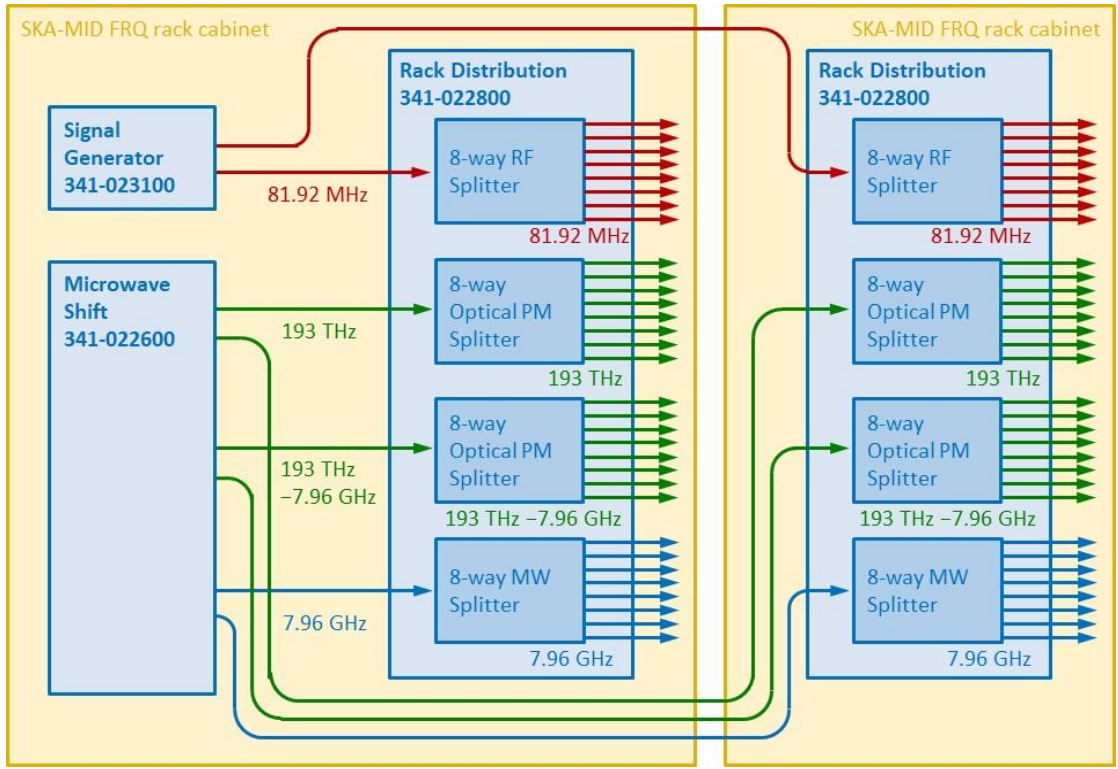

**Figure 32 SKA1-MID RD (341-022800) simplified schematic layout (internal)**

Photographs of the (internal) RD (341-022800) are shown in Figure 33.

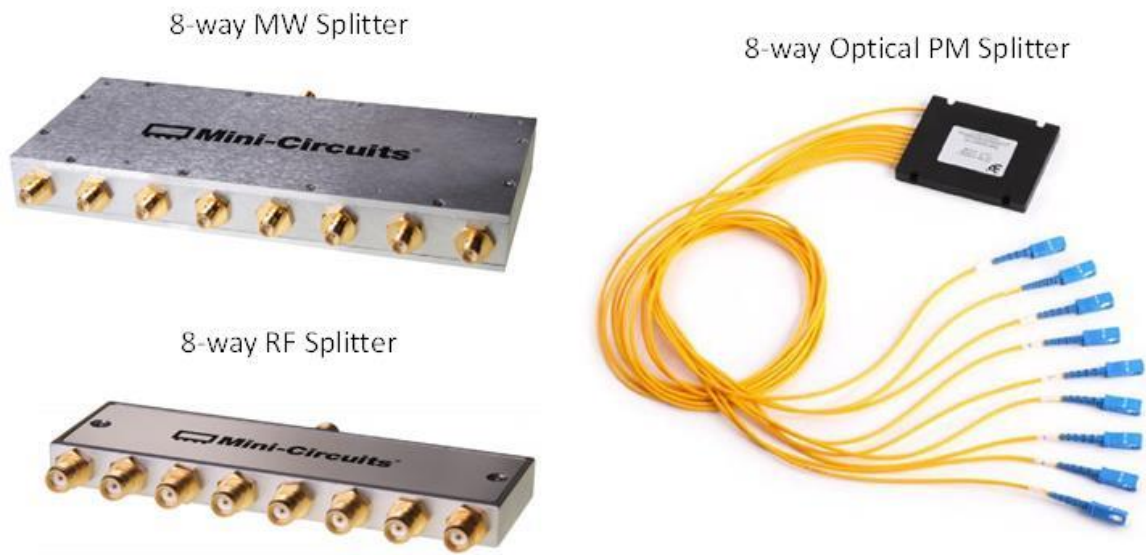

**Figure 33 SKA1-MID RD (341-022800) photograph collage (internal)**

The simplified schematics layout (external) for the RD (341-022800) is shown in Figure 34.

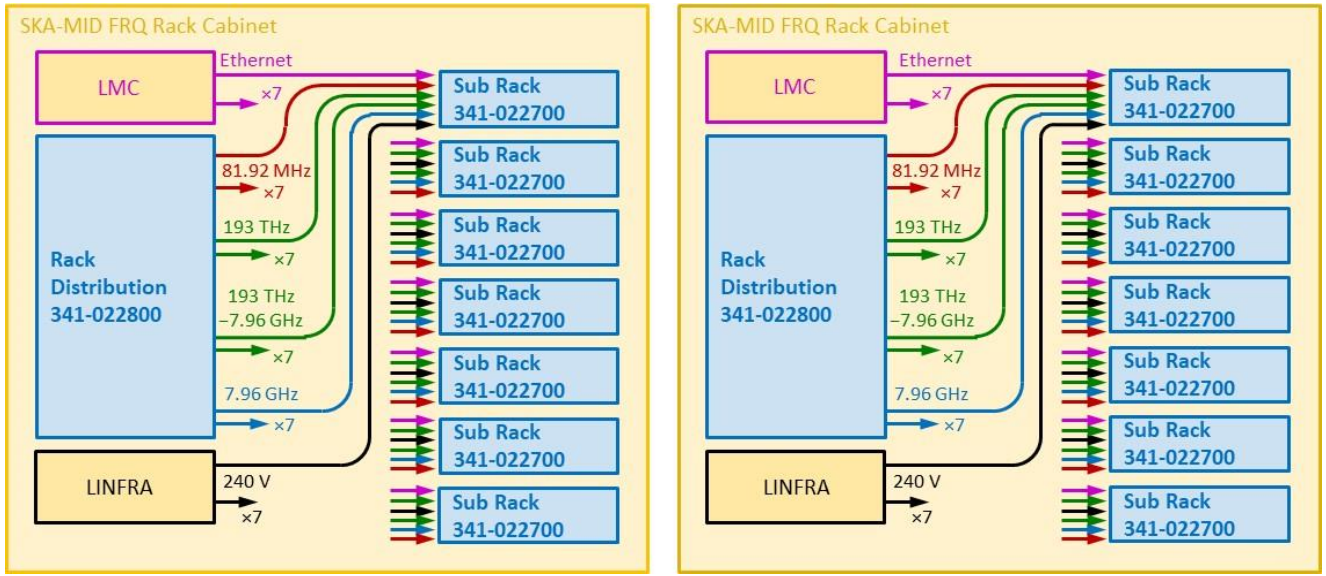

**Figure 34 SKA1-MID RD (341-022800) simplified schematic layout (external)**

A 3D render of the (external) RD (341-022800) is shown in Figure 35.

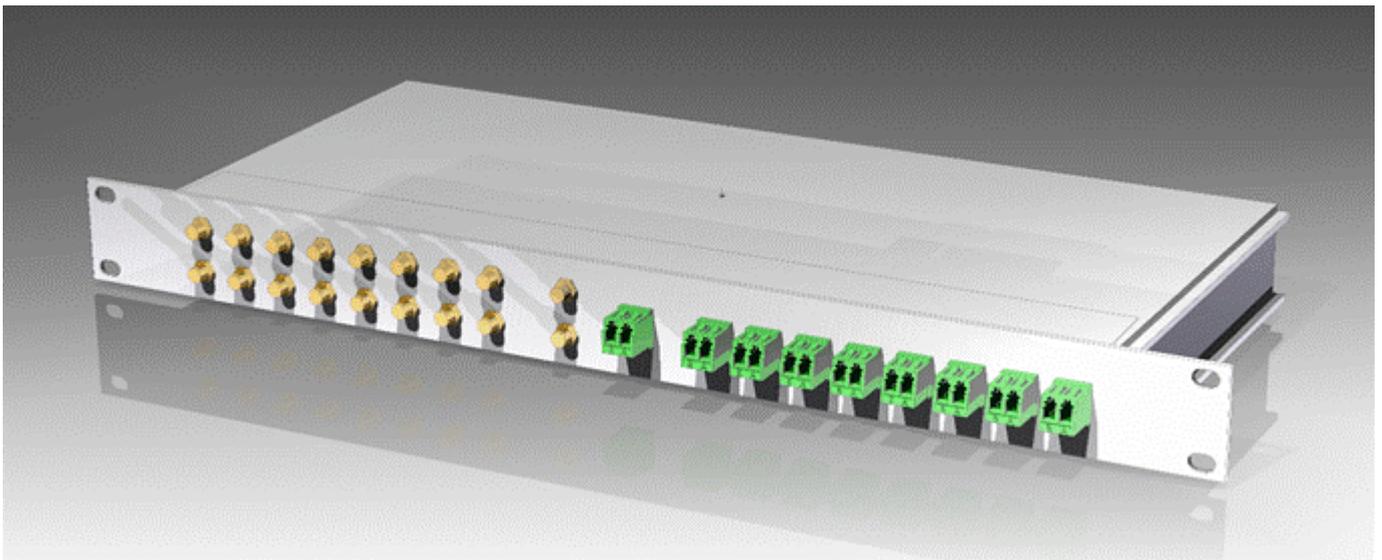

**Figure 35 SKA1-MID RD (341-022800) 3D render (external)**

#### 4.4.1.7 Sub Rack (341-022700)

The simplified schematics layout for the SR (341-022700) is shown in Figure 36.

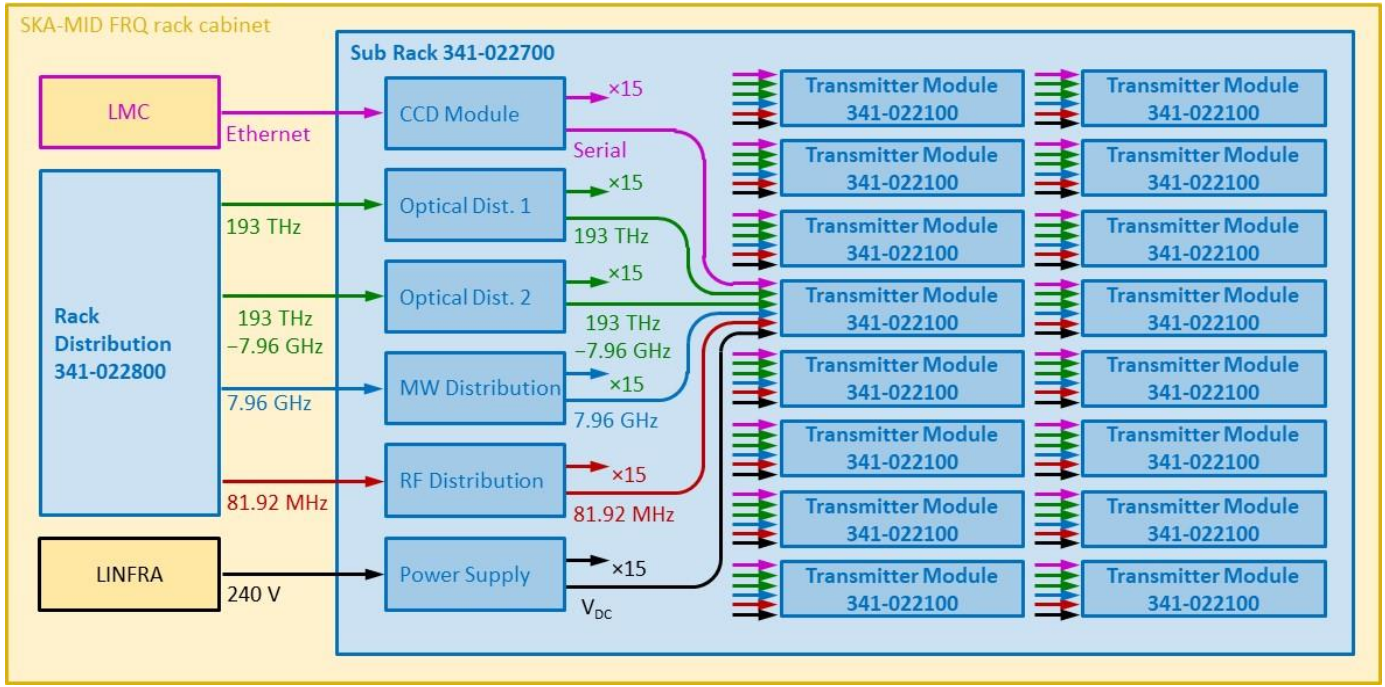

**Figure 36 SKA1-MID SR (341-022700) simplified schematic layout**

A 3D render of the SR (341-022700) is shown in Figure 37.

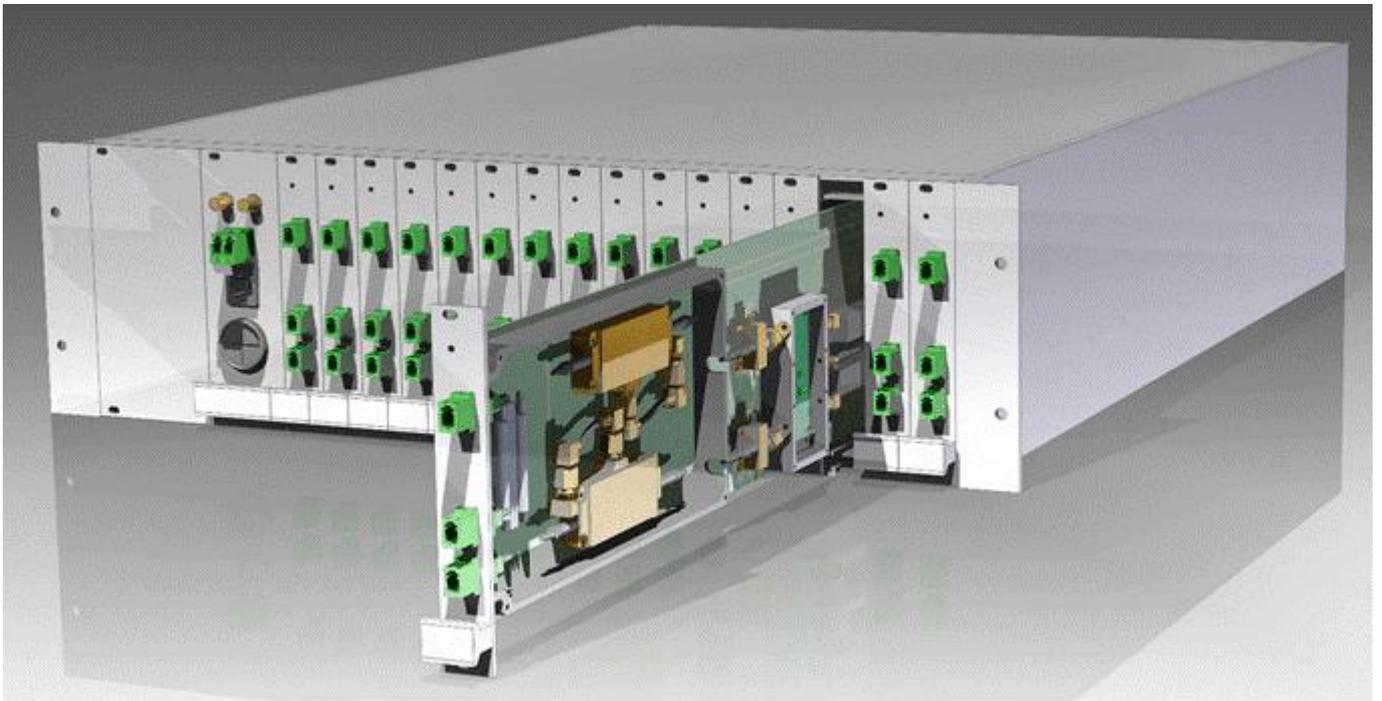

**Figure 37 SKA1-MID SR (341-022700) 3D render**

A photograph of the prototype of the SR (341-022700) is shown in Figure 38.

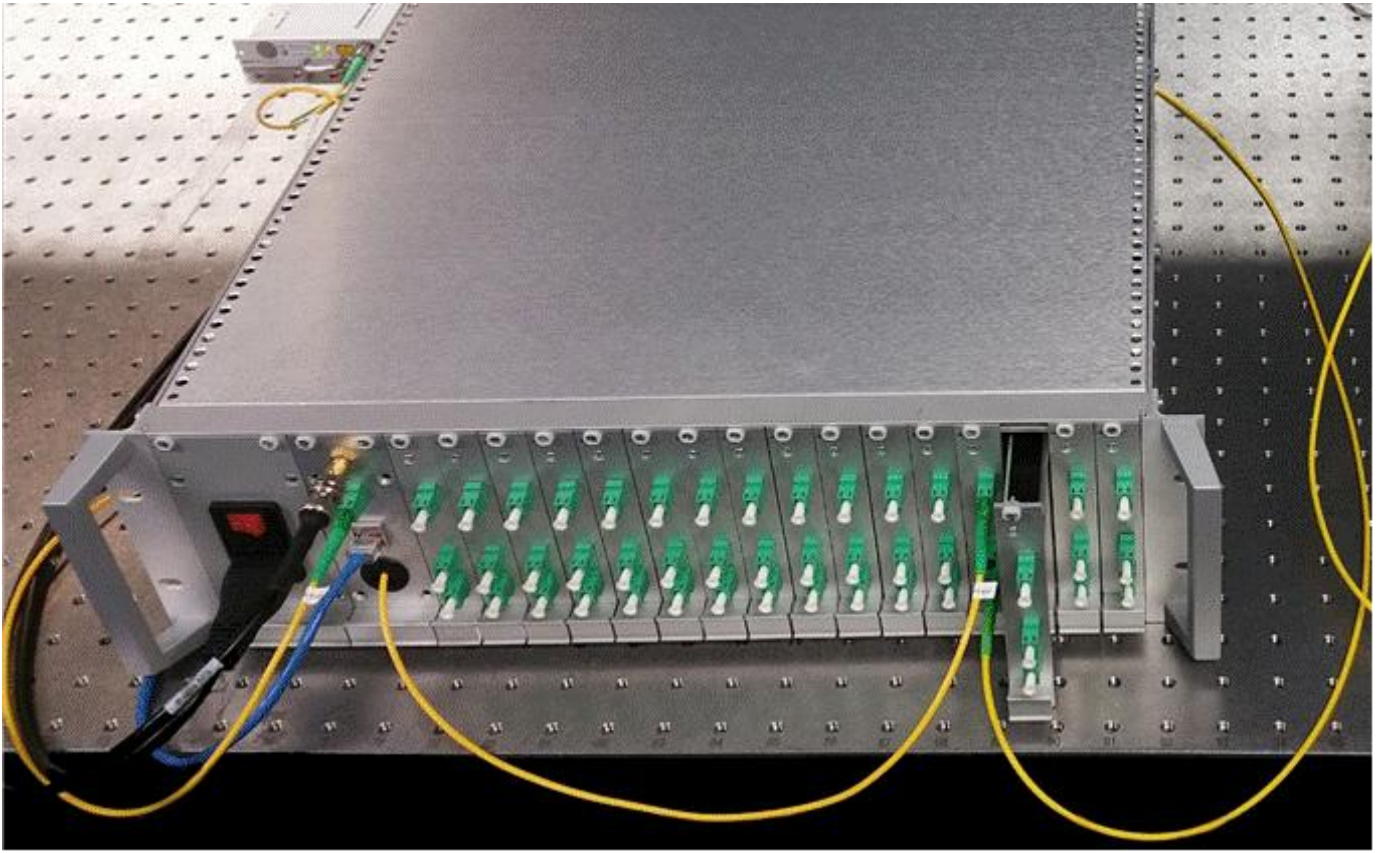

**Figure 38 Photograph of the SKA1-MID SR (341-022700) prototype**

### 4.4.1.8 Transmitter Module (341-022100)

The simplified schematics layout for the TM (341-022100) is shown in Figure 39.

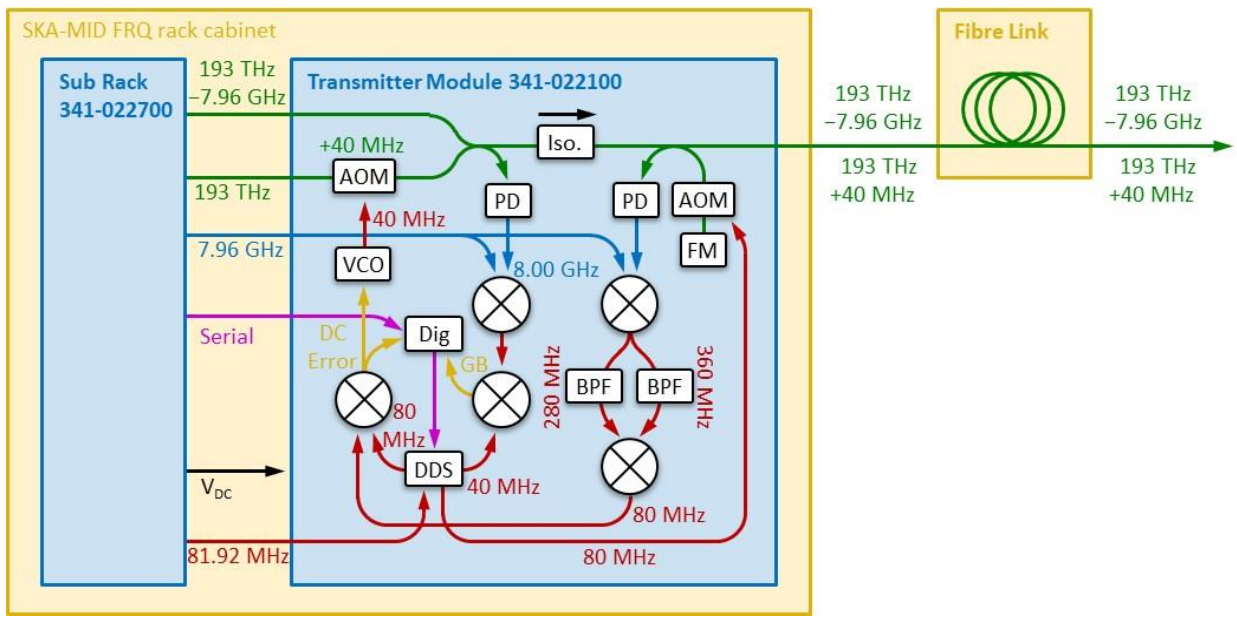

**Figure 39 SKA1-MID TM (341-022100) simplified schematic layout**

A 3D render of the TM (341-022100) is shown in Figure 40.

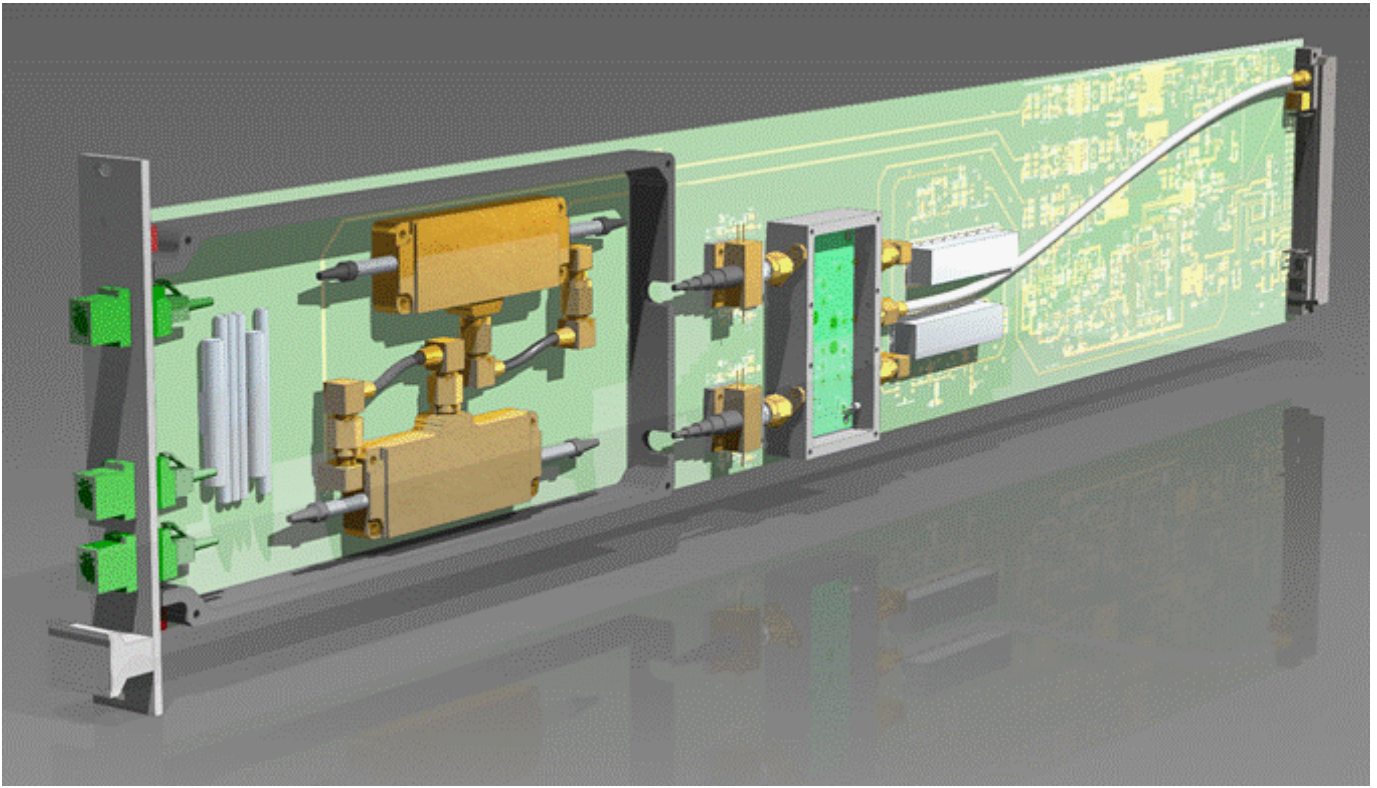

**Figure 40 SKA1-MID Transmitter Module (341-022100) 3D render**

A photograph of individual prototype PCBs of the TM (341-022100) is shown in Figure 41.

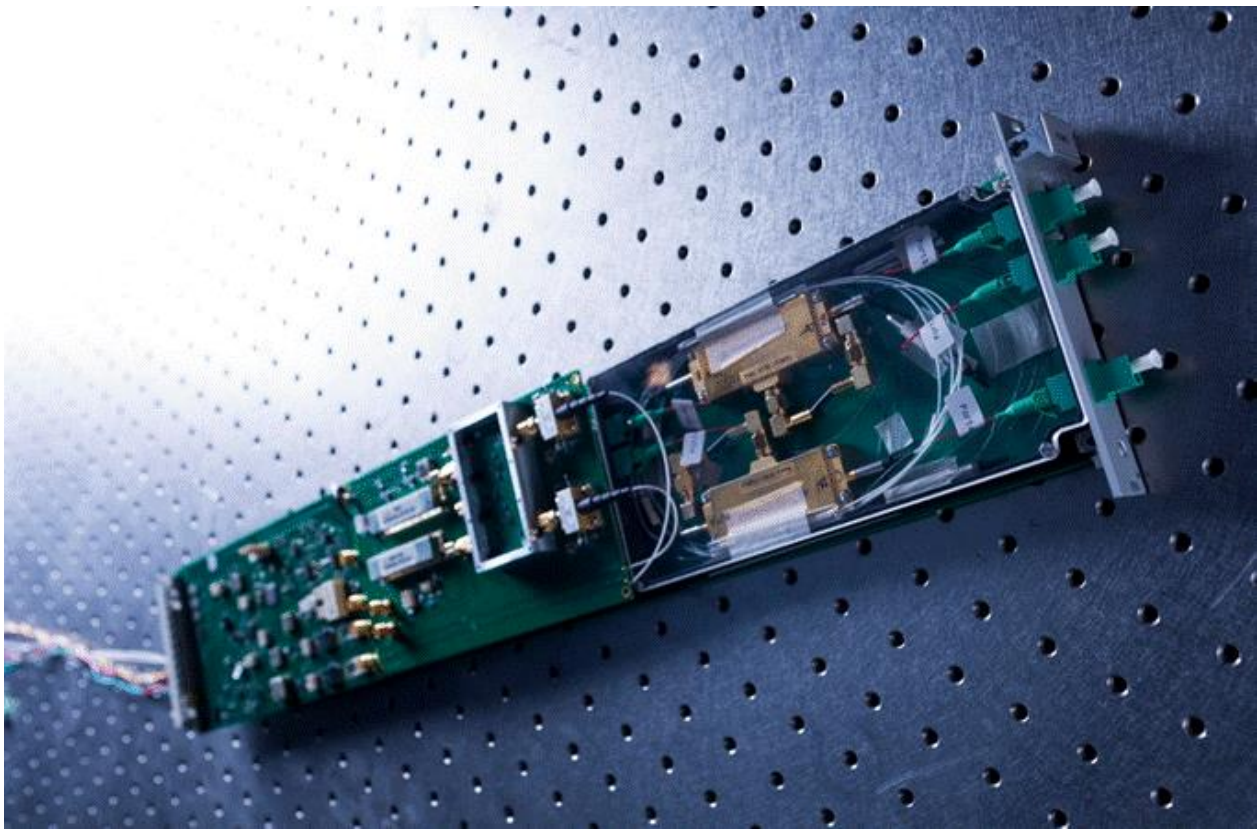

**Figure 41 Photograph of the SKA1-MID TM (341-022100) individual prototype PCBs**

#### 4.4.1.9 Fibre Patch Lead (341-023200)

This LRU is a simplex FP lead that joins the TM to a LINFRA patch-panel at the end of relevant rack cabinet row. The LINFRA patch-panel signifies the start of the fibre link.

#### 4.4.1.10 Optical Amplifier (341-022200)

The simplified schematics layout for the OA (341-022200) is shown in Figure 42.

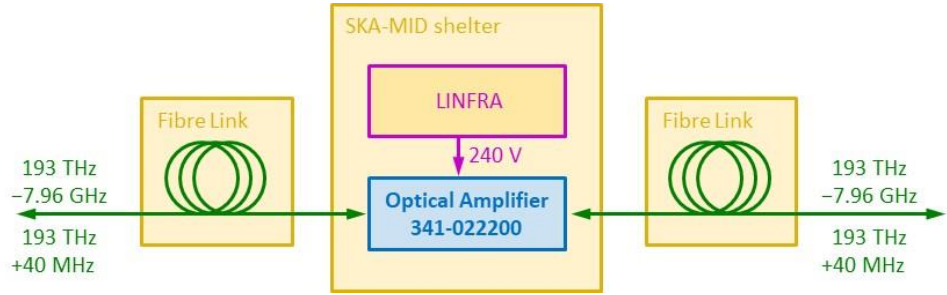

**Figure 42 SKA1-MID OA (341-022200) simplified schematic layout**

A photograph of the OA (341-022200) is shown in Figure 43.

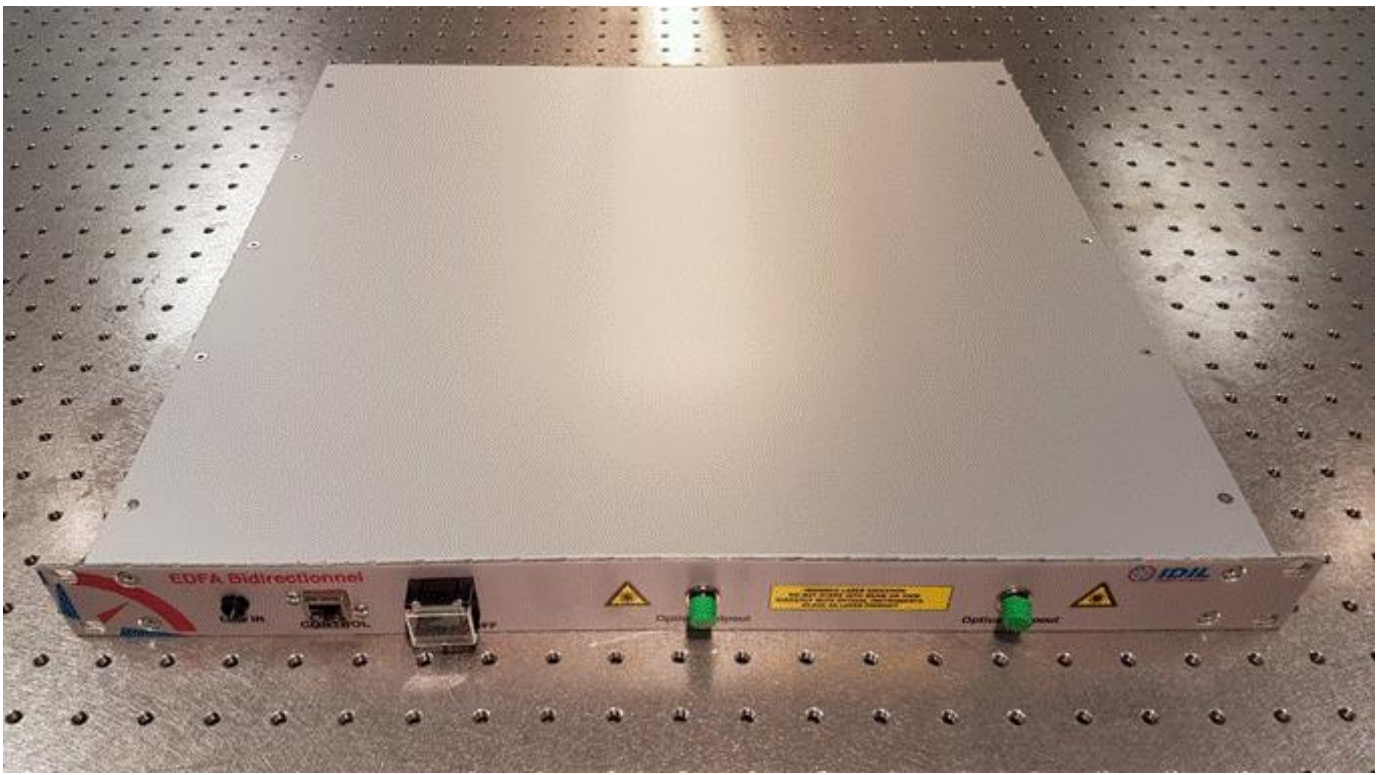

**Figure 43 Photograph of the SKA1-MID OA (341-022200)**

The network distribution of the 17 SKA1-MID OAs is shown in Figure 44.

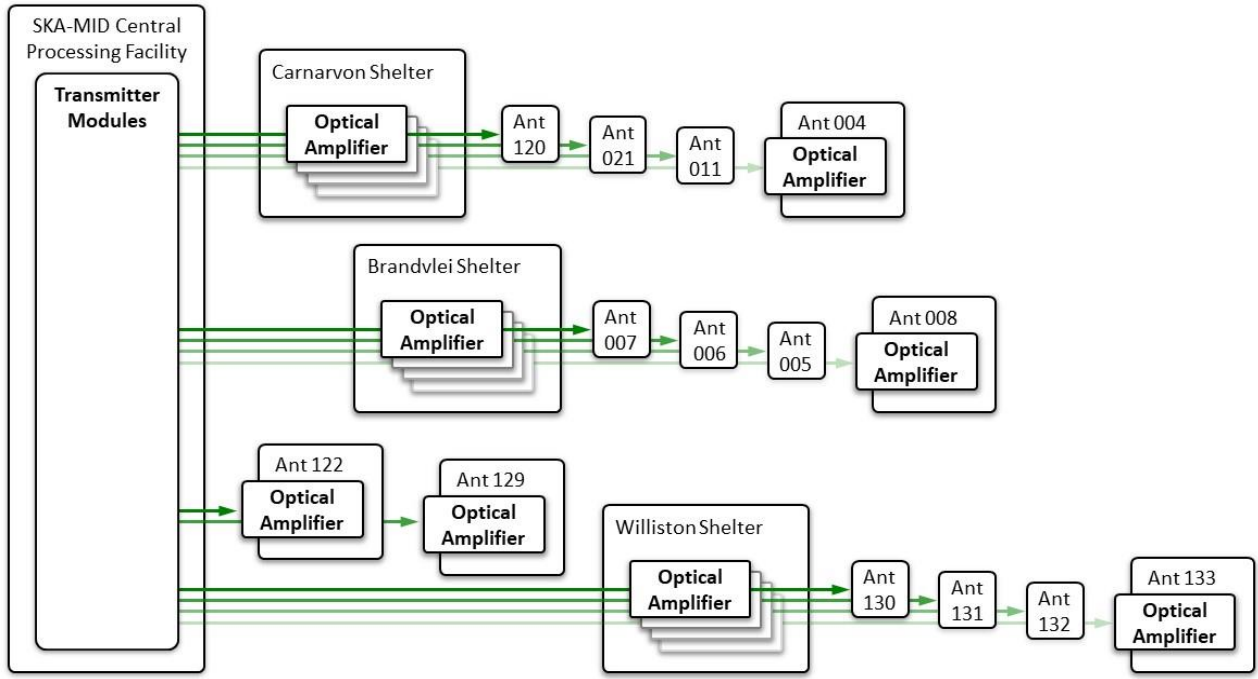

**Figure 44 SKA1-MID OA (341-022200) network distribution**

### 4.4.1.11 Receiver Module (341-022300)

The simplified schematic layout for the RM (341-022300) is shown in Figure 45.

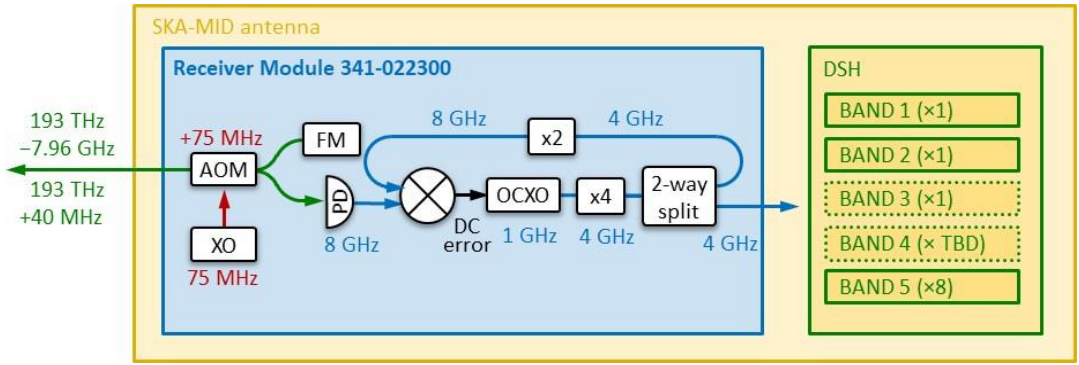

**Figure 45 SKA1-MID RM (341-022300) simplified schematic layout**

A photograph of the key RM (341-022300) mocked-up equipment inside a MeerKAT 'time and frequency' enclosure is shown in Figure 46.

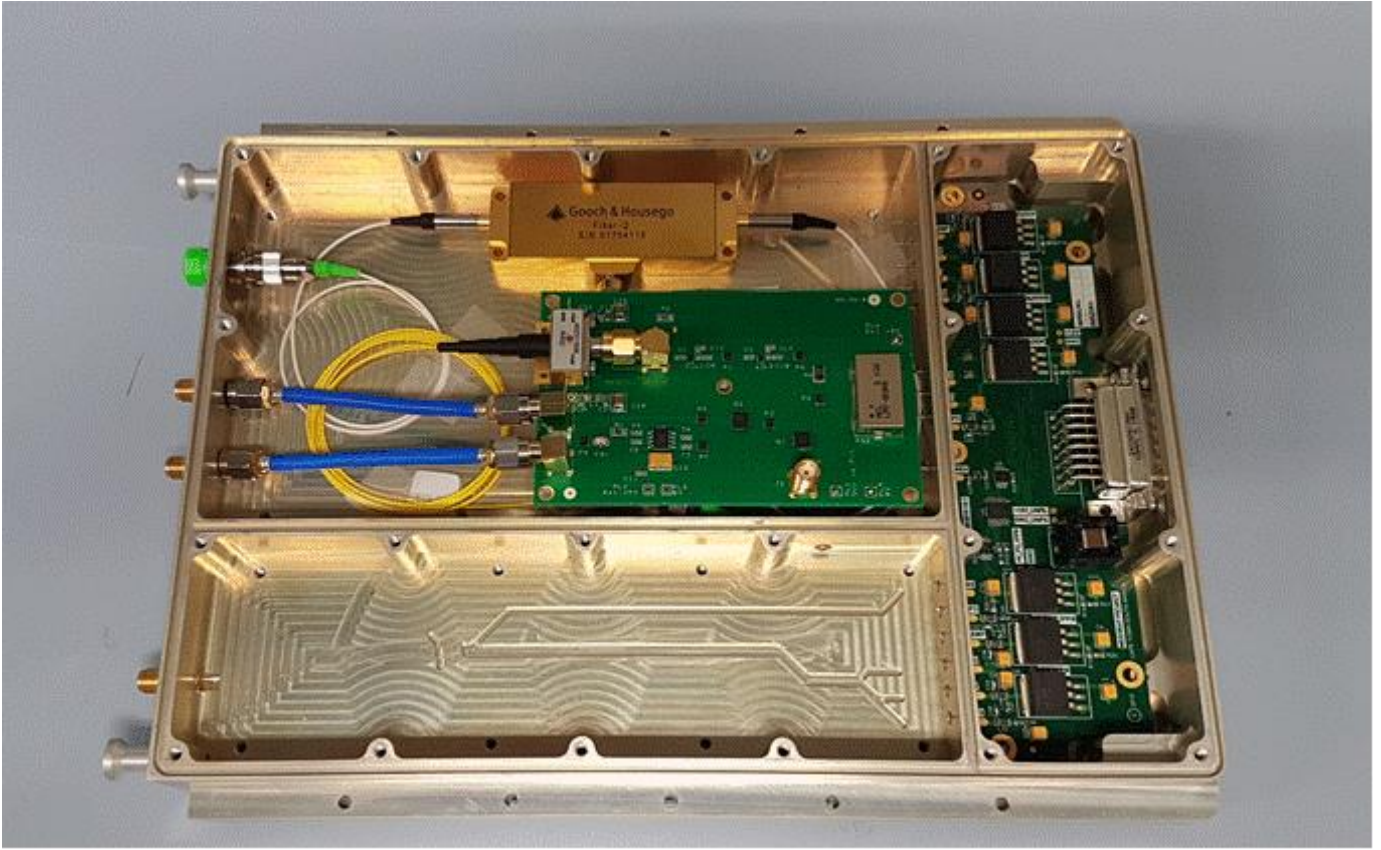

**Figure 46 Photograph of the SKA1-MID RM (341-022300) mock-up using MeerKAT enclosure**

A photograph of the RM (341-022300) prototype is shown in Figure 47.

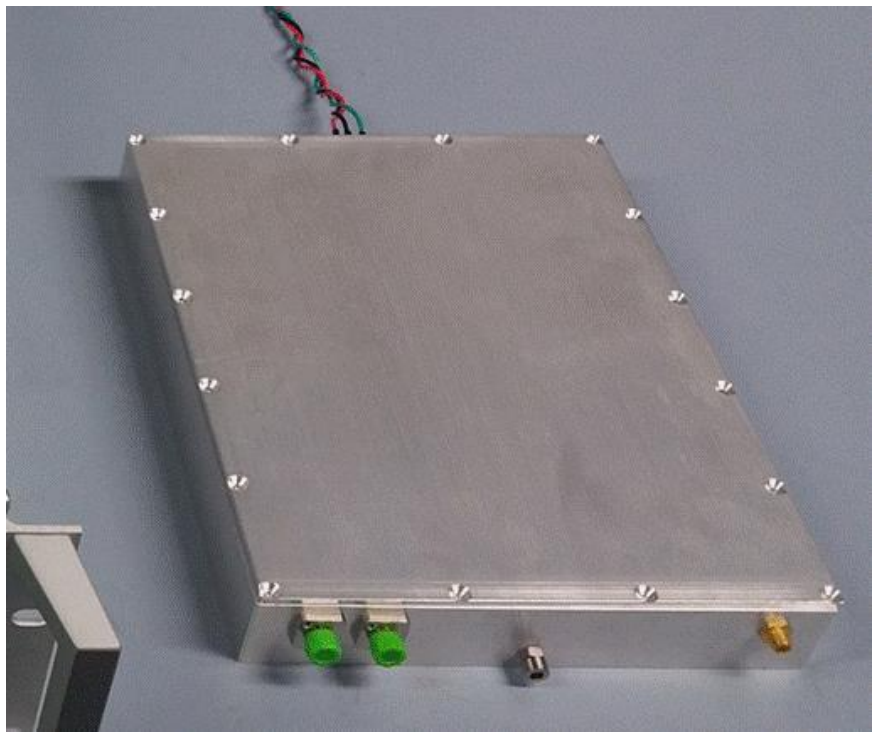

**Figure 47 Photograph of the SKA1-MID RM (341-022300) prototype**

### 4.4.2 Mechanical Detailed Design

#### 4.4.2.1 Rack Cabinet (341-022900)

The mechanical detailed design for rack cabinet (341-022900) is given in Table 20.

| Type | Fan Tray |
|---|---|
| Mounting | Rack |
| Rack Units | 1U |
| Number of Fans | 6 |
| Air Flow | 167m$^3$/h |
| Height | 1U |
| Width | 350mm |

**Table 20 SKA1-MID rack cabinet (341-022900) mechanical detailed design**

The related Solid Edge part is included in the Mechanical Detailed Design File pack in Appendix 8.5.1.

#### 4.4.2.2 Optical Source (341-022400)

The mechanical detailed design for the OS (341-022400) is a 3U, 19" rackmount enclosure, with a depth of 448 mm. The power socket is at the rear.

The related Solid Edge part is included in the Mechanical Detailed Design File pack in Appendix 8.5.1.

#### 4.4.2.3 Frequency Source (341-022500)

The mechanical detailed design for the FS (341-022500) is a 3U, 19" rackmount enclosure, with a depth of 489 mm. The power socket is at the rear.

The related Solid Edge part is included in the Mechanical Detailed Design File pack in Appendix 8.5.1.

### 4.4.2.4 Microwave Shift (341-022600)

The mechanical detailed design for the MS (341-022600) is shown in Figure 48.

**Figure 48 SKA1-MID MS (341-022600) mechanical detailed design**

The RD is a 1U, 19" rackmount enclosure, with a depth of 462.5 mm. The power socket is at the rear.

The related Solid Edge part is included in the Mechanical Detailed Design File pack in Appendix 8.5.1.

### 4.4.2.5 Signal Generator (341-023100)

The mechanical detailed design for the SG (341-023100) is a 3U, 19" rackmount enclosure, with a depth of 489 mm. The power socket is at the rear.

The related Solid Edge part is included in the Mechanical Detailed Design File pack in Appendix 8.5.1.

### 4.4.2.6 Rack Distribution (341-022800)

The mechanical detailed design for the RD (341-022800) is shown in Figure 49.

**Figure 49 SKA1-MID RD (341-022800) mechanical detailed design**

The RD is a 1U, 19" rackmount enclosure, with a depth of 178 mm. The power socket is at the rear.

The related Solid Edge part is included in the Mechanical Detailed Design File pack in Appendix 8.5.1.

### 4.4.2.7 Sub Rack (341-022700)

The mechanical detailed design for the SR (341-022700) is shown in Figure 50.

**Figure 50 SKA1-MID SR (341-022700) mechanical detailed design**

The SR is a 3U, 19" rackmount enclosure, with a depth of 600 mm (the maximum specified in [RD2]). The TM-power module is 12HP, and the TM-distribution module is 8HP. The power socket is at the rear.

The related Solid Edge part is included in the Mechanical Detailed Design File pack in Appendix 8.5.1.

### 4.4.2.8 Transmitter Module (341-022100)

#### 4.4.2.8.1 Transmitter Module (341-022100) PCB

The mechanical detailed design for the TM (341-022100) PCB is shown in Figure 51.

**Figure 51 SKA1-MID TM (341-022100) PCB mechanical detailed design**

The related Solid Edge part is included in the Mechanical Detailed Design File pack in Appendix 8.5.1.

#### 4.4.2.8.2 Transmitter Module (341-022100) Optics Enclosure

The mechanical detailed design for the TM (341-022100) optics enclosure is shown in Figure 52.

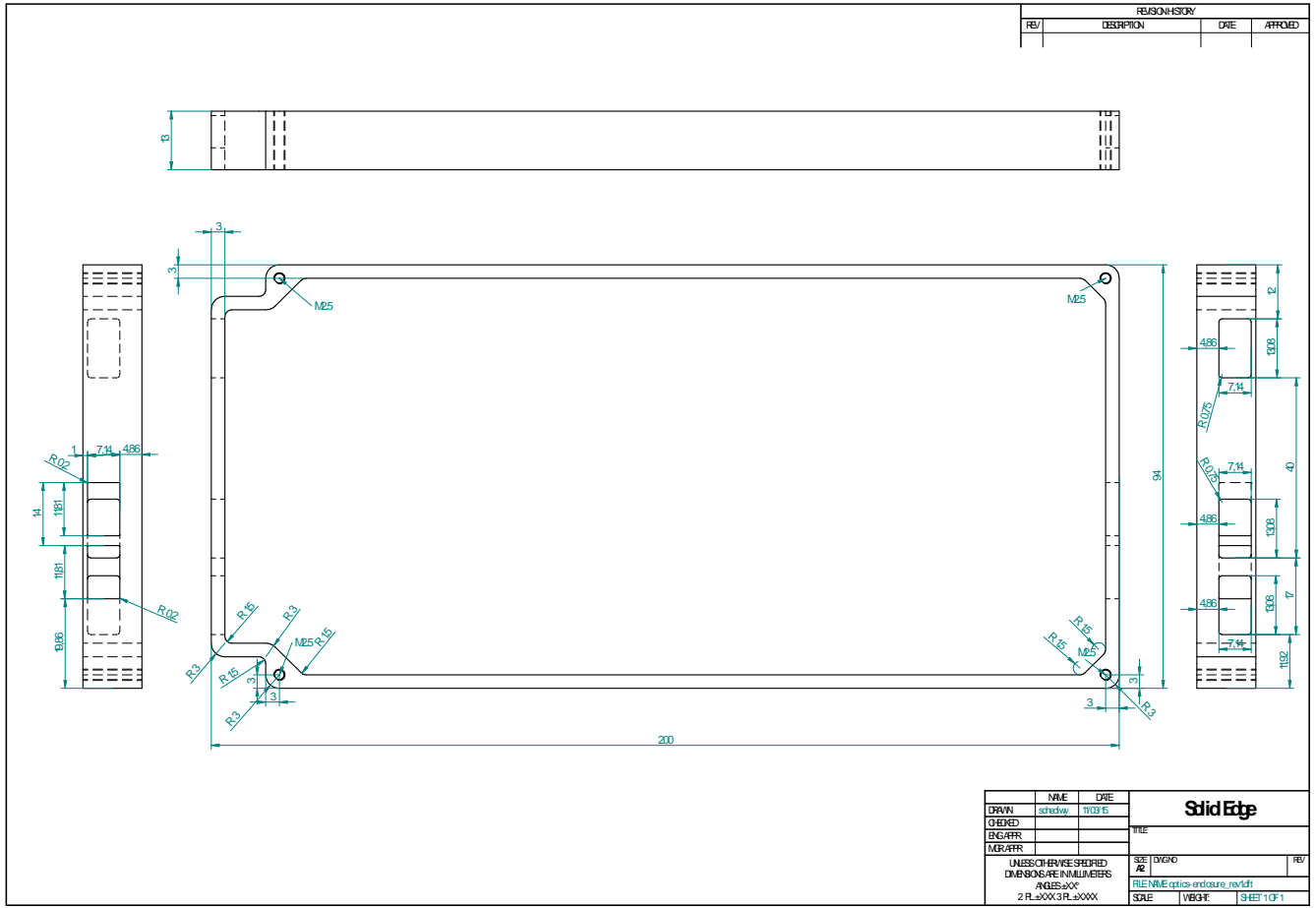

**Figure 52 SKA1-MID TM (341-022100) optics enclosure mechanical detailed design**

The related Solid Edge part is included in the Mechanical Detailed Design File pack in Appendix 8.5.1.

#### 4.4.2.8.3 Transmitter Module (341-022100) Optics Enclosure Lid

The mechanical detailed design for the TM (341-022100) optics enclosure lid is shown in Figure 53.

**Figure 53 SKA1-MID TM (341-022100) optics enclosure lid mechanical detailed design**

The related Solid Edge part is included in the Mechanical Detailed Design File pack in Appendix 8.5.1.

#### 4.4.2.8.4 Transmitter Module (341-022100) MW Enclosure Body

The mechanical detailed design for the TM (341-022100) MW enclosure body is shown in Figure 54.

**Figure 54 SKA1-MID TM (341-022100) MW enclosure body mechanical detailed design**

The related Solid Edge part is included in the Mechanical Detailed Design File pack in Appendix 8.5.1.

#### 4.4.2.8.5  Transmitter Module (341-022100) MW Enclosure Lid

The mechanical detailed design for the TM (341-022100) MW enclosure lid is shown in Figure 55.

**Figure 55 SKA1-MID TM (341-022100) MW enclosure lid mechanical detailed design**

The related Solid Edge part is included in the Mechanical Detailed Design File pack in Appendix 8.5.1.

#### 4.4.2.8.6    Transmitter Module (341-022100) TM PD Standoff

The mechanical detailed design for the TM (341-022100) TM PD standoff is shown in Figure 56.

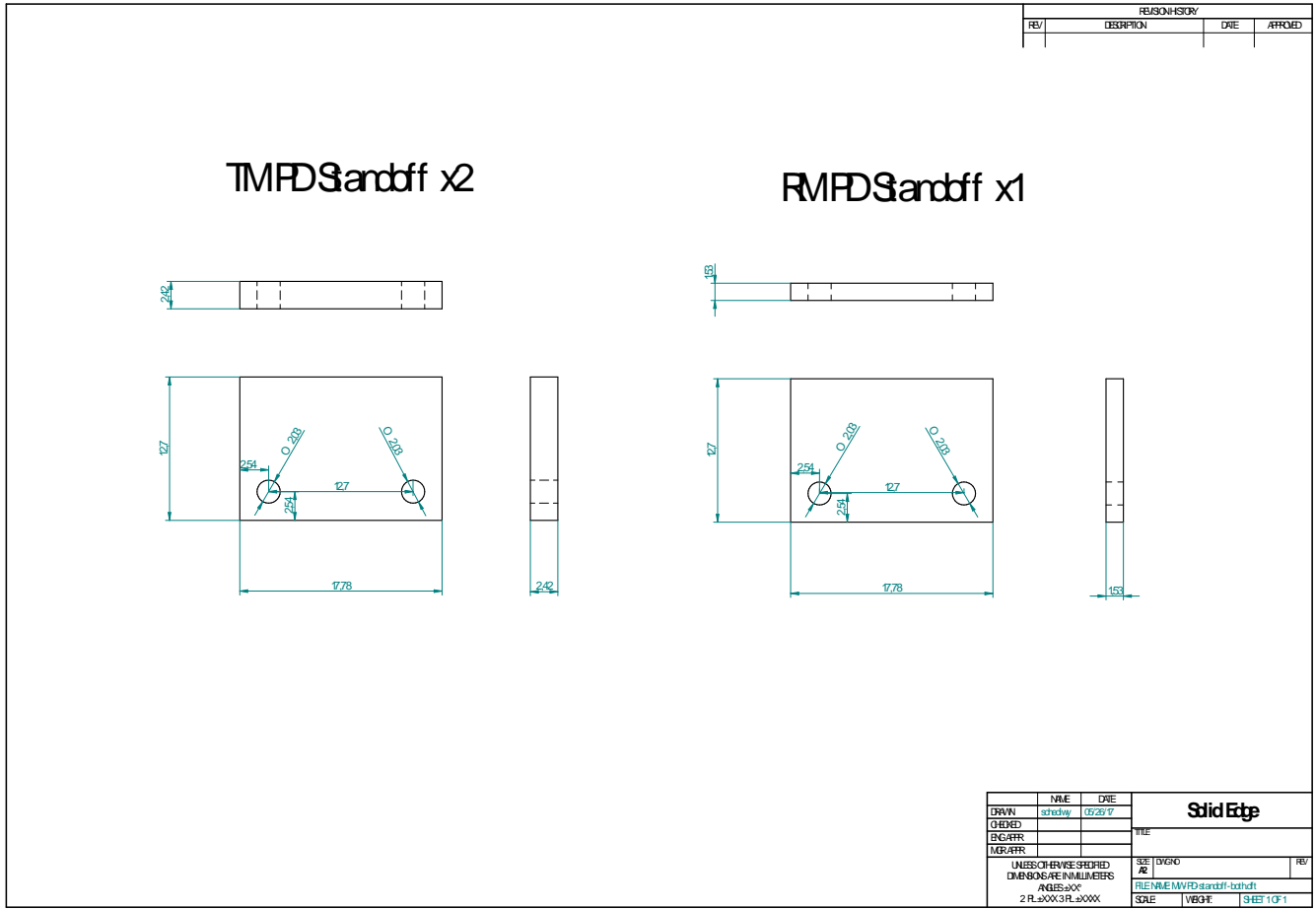

**Figure 56 SKA1-MID TM (341-022100) TM PD standoff mechanical detailed design**

The related Solid Edge part is included in the Mechanical Detailed Design File pack in Appendix 8.5.1.

#### 4.4.2.8.7 Transmitter Module (341-022100) PCB Front Panel

The mechanical detailed design for the TM (341-022100) PCB front panel is shown in Figure 57.

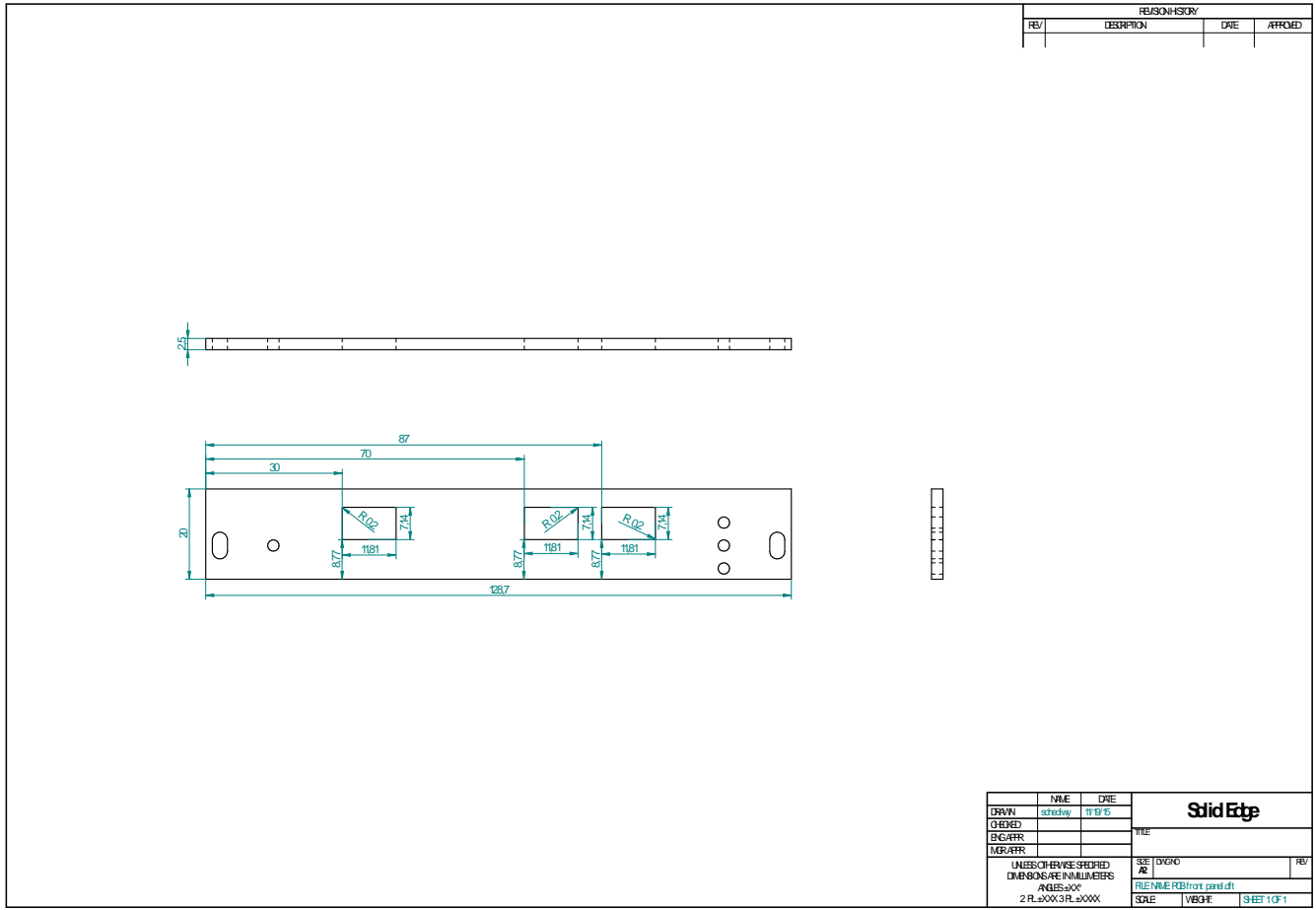

**Figure 57 SKA1-MID TM (341-022100) PCB front panel mechanical detailed design**

The related Solid Edge part is included in the Mechanical Detailed Design File pack in Appendix 8.5.1.

### 4.4.2.9  Fibre Patch Lead (341-023200)

The simplex FP lead joins the TM to a LINFRA patch-panel at the end of relevant rack cabinet row. The TM side has a simplex LC/APC fibre connector. The mechanical/physical attributes of the other side are provided in the LINFRA-SAT.STFR.FRQ Internal Interface Control Document (IICD).

### 4.4.2.10 Optical Amplifier (341-022200)

The mechanical detailed design for the OA (341-022200) is shown in Figure 58.

**FRONT VIEW**

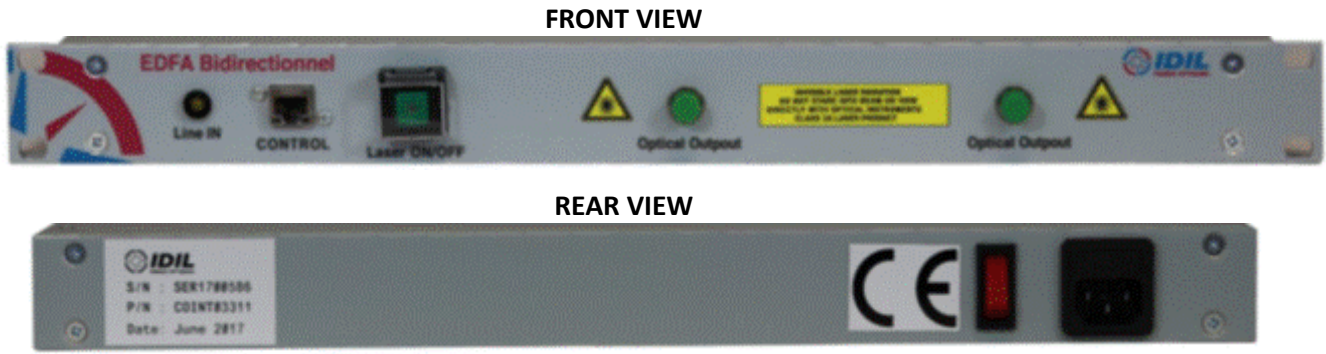

**REAR VIEW**

**Figure 58 SKA1-MID OA (341-022200) mechanical detailed design**

The depth of the OA (341-022200) is 450 mm.

**NOTE 11:** The purchased IEC C14 power socket for the specific model of OA is at the rear of the unit as this is convenient for the current work. The supplier however, has indicated that the design can be modified to include the connector at the front of the unit. This is required for operation in the receptor pedestal.

### 4.4.2.11 Receiver Module (341-022300)

#### 4.4.2.11.1 Receiver Module (341-022300) Enclosure Body

The mechanical detailed design for RM (341-022300) enclosure body is shown in Figure 59.

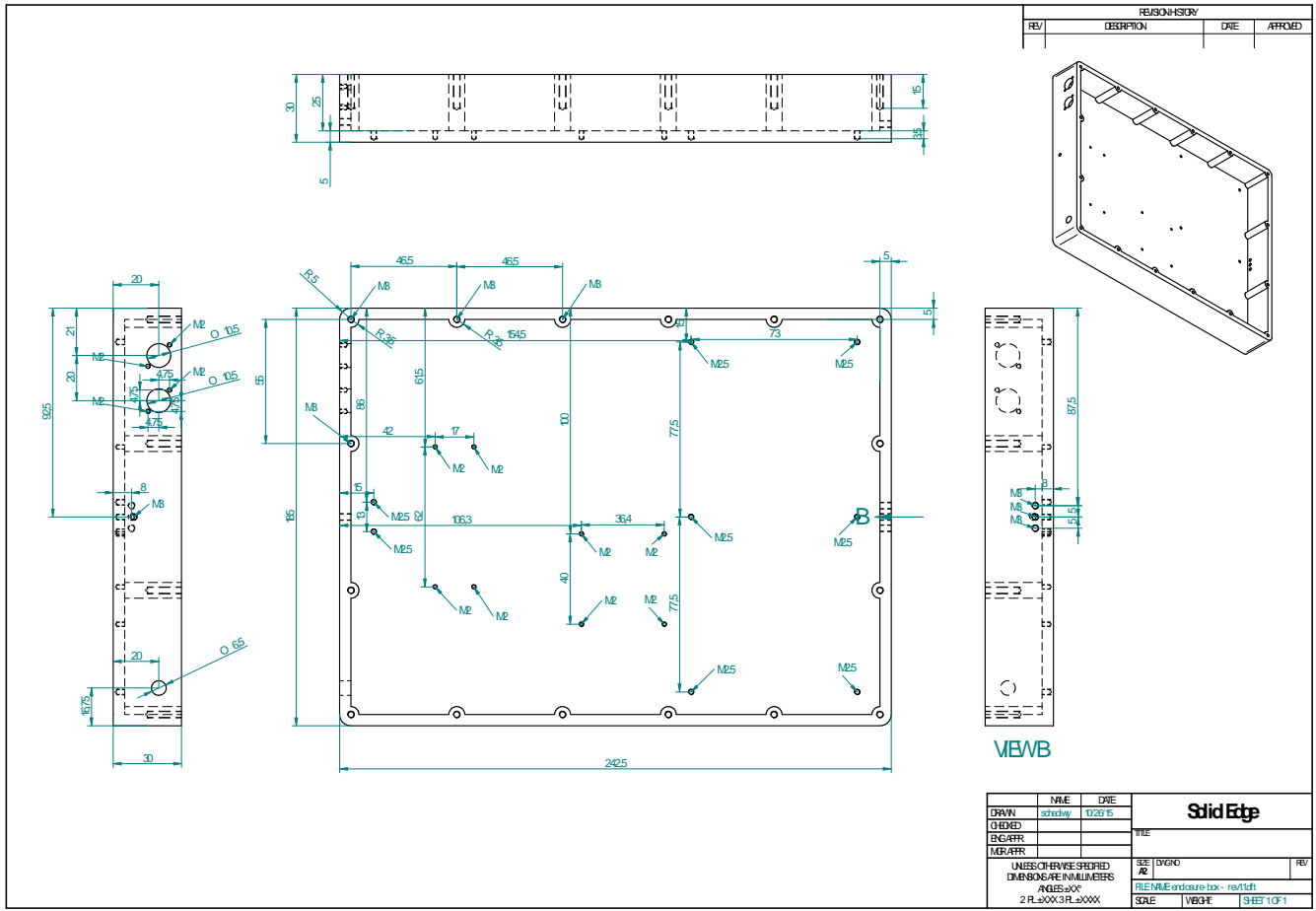

**Figure 59 SKA1-MID RM (341-022300) enclosure body mechanical detailed design**

The related Solid Edge part is included in the Mechanical Detailed Design File pack in Appendix 8.5.1.

The RM, and a DC power supply, is housed inside a 1U, 19" rackmount enclosure. The power socket is at the front as is required by [RD2] for equipment located at the receptor pedestal.

Alternatively, the RM optics and electronics can also be mounted inside a MeerKAT 'time and frequency' module (as shown in Figure 46) and therefore be physically compatible with the ADC receiver enclosure mounted on the indexer of the MeerKAT receptor. This was done in preparation for the anticipated ECP with the DISH consortium, to move the RM to the SKA1-MID receptor indexer. For more information refer to §6.2.2.

### 4.4.2.11.2 Receiver Module (341-022300) Enclosure Lid

The mechanical detailed design for the RM (341-022300) enclosure lid is shown in Figure 60.

**Figure 60 SKA1-MID RM (341-022300) enclosure lid mechanical detailed design**

The related Solid Edge part is included in the Mechanical Detailed Design File pack in Appendix 8.5.1.

### 4.4.2.11.3 Receiver Module (341-022300) PD Standoff

The mechanical detailed design for the RM (341-022300) PD standoff is shown in Figure 61.

**Figure 61 SKA1-MID RM (341-022300) PD standoff mechanical detailed design**

The related Solid Edge part is included in the Mechanical Detailed Design File pack in Appendix 8.5.1.

### 4.4.3 Optical Detailed Design

#### 4.4.3.1 Rack Cabinet (341-022900)

Not applicable.

#### 4.4.3.2 Optical Source (341-022400)

The optical detailed design for the OS (341-022400) is shown in Figure 62.

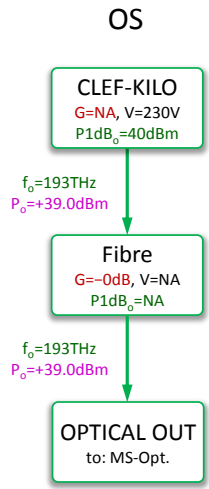

**Figure 62 SKA1-MID OS (341-022400) optical detailed design**

The related PowerPoint design is included in the Optical Detailed Design File Pack in Appendix 8.6.1.

#### 4.4.3.3 Frequency Source (341-022500)

Not applicable.

### 4.4.3.4 Microwave Shift (341-022600)

The optical detailed design for the MS (341-022600) is shown in Figure 63.

**MS-Opt.**

```
OPTICAL IN
from: OS
   │  f_o=193THz
   │  P_o=+39.0dBm
   ▼
E2000
G=−0.2dB, V=NA
P1dB_o=NA
   │  f_o=193THz
   │  P_o=+38.8dBm
   ▼
PN1550R1A1
G=−0.3/−20.3dB
P1dB_o=37dBm
   ├────────────────────────────┐
   │ f_o=193THz                 │ f_o=193THz
   │ P_o=+38.5dBm               │ P_o=+18.5dBm
   │                            ▼
   │                         PN1550R1A1
   │                         G=−20.3/−0.3dB
   │                         P1dB_o=37dBm
   │                            ├──────────────┐
   │                            │              │
   │              f_o=193THz    │  f_o=193THz  │         MW IN
   │              P_o=−1.8dBm   │  P_o=+18.2dBm│       from: MS-Elect.
   │                            ▼              ▼         f_o=7.96GHz
   │                         T-M040-PM      MXIQ-LN      P_o=+23dBm
   │                         G=−2.0dB,      G=−7dB, V=NA
   │                         Δf=+40         P1dB_o=20dBm  DC IN
   │                         P1dB_e=+23dBm                from: MS-Elect.
   │                                                      MW IN
   │                                                      from: MS-Elect.
   │                                                      f_o=7.96GHz
   │                                                      P_o=+23dBm
   │                            │              │
   │              f_o=193THz+40MHz  f_o=193THz−7.96GHz  f_o=193THz−7.96GHz
   │              P_o=−3.8dBm       P_o=+11.2dBm        P_o=+23dBm
   │                            │              ▼
   │                            │           PN1550R2A1
   │                            │           G=−10.3/−0.3dB
   │                            │           P1dB_o=NA
   │                            │              ├──────────────┐
   │                            │  f_o=193THz−7.96GHz   f_o=193THz−7.96GHz
   │                            │  P_o=+0.9dBm          P_o=+10.9dBm
   │                            ▼              ▼
   │                         PN1550R5A1       EDFA
   │                         G=−3.2dB, V=NA   G=+25.6dB, V=230V
   │                         P1dB_o=37dBm     P1dB_o=NA
   │                            ├──────┐         │
   │       f_o=193THz+40MHz  f_o=193THz−7.96GHz  f_o=193THz−7.96GHz
   │       P_o=−7.1dBm       P_o=−2.4dBm         P_o=+36.5dBm
   │                            │              ▼
   ▼                            ▼           PN1550R5A1
PN1550R5A1                  OPTICAL OUT     G=−3.3dB, V=NA
G=−3.3dB, V=NA              to: MS-Elect.   P1dB_o=37dBm
P1dB_o=37dBm
```

(branches to four OPTICAL OUT to: RD-Opt. via E2000 stages)

| | f_o=193THz | f_o=193THz | f_o=193THz−7.96GHz | f_o=193THz−7.96GHz |
| P_o=+35.2dBm | P_o=+35.2dBm | P_o=+33.2dBm | P_o=+33.2dBm |

E2000 (×4): G=−0.2dB, V=NA, P1dB_o=NA

Outputs: P_o=+35.0dBm, +35.0dBm, +33.0dBm, +33.0dBm → OPTICAL OUT to: RD-Opt.

**Figure 63 SKA1-MID MS (341-022600) optical detailed design**

The related PowerPoint design is included in the Optical Detailed Design File Pack in Appendix 8.6.1.

### 4.4.3.5 Signal Generator (341-023100)

Not applicable.

### 4.4.3.6 Rack Distribution (341-022800)

The optical detailed design for the RD (341-022800) is shown in Figure 64.

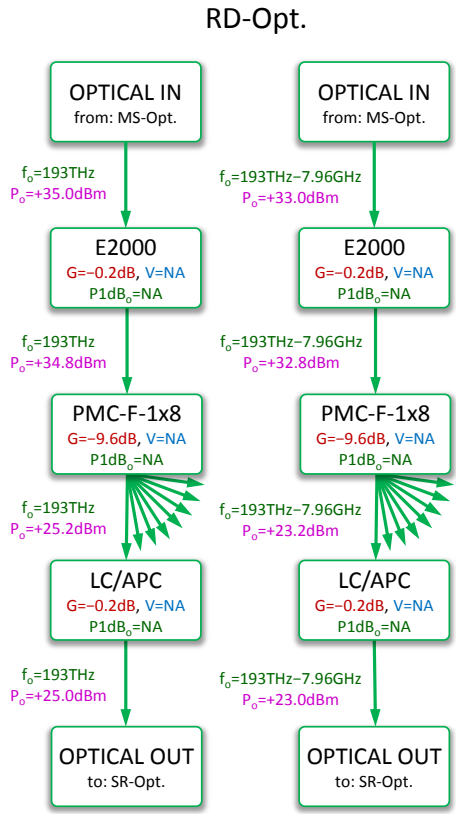

**Figure 64 SKA1-MID RD (341-022800) optical detailed design**

The related PowerPoint design is included in the Optical Detailed Design File Pack in Appendix 8.6.1.

### 4.4.3.7 Sub Rack (341-022700)

The optical detailed design for the SR (341-022700) is shown in Figure 65.

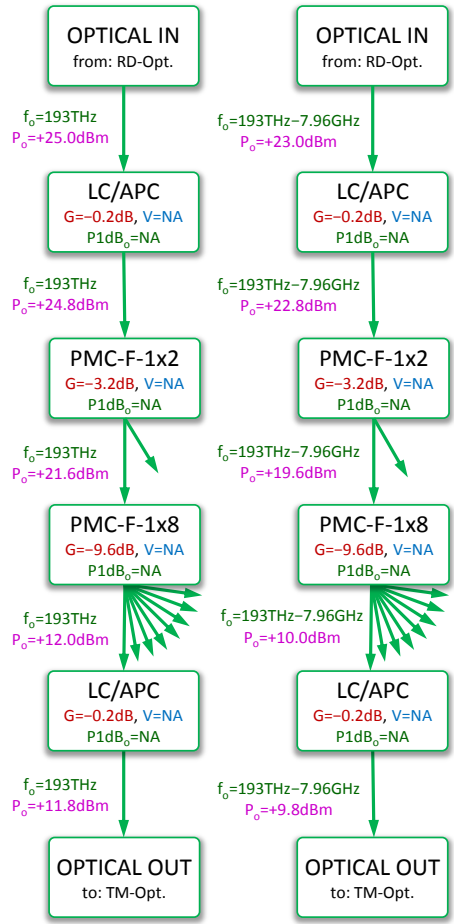

**Figure 65 SKA1-MID SR (341-022700) optical detailed design**

The related PowerPoint design is included in the Optical Detailed Design File Pack in Appendix 8.6.1.

#### 4.4.3.8 Transmitter Module (341-022100)

The optical detailed design for the TM (341-022100) is shown in Figure 66.

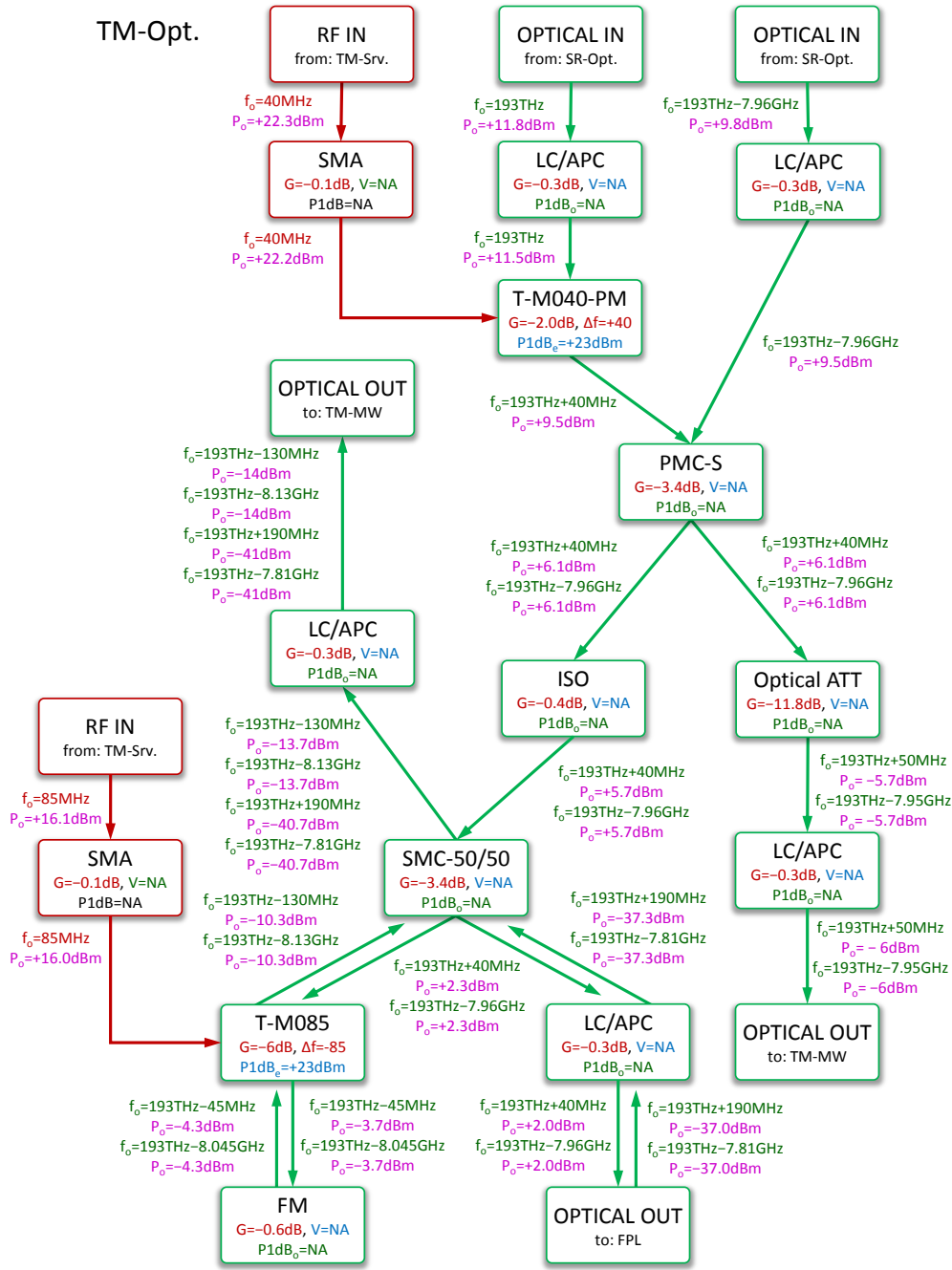

**Figure 66 SKA1-MID TM (341-022100) optical detailed design**

The related PowerPoint design is included in the Optical Detailed Design File Pack in Appendix 8.6.1.

### 4.4.3.9 Fibre Patch Lead (341-023200)

The optical detailed design for the FP lead (341-023200) is shown in Figure 67.

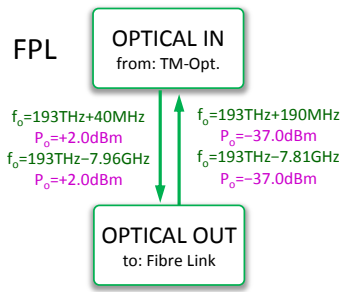

**Figure 67 SKA1-MID FP lead (341-023200) optical detailed design**

### 4.4.3.10 Optical Amplifier (341-022200)

The optical detailed design for the OA (341-022200) is shown in Figure 68.

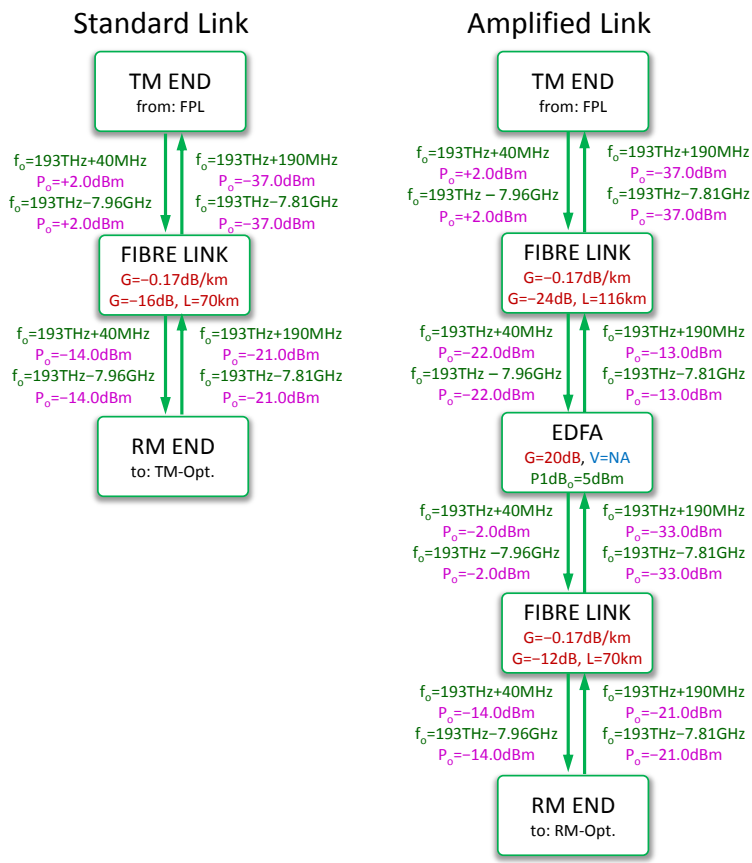

**Figure 68 SKA1-MID OA (341-022200) optical detailed design**

The related PowerPoint design is included in the Optical Detailed Design File Pack in Appendix 8.6.1.

### 4.4.3.11 Receiver Module (341-022300)

The optical detailed design for the RM (341-022300) is shown in Figure 69.

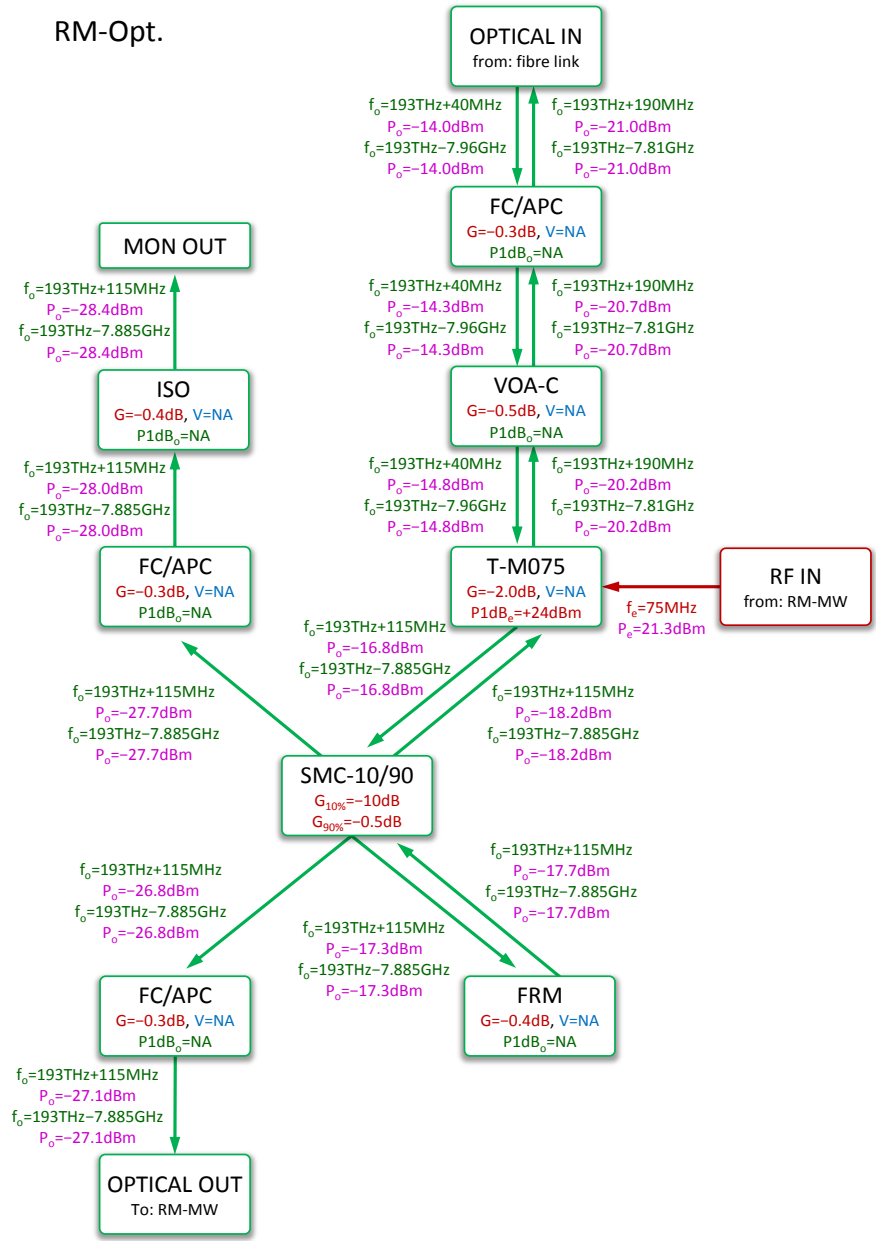

**Figure 69 SKA1-MID RM (341-022300) optical detailed design**

The related PowerPoint design is included in the Optical Detailed Design File Pack in Appendix 8.6.1.

### 4.4.4 Electronic Detailed Design

#### 4.4.4.1 Rack Cabinet (341-022900)

Not applicable.

#### 4.4.4.2 Optical Source (341-022400)

Not applicable.

#### 4.4.4.3 Frequency Source (341-022500)

The electronic detailed design for the FS (341-022500) is shown in Figure 70.

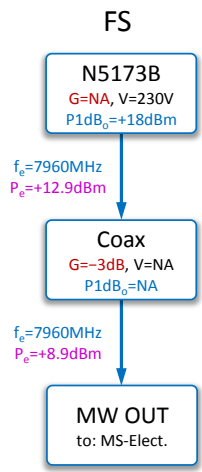

**Figure 70 SKA1-MID FS (341-022500) electronic detailed design**

The related PowerPoint design is included in the Electronic Detailed Design File Pack in Appendix 8.6.1.

#### 4.4.4.4 Microwave Shift (341-022600)

The electronic detailed design for the MS (341-022600) is shown in Figure 71.

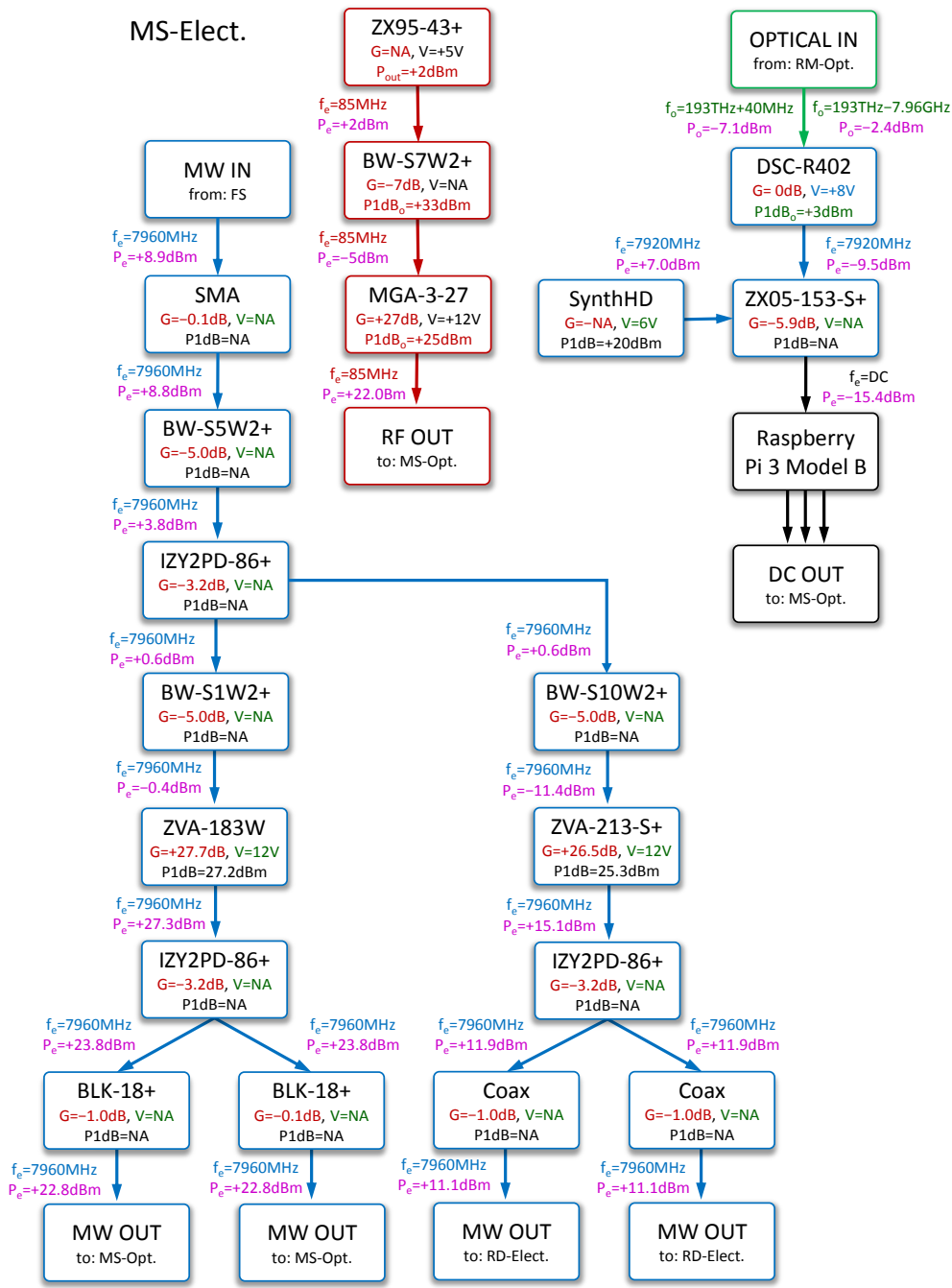

**Figure 71 SKA1-MID MS (341-022600) electronic detailed design**

The related PowerPoint design is included in the Electronic Detailed Design File Pack in Appendix 8.6.1.

### 4.4.4.5 Signal Generator (341-023100)

The electronic detailed design for the SG (341-023100) is shown in Figure 72.

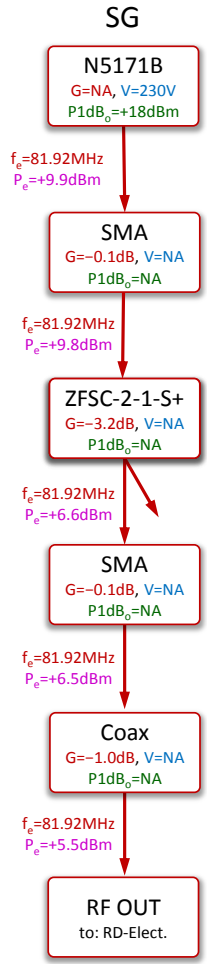

**Figure 72 SKA1-MID SG (341-023100) electronic detailed design**

The related PowerPoint design is included in the Electronic Detailed Design File Pack in Appendix 8.7.1.

#### 4.4.4.6 Rack Distribution (341-022800)

The electronic detailed design for the RD (341-022800) is shown in Figure 73.

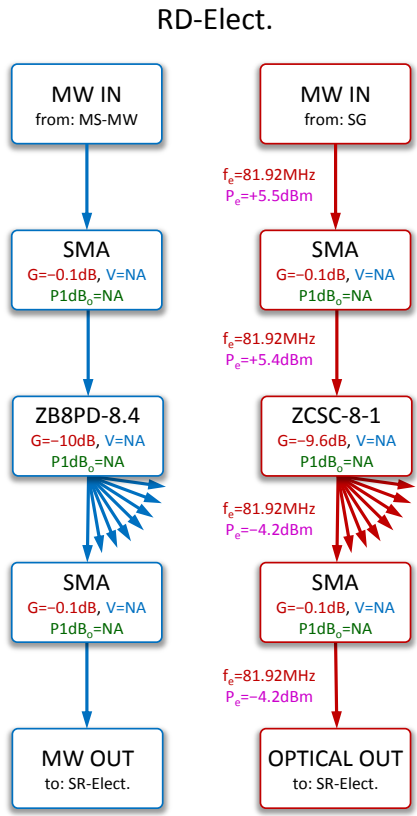

**Figure 73 SKA1-MID RD (341-022800) electronic detailed design**

The related PowerPoint design is included in the Electronic Detailed Design File Pack in Appendix 8.7.1.

### 4.4.4.7 Sub Rack (341-022700)

The electronic detailed design for the SR (341-022700) is shown in Figure 74.

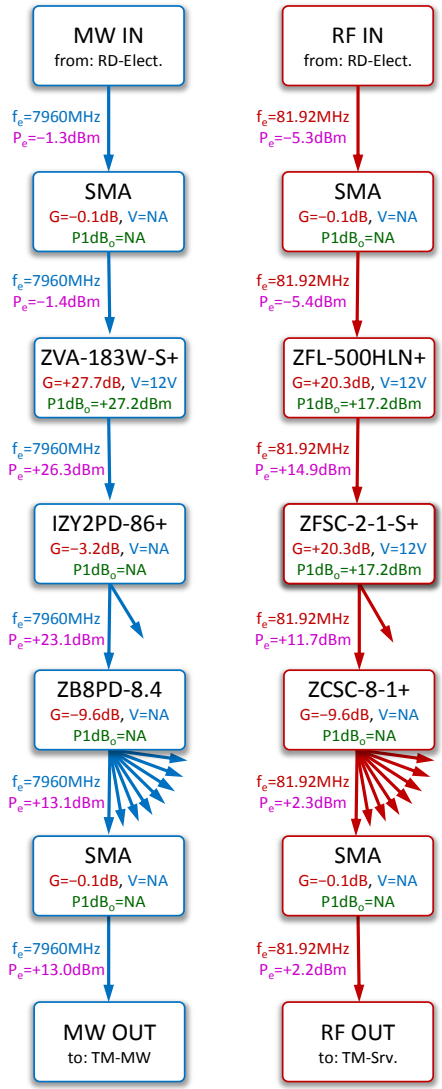

Figure 74 SKA1-MID SR (341-022700) electronic detailed design

The related PowerPoint design is included in the Electronic Detailed Design File Pack in Appendix 8.7.1.

### 4.4.4.8 Transmitter Module (341-022100)

The electronic detailed design for the TM (341-022100) (TM-MW PCB) is shown in Figure 75.

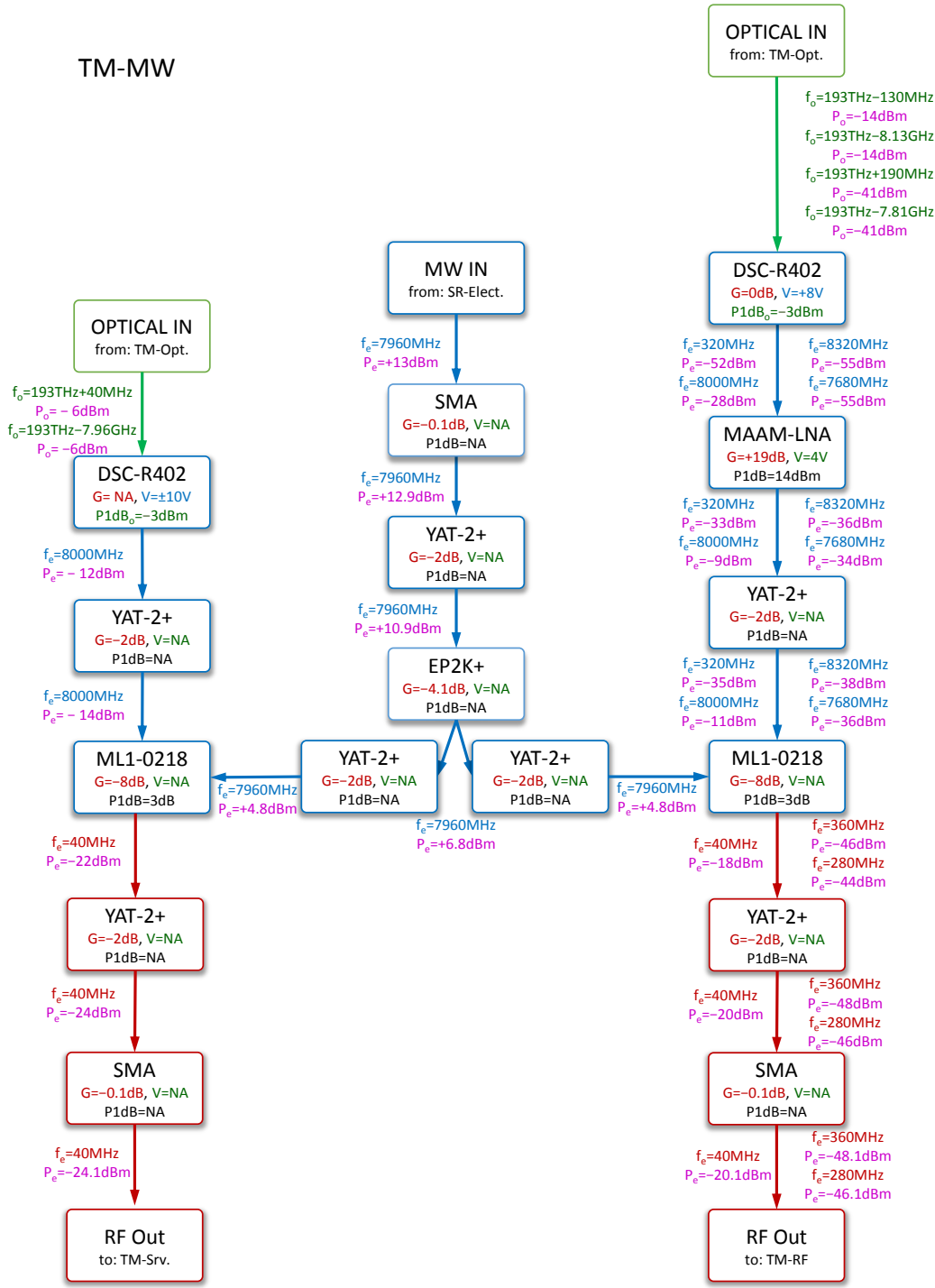

**Figure 75 SKA1-MID TM (341-022100) (TM-MW PCB) electronic detailed design**

The related PowerPoint design is included in the Electronic Detailed Design File Pack in Appendix 8.7.1.

The electronic circuit schematic for the TM (341-022100) (TM-MW PCB) is shown in Figure 76.

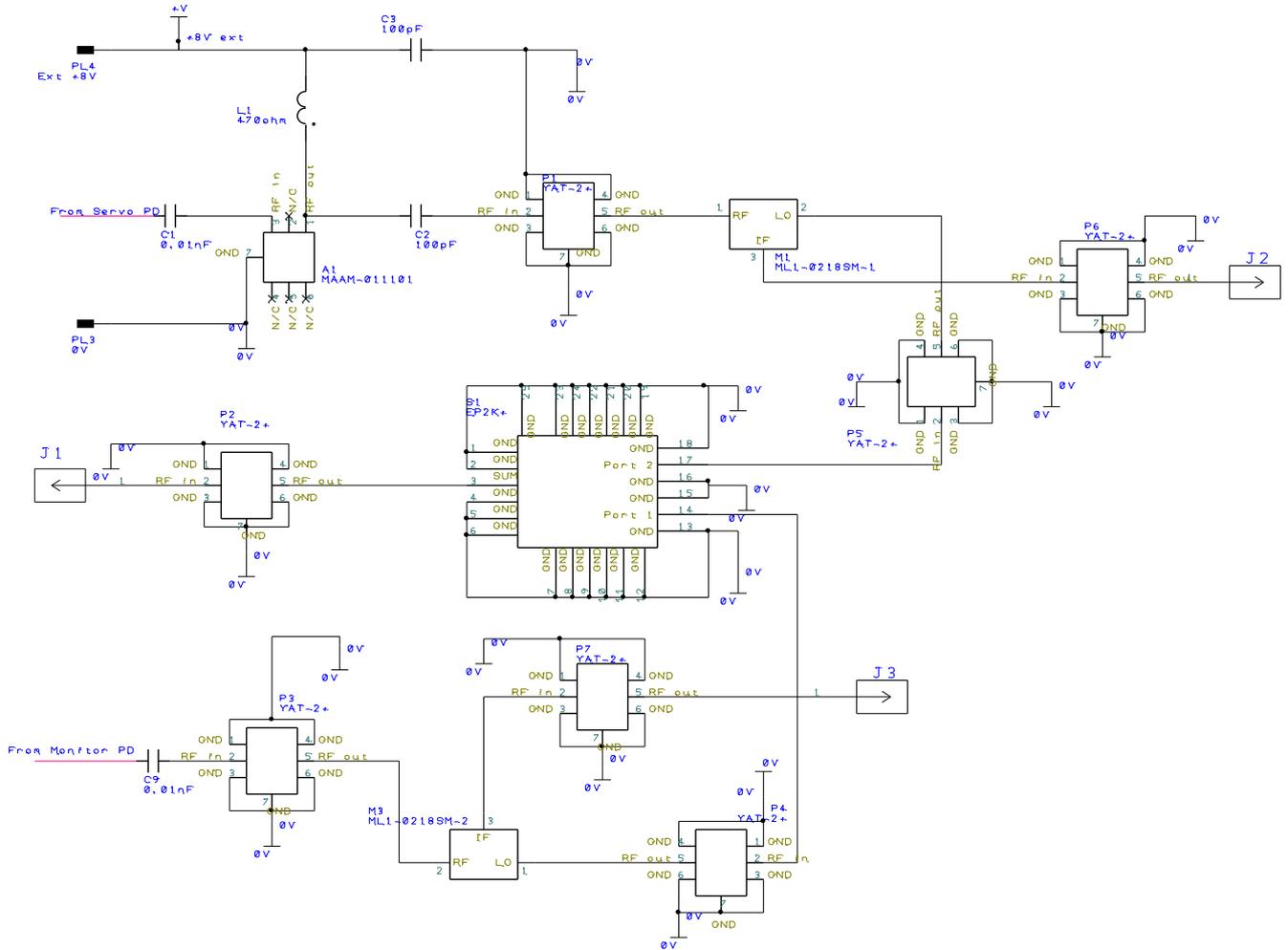

**Figure 76 SKA1-MID TM (341-022100) (TM-MW PCB) electronic circuit schematic**

The related DesignSpark circuit schematic is included in the Electronic Detailed Design File Pack in Appendix 8.7.2.

The PCB layout for the TM (341-022100) (TM-MW PCB) is shown in Figure 77.

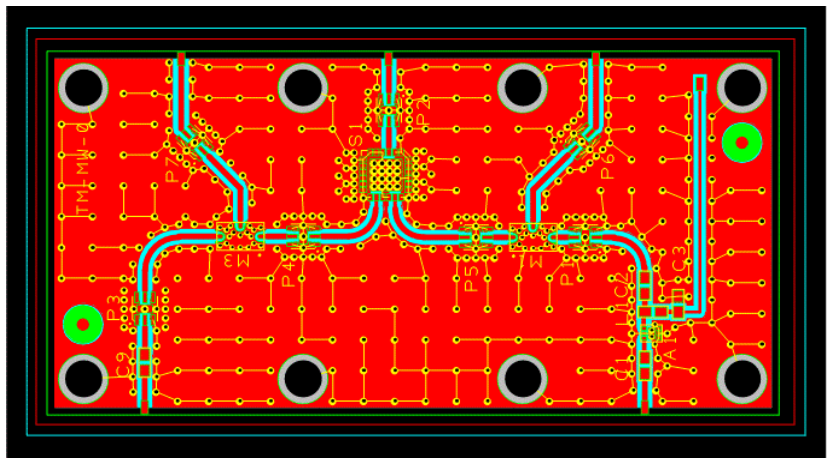

**Figure 77 SKA1-MID TM (341-022100) (TM-MW PCB) PCB layout**

The related DesignSpark PCB layout is included in the Electronic Detailed Design File Pack in Appendix 8.7.2.

The electronic detailed design for the TM (341-022100) (TM-RF PCB) is shown in Figure 78.

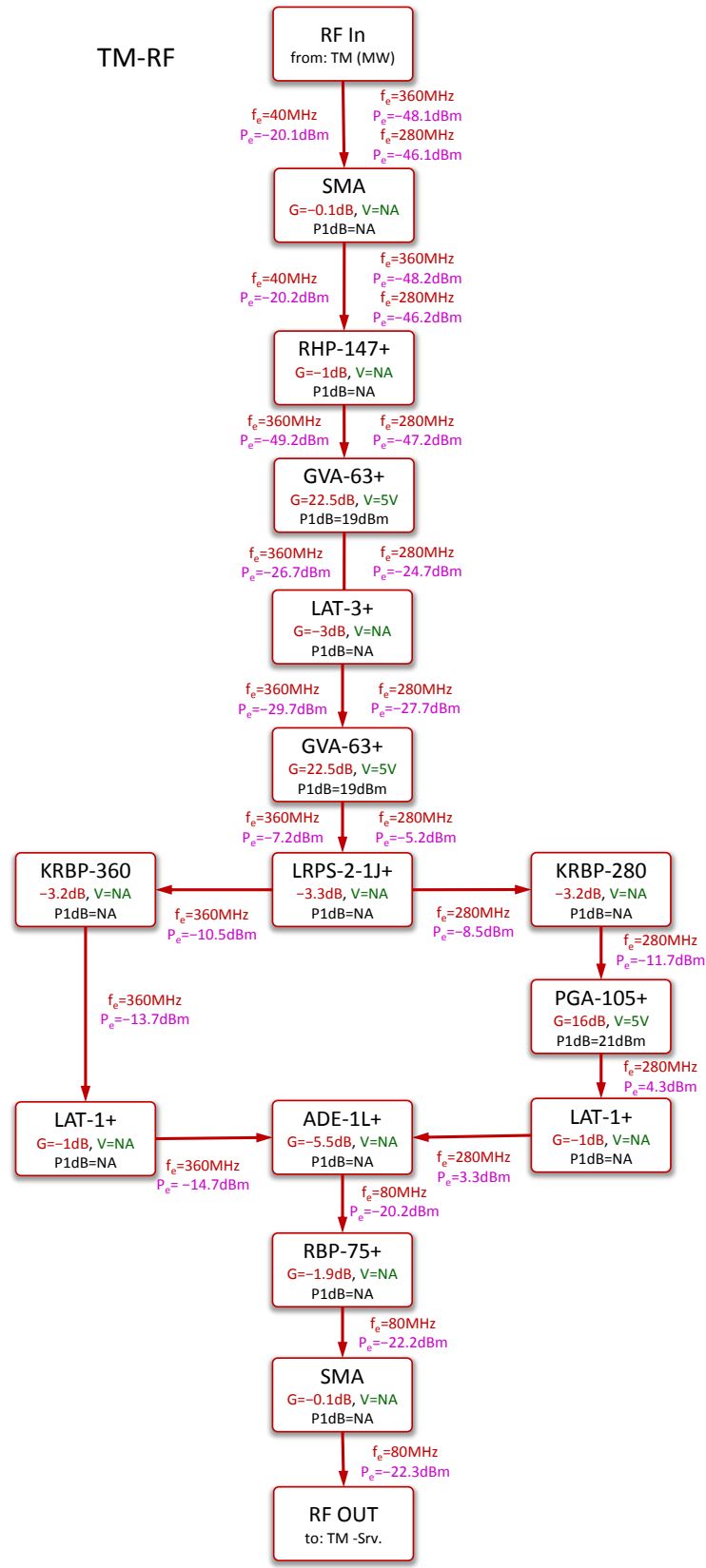

**Figure 78 SKA1-MID TM (341-022100) (TM-RF PCB) electronic detailed design**

The related PowerPoint design is included in the Electronic Detailed Design File Pack in Appendix 8.7.1.

The electronic circuit schematic for the TM (341-022100) (TM-RF PCB) is shown in Figure 79.

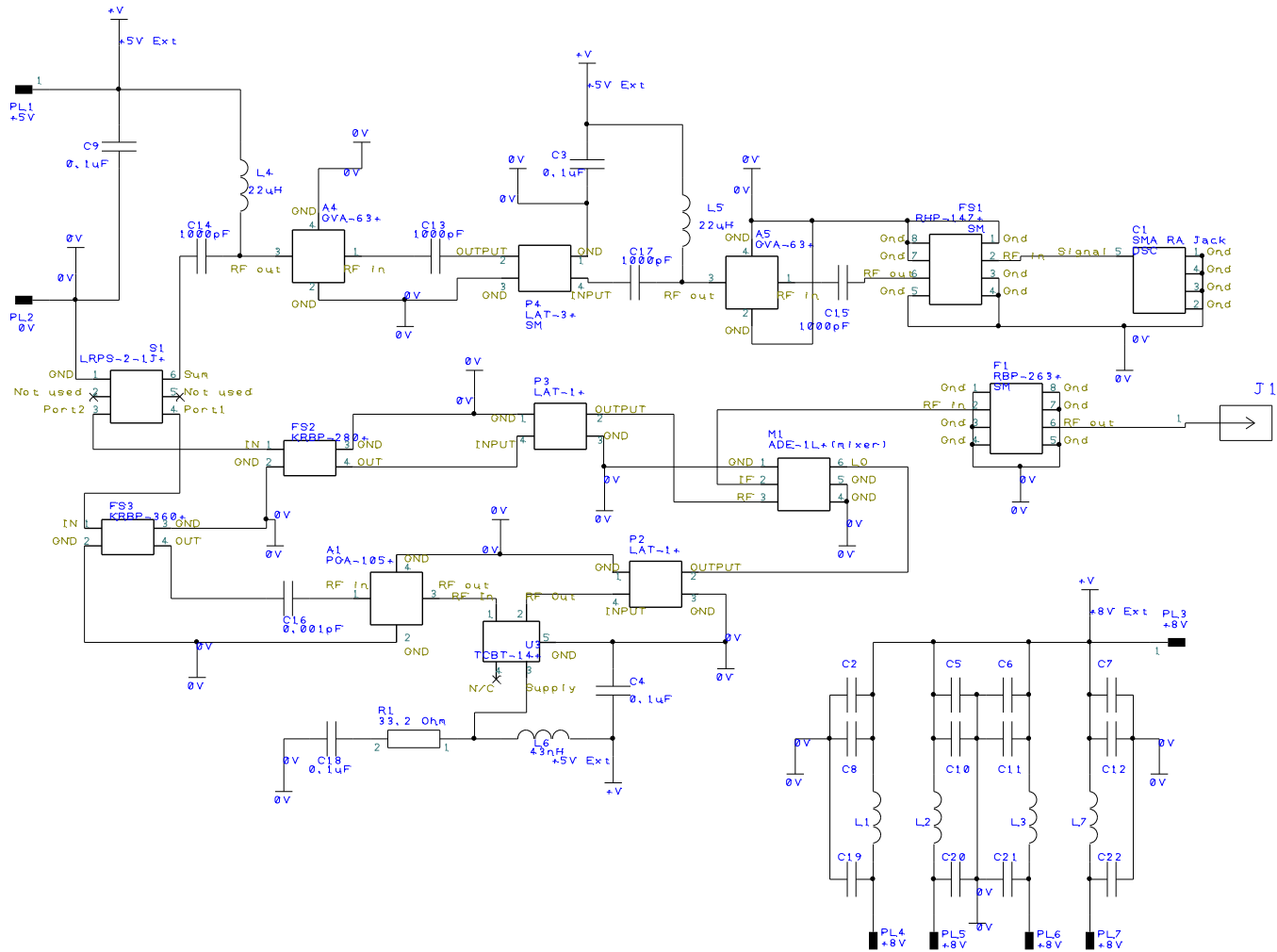

**Figure 79 SKA1-MID TM (341-022100) (TM-RF PCB) electronic circuit schematic**

The related DesignSpark circuit schematic is included in the Electronic Detailed Design File Pack in Appendix 8.7.2.

The electronic detailed design for the TM (341-022100) (TM-Srv. PCB) is shown in Figure 80.

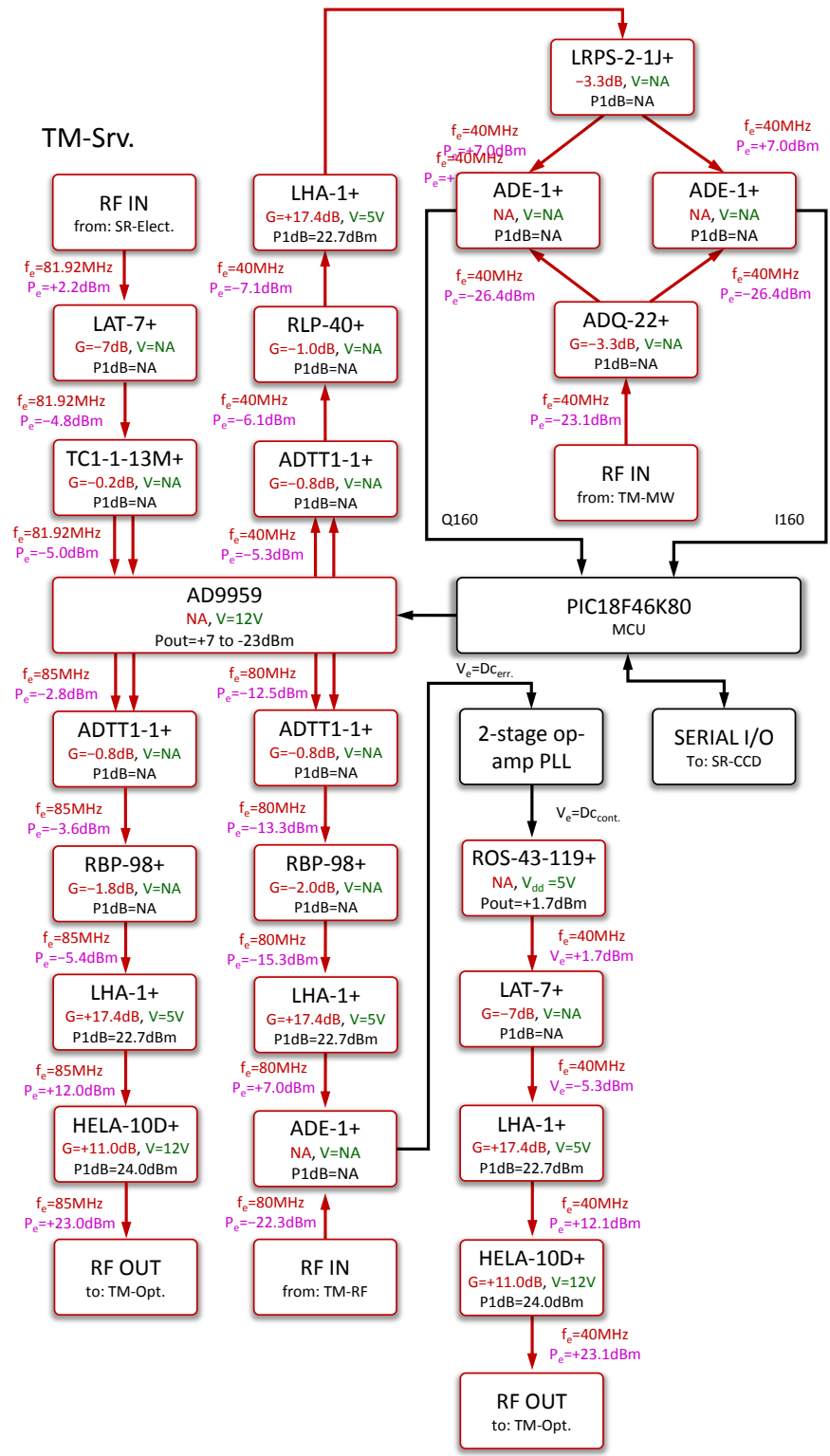

**Figure 80 SKA1-MID TM (341-022100) (TM-Srv. PCB) electronic detailed design**

The related PowerPoint design is included in the Electronic Detailed Design File Pack in Appendix 8.7.1.

The electronic circuit schematic for the TM (341-022100) (TM-Srv. PCB) is shown in Figure 81.

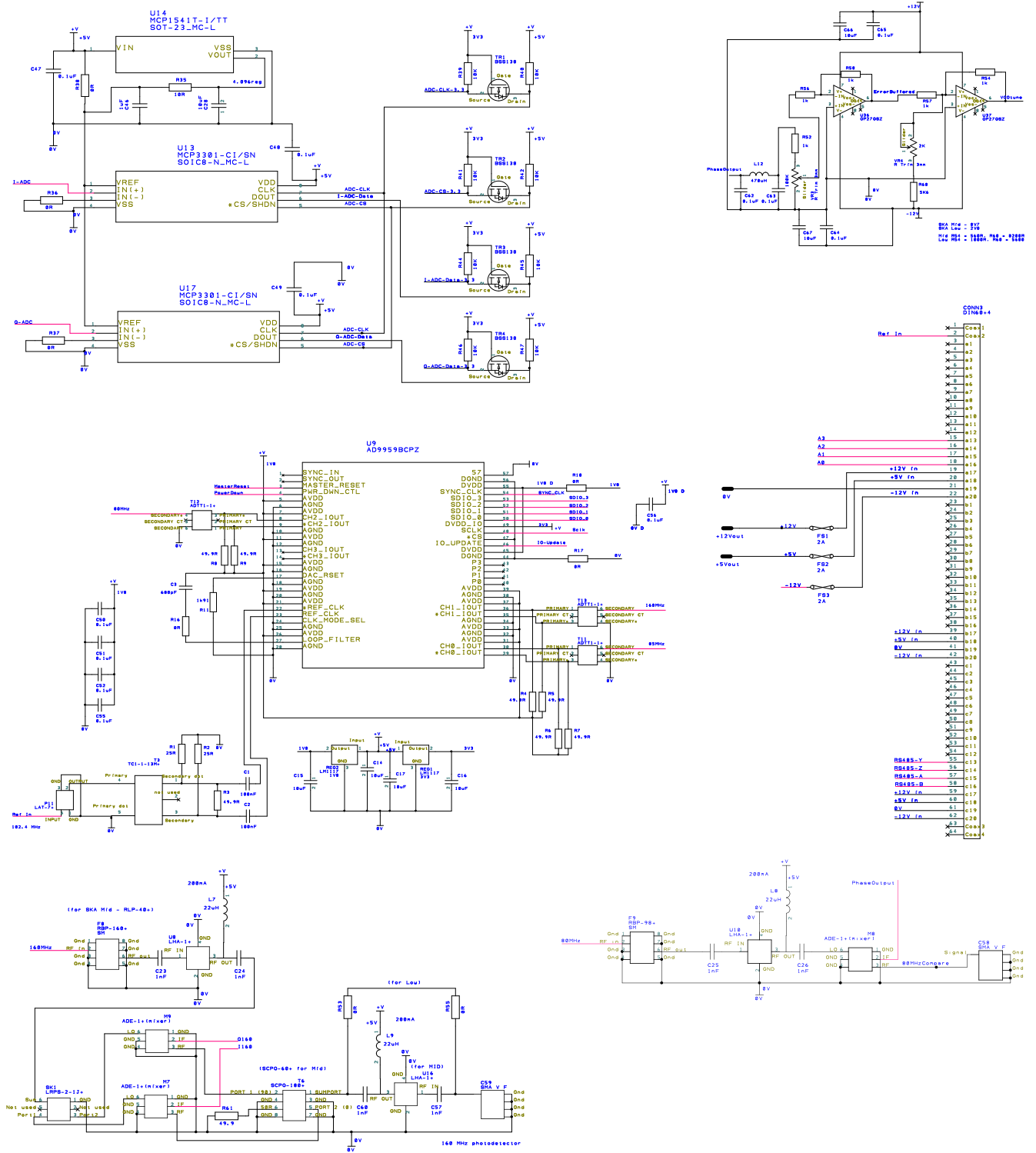

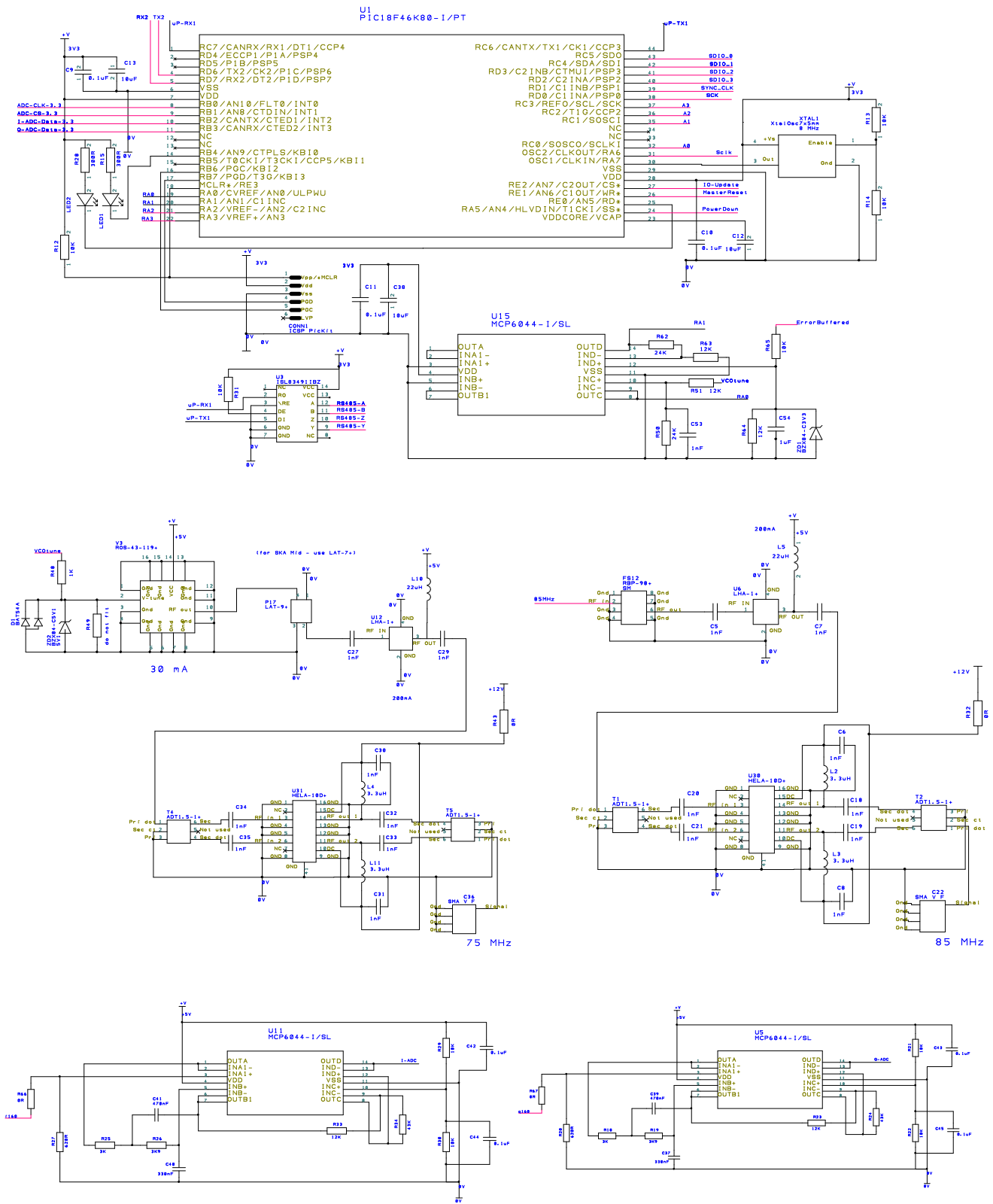

**Figure 81 SKA1-MID TM (341-022100) (TM-Srv. PCB) electronic circuit schematic**

The related DesignSpark circuit schematic is included in the Electronic Detailed Design File Pack in Appendix 8.7.2.

The PCB layout for the TM (341-022100) (TM-MW PCB) is shown in Figure 82.

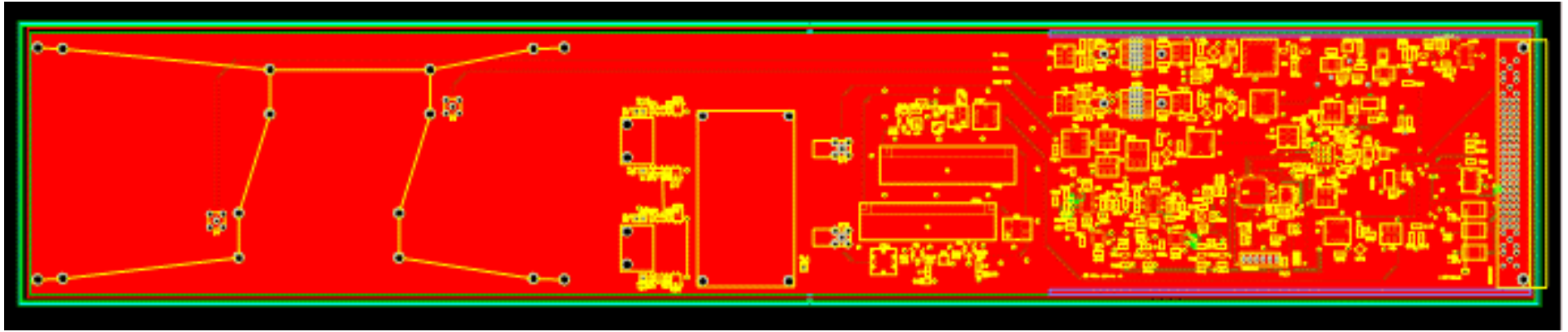

**Figure 82 SKA1-MID TM (341-022100) (TM-RF PCB) PCB layout**

The related DesignSpark PCB layout is included in the Electronic Detailed Design File Pack in Appendix 8.7.2.

### 4.4.4.9 Fibre Patch Lead (341-023200)

Not applicable.

### 4.4.4.10 Optical Amplifier (341-022200)

Not applicable.

#### 4.4.4.11 Receiver Module (341-022300)

The electronic detailed design for the RM (341-022300) is shown in Figure 83.

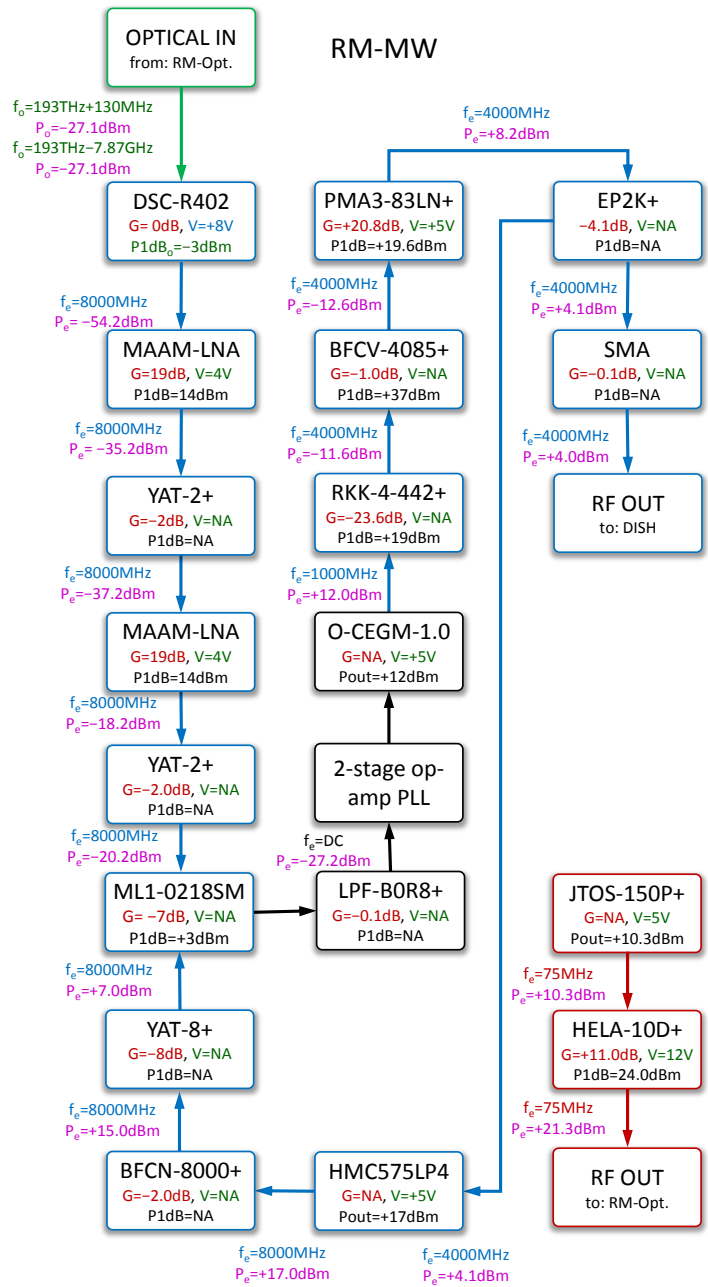

**Figure 83 SKA1-MID RM (341-022300) electronic detailed design**

The related PowerPoint design is included in the Electronic Detailed Design File Pack in Appendix 8.7.1.

The electronic circuit schematic and PCB layout for the RM (341-022300) are a work-in-progress.

While prototype PCBs have already been designed and manufactured as part of the design for mass manufacture work described in §6.2.3, the design details have since changed enough that the aforementioned electronic circuit schematic and PCB layout no longer accurately reflects the current design.

The related DesignSpark circuit schematic and DesignSpark PCB layout however, are still included for the record in the Electronic Detailed Design File Pack in Appendix 8.7.2.

## 4.5 Software

### 4.5.1 System Start-up

The following is the typical start-up procedure for the FRQ (UWA) system.

1. Power up equipment in the following order:

   1. OS (341-022400)
   2. SG (341-023100)
   3. FS (341-022500).

2. Wait a nominated period (a few minutes) for these devices to reach normal operating conditions, then power-up the MS (341-022600).
3. Wait for a nominated period (30 minutes) for the MS to reach thermal equilibrium and acquire a stable SSB-SC modulation state.
4. When complete, power up equipment in the following order:

   1. RD (341-022800)
   2. SRs (341-022700); this will also power-up the TMs (341-022100)
   3. OAs (341-022300)
   4. RMs (341-022300)

The frequency stabilisation servo system automatically engages lock.

5. Run an LMC system start up diagnostic programme that:

   - Confirms the RMS voltage of the servo-loop 'Lock signal' (see §4.3.4.2 for an explanation of this signal) for all Transmitter Modules (TMs) is below the specified threshold voltage.
   - Confirms the accumulated phase change as measured by the 'Phase measurement' signal (see §4.3.4.2 for an explanation of this signal) is within a set typical range.

If both these conditions are met, and no other faults are reported, then the LMC reports to the Telescope Manager that the FRQ (UWA) system is now operational.

### 4.5.2 Normal Operations

**NOTE 12:** The full details of all LMC FRQ interfaces are given in [RD51].

Engage/change/disengage the frequency offset by using the LMC to send a command to the specified TM(s), via the CCD in the host SR to either:

- Apply the new desired offset frequency value to the servo LO reference frequencies produced by the DDS, or
- Reset the servo LO reference frequencies produced by the DDS to the nominal frequency.

The CCD reports the new frequency offset value to SAT.LMC, which then passes this information to the Telescope Manager.

The mass manufacture prototype uses a proxy-SAT.LMC system to engage/change/disengage the frequency offset via the DDS using the process described here. A screen capture of the Graphical User Interface (GUI) for controlling and monitoring the DDS is shown in Figure 84.

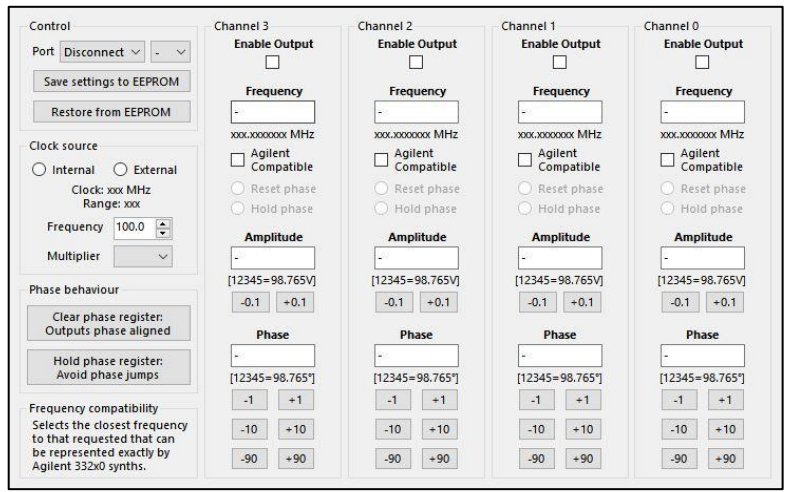

**Figure 84 GUI for controlling and monitoring the DDS**

**NOTE 13:** In a change from the draft DDD submitted in Q3, 2017, the SSB-SC modulation state 'optimisation' routine is now carried out continuously within the bounds of the MS LRU without any input required from SAT.LMC.

### 4.5.3 Off-Normal Operations

Table 21 shows the 'off-normal' operations.

| Scenario | Result | Reason | Diagnostic | Resolution |
|---|---|---|---|---|
| The antenna site power only is cut. | The FRQ (UWA) system no longer functions, no MW-frequency reference signals available at remote site. | The reflected signal is no longer available (the remote AOM does not transmit light if no RF drive signal is provided), the frequency stabilisation on servo-loop cannot lock. | The fault is detected by SAT.LMC monitor of Lock signal. The fault is trigger by SAT.LMC when the RMS voltage of servo-loop error signal the moves above the specified threshold voltage. | After the power is restored, the frequency stabilisation on servo system automatically re-engages. |
| The transmission through the fibre link is cut (break in fibre, or fault in the OA on long links) | The FRQ (UWA) system functions with limited long-term coherence. The clean-up OCXOs in the RM continue to function and provide short-term coherence for the array. | The reflected signal is no longer available, the frequency stabilisation on servo-loop cannot lock. | The fault is detected by LMC monitor of Lock signal (see §4.3.4.2): The fault is trigger by SAT.LMC when the RMS voltage of servo-loop error signal the moves above the specified threshold voltage (in the case of long links, the fault is also trigger by SAT.LMC from the photodetector power monitor in the OA, this will also indicate whether the fault is up-stream or down-stream of OA, or with the OA itself). | After the fibre continuity is restored, the frequency stabilisation on servo system automatically re-engages. |
| The CPF power only is cut. | The FRQ (UWA) system functions with limited long-term coherence. The clean-up OCXOs in the RM continue to function and provide short-term coherence for the array. | There is no power to transmit the reference signals, or frequency stabilisation on servo system. | Fault is detected by SAT.LMC when status communication ceases from any equipment located within the CPF. | After the power is restored, a full system start-up is required (as described in §4.6.1). |

**Table 21 Off-normal operations**

### 4.5.4 System Shutdown

Power down equipment in the following order:

1. RMs (341-022300)
2. OAs (341-022300)
3. SRs (341-022700); this will also power-down the TMs (341-022100)
4. RD (341-022800)
5. MS (341-022600)
6. FS (341-022500)
7. SG (341-023100)
8. OS (341-022400).

## 4.6 Security

At the time of writing there is no SKA wide security policy documentation available. The following list defines the security requirements included in [AD1]:

*SKA1-SYS_REQ-2791 The SKA must provide a security management system that includes:*

- *personnel security*
- *physical security (asset)*
- *security of information*

*SKA1-SYS_REQ-2478 The observatory must provide a secure environment for equipment including protection of generators, fuel, solar cells, equipment spare stores, and inter-station assets such as copper cables.*

*SKA1-SYS_REQ-2482 It must be possible to control on a per user basis which SKA1 facilities and resources (both hardware and software) may be accessed by the user.*

*SKA1-SYS_REQ-2479 Science Data Product security The SKA1_Low and SKA1_Mid telescopes shall provide a secure environment for all Science Data Products preserved long term*

In the absence of a security policy, security risk assessments were conducted for all FRQ assets. A risk assessment was completed per location (CPF, shelter, and pedestal), as the likelihood values and mitigating controls of each threat are different at each location. The FRQ assets were assessed against a number of threats and given a vulnerability score against Confidentiality, Integrity, and Availability (CIA). Each threat was initially assessed with no mitigation in place, which identified the uncontrolled risks. The mitigation fields were then completed, which included any mitigation that will be included as part of the design or rollout. As there is no security policy in place, and therefore no existing mitigating controls defined, the residual risks remained similar to the uncontrolled risks. The uncontrolled and residual risks are colour coded as follows:

- Red = Critical
- Amber = Significant
- Green = Low.

Refer to the following documents for the detailed results of the risk assessments:

- [RD52]
- [RD53]
- [RD54]
- [RD55].

§4.6.1 through §4.6.4 explain how the design meets the SKA security requirements listed above. §4.6.5 includes the results of the security risk assessments, and provides recommendations to include in the SKA security policies and during the procurement process.

### 4.6.1 Requirement SKA1-SYS_REQ-2791

It is understood that the CPF data room will have restricted access controls and CCTV cameras in place. It is also understood that the shelters and pedestals will be locked, but any other physical security controls are unknown. It is expected that physical and personnel security controls will be included within SKA Hosting Agreements, and is out of scope for this design document.

This FRQ design document only addresses the security of information. Information in the FRQ context is defined as:

- Management data, and access to and from the LMC;
- Configuration data held on the FRQ assets.

It is expected that physical and logical access to the control networks, where FRQ assets will reside, will be restricted and controlled, and that firewalls will be positioned between the SKA networks and the internet. These security controls are out of scope for this design document and are considered Enterprise functions.

The SKA organisation has adopted the Tango Framework for transmitting monitoring and control data over the network. This software has limited security controls and is out of scope for this work package.

Access control features of COTS equipment is vendor specific, and any requirements for authentication should be included in the procurement process.

Due to the limited security controls that can be put in place for the FRQ equipment, the overall SADT design places this equipment on its own VLAN. It is recommended that access to this VLAN is restricted with access control lists that only allow the required traffic to traverse between networks.

### 4.6.2 Requirement SKA1-SYS_REQ-2478

This requirement is not applicable to the FRQ assets.

### 4.6.3 Requirement SKA1-SYS_REQ-2482

The FRQ solution is made up of a number of components. The bespoke equipment does not have any authentication functionality, and is unable to meet this requirement. Authentication features for COTS equipment is vendor specific, and any authentication requirements would need to be included as part of the procurement process. Security considerations for each component are defined below:

- The SR (CIN 341-022700), containing the CCD Module, does not have any authentication features and can be accessed and modified by anyone who has access to the network that this equipment resides on. Any unauthorised modification of each SR would affect up to 16 receptors. This equipment will be located in the CPF data room and will be protected by the physical security in place.

- The RM (CIN 341-022300) does not have network connectivity and can only be modified by someone with physical access to this asset. This equipment will be located in a pedestal and will be protected by the physical security in place.

- The following assets are COTS equipment:

    - FS (CIN 341-022500)
    - OS (CIN 341-022400)
    - MS (CIN 341-022600)

- SG (CIN 341-023100)

The inclusion of authentication features in the COTS equipment will be vendor dependent. As this design document must be vendor agnostic, requirements for any authentication features should be included as part of the procurement process. All of these assets will reside in the CPF data room and will be protected by the physical security in place. All of these devices will have network connectivity. Any unauthorised modification of these assets could potentially affect the entire telescope.

The OAs will reside in three shelters and five pedestals, and will be protected by the physical security in place at these locations. The inclusion of authentication features will be vendor dependent. As this design document must be vendor agnostic, requirements for any authentication features should be included as part of the procurement process. All of these devices will have network connectivity. Any unauthorised modification of these assets would affect the receptor that is attached to it.

In order to reduce the risk of unauthorised access or modification of these assets in the absence of any authentication features, these assets should be placed on their own VLAN. See [RD56] for further information.

This VLAN should have its access restricted to authorised and trained personnel only. Access control lists on routers/firewalls should restrict network access to this VLAN by only allowing communication between specific devices and specific ports. Remote access is not required, and these assets do not require any internet access.

### 4.6.4 Requirement SKA1-SYS_REQ-2479

The FRQ assets do not store any science data. The COTS equipment may store configuration and log information. This information may be protected if the COTS equipment includes authentication functionality that requires a username and password to be entered in order to gain access to the device. This functionality may be vendor dependent and should be included as part of the requirements during the procurement process. The SAT.LMC server will make provision for received CCD Module information to be stored. Refer to [RD57] for further details of how this information is stored.

### 4.6.5 Security Recommendations

The security risk assessment highlighted three threats that the FRQ assets were susceptible to prior to implementing any security controls (uncontrolled risks). Once the mitigating controls had been completed, only one residual risk remained. Table 22 lists the three threats that were identified, the mitigating controls implemented as part of the design, the residual risk status, plus further recommendations for risk reduction.

| Threat | Initial Severity | Mitigation included in the Design | Residual Risk | Further Recommendations |
|---|---|---|---|---|
| Unauthorised physical access/tampering | Significant | Equipment will reside in either a data room with access control and CCTV, a locked shelter, or a locked pedestal. | Low | Configure the COTS equipment to require a username and password to access the device. Ensure there are disciplinary procedures in place. Ensure all visitors are escorted and not left unattended (this process may already be in place). |
| Maintenance error/operator error/ misconfiguration | Significant | None | Significant | Only allow trained operators to maintain equipment. Have documented procedures in place for regular maintenance activities. Configure the COTS equipment to require a |

| Threat | Initial Severity | Mitigation included in the Design | Residual Risk | Further Recommendations |
|---|---|---|---|---|
| | | | | username and password to access the device.<br>Implement a change control procedure with roll back options.<br>Back up or document the configuration prior to making changes.<br>Test the changes once implemented to ensure the device still functions correctly. |
| Communications infiltration | Significant | The equipment will not have internet access inbound or outbound and will reside on its own VLAN.<br>It is expected that firewalls will be in place at the perimeter and access to the control networks will be restricted.<br>The equipment will be located in either a shielded data room that has access control and CCTV, a locked shelter, or a locked pedestal.<br>The equipment does not contain any sensitive information. | Low | Apply any vendor supplied patches in a timely manner.<br>Restrict access to the VLAN using access control lists and only allow the required communication.<br>Restrict access to the equipment to trained and authorised personnel only.<br>Where possible, configure the equipment to require a username and password to gain access to the device.<br>Restrict the use of removable storage.<br>Implement an IDS\IPS to monitor for unusual traffic. |

**Table 22 SKA1-MID security recommendations**

Other security recommendations that were identified during the risk assessment were:

- Back up the configuration, so that it can be applied to a replacement device in the event of a device failure, equipment misconfiguration, or unauthorised change.

- Purchase COTS equipment that includes authentication and allows usernames and passwords to be configured in order to access the device (either local authentication or via Authentication, Authorisation, and Accounting (AAA) servers).

- Ensure that any usernames and passwords are not transmitted or stored in clear text.

- Ensure that default usernames and passwords are not used.

- Ensure the vendors of the COTS equipment regularly distribute security patches, and implement a patch management policy.

- Only allow trained individuals to maintain/operate the equipment.

- Ensure all cables are clearly labelled.

- Implement a change control procedure.

- Implement an asset disposal policy.

- Implement an incident response procedure.

- Implement a disaster recovery plan.

- Ensure all FRQ equipment is contained within an asset register.

## 4.7 Safety

### 4.7.1 Hazardous Items

Applicable to:

- OS (341-022400)
- MS (341-022600).

The 1552 nm infrared fibre laser that comprises an OS has output power of +40 dBm (10 W). According to the definitions of the IEC 60825-1 standard, this laser is rated as Class 1, given that the laser light is entirely contained within optical fibre. However, should a fibre be accidently disconnected or broken, the laser becomes Class 4. As shown in Figure 20, this light is then transferred directly to the MS. Its four outputs, are between +33 and +35 dBm (see Figure 63), are then routed to the RD. At the output of the RD the power is below +25 dBm (0.32 W), which is below the 0.5 W, Class 3B limit and therefore considered eye safe (as Class 3R).

The optical power being sent down the fibre links is below 10 mW and is therefore Class 1.

### 4.7.2 Training

Applicable to:

- RM (341-022300)
- OS (341-022400)*
- MS (341-022600)*.

* Due to the high optical power within the OS and MS, certified laser training is required for any personnel working or interacting with the optical fibre coming out of either of these units.

The SKA phase synchronisation system has been designed such that the commissioning process requires the optimisation of only one free parameter per link; no other adjustments or optimisations are required across the entire system. The optimisation involves making the total power loss of all optical fibre links (regardless of length) equal. This ensures that the optical power levels all TMs and RMs are equal across the array. This means that resulting electronic signal levels are like-wise optimised and equal between all systems. The commissioning personnel will require training to conduct this optimisation process.

The process is very simple, takes less than a minute to execute, and involves only a simple optical power meter. After installation of all necessary equipment on a particular link, the optical power meter is temporarily connected to the 'monitor port' of the RM (see Figure 69). A screwdriver is then used to adjust the trim of a variable optical attenuator which is accessed via recess on the front of the RM (see Figure 59) until the power meter indicates the nominal optical power value.

## 4.8 Integration

### 4.8.1 Component and System Integration

UWA researchers in partnership with MeerKAT and UoM electronic engineers, progressed the detailed designs described in this document, into a set of mass manufacture archetypes, effectively getting a head-start at addressing manufacturing issues that may be encounter by contractors during SKA construction. This includes construction-ready sets of optical, electronic, and mechanical design files, and a Bill of Materials (BOM) based on components purchased from vendors to construct these archetypes. This forms the basis of the UWA Detailed Cost Model (§4.10). The first set of mass manufacture archetypes for SKA1-LOW were completed in Q2, 2016 [RD21]; and for SKA1-MID in Q1, 2017 [RD22]. The assembly and testing of the completed archetypes was performed at UWA. To get an independent evaluation of labour costs associated with assembly and testing, a local optical manufacturing technology consultancy company was employed to provide an independent review of the costings (see Appendix 8.9.3).

### 4.8.2 Wider System-of-interest Integration

The SKA phase synchronisation system has interfaces with several other SKA elements (Wider System-of-interest (WSOI)), as outlined in §4.8.4 and §4.8.5, however, it is fundamentally self-contained. More information is provided in §5.1.1.

### 4.8.3 Precursor Integration

Previous astronomical verification work with ASKAP [RD19] and ATCA [RD18], [RD19] and [RD20] is described in §5.1.1. There is €14.4k of funding to continue SKA1-MID integration tests with MeerKAT. UWA and SKA SA have arranged a joint follow-up South African overhead fibre field trial. The plan calls to integrate the SKA phase synchronisation system with the SKA1-MID precursor telescope, MeerKAT. The RM will be installed in the MeerKAT L-band receiver enclosure mounted on the receptor indexer. The reason for 6.8 GHz transfer is that L-band receiver clock is 1.72 GHz, and so 4× 1.72 GHz produces the closest frequency to the nominal SKA1-MID 8 GHz transmission frequency.

### 4.8.4 SADT Interdependencies

As shown in detail in §4.4.1, the SKA1-MID variant of The SKA phase synchronisation system has interfaces with the following SADT elements:

- CLOCK
- LMC
- LINFRA
- Non-Science Data Network (NSDN).

Further information on the interfaces with these elements is provided in the relevant IICD.

### 4.8.5 Non-SADT (External) Dependencies

As shown in detail in §4.4.1, the SKA phase synchronisation system has interfaces with only the DISH Consortium hardware. Further information on this interface is provided in the relevant EICD.

## 4.9 Interoperability

### 4.9.1 SADT Interoperability

The SKA phase synchronisation system has been extensive tested with elements that simulate or replicate the various SADT interfaces outlined in §4.8.4. Specifically, the software interfaces with SAT.LMC (and hardware interfaces with the NSDN) was replicated using a mock-up implementation of both the Lock signal and Phase measurement monitoring, and this was demonstrated to the SADT Consortium during the face-to-face meeting in Perth 2016 (see §4.3.4.2). The SAT.CLOCK interface (receiving a 10 MHz reference signal from a H-Maser) is daily practice. All of the evaluation measurements outlined in §5.1.1 were conducted with this in place. The design for mass manufacture archetypes have been installed in rack mounted enclosures for interfacing with the LINFRA rack cabinet.

### 4.9.2 Non-SADT (External) Interoperability

The SKA phase synchronisation system has been extensive tested with an interface that is closely resembling of the DISH interface outlined in §4.8.5 as is possible. During the astronomical verification test with ATCA [RD18], [RD19] and [RD20], the output of the RMs were used to provide the reference clock for the ATCA receivers. As described in §4.3.2, the outcome of these tests were used to verify the functional performance requirements of the system.

## 4.10 Costing

### 4.10.1 Development Costs

The development costs for both the SKA1-MID and SKA1-LOW variants of the SKA phase synchronisation system has been €482.0k. These expenses were incurred between the period start-February 2014 and end-June 2017. This cost is divided as follows:

- Salaries and on-costs: €354.4k
- Laboratory equipment: €64.8k
- Consumables: €17.1k
- Collaborative travel: €25.1k
- Field work: €20.6k.

### 4.10.2 Capital Expenditure Equipment Costs

The summary of capital expenditure cost for SKA1-MID equipment as determined by the UWA Detailed Cost Model (see Appendix 8.9.1) is given in Table 23.

**SUMMARY**  
**Capex Equipment Cost**

| SKA-MID UWA STFR.FRQ LRUs (-) | Per Unit Equipment (€) | Totals Quantity (#) | Cost (€) |
|---|---|---|---|
| Rack Cabinet (341-022900) | € 2,134 | 1 | € 2,134 |
| Optical Source (341-022400) | € 33,486 | 1 | € 33,486 |
| Frequency Synthesiser (341-022500) | € 15,507 | 1 | € 15,507 |
| Microwave Shift (341-022600) | € 20,085 | 1 | € 20,085 |
| Signal Generator (341-023100) | € 6,154 | 1 | € 6,154 |
| Rack Distribution (341-022800) | € 2,848 | 2 | € 5,696 |
| Sub Rack (341-022700) | € 4,634 | 13 | € 60,245 |
| Transmitter Module (341-022100) | € 5,782 | 197 | € 1,138,993 |
| Fibre Patch Lead (341-023200) | € 9 | 197 | € 1,812 |
| Receiver Module (341-022300) | € 2,866 | 197 | € 564,534 |
| Optical Amplifier (341-022200) | € 5,700 | 17 | € 96,900 |
| **Total Cost** | | | € 1,945,546 |
| **Cost Per Link** | | | € 9,876 |

**Table 23 SKA1-MID summary of the capex equipment Detailed Cost Model Database**

The 'per unit' values in the table above are derived from the detailed bill of materials outlined in the following eleven Capital Expenditure Equipment Costs sub-sections, one for each LRU. In these sub-sections, the cost of each LRU is given on an individual basis, with the cost of each item based on real-world quotes at the quantities required to build one-off units (e.g. to purchase just two AOMs for the TM). The summary table above, presents the cost for the entire system with an appropriate volume discount applied (e.g. purchasing a total of 600 AOMs). I used a simple algorithm to apply a 'volume discount rate' to the equipment based on the quantity of each LRU. The rate and algorithm form were chosen based on a best fit from real-world quoted volume discounts for the cost dominant items (i.e. I requested several high volume quotes for items including AOMs, PDs, optical assemblies, and some electronics). The ancillary parameters, used in capital expenditure cost for SKA1-MID equipment Detailed Cost Model Database, including the 'volume discount rate', are given in Table 24.

| Exchange Rate | AUD | EUR | GBP | USD |
|---|---|---|---|---|
| 1 EUR | 1.46000 | 1.00000 | 0.77483 | 1.12870 |
| Inverse (2016/04/26) | 0.68493 | 1.00000 | 1.29061 | 0.88598 |

| Volume Discount Rate | 0.0025 | per unit |
|---|---|---|

**Table 24 SKA1-MID ancillary parameters used in the capex equipment Detailed Cost Model Database**

This approach was found especially useful when the final number of links, and required sparing quantities, were still in flux. The process of choosing and applying the algorithm was discussed at length with Mike Pearson from SADT.

#### 4.10.2.1 Rack Cabinet (341-022900)

The Detailed Cost Model Database for the rack cabinet (341-022900) is shown in Table 25.

| Rack Cabinet (341-022900) | | | | | | | | | | | |
|---|---|---|---|---|---|---|---|---|---|---|---|
| Equipment Cost | € 2,134.03 | | | | | | | | | | |
| Total Power | 240.00 | | | | | | | | | | |

| Identifier | Company | Part Number | Part Description | Case Style | CAD/PCE | URL | Q.Each | Min.Q. | Unit Price | Currency | Exg | Total |
|---|---|---|---|---|---|---|---|---|---|---|---|---|
| RC-Equipment | RS | 118-3279 | 6 Fan Rack Mount Fan Tray, 1U | 19", 1U rack | Y | http://au | 8 | 1 | $ 389.46 | AUD | 0.685 | € 2,134.03 |

**Table 25 SKA1-MID rack cabinet (341-022900) capex equipment Detailed Cost Model Database**

#### 4.10.2.2 Optical Source (341-022400)

The Detailed Cost Model Database for the OS (341-022400) is shown in Table 26.

| Optical Source (341-022400) | | | | | | | | | | | |
|---|---|---|---|---|---|---|---|---|---|---|---|
| Equipment Cost | € 33,486.30 | | | | | | | | | | |
| Total Power | 120.00 | | | | | | | | | | |

| Identifier | Company | Part Number | Part Description | Case Style | CAD/PCE | URL | Q.Each | Min.Q. | Unit Price | Currency | Exg | Total |
|---|---|---|---|---|---|---|---|---|---|---|---|---|
| OS-Equipment | Keopsys | CEFL-KILO-10-LP-W31- | CONTINUOUS WAVE ERBIUM FIBER LASER | 19", 3U rack | N | https:// | 1 | 1 | $ 48,890.00 | AUD | 0.685 | € 33,486.30 |

**Table 26 SKA1-MID OS (341-022400) capex equipment Detailed Cost Model Database**

#### 4.10.2.3 Frequency Source (341-022500)

The Detailed Cost Model Database for the FS (341-022500) is shown in Table 27.

| Frequency Synthesiser (341-022500) | | | | | | | | | | | |
|---|---|---|---|---|---|---|---|---|---|---|---|
| Equipment Cost | € 15,506.68 | | | | | | | | | | |
| Total Power | 75.00 | | | | | | | | | | |

| Identifier | Company | Part Number | Part Description | Case Style | CAD/PCE | URL | Q.Each | Min.Q. | Unit Price | Currency | Exg | Total |
|---|---|---|---|---|---|---|---|---|---|---|---|---|
| FS-Equipment | Keysight | N5173B-513 | EXG, Frequency Synthesiser 9 kHz to 13 GHz | 19", 2U rack | Y | http://w | 1 | 1 | $ 22,639.75 | AUD | 0.685 | € 15,506.68 |

**Table 27 SKA1-MID FS (341-022500) capex equipment Detailed Cost Model Database**

### 4.10.2.4 Microwave Shift (341-022600)

The Detailed Cost Model Database for the MS (341-022600) is shown in Table 28.

| Microwave Shift (341-022600) | | |
|---|---|---|
| Equipment Cost | € | 20,084.88 |
| Total Power | | 0.00 |

| Identifier | Company | Part Number | Part Description | Case Style | CAD/PCE | URL | Q.Each | Min.Q. | Unit Price | Currency | Exg | Total |
|---|---|---|---|---|---|---|---|---|---|---|---|---|
| MS-Equipment | Jaycar | HB5125 | 2U 19" rack mount equipment enclosure | --- | N | http://v | 1 | 1 | $ 109.00 | AUD | 0.685 | € 74.66 |
| MS-Equipment | *various* | *various* | cables, connectors, adaptors, sundries | --- | N | --- | 1 | 1 | $ 1,000.00 | AUD | 0.685 | € 684.93 |
| MS-Equipment | Benchmark Enginne | --- | Custom aluminium enclosure | --- | N | --- | 1 | 1 | $ 1,000.00 | AUD | 0.685 | € 684.93 |
| MS-Equipment | Lektronics | --- | Ribox 2U x 350mm deep | --- | N | http://v | 1 | 1 | $ 128.00 | AUD | 0.685 | € 87.67 |
| MS-Opt. | Thor Labs | PN1550R1A1 | PM Optical Coupler (99:01) | | | https:// | 2 | 1 | $ 536.00 | USD | 0.886 | € 949.77 |
| MS-Opt. | Thor Labs | PN1550R2A1 | PM Optical Coupler (90:10) | | | https:// | 1 | 1 | $ 536.00 | USD | 0.886 | € 474.88 |
| MS-Opt. | Thor Labs | PN1550R5A1 | PM Optical Coupler (50:50) | | | https://w | 3 | 1 | $ 536.00 | USD | 0.886 | € 1,424.65 |
| MS-Opt. | Gooch & Housego | T-M040-PM | AOM | | | https:// | 1 | 1 | $ 2,255.00 | GBP | 1.291 | € 2,910.32 |
| MS-Opt. | Photline | MXIQ-LN-40-P-P-PD-FA- | Dual Parallel Mach-Zehnder Modulator | | | https:// | 1 | 1 | $ 3,900.00 | EUR | 1.000 | € 3,900.00 |
| MS-Opt. | oeMarket | EDFA-D | Two-Stage Amplifier 25dB | | | http://w | 1 | 1 | $ 3,890.00 | USD | 0.886 | € 3,446.44 |
| MS-Elect. | Minicircuits | BW-S5W2+ | Attenuator | | | https://w | 1 | 1 | $ 31.95 | USD | 0.886 | € 28.31 |
| MS-Elect. | Minicircuits | IZY2PD-86+ | Splitter | | | https://w | 3 | 1 | $ 94.95 | USD | 0.886 | € 252.37 |
| MS-Elect. | Minicircuits | BW-S1W2+ | Attenuator | | | https://w | 1 | 1 | $ 31.95 | USD | 0.886 | € 28.31 |
| MS-Elect. | Minicircuits | ZVA-183W | Wideband Amplifier | | | https://w | 1 | 1 | $ 1,534.95 | USD | 0.886 | € 1,359.93 |
| MS-Elect. | Minicircuits | BLK-18+ | DC Block | | | https://w | 2 | 1 | $ 21.95 | USD | 0.886 | € 38.89 |
| MS-Elect. | Minicircuits | BW-S10W2+ | Attenuator | | | https://w | 1 | 1 | $ 31.95 | USD | 0.886 | € 28.31 |
| MS-Elect. | Minicircuits | ZVA-213-S+ | Wideband Amplifier | | | https://w | 1 | 1 | $ 1,094.95 | USD | 0.886 | € 970.10 |
| MS-Elect. | Discovery Semicond | DSC-R402 | Low Noise, High Gain Optical Receiver | | | http://w | 1 | 1 | $ 1,350.00 | USD | 0.886 | € 1,196.07 |
| MS-Elect. | Windfreak | SynthHD | Dual channel Microwave Generator | | | https://w | 1 | 1 | $ 1,299.00 | USD | 0.886 | € 1,150.88 |
| MS-Elect. | Minicircuits | ZX05-153-S+ | Coaxial Frequency Mixer | | | https://w | 1 | 1 | $ 48.95 | USD | 0.886 | € 43.37 |
| MS-Elect. | Element14 | Raspberry Pi 3 Model B | Raspberry Pi (Microcomputer) | | | https:// | 1 | 1 | $ 49.50 | AUD | 0.685 | € 33.90 |
| MS-Elect. | Minicircuits | ZX95-43+ | VCO | | | https://w | 1 | 1 | $ 44.95 | USD | 0.886 | € 39.82 |
| MS-Elect. | Minicircuits | BW-S7W2+ | Attenuator | | | https://w | 1 | 1 | $ 31.95 | USD | 0.886 | € 28.31 |
| MS-Elect. | RF Bay | MGA-3-27 | RF Amplifier | | | http://rfb | 1 | 1 | $ 280.00 | USD | 0.886 | € 248.07 |

**Table 28 SKA1-MID MS (341-022600) capex equipment Detailed Cost Model Database**

### 4.10.2.5 Signal Generator (341-023100)

The Detailed Cost Model Database for the SG (341-023100) is shown in Table 29.

| Signal Generator (341-023100) | | |
|---|---|---|
| Equipment Cost | € | 6,154.05 |
| Total Power | | 75.00 |

| Identifier | Company | Part Number | Part Description | Case Style | CAD/PCE | URL | Q.Each | Min.Q. | Unit Price | Currency | Exg | Total |
|---|---|---|---|---|---|---|---|---|---|---|---|---|
| SG-Equipment | Keysight | N5171B-501 | EXG Signal Generator, 9 kHz to 1 GHz | 19", 1U rack | Y | http:// | 1 | 1 | $ 8,921.60 | AUD | 0.685 | € 6,110.68 |
| SG-Equipment | Minicircuits | ZFSC-2-1+ | PWR SPLTR CMBD / BNC / RoHS | K18 | N | https:// | 1 | 1 | $ 48.95 | USD | 0.886 | € 43.37 |

**Table 29 SKA1-MID SG (341-023100) capex equipment Detailed Cost Model Database**

### 4.10.2.6 Rack Distribution (341-022800)

The Detailed Cost Model Database for the RD (341-022800) is shown in Table 30.

| Rack Distribution (341-022800) | | |
|---|---|---|
| Equipment Cost | € | 2,855.37 |
| Total Power | | 0.00 |

| Identifier | Company | Part Number | Part Description | Case Style | CAD/PCE | URL | Q.Each | Min.Q. | Unit Price | Currency | Exg | Total |
|---|---|---|---|---|---|---|---|---|---|---|---|---|
| RD-Equipment | oeMarket | PMCM-1550-1x8-1-FAF | PM fibre 1/8 splitters | --- | N | http:// | 2 | 1 | $ 1,427.00 | USD | 0.886 | € 2,528.57 |
| RD-Equipment | Minicircuits | ZB8PD-8.4 | microwave-frequency splitter 1/8 | Z41 | N | https:// | 1 | 1 | $ 149.95 | USD | 0.886 | € 132.85 |
| RD-Equipment | Minicircuits | ZB8PD-8.4 | radio-frequency splitter 1/8 | UU215 | N | https:// | 1 | 1 | $ 119.95 | USD | 0.886 | € 106.27 |
| RD-Equipment | Lektronics | 3687818.000 | Ribox 1U x 350mm deep | 19", 1U rack | Y | http:// | 1 | 1 | $ 128.00 | AUD | 0.685 | € 87.67 |

**Table 30 SKA1-MID RD (341-022800) capex equipment Detailed Cost Model Database**

### 4.10.2.7 Sub Rack (341-022700)

The Detailed Cost Model Database for the SR (341-022700) is shown in Table 31.

| Sub Rack (341-022700) | | |
|---|---|---|
| Equipment Cost | € | 4,777.58 |
| Total Power | | 0.00 |

| Identifier | Company | Part Number | Part Description | Case Style | CAD/PCE | URL | Q.Each | Min.Q. | Unit Price | Currency | Exg | Total |
|---|---|---|---|---|---|---|---|---|---|---|---|---|
| SR-Eurocard frame | Lektronics | 3684.026 | Rittal Ripac Vario 3Ux84HPx465 sub-rack | --- | Y | | 1 | 1 | $ 190.00 | AUD | 0.685 | € 130.14 |
| SR-Eurocard frame | Lektronics | 3636.010 | Rittal Pkt2 3U handles | --- | Y | | 1 | 1 | $ 18.00 | AUD | 0.685 | € 12.33 |
| SR-Eurocard frame | Lektronics | 3652.050 | Rittal 3Ux8HP standard front panel kit (with pl --- | Y | | 1 | 1 | $ 19.45 | AUD | 0.685 | € 13.32 |
| SR-Eurocard frame | Lektronics | 3652.070 | Rittal 3Ux12HP standard front panel kit (with --- | Y | | 1 | 1 | $ 22.85 | AUD | 0.685 | € 15.65 |
| SR-Eurocard frame | Lektronics | 3684.656 | Rittal 280mm Plastic snap in card guide | --- | N | | 32 | 1 | $ 0.90 | AUD | 0.685 | € 19.73 |
| SR-Eurocard frame | Lektronics | 3684.691 | Rittal 84HPx465 solid unvented lid | --- | Y | | 2 | 1 | $ 39.80 | AUD | 0.685 | € 54.52 |
| SR-Eurocard frame | Lektronics | 3684.234 | Rittal mounting blocks (Pkt10) | --- | N | | 2 | 1 | $ 19.45 | AUD | 0.685 | € 26.64 |
| SR-Eurocard frame | Lektronics | 3684.233 | Rittal screws (Pkt100) | --- | N | | 1 | 1 | $ 12.00 | AUD | 0.685 | € 8.22 |
| SR-Eurocard frame | Lektronics | --- | Shipping | --- | N/A | | 1 | 1 | $ 25.00 | AUD | 0.685 | € 17.12 |
| SR-CCD Module PCB | Advanced Technolog | --- | FR4 PCB manufacture and assembly | | | | 1 | 1 | $ 200.00 | AUD | 0.685 | € 136.99 |
| SR-CCD Module PCB | *various* | *various* | CCD components and Ethernet controller | | | | 1 | 1 | $ 100.00 | EUR | 1.000 | € 100.00 |
| SR-Backplane PCB | Advanced Technolog | --- | FR4 PCB manufacture and assembly | | | | 1 | 1 | $ 200.00 | AUD | 0.685 | € 136.99 |
| SR-Backplane PCB | *various* | *various* | Passive RF splits and connectors | | | | 1 | 1 | $ 50.00 | EUR | 1.000 | € 50.00 |
| SR-Optical Distribution | oeMarket | PMC-F-1550-1x2-50/50 | fiberised PM optical coupler (50/50) | --- | N | http:// | 1 | 1 | $ 299.00 | USD | 0.886 | € 264.91 |
| SR-Optical Distribution | oeMarket | PMCM-1550-1x8-1-FAF | Secondary PM fibre 1/8 splitters | --- | N | http:// | 2 | 1 | $ 1,620.00 | USD | 0.886 | € 2,870.56 |
| SR-Power Supplies | Lektronics | 3686.682 | CPCI Power Supply Unit | 3U, 12HP | Y | http://im | 1 | 1 | $ 550.00 | AUD | 0.685 | € 376.71 |
| SR-Power Supplies | Lektronics | 3685.33 | CPCI Power Supply Faceplate | 3U, 12HP | Y | http://im | 1 | 1 | $ 10.00 | AUD | 0.685 | € 6.85 |
| SR-Connectors | RS | | Ethernet bulkhead | | Y | | 1 | 1 | $ 10.20 | AUD | 0.685 | € 6.99 |
| SR-Connectors | RS | | SMA bulkhead | | Y | | 2 | 1 | $ 6.80 | AUD | 0.685 | € 9.32 |
| SR-Connectors | oeMarket | | SC/APC bulkhead | | Y | | 1 | 1 | $ 3.70 | AUD | 0.685 | € 2.53 |
| SR-MW Distribution | Minicircuits | ZB8PD-8.4 | microwave-frequency splitter 1/8 | Z41 | N | https:// | 2 | 1 | $ 149.95 | USD | 0.886 | € 265.70 |
| SR-MW Distribution | Minicircuits | IZY2PD-86+ | Splitter | | | https://w | 3 | 1 | $ 94.95 | USD | 0.886 | € 252.37 |
| SR-MW Distribution | Rojone | | | | | | | | | | | |

**Table 31 SKA1-MID SR (341-022700) capex equipment Detailed Cost Model Database**

### 4.10.2.8 Transmitter Module (341-022100)

The Detailed Cost Model Database for the TM (341-022100) is shown in Table 32.

| Transmitter Module (341-022100) | | |
|---|---|---|
| Equipment Cost | € | 11,336.65 |
| Total Power | | 0.00 |

| Identifier | Company | Part Number | Part Description | Case Style | CAD/PCE | URL | Q.Each | Min.Q. | Unit Price | Currency | Exg | Total |
|---|---|---|---|---|---|---|---|---|---|---|---|---|
| TM-Opt. | Gooch&Housego | T-M080-0.5C8J-3-F2S | -85MHz SM AOM | --- | Y | http://go | 1 | | $ 1,850.00 | GBP | 1.291 | € 2,387.62 |
| TM-Opt. | Gooch&Housego | T-M040-0.5C8J-3-F2P | +40MHz PM AOM | --- | Y | http://go | 1 | | $ 2,255.00 | GBP | 1.291 | € 2,910.32 |
| TM-Opt. | OEMarket | PMC-S-1550-2x2-50/50 | PM Splitter | --- | Y | http://w | 1 | | $ 525.00 | USD | 0.886 | € 465.14 |
| TM-Opt. | OEMarket | SMC-1550-2x2-P-50/50 | 50/50 SM Splitter | --- | Y | http://w | 1 | | $ 17.90 | USD | 0.886 | € 15.86 |
| TM-Opt. | OEMarket | FRM-1550-0-N | Faraday Mirror | --- | Y | http://w | 1 | | $ 69.95 | USD | 0.886 | € 61.97 |
| TM-Opt. | OEMarket | ISO-MI-1550-S-0-N | Miniature Opt. Fibre Isolator (Single Stage) | --- | Y | http://w | 1 | | $ 89.00 | USD | 0.886 | € 78.85 |
| TM-Opt. | OEMarket | FOA-IL-1550-12.5-0-N | Optical Attenuator | --- | Y | http://w | 1 | | $ 17.99 | USD | 0.886 | € 15.94 |
| TM-Opt. | OEMarket | ADP-MF-FC-PMAF | Optical Connector | --- | Y | http://w | 4 | | $ 18.90 | USD | 0.886 | € 66.98 |
| TM-Opt. | OEMarket | Optical sub-assembly | Splicing assembly | --- | --- | | 1 | | $ 300.00 | USD | 0.886 | € 265.79 |
| TM-Opt. | Element 14 | 124 8989 | SMA Right-angle female | SMA Throug | Y | http://au | 2 | | $ 2.56 | USD | 0.886 | € 4.54 |
| TM-Opt. | Lektronics | 3652.010 | Rittal 3Ux4HP standard front panel kit (with pl | --- | Y | | 1 | | $ 18.65 | AUD | 0.685 | € 12.77 |
| TM-Opt. | Benchmark Engineer | Uwa0007 | Optics Enclosure Body | --- | Y | | 1 | | $ 281.80 | AUD | 0.685 | € 193.01 |
| TM-Opt. | Benchmark Engineer | Uwa0008 | Optics Enclosure Lid | | Y | | 1 | | $ 33.90 | AUD | 0.685 | € 23.22 |
| TM-MW | Discovery Semicond | DSC-R402AC-LW-39-LC | Photodetector | K-Connector | Y | http://w | 2 | | $ 1,350.00 | USD | 0.886 | € 2,392.13 |
| TM-MW | Marki Microwave | ML1-0218LSM-1 | Microwave Mixer | SM Pad | Y | http://w | 2 | | $ 99.00 | USD | 0.886 | € 175.42 |
| TM-MW | Marki Microwave | ML1-0218LSM-2 | Microwave Mixer | SM Pad | Y | http://w | 2 | | $ 99.00 | USD | 0.886 | € 175.42 |
| TM-MW | Minicircuits | EP2K+ | Power Splitter | SM Pad | Y | https:// | 1 | | $ 36.95 | USD | 0.886 | € 32.74 |
| TM-MW | Minicircuits | YAT-2+ | Microwave Attenuator | SM Pad | Y | http://w | 7 | | $ 2.99 | USD | 0.886 | € 18.54 |
| TM-MW | MACOM | MAAM-011101 | Low Noise Amplifier | SM Pad | Y | http://w | 1 | | $ 13.24 | AUD | 0.685 | € 9.07 |
| TM-MW | Advanced Technolog | 161028UWA | TM-MW PCB assembly | | Y | | 1 | | $ 66.50 | AUD | 0.685 | € 45.55 |
| TM-MW | Gigatronix | JP3ZXJP3ZXT4F2-300 | Coax | | | https://w | 1 | | $ 10.42 | GBP | 1.291 | € 13.45 |
| TM-MW | UWA Physics works | --- | MW Enclosure Body and Lid | | Y | | 1 | | $ 525.00 | AUD | 0.685 | € 359.59 |
| | | | | | | | | | | | | € - |
| TM-RF | KR Electronics | 3032-360BPF | 360MHz Band Pass Filter | SM Pad | Y | http://kr | 1 | | $ 542.00 | USD | 0.886 | € 480.20 |
| TM-RF | KR Electronics | 3032-280BPF | 280MHz Band Pass Filter | SM Pad | Y | http://kr | 1 | | $ 542.00 | USD | 0.886 | € 480.20 |
| TM-RF | Minicircuits | RHP-147+ | 250 MHz High Pass Filter | SM Pad | Y | https:// | 1 | | $ 12.70 | USD | 0.886 | € 11.25 |
| TM-RF | Minicircuits | PGA-105+ | Low Noise Amplifier | SM Pad | Y | https:// | 1 | | $ 1.99 | USD | 0.886 | € 1.76 |
| TM-RF | Minicircuits | GVA-63+ | Amplifier | SM Pad | Y | http://w | 2 | | $ 0.99 | USD | 0.886 | € 1.75 |
| TM-RF | Minicircuits | TCBT-14+ | Amp Bias Tee | SM Pad | Y | https:// | 1 | | $ 8.45 | USD | 0.886 | € 7.49 |
| TM-RF | Minicircuits | LRPS-2-1J+ | Power Splitter | SM Pad | Y | http://w | 1 | | $ 8.70 | USD | 0.886 | € 7.71 |
| TM-RF | Minicircuits | ADE-1L+ | RF Mixer | SM Pad | Y | http://w | 1 | | $ 15.95 | USD | 0.886 | € 14.13 |
| TM-RF | Minicircuits | LAT-1+ | RF Attenuator | SM Pad | Y | https:// | 2 | | $ 1.95 | USD | 0.886 | € 3.46 |
| TM-RF | Minicircuits | LAT-3+ | RF Attenuator | SM Pad | Y | https:// | 1 | | $ 1.95 | USD | 0.886 | € 1.73 |
| TM-RF | Element 14 | 1248989 | SMA Right-angle female | SMA Throug | Y | http://au | 2 | | $ 2.46 | AUD | 0.685 | € 3.37 |
| TM-RF | Amphenol RF | 132135 | SMA Right-angle male | SMA Throug | Y | http://au | 2 | | $ 14.24 | AUD | 0.685 | € 19.51 |
| TM-RF | Radiall | 125.484.000 | SMA Flange Connector | | Y | | 5 | | $ 4.99 | AUD | 0.685 | € 17.09 |
| TM-RF | Radiall | 4141379 | SMA Straight Female | SMA Throug | Y | http://au | 2 | | $ 4.65 | AUD | 0.685 | € 6.37 |
| TM-RF | Minicircuits | RBP-75+ | 80MHz Band Pass Filter | | | https://w | 1 | | $ 13.70 | USD | 0.886 | € 12.14 |
| TM-RF | Advanced Technolog | 161028UWA | TM-RF PCB assembly | | Y | -- | 1 | | $ 69.90 | AUD | 0.685 | € 47.88 |

| | | | | | | | | | | | |
|---|---|---|---|---|---|---|---|---|---|---|---|
| TM-Srv. | RS | 9030006262 | Female Copper Alloy DIN Connector | | | http://do | 1 | $ | 9.91 GBP | 1.291 € | 12.79 |
| TM-Srv. | RS | 9031606901 | Right Angle DIN 41612 Connector | | | https://b | 1 | $ | 8.63 GBP | 1.291 € | 11.14 |
| TM-Srv. | Minicircuits | HELA-10D+ | Power Amplifier | | Y | https://w | 2 | $ | 19.95 USD | 0.886 € | 35.35 |
| TM-Srv. | Minicircuits | ADE-1+ | Mixer | | Y | http://w | 2 | $ | 15.00 USD | 0.886 € | 26.58 |
| TM-Srv. | Minicircuits | LHA-1+ | Amplifier | | Y | http://w | 5 | $ | 1.64 USD | 0.886 € | 7.26 |
| TM-Srv. | Minicircuits | RBP-98+ | Filter 100 MHz | | Y | http://w | 1 | $ | 13.70 USD | 0.886 € | 12.14 |
| TM-Srv. | Minicircuits | ADTT1-1+ | Transformer | | Y | http:// | 4 | $ | 7.88 USD | 0.886 € | 27.93 |
| TM-Srv. | RS | OP27EP | DC amplifier | | Y | | 1 | $ | 6.34 USD | 0.886 € | 5.62 |
| TM-Srv. | RS | OP27EP | Filter DC | | Y | | 1 | $ | 6.34 USD | 0.886 € | 5.62 |
| TM-Srv. | Analog Devices | AD9959 | DDS | | Y | http://w | 1 | $ | 87.36 USD | 0.886 € | 77.40 |
| TM-Srv. | Microchip | PIC18F46K80-I/PT | uProcessor | | Y | http://v | 1 | $ | 5.80 USD | 0.886 € | 5.14 |
| TM-Srv. | Element 14 | 73100-0114 | SMA PCB launcher | | Y | http://a | 5 | $ | 9.08 AUD | 0.685 € | 31.10 |
| TM-Srv. | Newbury Electronics | --- | FR4 4-layer PCB | | Y | | 1 | $ | 50.00 GBP | 1.291 € | 64.53 |
| TM-Srv. | Newbury Electronics | --- | Assembly | | Y | | 1 | $ | 100.00 GBP | 1.291 € | 129.06 |
| TM-Srv. | Minicircuits | LAT-7+ | Attenuator | MMM168 | | https:// | 1 | $ | 2.15 USD | 0.886 € | 1.90 |
| TM-Srv. | Minicircuits | TC-1-13M+ | Transformer | AT224-1A | | https:// | 1 | $ | 2.50 USD | 0.886 € | 2.21 |
| TM-Srv. | Minicircuits | RLP-40+ | 40MHz Low Pass Filter | GP731 | | https:// | 1 | $ | 7.95 USD | 0.886 € | 7.04 |
| TM-Srv. | Minicircuits | LRPS-2-1J+ | Power Splitter | QQQ569 | | https:// | 1 | $ | 8.95 USD | 0.886 € | 7.93 |
| TM-Srv. | Minicircuits | ADQ-22+ | Power Splitter | CJ725 | | https:// | 1 | $ | 9.35 USD | 0.886 € | 8.28 |
| TM-Srv. | Minicircuits | ROS-43-119+ | VCO 40 MHz | CK-605 | | https:// | 1 | $ | 19.95 USD | 0.886 € | 17.68 |

**Table 32 SKA1-MID TM (341-022100) capex equipment Detailed Cost Model Database**

### 4.10.2.9 Fibre Patch Lead (341-023200)

The Detailed Cost Model Database for the FP lead (341-023200) is shown in Table 33.

| Fibre Patch Lead (341-023200) | | | | | | | | | | | |
|---|---|---|---|---|---|---|---|---|---|---|---|
| Equipment Cost | € 9.20 | | | | | | | | | | |
| Total Power | 0.00 | | | | | | | | | | |

| Identifier | Company | Part Number | Part Description | Case Style | CAD/PCE | URL | Q.Each | Min.Q. | Unit Price | Currency | Exg | Total |
|---|---|---|---|---|---|---|---|---|---|---|---|---|
| FPL-Equipment | go4fiber | S1-F6L6-01-0150-GF | FC/APC to LC/APC Patchcord, 9/125um | NA | Y | https://w | 1 | 1 | $ 10.38 | USD | 0.886 | € 9.20 |

**Table 33 SKA1-MID FP lead (341-023200) capex equipment Detailed Cost Model Database**

### 4.10.2.10 Optical Amplifier (341-022200)

The Detailed Cost Model Database for the OA (341-022200) is shown in Table 34.

| Optical Amplifier (341-022200) | | | | | | | | | | | |
|---|---|---|---|---|---|---|---|---|---|---|---|
| Equipment Cost | € 5,937.50 | | | | | | | | | | |
| Total Power | 50.00 | | | | | | | | | | |

| Identifier | Company | Part Number | Part Description | Case Style | CAD/PCE | URL | Q.Each | Min.Q. | Unit Price | Currency | Exg | Total |
|---|---|---|---|---|---|---|---|---|---|---|---|---|
| OA-Equipment | IDIL | COINT03311 | Erbium doped bidirectionnal amplifier (AC | 19", 2U rack | N | | 1 | 1 | $ 5,937.50 | EUR | 1.000 | € 5,937.50 |

**Table 34 SKA1-MID OA (341-022200) capex equipment Detailed Cost Model Database**

### 4.10.2.11  Receiver Module (341-022300)

The Detailed Cost Model Database for the RM (341-022300) is shown in Table 35.

| Receiver Module (341-022300) | |
|---|---|
| Equipment Cost | € 5,618.93 |
| Total Power | 0.00 |

| Identifier | Company | Part Number | Part Description | Case Style | CAD/PCE | URL | Q.Each | Min.Q. | Unit Price | Currency | Exg | Total |
|---|---|---|---|---|---|---|---|---|---|---|---|---|
| RM-Opt. | Gooch&Housego | T-M080-0.5C8J-3-F2S | 75MHz SM AOM | | | https://g | 1 | 1 | $ 1,850.00 | GBP | 1.291 | € 2,387.62 |
| RM-Opt. | oeMarket | FRM-1550-0-N | SM Faraday Mirror | | | http://w | 1 | 1 | $ 69.95 | USD | 0.886 | € 61.97 |
| RM-Opt. | oeMarket | VOA-C-1550-9-0-N | Variable Optical Attenuator | | | http://w | 1 | 1 | $ 179.95 | USD | 0.886 | € 159.43 |
| RM-Opt. | oeMarket | SMC-1550-2x2-P-10/90 | 10/90 SM Splitter | | | http://w | 1 | 1 | $ 17.90 | USD | 0.886 | € 15.86 |
| RM-Opt. | oeMarket | ADP-MF-FC-PMAF | Optical connector | | | http://w | 3 | 1 | $ 18.90 | USD | 0.886 | € 50.23 |
| RM-MW | Discovery Semicond | DSC-R402AC-LW-39-LC | Photodetector | K-Connector | Y | http://w | 1 | 1 | $ 1,350.00 | USD | 0.886 | € 1,196.07 |
| RM-MW | Marki Microwave | ML1-0218LSM-1 | Microwave Mixer | SM Pad | | http://w | 1 | 1 | $ 99.00 | USD | 0.886 | € 87.71 |
| RM-MW | Minicircuits | EP2K+ | Power Splitter | SM Pad | | https://w | 1 | 1 | $ 36.95 | USD | 0.886 | € 32.74 |
| RM-MW | Minicircuits | YAT-2+ | Microwave Attenuator | SM Pad | | http://w | 2 | 1 | $ 2.99 | USD | 0.886 | € 5.30 |
| RM-MW | MACOM | MAAM-011101 | Low Noise Amplifier | SM Pad | | http://w | 2 | 1 | $ 13.24 | AUD | 0.685 | € 18.14 |
| RM-MW | Minicircuits | YAT-8+ | Microwave Attenuator | SM Pad | | https://w | 1 | 1 | $ 2.99 | USD | 0.886 | € 2.65 |
| RM-MW | Minicircuits | BFCN-8000+ | Bandpass Filter | SM Pad | | https://w | 1 | 1 | $ 3.95 | USD | 0.886 | € 3.50 |
| RM-MW | Analog Devices | HMC575LP4 | Divide by 2 | SM Pad | | http://w | 1 | 1 | $ 29.63 | USD | 0.886 | € 26.25 |
| RM-MW | NEL FC Inc | 1 GHz MHz O-CEGM-X | OCXO | Europack | | http://w | 1 | 1 | $ 1,200.00 | USD | 0.886 | € 1,063.17 |
| RM-MW | Minicircuits | RKK-4-442+ | Multiply by 4 | SM Pad | | https://w | 1 | 1 | $ 13.95 | USD | 0.886 | € 12.36 |
| RM-MW | Minicircuits | BFCV-4085+ | Bandpass Filter | SM Pad | | https://w | 1 | 1 | $ 4.95 | USD | 0.886 | € 4.39 |
| RM-MW | Minicircuits | PMA3-83LN+ | Low Noise Amplifier | SM Pad | | https://w | 1 | 1 | $ 11.95 | USD | 0.886 | € 10.59 |
| RM-MW | RS | OP27EP | DC amplifier | | | | 1 | 1 | $ 6.34 | USD | 0.886 | € 5.62 |
| RM-MW | RS | OP27EP | Filter DC | | | | 1 | 1 | $ 6.34 | USD | 0.886 | € 5.62 |
| RM-MW | Minicircuits | JTOS-150P+ | VCO | SM Pad | | https://w | 1 | 1 | $ 14.95 | USD | 0.886 | € 13.25 |
| RM-MW | Minicircuits | HELA-10D+ | High-powered Amplifier | SM Pad | | https://w | 1 | 1 | $ 19.95 | USD | 0.886 | € 17.68 |
| RM-MW | Advanced Technolog | --- | MW Rogers PCB manufacture and assembly | | Y | | 1 | 1 | $ 200.00 | AUD | 0.685 | € 136.99 |
| RM-Enclosure | Manchester | --- | Bespoke aluminium box and lid | --- | Y | http://v | 1 | 1 | $ 40.00 | GBP | 1.291 | € 51.62 |
| RM-Enclosure | RS | 885-9960 | SMA feedthrough | --- | Y | http://au | 1 | 1 | $ 19.93 | AUD | 0.685 | € 13.65 |
| RM-Enclosure | oeMarket | --- | FC/APC bulkhead adaptor | --- | Y | http://v | 3 | 1 | $ 2.19 | USD | 0.886 | € 5.82 |
| RM-Enclosure | Element 14 | 1186435 | Tusonix 4400-093, EMC filter, Type C, Style 1, | --- | Y | http://w | 3 | 1 | $ 14.95 | AUD | 0.685 | € 30.72 |
| RM-Power | MeerKAT | --- | MeerKAT power board | | | | 1 | 1 | $ 200.00 | EUR | 1.000 | € 200.00 |

**Table 35 SKA1-MID RM (341-022300) capex equipment Detailed Cost Model Database**

### 4.10.3 Capital Expenditure Labour Costs

The summary of SKA1-MID capital expenditure labour cost as determined by the UWA Detailed Cost Model (Appendix 8.9.2) is given in Table 36.

| Item | Per unit | | | | | For all units | | | | |
|---|---|---|---|---|---|---|---|---|---|---|
| | Labour time at grade 40 (hours) | Labour time at grade 42 (hours) | Labour time at grade 44 (hours) | Total labour time (hours) | Labour base salary cost, excluding overhead (EUR) | Quantity of item (including 10% spares) | Labour time at grade 40 (hours) | Labour time at grade 42 (hours) | Labour time at grade 44 (hours) | Total labour time (hours) | Labour base salary cost, excluding overhead (EUR) |
| Rack Cabinet (341-022900) | 14 | 32 | 3.25 | 49.25 | € 1,719.54 | 2 | 28 | 64 | 6.5 | 98.5 | € 3,439.08 |
| Optical Source (341-022400) | 0.5 | 1.2 | 7.75 | 9.45 | € 478.55 | 2 | 1 | 2.4 | 15.5 | 18.9 | € 957.09 |
| Freq. Synthesiser (341-022500) | 0.5 | 1.56667 | 3.25 | 5.3167 | € 244.86 | 2 | 1 | 3.13333 | 6.5 | 10.633 | € 489.72 |
| Microwave Shift (341-022600) | 3 | 17.05 | 18.25 | 38.3 | € 1,634.31 | 2 | 6 | 34.1 | 36.5 | 76.6 | € 3,268.61 |
| Signal Generator (341-023100) | 0.5 | 1.8 | 3.25 | 5.55 | € 251.97 | 2 | 1 | 3.6 | 6.5 | 11.1 | € 503.94 |
| Rack Distribution (341-022800) | 2 | 15.8 | 4.25 | 22.05 | € 793.89 | 3 | 6 | 47.4 | 12.75 | 66.15 | € 2,381.66 |
| Sub Rack (341-022700) | 1 | 20.5 | 1.25 | 22.75 | € 733.37 | 10 | 10 | 205 | 12.5 | 227.5 | € 7,333.71 |
| Transmitter Module (341-022100) | 1 | 6.26667 | 2.75 | 10.017 | € 381.17 | 147 | 147 | 921.2 | 404.25 | 1472.5 | € 56,031.72 |
| Receiver Module (341-022300) | 1 | 11.1 | 1.75 | 13.85 | € 474.07 | 147 | 147 | 1631.7 | 257.25 | 2036 | € 69,688.52 |
| Optical Amplifier (341-022200) | 5 | 8.21667 | 2.25 | 15.467 | € 575.50 | 19 | 95 | 156.117 | 42.75 | 293.87 | € 10,934.43 |
| TOTALS | | | | | | | 442 | 3068.65 | 801 | 4311.7 | € 155,028.49 (excluding overheads) |

Table 36 SKA1-MID summary of the capex labour Detailed Cost Model Database

These values were independently determined by the consultancy firm Light Touch Solutions, and their Capex Labour Costing Analysis report is attached as Appendix 8.9.3.

The ancillary parameters used in the SKA1-MID capital expenditure labour cost Detailed Cost Model Database are given in Table 37.

| | | Yearly salary (AUD) | Hourly rate (AUD) | Hourly rate (EUR) |
|---|---|---|---|---|
| Technician, Grade = | 40 | $ 80,100 | $ 44.50 | € 30.48 |
| Assistant engineer, Grade = | 42 | $ 106,500 | $ 59.17 | € 40.53 |
| Senior engineer, Grade = | 44 | $ 143,000 | $ 79.44 | € 54.41 |
| Time per splice | 0.25 | | | |
| Time per fibre secure | 0.166666667 | | | |
| Time per RF connection | 0.083333333 | | | |
| Time per optical connection | 0.033333333 | | | |
| Time per RF connection | 0.033333333 | | | |
| Time per ethernet connection | 0.033333333 | | | |
| Time per power connection | 0.033333333 | | | |
| Time per cable secure | 0.083333333 | | | |

Table 37 SKA1-MID ancillary parameters used in the capex labour Detailed Cost Model Database

The colour key used in the SKA1-MID capital expenditure labour cost Detailed Cost Model Database is presented in Table 38.

| |
|---|
| * Indicated building of receiver/transmitter/Microwave shift/Rack distribution |
| Requires Check |
| Requires input |
| Do not change |

Table 38 SKA1-MID key to shading colours used in the capex labour Detailed Cost Model Database

### 4.10.3.1 Rack Cabinet (341-022900)

The Detailed Cost Model Database for the rack cabinet (341-022900) is shown in Table 39.

| Activity | Quantity | Time (hours) | Grade | Hourly rate for grade (EUR) | Cost (EUR) | | Hourly rate (EUR) |
|---|---|---|---|---|---|---|---|
| | | | | | | Technician, Grade = 40 | $ 30.48 |
| At sub-contractor's site: | | | | | | Assistant engineer, Grade = 42 | $ 40.53 |
| Receiving at sub-contractor site | | 4 | 40 | $ 40.53 | $ 162.10 | Senior engineer, Grade = 44 | $ 54.41 |
| * Preparation for building of RM/TM/MS/RD | | 0 | 42 | $ 30.48 | $ - | | |
| * Fibre splices | 0 | 0 | 42 | $ 30.48 | $ - | Time per splice | 0.25 |
| * Securing fibre component and fibres | 0 | 0 | 42 | $ 30.48 | $ - | Time per fibre secure | 0.166667 |
| * RF or microwave connections | 0 | 0 | 42 | $ 30.48 | $ - | Time per RF connection | 0.083333 |
| * Testing | | 0 | 44 | $ 54.41 | $ - | | |
| Assembly of rack-mounting | | 6 | 42 | $ 30.48 | $ 182.88 | | |
| Mounting onto rack | | 3 | 42 | $ 30.48 | $ 91.44 | | |
| Optical connections | 0 | 0 | 42 | $ 30.48 | $ - | Time per optical connection | 0.033333 |
| RF or microwave connections | 0 | 0 | 42 | $ 30.48 | $ - | Time per RF connection | 0.033333 |
| Ethernet connections | 0 | 0 | 42 | $ 30.48 | $ - | Time per ethernet connection | 0.033333 |
| Power connections | 0 | 0 | 42 | $ 30.48 | $ - | Time per power connection | 0.033333 |
| Fastening of cables | 0 | 0 | 42 | $ 30.48 | $ - | Time per cable secure | 0.083333 |
| Testing/verification including documentation | | 3 | 44 | $ 54.41 | $ 163.24 | | |
| Outgoing inspection | | 0.25 | 44 | $ 54.41 | $ 13.60 | | |
| packaging for transport | | 5 | 40 | $ 40.53 | $ 202.63 | | |
| Transport logistics | | 5 | 40 | $ 40.53 | $ 202.63 | | |
| At in-situ site: | | | | | | | |
| Travel time to in-situ site | | 7.5 | 42 | $ 30.48 | $ 228.60 | Total Labour time at grade 40: | 14 |
| Receiving at in-situ site | | 3.5 | 42 | $ 30.48 | $ 106.68 | Total Labour time at grade 42: | 32 |
| installation at in-situ site | | 3.5 | 42 | $ 30.48 | $ 106.68 | Total Labour time at grade 44: | 3.25 |
| Testing/verification | | 1 | 42 | $ 30.48 | $ 30.48 | | |
| Travel time return from in-situ site | | 7.5 | 42 | $ 30.48 | $ 228.60 | Total labour time at all grades: | 49.25 |
| | | | Labour base salary cost per module (EUR): | | $ 1,719.54 | (Excluding overheads) | |

**Table 39 SKA1-MID rack cabinet (341-022900) capex labour Detailed Cost Model Database**

### 4.10.3.2 Optical Source (341-022400)

The Detailed Cost Model Database for the OS (341-022400) is shown in Table 40.

| Activity | Quantity | Time (hours) | Grade | Hourly rate for grade (EUR) | Cost (EUR) | | | Hourly rate (EUR) |
|---|---|---|---|---|---|---|---|---|
| | | | | | | Technician, Grade = | 40 | $ 30.48 |
| **At sub-contractor's site:** | | | | | | Assistant engineer, Grade = | 42 | $ 40.53 |
| Receiving at sub-contractor site | | 0.5 | 40 | $ 40.53 | $ 20.26 | Senior engineer, Grade = | 44 | $ 54.41 |
| * Preparation for building of RM/TM/MS/RD | | 0 | 42 | $ 30.48 | $ - | | | |
| * Fibre splices | 0 | 0 | 42 | $ 30.48 | $ - | Time per splice | 0.25 | |
| * Securing fibre component and fibres | 0 | 0 | 42 | $ 30.48 | $ - | Time per fibre secure | 0.166667 | |
| * RF or microwave connections | 0 | 0 | 42 | $ 30.48 | $ - | Time per RF connection | 0.083333 | |
| * Testing | | 0 | 44 | $ 54.41 | $ - | | | |
| Assembly of rack-mounting | | 0.25 | 42 | $ 30.48 | $ 7.62 | | | |
| Mounting onto rack | | 0.5 | 42 | $ 30.48 | $ 15.24 | | | |
| Optical connections | 1 | 0.0333333 | 42 | $ 30.48 | $ 1.02 | Time per optical connection | 0.033333 | |
| RF or microwave connections | 0 | 0 | 42 | $ 30.48 | $ - | Time per RF connection | 0.033333 | |
| Ethernet connections | 1 | 0.0333333 | 42 | $ 30.48 | $ 1.02 | Time per ethernet connection | 0.033333 | |
| Power connections | 1 | 0.0333333 | 42 | $ 30.48 | $ 1.02 | Time per power connection | 0.033333 | |
| Fastening of cables | 3 | 0.25 | 42 | $ 30.48 | $ 7.62 | Time per cable secure | 0.083333 | |
| Testing/verification including documentation | | 7.5 | 44 | $ 54.41 | $ 408.11 | | | |
| Outgoing inspection | | 0.25 | 44 | $ 54.41 | $ 13.60 | | | |
| packaging for transport | | 0 | 40 | $ 40.53 | $ - | | | |
| Transport logistics | | 0 | 40 | $ 40.53 | $ - | | | |
| **At in-situ site:** | | | | | | | | |
| Travel time to in-situ site | | 0 | 42 | $ 30.48 | $ - | Total Labour time at grade 40: | 0.5 | |
| Receiving at in-situ site | | 0 | 42 | $ 30.48 | $ - | Total Labour time at grade 42: | 1.2 | |
| installation at in-situ site | | 0 | 42 | $ 30.48 | $ - | Total Labour time at grade 44: | 7.75 | |
| Testing/verification | | 0.1 | 42 | $ 30.48 | $ 3.05 | | | |
| Travel time return from in-situ site | | 0 | 42 | $ 30.48 | $ - | Total labour time at all grades: | 9.45 | |
| | | Labour base salary cost per module (EUR): | | $ | 478.55 | (Excluding overheads) | | |

**Table 40 SKA1-MID OS (341-022400) capex labour Detailed Cost Model Database**

### 4.10.3.3 Frequency Source (341-022500)

The Detailed Cost Model Database for the FS (341-022500) is shown in Table 41.

| Activity | Quantity | Time (hours) | Grade | Hourly rate for grade (EUR) | Cost (EUR) | | | Hourly rate (EUR) |
|---|---|---|---|---|---|---|---|---|
| | | | | | | Technician, Grade = | 40 | $ 30.48 |
| **At sub-contractor's site:** | | | | | | Assistant engineer, Grade = | 42 | $ 40.53 |
| Receiving at sub-contractor site | | 0.5 | 40 | $ 40.53 | $ 20.26 | Senior engineer, Grade = | 44 | $ 54.41 |
| * Preparation for building of RM/TM/MS/RD | | 0 | 42 | $ 30.48 | $ - | | | |
| * Fibre splices | 0 | 0 | 42 | $ 30.48 | $ - | Time per splice | 0.25 | |
| * Securing fibre component and fibres | 0 | 0 | 42 | $ 30.48 | $ - | Time per fibre secure | 0.166667 | |
| * RF or microwave connections | 0 | 0 | 42 | $ 30.48 | $ - | Time per RF connection | 0.083333 | |
| * Testing | | 0 | 44 | $ 54.41 | $ - | | | |
| Assembly of rack-mounting | | 0 | 42 | $ 30.48 | $ - | | | |
| Mounting onto rack | | 1 | 42 | $ 30.48 | $ 30.48 | | | |
| Optical connections | 0 | 0 | 42 | $ 30.48 | $ - | Time per optical connection | 0.033333 | |
| RF or microwave connections | 2 | 0.0666667 | 42 | $ 30.48 | $ 2.03 | Time per RF connection | 0.033333 | |
| Ethernet connections | 1 | 0.0333333 | 42 | $ 30.48 | $ 1.02 | Time per ethernet connection | 0.033333 | |
| Power connections | 1 | 0.0333333 | 42 | $ 30.48 | $ 1.02 | Time per power connection | 0.033333 | |
| Fastening of cables | 4 | 0.3333333 | 42 | $ 30.48 | $ 10.16 | Time per cable secure | 0.083333 | |
| Testing/verification including documentation | | 3 | 44 | $ 54.41 | $ 163.24 | | | |
| Outgoing inspection | | 0.25 | 44 | $ 54.41 | $ 13.60 | | | |
| packaging for transport | | 0 | 40 | $ 40.53 | $ - | | | |
| Transport logistics | | 0 | 40 | $ 40.53 | $ - | | | |
| **At in-situ site:** | | | | | | | | |
| Travel time to in-situ site | | 0 | 42 | $ 30.48 | $ - | Total Labour time at grade 40: | 0.5 | |
| Receiving at in-situ site | | 0 | 42 | $ 30.48 | $ - | Total Labour time at grade 42: | 1.566667 | |
| installation at in-situ site | | 0 | 42 | $ 30.48 | $ - | Total Labour time at grade 44: | 3.25 | |
| Testing/verification | | 0.1 | 42 | $ 30.48 | $ 3.05 | | | |
| Travel time return from in-situ site | | 0 | 42 | $ 30.48 | $ - | Total labour time at all grades: | 5.316667 | |
| | | | | Labour base salary cost per module (EUR): | $ 244.86 | (Excluding overheads) | | |

**Table 41 SKA1-MID FS (341-022500) capex labour Detailed Cost Model Database**

### 4.10.3.4 Microwave Shift (341-022600)

The Detailed Cost Model Database for the MS (341-022600) is shown in Table 42.

| Activity | Quantity | Time (hours) | Grade | Hourly rate for grade (EUR) | Cost (EUR) | | | Hourly rate (EUR) |
|---|---|---|---|---|---|---|---|---|
| | | | | | | Technician, Grade = | 40 | $ 30.48 |
| **At sub-contractor's site:** | | | | | | Assistant engineer, Grade = | 42 | $ 40.53 |
| Receiving at sub-contractor site | | 3 | 40 | $ 40.53 | $ 121.58 | Senior engineer, Grade = | 44 | $ 54.41 |
| * Preparation for building of RM/TM/MS/RD | | 4 | 42 | $ 30.48 | $ 121.92 | | | |
| * Fibre splices | 4 | 1 | 42 | $ 30.48 | $ 30.48 | Time per splice | 0.25 | |
| * Securing fibre component and fibres | 8 | 1.3333333 | 42 | $ 30.48 | $ 40.64 | Time per fibre secure | 0.166667 | |
| * RF or microwave connections | 8 | 0.6666667 | 42 | $ 30.48 | $ 20.32 | Time per RF connection | 0.083333 | |
| * Testing (inc. calibration of voltage) | | 15 | 44 | $ 54.41 | $ 816.21 | | | |
| Assembly of rack-mounting | | 0.5 | 42 | $ 30.48 | $ 15.24 | | | |
| Mounting onto rack | | 0.5 | 42 | $ 30.48 | $ 15.24 | | | |
| Optical connections | 4 | 0.1333333 | 42 | $ 30.48 | $ 4.06 | Time per optical connection | 0.033333 | |
| RF or microwave connections | 3 | 0.1 | 42 | $ 30.48 | $ 3.05 | Time per RF connection | 0.033333 | |
| Ethernet connections | 1 | 0.0333333 | 42 | $ 30.48 | $ 1.02 | Time per ethernet connection | 0.033333 | |
| Power connections | 1 | 0.0333333 | 42 | $ 30.48 | $ 1.02 | Time per power connection | 0.033333 | |
| Fastening of cables | 9 | 0.75 | 42 | $ 30.48 | $ 22.86 | Time per cable secure | 0.083333 | |
| Testing/verification including documentation | | 3 | 44 | $ 54.41 | $ 163.24 | | | |
| Outgoing inspection | | 0.25 | 44 | $ 54.41 | $ 13.60 | | | |
| packaging for transport | | 0 | 40 | $ 40.53 | $ - | | | |
| Transport logistics | | 0 | 40 | $ 40.53 | $ - | | | |
| **At in-situ site:** | | | | | | | | |
| Travel time to in-situ site | | 0 | 42 | $ 30.48 | $ - | Total Labour time at grade 40: | 3 | |
| Receiving at in-situ site | | 0 | 42 | $ 30.48 | $ - | Total Labour time at grade 42: | 17.05 | |
| installation at in-situ site | | 0 | 42 | $ 30.48 | $ - | Total Labour time at grade 44: | 18.25 | |
| Testing/verification (inc. calibration of voltage) | | 8 | 42 | $ 30.48 | $ 243.84 | | | |
| Travel time return from in-situ site | | 0 | 42 | $ 30.48 | $ - | Total labour time at all grades: | 38.3 | |
| | | | | Labour base salary cost per module (EUR): | $ 1,634.31 | (Excluding overheads) | | |

**Table 42 SKA1-MID MS (341-022600) capex labour Detailed Cost Model Database**

### 4.10.3.5 Signal Generator (341-023100)

The Detailed Cost Model Database for the SG (341-023100) is shown in Table 43.

| Activity | Quantity | Time (hours) | Grade | Hourly rate for grade (EUR) | Cost (EUR) | | | Hourly rate (EUR) |
|---|---|---|---|---|---|---|---|---|
| | | | | | | Technician, Grade = | 40 | $ 30.48 |
| **At sub-contractor's site:** | | | | | | Assistant engineer, Grade = | 42 | $ 40.53 |
| Receiving at sub-contractor site | | 0.5 | 40 | $ 40.53 | $ 20.26 | Senior engineer, Grade = | 44 | $ 54.41 |
| * Preparation for building of RM/TM/MS/RD | | 0 | 42 | $ 30.48 | $ - | | | |
| * Fibre splices | 0 | 0 | 42 | $ 30.48 | $ - | Time per splice | 0.25 | |
| * Securing fibre component and fibres | 0 | 0 | 42 | $ 30.48 | $ - | Time per fibre secure | 0.166667 | |
| * RF or microwave connections | 0 | 0 | 42 | $ 30.48 | $ - | Time per RF connection | 0.083333 | |
| * Testing | | 0 | 44 | $ 54.41 | $ - | | | |
| Assembly of rack-mounting | | 0 | 42 | $ 30.48 | $ - | | | |
| Mounting onto rack | | 1 | 42 | $ 30.48 | $ 30.48 | | | |
| Optical connections | 0 | 0 | 42 | $ 30.48 | $ - | Time per optical connection | 0.033333 | |
| RF or microwave connections | 4 | 0.1333333 | 42 | $ 30.48 | $ 4.06 | Time per RF connection | 0.033333 | |
| Ethernet connections | 1 | 0.0333333 | 42 | $ 30.48 | $ 1.02 | Time per ethernet connection | 0.033333 | |
| Power connections | 1 | 0.0333333 | 42 | $ 30.48 | $ 1.02 | Time per power connection | 0.033333 | |
| Fastening of cables | 6 | 0.5 | 42 | $ 30.48 | $ 15.24 | Time per cable secure | 0.083333 | |
| Testing/verification including documentation | | 3 | 44 | $ 54.41 | $ 163.24 | | | |
| Outgoing inspection | | 0.25 | 44 | $ 54.41 | $ 13.60 | | | |
| packaging for transport | | 0 | 40 | $ 40.53 | $ - | | | |
| Transport logistics | | 0 | 40 | $ 40.53 | $ - | | | |
| **At in-situ site:** | | | | | | | | |
| Travel time to in-situ site | | 0 | 42 | $ 30.48 | $ - | Total Labour time at grade 40: | 0.5 | |
| Receiving at in-situ site | | 0 | 42 | $ 30.48 | $ - | Total Labour time at grade 42: | 1.8 | |
| installation at in-situ site | | 0 | 42 | $ 30.48 | $ - | Total Labour time at grade 44: | 3.25 | |
| Testing/verification | | 0.1 | 42 | $ 30.48 | $ 3.05 | | | |
| Travel time return from in-situ site | | 0 | 42 | $ 30.48 | $ - | Total labour time at all grades: | 5.55 | |
| | | Labour base salary cost per module (EUR): | | | $ 251.97 | (Excluding overheads) | | |

**Table 43 SKA1-MID SG (341-023100) capex labour Detailed Cost Model Database**

### 4.10.3.6 Rack Distribution (341-022800)

The Detailed Cost Model Database for the RD (341-022800) is shown in Table 44.

| Activity | Quantity | Time (hours) | Grade | Hourly rate for grade (EUR) | Cost (EUR) | | | Hourly rate (EUR) |
|---|---|---|---|---|---|---|---|---|
| | | | | | | Technician, Grade = | 40 | $ 30.48 |
| **At sub-contractor's site:** | | | | | | Assistant engineer, Grade = | 42 | $ 40.53 |
| Receiving at sub-contractor site | | 2 | 40 | $ 40.53 | $ 81.05 | Senior engineer, Grade = | 44 | $ 54.41 |
| * Preparation for building of RM/TM/MS/RD | | 5 | 42 | $ 30.48 | $ 152.40 | | | |
| * Fibre splices | 0 | 0 | 42 | $ 30.48 | $ - | Time per splice | 0.25 | |
| * Securing fibre component and fibres | 18 | 3 | 42 | $ 30.48 | $ 91.44 | Time per fibre secure | 0.166667 | |
| * RF or microwave connections | 18 | 1.5 | 42 | $ 30.48 | $ 45.72 | Time per RF connection | 0.083333 | |
| * Testing | | 2 | 44 | $ 54.41 | $ 108.83 | | | |
| Assembly of rack-mounting | | 1 | 42 | $ 30.48 | $ 30.48 | | | |
| Mounting onto rack | | 1 | 42 | $ 30.48 | $ 30.48 | | | |
| Optical connections | 18 | 0.6 | 42 | $ 30.48 | $ 18.29 | Time per optical connection | 0.033333 | |
| RF or microwave connections | 18 | 0.6 | 42 | $ 30.48 | $ 18.29 | Time per RF connection | 0.033333 | |
| Ethernet connections | 0 | 0 | 42 | $ 30.48 | $ - | Time per ethernet connection | 0.033333 | |
| Power connections | 0 | 0 | 42 | $ 30.48 | $ - | Time per power connection | 0.033333 | |
| Fastening of cables | 36 | 3 | 42 | $ 30.48 | $ 91.44 | Time per cable secure | 0.083333 | |
| Testing/verification including documentation | | 2 | 44 | $ 54.41 | $ 108.83 | | | |
| Outgoing inspection | | 0.25 | 44 | $ 54.41 | $ 13.60 | | | |
| packaging for transport | | 0 | 40 | $ 40.53 | $ - | | | |
| Transport logistics | | 0 | 40 | $ 40.53 | $ - | | | |
| **At in-situ site:** | | | | | | | | |
| Travel time to in-situ site | | 0 | 42 | $ 30.48 | $ - | Total Labour time at grade 40: | 2 | |
| Receiving at in-situ site | | 0 | 42 | $ 30.48 | $ - | Total Labour time at grade 42: | 15.8 | |
| installation at in-situ site | | 0 | 42 | $ 30.48 | $ - | Total Labour time at grade 44: | 4.25 | |
| Testing/verification | | 0.1 | 42 | $ 30.48 | $ 3.05 | | | |
| Travel time return from in-situ site | | 0 | 42 | $ 30.48 | $ - | Total labour time at all grades: | 22.05 | |
| | | | | Labour base salary cost per module (EUR): | $ 793.89 | (Excluding overheads) | | |

**Table 44 SKA1-MID RD (341-022800) capex labour Detailed Cost Model Database**

### 4.10.3.7 Sub Rack (341-022700)

The Detailed Cost Model Database for the SR (341-022700) is shown in Table 45.

| Activity | Quantity | Time (hours) | Grade | Hourly rate for grade (EUR) | Cost (EUR) | | | Hourly rate (EUR) |
|---|---|---|---|---|---|---|---|---|
| | | | | | | Technician, Grade = | 40 | $ 30.48 |
| **At sub-contractor's site:** | | | | | | Assistant engineer, Grade = | 42 | $ 40.53 |
| Receiving at sub-contractor site | | 1 | 40 | $ 40.53 | $ 40.53 | Senior engineer, Grade = | 44 | $ 54.41 |
| * Preparation for building of RM/TM/MS/RD | | 0 | 42 | $ 30.48 | $ - | | | |
| * Fibre splices | 0 | 0 | 42 | $ 30.48 | $ - | Time per splice | 0.25 | |
| * Securing fibre component and fibres | 0 | 0 | 42 | $ 30.48 | $ - | Time per fibre secure | 0.166667 | |
| * RF or microwave connections | 0 | 0 | 42 | $ 30.48 | $ - | Time per RF connection | 0.083333 | |
| * Testing | | 0 | 44 | $ 54.41 | $ - | | | |
| Assembly of rack-mounting | | 7.5 | 42 | $ 30.48 | $ 228.60 | | | |
| Mounting onto rack | | 1 | 42 | $ 30.48 | $ 30.48 | | | |
| Optical connections | 34 | 1.1333333 | 42 | $ 30.48 | $ 34.54 | Time per optical connection | 0.033333 | |
| RF or microwave connections | 34 | 1.1333333 | 42 | $ 30.48 | $ 34.54 | Time per RF connection | 0.033333 | |
| Ethernet connections | 17 | 0.5666667 | 42 | $ 30.48 | $ 17.27 | Time per ethernet connection | 0.033333 | |
| Power connections | 17 | 0.5666667 | 42 | $ 30.48 | $ 17.27 | Time per power connection | 0.033333 | |
| Fastening of cables | 102 | 8.5 | 42 | $ 30.48 | $ 259.08 | Time per cable secure | 0.083333 | |
| Testing/verification including documentation | | 1 | 44 | $ 54.41 | $ 54.41 | | | |
| Outgoing inspection | | 0.25 | 44 | $ 54.41 | $ 13.60 | | | |
| packaging for transport | | 0 | 40 | $ 40.53 | $ - | | | |
| Transport logistics | | 0 | 40 | $ 40.53 | $ - | | | |
| **At in-situ site:** | | | | | | | | |
| Travel time to in-situ site | | 0 | 42 | $ 30.48 | $ - | Total Labour time at grade 40: | 1 | |
| Receiving at in-situ site | | 0 | 42 | $ 30.48 | $ - | Total Labour time at grade 42: | 20.5 | |
| installation at in-situ site | | 0 | 42 | $ 30.48 | $ - | Total Labour time at grade 44: | 1.25 | |
| Testing/verification | | 0.1 | 42 | $ 30.48 | $ 3.05 | | | |
| Travel time return from in-situ site | | 0 | 42 | $ 30.48 | $ - | Total labour time at all grades: | 22.75 | |
| | | | Labour base salary cost per module (EUR): | $ | 733.37 | (Excluding overheads) | | |

**Table 45 SKA1-MID SR (341-022700) capex labour Detailed Cost Model Database**

### 4.10.3.8 Transmitter Module (341-022100)

The Detailed Cost Model Database for the TM (341-022100) is shown in Table 46.

| Activity | Quantity | Time (hours) | Grade | Hourly rate for grade (EUR) | Cost (EUR) | | | Hourly rate (EUR) |
|---|---|---|---|---|---|---|---|---|
| | | | | | | Technician, Grade = | 40 | $ 30.48 |
| **At sub-contractor's site:** | | | | | | Assistant engineer, Grade = | 42 | $ 40.53 |
| Receiving at sub-contractor site | | 1 | 40 | $ 40.53 | $ 40.53 | Senior engineer, Grade = | 44 | $ 54.41 |
| * Preparation for building of RM/TM/MS/RD | | 2 | 42 | $ 30.48 | $ 60.96 | | | |
| * Fibre splices | 4 | 1 | 42 | $ 30.48 | $ 30.48 | Time per splice | 0.25 | |
| * Securing fibre component and fibres | 6 | 1 | 42 | $ 30.48 | $ 30.48 | Time per fibre secure | 0.166667 | |
| * RF or microwave connections | 7 | 0.5833333 | 42 | $ 30.48 | $ 17.78 | Time per RF connection | 0.083333 | |
| * Testing | | 2 | 44 | $ 54.41 | $ 108.83 | | | |
| Assembly of rack-mounting | | 0.5 | 42 | $ 30.48 | $ 15.24 | | | |
| Mounting onto rack | | 0.5 | 42 | $ 30.48 | $ 15.24 | | | |
| Optical connections | 1 | 0.0333333 | 42 | $ 30.48 | $ 1.02 | Time per optical connection | 0.033333 | |
| RF or microwave connections | 2 | 0.0666667 | 42 | $ 30.48 | $ 2.03 | Time per RF connection | 0.033333 | |
| Ethernet connections | 1 | 0.0333333 | 42 | $ 30.48 | $ 1.02 | Time per ethernet connection | 0.033333 | |
| Power connections | 1 | 0.0333333 | 42 | $ 30.48 | $ 1.02 | Time per power connection | 0.033333 | |
| Fastening of cables | 5 | 0.4166667 | 42 | $ 30.48 | $ 12.70 | Time per cable secure | 0.083333 | |
| Testing/verification including documentation | | 0.5 | 44 | $ 54.41 | $ 27.21 | | | |
| Outgoing inspection | | 0.25 | 44 | $ 54.41 | $ 13.60 | | | |
| packaging for transport | | 0 | 40 | $ 40.53 | $ - | | | |
| Transport logistics | | 0 | 40 | $ 40.53 | $ - | | | |
| **At in-situ site:** | | | | | | | | |
| Travel time to in-situ site | | 0 | 42 | $ 30.48 | $ - | Total Labour time at grade 40: | 1 | |
| Receiving at in-situ site | | 0 | 42 | $ 30.48 | $ - | Total Labour time at grade 42: | 6.266667 | |
| installation at in-situ site | | 0 | 42 | $ 30.48 | $ - | Total Labour time at grade 44: | 2.75 | |
| Testing/verification | | 0.1 | 42 | $ 30.48 | $ 3.05 | | | |
| Travel time return from in-situ site | | 0 | 42 | $ 30.48 | $ - | Total labour time at all grades: | 10.01667 | |
| | | | | Labour base salary cost per module (EUR): | $ 381.17 | (Excluding overheads) | | |

**Table 46 SKA1-MID TM (341-022100) capex labour Detailed Cost Model Database**

### 4.10.3.9 Fibre Patch Lead (341-023200)

The labour base cost for a FP lead (341-023200) module is estimated to be 30 EUR (excluding overheads).

### 4.10.3.10 Optical Amplifier (341-022200)

The Detailed Cost Model Database for the OA (341-022200) is shown in Table 47.

| Activity | Quantity | Time (hours) | Grade | Hourly rate for grade (EUR) | Cost (EUR) | | | |
|---|---|---|---|---|---|---|---|---|
| | | | | | | Technician, Grade = | 40 | $ 30.48 |
| At sub-contractor's site: | | | | | | Assistant engineer, Grade = | 42 | $ 40.53 |
| Receiving at sub-contractor site | | 1 | 40 | $ 40.53 | $ 40.53 | Senior engineer, Grade = | 44 | $ 54.41 |
| * Preparation for building of RM/TM/MS/RD | | 0 | 42 | $ 30.48 | $ - | | | |
| * Fibre splices | 0 | 0 | 42 | $ 30.48 | $ - | Time per splice | 0.25 | |
| * Securing fibre component and fibres | 0 | 0 | 42 | $ 30.48 | $ - | Time per fibre secure | 0.166667 | |
| * RF or microwave connections | 0 | 0 | 42 | $ 30.48 | $ - | Time per RF connection | 0.083333 | |
| * Testing | | 0 | 44 | $ 54.41 | $ - | | | |
| Assembly of rack-mounting | | 0.5 | 42 | $ 30.48 | $ 15.24 | | | |
| Mounting onto rack | | 0.5 | 42 | $ 30.48 | $ 15.24 | | | |
| Optical connections | 2 | 0.0666667 | 42 | $ 30.48 | $ 2.03 | Time per optical connection | 0.033333 | |
| RF or microwave connections | 0 | 0 | 42 | $ 30.48 | $ - | Time per RF connection | 0.033333 | |
| Ethernet connections | 1 | 0.0333333 | 42 | $ 30.48 | $ 1.02 | Time per ethernet connection | 0.033333 | |
| Power connections | 1 | 0.0333333 | 42 | $ 30.48 | $ 1.02 | Time per power connection | 0.033333 | |
| Fastening of cables | 4 | 0.3333333 | 42 | $ 30.48 | $ 10.16 | Time per cable secure | 0.083333 | |
| Testing/verification including documentation | | 2 | 44 | $ 54.41 | $ 108.83 | | | |
| Outgoing inspection | | 0.25 | 44 | $ 54.41 | $ 13.60 | | | |
| packaging for transport | | 2 | 40 | $ 40.53 | $ 81.05 | | | |
| Transport logistics | | 2 | 40 | $ 40.53 | $ 81.05 | | | |
| At in-situ site: | | | | | | | | |
| Travel time to in-situ site | | 2.5 | 42 | $ 30.48 | $ 76.20 | Total Labour time at grade 40: | 5 | |
| Receiving at in-situ site | | 0.25 | 42 | $ 30.48 | $ 7.62 | Total Labour time at grade 42: | 8.216667 | |
| installation at in-situ site | | 1 | 42 | $ 30.48 | $ 30.48 | Total Labour time at grade 44: | 2.25 | |
| Testing/verification | | 0.5 | 42 | $ 30.48 | $ 15.24 | | | |
| Travel time return from in-situ site | | 2.5 | 42 | $ 30.48 | $ 76.20 | Total labour time at all grades: | 15.46667 | |
| | | | | Labour base salary cost per module (EUR): | $ 575.50 | (Excluding overheads) | | |

**Table 47 SKA1-MID OA (341-022200) capex labour Detailed Cost Model Database**

### 4.10.3.11  Receiver Module (341-022300)

The Detailed Cost Model Database for the RM (341-022300) is shown in Table 48.

| Activity | Quantity | Time (hours) | Grade | Hourly rate for grade (EUR) | Cost (EUR) | | | Hourly rate (EUR) |
|---|---|---|---|---|---|---|---|---|
| | | | | | | Technician, Grade = | 40 | $ 30.48 |
| **At sub-contractor's site:** | | | | | | Assistant engineer, Grade = | 42 | $ 40.53 |
| Receiving at sub-contractor site | | 1 | 40 | $ 40.53 | $ 40.53 | Senior engineer, Grade = | 44 | $ 54.41 |
| * Preparation for building of RM/TM/MS/RD | | 2 | 42 | $ 30.48 | $ 60.96 | | | |
| * Fibre splices | 2 | 0.5 | 42 | $ 30.48 | $ 15.24 | Time per splice | 0.25 | |
| * Securing fibre component and fibres | 2 | 0.3333333 | 42 | $ 30.48 | $ 10.16 | Time per fibre secure | 0.166667 | |
| * RF or microwave connections | 2 | 0.1666667 | 42 | $ 30.48 | $ 5.08 | Time per RF connection | 0.083333 | |
| * Testing | | 1 | 44 | $ 54.41 | $ 54.41 | | | |
| Assembly of rack-mounting | | 0.5 | 42 | $ 30.48 | $ 15.24 | | | |
| Mounting onto rack | | 0.5 | 42 | $ 30.48 | $ 15.24 | | | |
| Optical connections | 1 | 0.0333333 | 42 | $ 30.48 | $ 1.02 | Time per optical connection | 0.033333 | |
| RF or microwave connections | 1 | 0.0333333 | 42 | $ 30.48 | $ 1.02 | Time per RF connection | 0.033333 | |
| Ethernet connections | 0 | 0 | 42 | $ 30.48 | $ - | Time per ethernet connection | 0.033333 | |
| Power connections | 1 | 0.0333333 | 42 | $ 30.48 | $ 1.02 | Time per power connection | 0.033333 | |
| Fastening of cables | 3 | 0.25 | 42 | $ 30.48 | $ 7.62 | Time per cable secure | 0.083333 | |
| Testing/verification including documentation | | 0.5 | 44 | $ 54.41 | $ 27.21 | | | |
| Outgoing inspection | | 0.25 | 44 | $ 54.41 | $ 13.60 | | | |
| packaging for transport | | 0 | 40 | $ 40.53 | $ - | | | |
| Transport logistics | | 0 | 40 | $ 40.53 | $ - | | | |
| **At in-situ site:** | | | | | | | | |
| Travel time to in-situ site | | 2.5 | 42 | $ 30.48 | $ 76.20 | Total Labour time at grade 40: | 1 | |
| Receiving at in-situ site | | 0.25 | 42 | $ 30.48 | $ 7.62 | Total Labour time at grade 42: | 11.1 | |
| installation at in-situ site | | 1 | 42 | $ 30.48 | $ 30.48 | Total Labour time at grade 44: | 1.75 | |
| Testing/verification | | 0.5 | 42 | $ 30.48 | $ 15.24 | | | |
| Travel time return from in-situ site | | 2.5 | 42 | $ 30.48 | $ 76.20 | Total labour time at all grades: | 13.85 | |
| | | | | Labour base salary cost per module (EUR): | $ 474.07 | (Excluding overheads) | | |

**Table 48 SKA1-MID RM (341-022300) capex labour Detailed Cost Model Database**

# 5 EVALUATION

## 5.1 Evaluation

Between February 2014 and July 2017, the SKA phase synchronisation system was evaluated using a total of 17 test campaigns, as defined in [RD58], and outlined in §5.1.1. These tests successfully managed to verify by demonstration the functionality of the system against all SKA requirements, as well as many other more rigorous practical considerations.

In addition, UWA researchers in partnership with MeerKAT and UoM electronic engineers, progressed the detailed designs described in this document, into a set of mass manufacture archetypes as described in §4.8.1.

The report on the construction of the SKA phase synchronisation system mass manufacture archetype [RD22] is summarised in §5.1.2.

### 5.1.1 Test and Verification by Demonstration

The following tests are relevant to SKA1-MID:

- Test(1) – SKA STFR.FRQ (UWA) v1 – Preliminary Laboratory Test (§5.1.1.1).
- Test(3) – SKA-MID STFR.FRQ (UWA) v1 – Laboratory Demonstration (§5.1.1.2).
- Test(5) – SKA STFR.FRQ (UWA) v2 – TM Stabilisation (§5.1.1.3).
- Test(7) – SKA1 Overhead Optical Fibre Characterisation Field Trial (§5.1.1.4).
- Test(8) – SKA1 STFR.FRQ (UWA) v2 – Long-Haul Overhead Fibre Verification (§5.1.1.5).
- Test(9) – SKA-MID STFR.FRQ (UWA) v2 – Astronomical Verification (§5.1.1.6).
- Test(11) – SKA-MID STFR.FRQ (UWA) v2 – Laboratory Demonstration (§5.1.1.7).
- Test(13) – SKA-MID STFR.FRQ (UWA) v2 – Temperature and Humidity (§5.1.1.8).
- Test(15) – SKA-MID STFR.FRQ (UWA) v2 – Seismic Resilience (§5.1.1.9).
- Test(17) – SKA-MID STFR.FRQ (UWA) v2 – Electromagnetic Compatibility (§5.1.1.10).

#### 5.1.1.1 Test(1) – SKA STFR.FRQ (UWA) v1 – Preliminary Laboratory Test

Refer to [RD12].

This document reports on the laboratory verification of the dual-servo version of the SKA frequency transfer system STFR.FRQ (UWA) for stabilising the RF or MW frequency separation between two optical signals. The dual-servo version is indicated by "v1" to distinguish it from the latter version that only uses a single servo; that is, the single-servo version, designated "v2".

The transition from "v1" to "v2" occurred following the first SKA-LOW astronomical verification field trail with the ASKAP in November 2014. This period was after the of the Preliminary Design Review (PDR) documentation submission, so the PDR 'SADT.SAT.STFR Design Report' contains information about the v1 system only. By the PDR assessment panel meeting, the transition to v2 was complete. The full explanation of the transition including the motivation and process are described in the UWA 'SKA-LOW Astronomical Verification' report.

Therefore, the technical details of the system described in this report are strictly not applicable to the current design ("v2") that UWA adopted following the PDR. The changes in technical details between "v1" and "v2" however, are minor enough that all of the key system characteristics described in §5.1.1.2 are unchanged.

The tests described in this section of the report were performed in the UWA SADT laboratory over 0 km (short patch leads) and a 1 km spool of fibre. A 70 MHz RF signal was transferred using an AOM to produce the frequency separation between two optical frequencies.

The results of this series of tests confirmed that the dual-servo system worked as planned and has provided confidence that its performance would also be sufficient to meet the SKA performance requirements at the higher MW frequencies. The results were presented at the SADT face-to-face meeting in June 2014 and at SAT Kick-Off Meeting in Beijing.

The aim of this series of tests was to confirm that phase-stabilising two optical frequencies independently would result in effective phase-stabilisation of the frequency difference between the two optical signals. These tests also provided preliminary data sufficient to confirm that, in principle, the performance of this frequency transfer technique is sufficient to meet SKA performance requirements.

Key FRQ (UWA) system characteristics specifically investigated as part of this test:

- Transfer stability
- Transfer accuracy.

### 5.1.1.2 Test(3) – SKA-MID STFR.FRQ (UWA) v1 – Laboratory Demonstration

Refer to [RD24] and [RD12].

The SKA FRQ (UWA) preliminary test described in §1 and Laboratory Prototype described in §4 demonstrate the ability of the SKA frequency transfer system to provide a stabilised RF frequency at a remote site using a constant RF frequency shift in one arm of the interferometer. To provide the higher reference frequencies required by SKA1-MID, the SKA frequency transfer system must also be demonstrated to work with a MW frequency shift. The MW version of the UWA system is referred to as SKA1-MID STFR.FRQ (UWA).

This section describes the construction and characterization of the SKA1-MID Laboratory Prototype transferring frequencies of 2 GHz and 10 GHz over distances of up to 30 km (with additional optical attenuation simulating the power loss of a further 60 km of optical fibre for the 2 GHz tests).

The SKA1-MID prototype was initially tested at a transfer frequency of 2 GHz, as this was the upper frequency limit imposed by the bandwidth of the ancillary electronics used the drive the MW frequency shifting device. The 2 GHz tests were performed between 23$^{rd}$ April and 2$^{nd}$ May 2014. Electronics capable of handling higher frequencies were purchased and system tests at 10 GHz were completed between 16$^{th}$ May and 2$^{nd}$ September 2014. 10 GHz was selected as it was more technologically representative of the frequencies required by different SKA1-MID receiver bands, and for practical reasons, as the 20 GHz second-harmonic of this signal would be visible to the 20 GHz signal analysis equipment for diagnostic purposes.

The stability of the individual optical signals was measured in parallel with the stability of the MW signal for a selection of the 10 GHz tests. This allowed the comparison of optical and MW performance and characteristics of the system. An extended set of results for these dual optical-MW measurements were obtained between 1$^{st}$ July and 25$^{th}$ August 2014.

The aim of this series of tests was to demonstrate the performance and phase stability of the system at MW frequencies as well as verify additional system characteristics that could not be tested with the previous RF setups. These characteristics include selectable transmission frequency, the effective transmission distance, efficient sub-arraying for distribution to multiple receiver sites, and the fact that the UWA system does not suffer from signal fading apparent in typical transmission systems that use dual-sideband modulation.

The aim of making parallel measurements of the stability of the two independent optical signals and the MW frequency signal resulting from the frequency separation of the two optical signals was to experimentally investigate the relationship between the stability of the optical signals and the resulting

MW signal.

Key FRQ (UWA) system characteristics specifically investigated as part of this test were:

- Transfer stability
- Optical amplification
- Fibre bandwidth
- Signal fading
- Efficient sub-arraying
- Immunity from reflections
- Polarisation alignment
- Fast frequency switching
- Selectable transmission frequency
- Frequency offset
- OS
- Fibre topology
- Specialty fibre.

### 5.1.1.3 Test(5) – SKA STFR.FRQ (UWA) v2 – TM Stabilisation

Refer to [RD19].

A key issue encountered when initially testing the dual-servo version (v1) of the SKA frequency transfer system was that acoustic vibrations in the TM were significantly degrading the stability of the transmitted 16 MHz signal. This was attributed to the fact that:

- Approximately 0.5 m of fibre in the beginning of each arm of the MZI was outside the servo loop, and therefore any relative thermal and/or force gradients between the MZI arms led to frequency fluctuations which could not be compensated for by the servos.
- The output frequency of the MZI was especially sensitive to thermal/acoustic fluctuations in the TM since each arm was spatially separated.

The sensitivity of the MZI was characterised using a vibration rig at acoustic frequencies, and analysing the spectrum of an error signal representing the relative phase fluctuations in each arm of the interferometer.

The aims of this test are to eliminate the sensitivity of the TM to environmental conditions at the local site.

The key FRQ (UWA) system characteristic specifically investigated as part of this test was the environment at the local site.

### 5.1.1.4 Test(7) – SKA1 Overhead Optical Fibre Characterisation Field Trial

Refer to [RD15] and [RD14].

The UWA High-Precision Fibre Characterisation System revealed that overhead optical fibre links are significantly more susceptible to environmental impacts, when compared to non-overhead links of comparable length. In the environmental conditions present during the trials, this resulted in frequency fluctuations over 1,000 times greater (at one second of integration) than those previously measured using the same system on conventional links (buried, conduit, spool). To put this into perspective, the 153 km South African overhead fibre link exhibits about 500 times more frequency noise than the 1840 km world-record link across Germany [RD59]. This additional noise was measured to be related to the severity of the environmental conditions (specifically, wind speed), and so it can be expected that more severe weather conditions will result in frequency fluctuations greater than the 500 times increase measured.

Two optical fibre reticulation options are being considered for the spiral arm component of SKA1-MID's SAT system. The default option is a trenched route outside the core region that has to avoid physical obstructions. This results in a route with a total length of 604 km across the three spiral arms, or an average of around 200 km per spiral arm. This is approximately 50% longer than a route that directly follows the layout of the spiral arms determined to be 393 km, or about 130 km per arm. With the additional length of the trenched fibre in the core, the maximum SKA1-MID arm length for this configuration is about 150 km.

The total SADT local infrastructure costs of the trenched route based on the default option is currently calculated to be around 11 million euros. However, if the SAT system can be shown to be able to operate on overhead fibre, then all of SADT fibre cable could simply be suspended on the same overhead poles that distribute power along the spiral arms. This route has a total distance of only 394 km across the three spiral arms. Therefore, the only infrastructure cost specific for SAT are the additional (now shorter) SAT fibre cables, estimated to be less than 300 thousand euros. Detailed cost analysis studies for moving all SADT fibre onto the overhead route are ongoing, but it is expected that not having to trench the SKA1-MID spiral arms could save the SKA project several million euros.

To quantify the difference between overhead and conventional fibre links in terms of SAT frequency transfer, several members of the SADT Consortium deployed the UWA High-Precision Fibre Characterisation System over various overhead fibre links on the South African SKA site between Sunday 31$^{st}$ of May and Wednesday the 3$^{rd}$ of June, 2015. These were compared to frequency transfer data taken in Australia on conventional links of comparable distance as measured during the previous few weeks. The overhead optical fibre characterisation field trials included deployment over a 153 km overhead fibre link which is an ideal analogue for the longest planned SKA1-MID fibre links, if an overhead route can be accommodated.

The aims and objectives of this work was to:

- Conduct a highly-precise comparison measurements of overhead fibre links compared to conventional links (buried, conduit, spool) of similar length, in order to ascertain the feasibility of using overhead fibre frequency transfer for the SKA.

- Test the robustness of AOM-based servo actuation systems in conditions that required a factor of 6,000 greater actuation range and speed than the maximum required for SKA frequency transfer (193 THz frequency transfer compared with the 32 GHz transfer required SKA1-MID Band 5).

- Confirm that the coherence length of a moderately priced, COTS diode laser (Koheras X15) is suitable for SKA long-haul frequency transfer.

- Test component reliability by exposing them to severe physical punishment during air transport to South Africa (objective applied retroactively after it was discovered the shock-proof equipment case failed during transport resulting in external damage to the equipment).

### 5.1.1.5  Test(8) – SKA1 STFR.FRQ (UWA) v2 – Long-Haul Overhead Fibre Verification

Refer to [RD16] and [RD14].

The SKA frequency transfer system was demonstrated to be capable of frequency transfer over the longest SKA1 overhead fibre links.

Two optical fibre reticulation options are being considered for the spiral arm component of SKA1-MID's SAT system. The default option is a trenched route outside the core region that has to avoid physical obstructions. This results in a route with a total length of 604 km across the three spiral arms, or an overage of about 200 km per spiral arm. This is approximately 50% longer than a route that directly follows the layout of the spiral arms determined to be 393 km, or about 130 km per arm. With the additional length of the trenched fibre in the core, the maximum SKA1-MID arm length for this configuration is about 150 km.

The total SADT local infrastructure costs of the trenched route based on the default option is currently calculated to be around 11 million euros. However, if the SAT system can be shown to be able to operate on overhead fibre, then all of SADT fibre cable could simply be suspended on the same overhead poles that distribute power along the spiral arms. This route has a total distance of only 394 km across the three spiral arms. Therefore, the only infrastructure cost specific for SAT are the additional (now shorter) SAT fibre cables, estimated to be less than 300 thousand euros. Detailed cost estimate studies for moving all SADT fibre onto the overhead route are ongoing, but it is expected that not having to trench the SKA1-MID spiral arms could save the SKA project several million euros.

The SAT system comprises two independent systems; frequency transfer and absolute time transfer, with the absolute time transfer system being developed by other members of the SADT Consortium. Given its much lower susceptibility to optical link fluctuations than the frequency transfer system, it was expected that the SKA absolute time transfer system to be able to operate nominally on overhead fibre. Nonetheless, the SADT Consortium has put forward plans to also verify this system on overhead fibre in the near future.

The SKA frequency transfer system was used to transmit a stabilised RF signal over several different lengths of overhead optical fibre at the South African SKA site. An out-of-loop measurement system was used to independently measure the frequency stability of the frequency transfer. The system was shown to operate reliably for continuous periods up to 48 hours, over links up to 186 km, and during average wind speeds up to 56 km/h.

The aims and objectives of this work was to:

- Confirm that frequency signals transferred using the SKA frequency transfer system meets the SKA phase stability requirements when deployed on overhead fibre links, over the longest planned SKA fibre links, and during extreme environmental conditions.

- Confirm that the AOM-based active stabilisation servo used in the SKA frequency transfer system remains robust while deployed on overhead fibre links, over the longest planned SKA fibre links, and during extreme environmental conditions.

- Confirm that the SKA frequency transfer system can utilise optical amplification to extend fibre-based frequency transfer beyond the longest planned SKA fibre links.

5.1.1.6   Test(9) – SKA-MID STFR.FRQ (UWA) v2 – Astronomical Verification

Refer to [RD17] and [RD18].

Astronomical verification trails of the SKA1-MID frequency transfer system field-deployable prototype were carried out at the ATCA during May and June of 2016. The UWA SKA1-MID frequency transfer system prototype performed as designed, easily meeting SKA Level 1 coherence requirements, REQ-2268 and REQ-2693, under a broad range of observing conditions. Extrapolating the results of the field trials to SKA worst-case scenario performance predictions indicate that the SKA1-MID frequency transfer system exceed REQ-2268 (maximum 2% coherence loss after 1 s) by a factor of 2560. Over the 77 km test link at the ATCA, the UWA SKA1-MID frequency transfer system prototype exceeded REQ-2693 (phase drift over 600 s less than 1 rad) by an order of magnitude. The receiver architecture and the test methodology employed during these trials also enabled an investigation into the effects of other sources of phase-noise such as electronics, amplifiers, E/O-O/E systems, and FSs. These further tests strongly suggest that SKA frequency transfer systems that propose to use multiple, separately located synthesisers must be thoroughly vetted by astronomical verification employing the proposed synthesiser design under actual observing conditions in order to highlight any potential vulnerabilities of the synthesisers to ambient temperature fluctuations at the receiver site.

### 5.1.1.7 Test(11) – SKA-MID STFR.FRQ (UWA) v2 – Laboratory Demonstration

Refer to [RD13].

This report presents the experimental methods and frequency transfer performance results of the SKA-MID variant of the SKA phase synchronisation system (FRQ (UWA)). A subset of this information as well as the technical details of the frequency transfer technique are reported in the peer-reviewed publication [RD10].

The aims of the test were to describe the basic design and operation of the SKA-MID FRQ (UWA) system and demonstrate its compliance with the applicable SKA engineering requirements. The key SKA-MID FRQ (UWA) system characteristics specifically investigated in this report are:

- Frequency transfer stability — demonstrate the frequency transfer performance of the system and that it is compliant with the 1.9% coherence loss requirements at 1 s (REQ-2268) and 60 s (REQ-2692).

- Frequency transfer accuracy — demonstrate the frequency accuracy achieved with the system.

- Phase drift — demonstrate the phase drift of the transferred frequency and that it is compliant with REQ-2693 by being less than 1 rad over 600 s.

- Jitter — demonstrate that the jitter of the transferred signal is compliant with the requirement defined by EICD 300-0000000-026_02-SADT-DSH_ICD.

### 5.1.1.8 Test(13) – SKA-MID STFR.FRQ (UWA) v2 – Temperature and Humidity

Refer to [RD13].

SKA subsystems are expected to operate at their specified performance levels throughout a range of climatic conditions that they might reasonably be subjected to in their operational environments, including expected temperature and humidity ranges. The tests and results presented below demonstrate that the SKA-MID FRQ (UWA) system meets the climatic requirements by maintaining the required coherence loss limits throughout the specified climate range.

The aims of these measurements were to demonstrate compliance of the SKA-MID FRQ (UWA) system with the one-second (REQ-2268) and 60-second (REQ-2692) SKA coherence requirements under the following conditions:

- SAT.STFR.FRQ_REQ-2798-01-01 — for all ambient temperatures of non-weather protected fibres and equipment between −5°C and 50°C.

- SAT.STFR.FRQ_REQ-2798-01-01 — for rates of change of ambient temperature of non-weather protected fibres and equipment of up to ±3°C every 10 minutes.

- SAT.STFR.FRQ_REQ-305-075 — for all ambient temperatures in the EMI shielded cabinet in the pedestal between −5°C and 50°C.

- SAT.STFR.FRQ_REQ-305-075 — for all rates of change of ambient temperature temperatures in the EMI shielded cabinet in the pedestal up to a maximum of ±3°C every 10 minutes.

- SAT.STFR.FRQ_REQ-305-076 — for a range of non-condensing relative humidities in the EMI shielded cabinet in the pedestal between 40% and 60%.

- SAT.STFR.FRQ_REQ-305-077 — for all ambient temperatures in the inner/outer repeater shelters between 18°C and 26°C.

- SAT.STFR.FRQ_REQ-305-078 —for a range of non-condensing relative humidities in the outer repeater shelter or inner repeater shelter between 40% and 60%.

- SAT.STFR.FRQ_REQ-305-079 — for all ambient temperatures in the CPF between 18°C and 26°C.

- SAT.STFR.FRQ_REQ-305-080 — for all relative humidities in the CPF between 40% and 60%.

### 5.1.1.9 Test(15) – SKA-MID STFR.FRQ (UWA) v2 – Seismic Resilience

Refer to [RD13].

SKA subsystems are expected to continue to function at the required performance levels after a minor Earth tremor event. The tests and results presented below show that the SKA-MID FRQ (UWA) system meets the requirements by demonstrating continued operation at the required coherence loss limits after experiencing ground accelerations up to 1 m/s.

The aims of these measurements were to demonstrate compliance of the SKA-MID FRQ (UWA) system with the one-second (REQ-2268) and 60-second (REQ-2692) SKA coherence requirements after experiencing a seismic event resulting in a maximum peak ground acceleration of 1 m/s$^2$ (SADT.SAT.FRQ_REQ-2650-01-01).

### 5.1.1.10 Test(17) – SKA-MID STFR.FRQ (UWA) v2 – Electromagnetic Compatibility

Refer to [RD13].

The design considerations for the SKA aim to minimise the telescope's exposure and sensitivity to RFI that could impair the SKA's observational capabilities. This includes self-generated RFI, which is denoted as internal EMI. The tests and results presented below show that the EMI emitted by the SKA-MID FRQ (UWA) system is within the designed limits and the system meet the SKA electromagnetic compatibility (EMC) requirements.

The aims of this test were to demonstrate compliance of the SKA-MID FRQ (UWA) system with the RFI/EMC requirements REQ-2462 and SADT.SAT.STFR.FRQ_REQ-2462-01-01 that FRQ components emitting electromagnetic radiation within frequency intervals for broad and narrow band cases shall be within the SKA RFI/EMI Threshold Levels as defined in SKA-TEL-SKO-0000202-AG-RFI-ST-01.

### 5.1.2 Construction of a Mass Manufacture Archetype

UWA researchers in partnership with MeerKAT and UoM electronic engineers, progressed the detailed designs described in this document, into a set of mass manufacture archetypes as described in §4.8.1.

The SKA1-MID frequency transfer system previously consisted of a Field-deployable Prototype (FDP). This prototype was used to test the performance of the system while maintaining the capacity for iteration through the use of discrete optical and electronic components. The performance of the system was astronomically verified through testing at ATCA in 2017. As a result, the FDP can now progress to the Design for Manufacturability (DfM) stage. The DfM process must work within the constraints placed on the SKA1-MID by its MW frequency reference to develop a miniaturised 'DfM archetype' of the SKA1-MID frequency transfer system. Throughout the system's development, the design will be continually tested to ensure the new system maintains the stability level established by the FDP.

This report outlines the development and testing of this DfM archetype. First, it gives an overview of the relevant principles of frequency transfer, and describes the role of frequency transfer in the SKA. A summary of the previous work done at UWA on the SKA1-MID frequency transfer system is given, including the successful astronomical verification of the FDP carried out earlier this year. The report then covers the DfM process. This design section begins with the layout and redesign of the system on a schematic level, followed by the creation of several PCBs to be incorporated into the encompassing DfM system for testing. The results of this testing is then summarised, and the report concludes with an outline of the current state of the DfM archetype, as well as a discussion of further applications for the system.

## 5.2 Pre-technology Down-select Independent Assessment of Solution

The detailed design of the SKA phase synchronisation system as described in a draft version of this report, as well in the journal papers and SADT reports provided in §2.4, was critically assessed by three independent domain experts in phase synchronisation systems for radio telescopes arrays.

The primary purpose of the review was to conduct a critical assessment of the principles of the SKA phase synchronisation system and comment on its suitability as a system for SKA. The terms of reference were to attempt to identify any weaknesses/risks/limitations with the system that were either unknown (in which case the detailed design may be able to be updated), or look like weaknesses/risks/limitations because the material was not covered sufficiently well enough in the draft documentation (in which case the documentation could be improved).

The deadline for the reviews was Monday the 17$^{th}$ of July (two weeks prior to the submission of this document to the SADT office). No major issues were raised and no significant changes to the detailed design had to be implemented. This report however, was significantly updated in response to the many comments made by all three reviewers. In addition, some key suggestions were incorporated into recommendation and further work listed in §6.2. These three independent assessments helped build confidence in the detailed design but also helped to improve the design so that the best possible phase synchronisation system for the SKA telescope could be delivered.

### 5.2.1 ASTRON Netherlands Institute for Radio Astronomy – Gijs Schoonderbeek

This review focused on two elements of the SKA phase synchronisation system:

- The short term stability, dominated by the clean-up OCXO located within the RM at the antenna site.
- The long term stability, which is driven by the quality of the control loop of the phase synchronisation system.

The report is attached as Appendix 8.10.1.

### 5.2.2 Jet Propulsion Laboratory – Larry D'Addario

The comments in this review concerned three main areas:

- The review called for a budget of noise limits for the transfer performance, to help inform whether the performance could be improved, or to help identify any components that could be degraded without degrading the transfer performance (unfortunately, in the time available the detailed measurements required to support such a noise budget were unable to be conducted, but is something that will be explored outside the scope of SKA in the future).
- To provide more detailed explanations of design choices relating to selection of specific signal frequencies and power levels of the various components within the system (notably the AOMs and DDS reference frequencies).
- The review called for a discussion on robustness of the system during distributions in normal operation (for example, from power outages).

The report is attached as Appendix 8.10.2.

### 5.2.3 Square Kilometre Array South Africa – Johan Burger

The comments from the final review can be classed into three main areas:

- Comments relating to the minimising risk and complexity by locating complex equipment at the central site, and having only simple and robust equipment located at the antenna site.
- Comments relating to the signal-to-noise budget of the entire system and how this impacts the dynamic range, with the view to increase robustness through an increase in operational dynamic range.
- The review called for more information relating to the key component of the MS, the DPM.

The report is attached as Appendix 8.10.3.

## 5.3 Technology Down-select

As part of the SADT Consortium-led technology down-select process, this document was reviewed further by the following independent domain experts:

- William Shillue of the National Radio Astronomy Observatory;
- Miho Fujieda of the National Institute of Information and Communications Technology;
- Sven-Christian Ebenhag of RISE.

The document was updated taking into account the feedback provided.

### 5.3.1 Functional Performance Requirements

#### 5.3.1.1 Phase Drift Requirement

**§5.1.1, p.7** *"it would have been better to see phase drift vs. temperature directly, as this would be more convincing. …. The frequency counter data is unintelligible, it looks like noise again."*

One can only see noise because any effects of temperature change are removed by the stabilisation system; **exactly** what the stabilisation system is designed to do. The residual noise limit in this case is likely the amplitude noise-to-phase noise coupling in the servo system's RF mixer. If any physically-attributable trend were evident in these data, the system could (and should) be improved to address it and remove it.

#### 5.3.1.2 Jitter Requirements

**§4.3, p.4** *"need measurements for the whole system using the same oscillator that will be used in the final product"*

This is a valid point, however, this topic is complex and involves two related issues:

- The 74 fs jitter requirement has not been appropriately specified; it has since been acknowledged (even by the SKAO) that a less stringent value (perhaps 288 fs) is much closer to the actual required specification.

- The ridiculous situation of having the FRQ-DISH interface in the pedestal, while the ADCs are in the indexer, thereby requiring a secondary frequency transfer system (with its own clean-up OCXO). Clearly, the logical conclusion is to move the FRQ RM to the indexer and combine these two systems. SADT and DISH have agreed to raise an ECP to this effect after CDR.

This, therefore, brings the expectation that space (and power/dissipated heat) will be much more tightly constrained that the current allowed 3U 19 rackmount in the pedestal; however, it is not clear how much space will be available. Therefore, selecting a compact clean-up OCXO (to go with the otherwise compact RM) has been a choice for UWA since commencing the detailed design. Having such small form factor however, significantly reduces the available products and restricts the range of specifications. UWA has purchased and experimentally tested two compact oscillators:

- An oscillator that is capable of tuning over the required frequency offset range, but does not meet the stringent 74 fs jitter specification.

- An oscillator that does meet the specification (28.3 fs) but cannot tune over the whole range (plus is not at the ideal frequency, so requires additional complication of frequency multiplication circuitry).

Relaxing the jitter specification will drastically increase the range of available suitable compact oscillators to be used in the case of the RM moving to the indexer (alternatively, if it is confirmed that the interfaces will remain in the pedestal, a much larger oscillator can be selected). Only once this intra-consortium architecture design choice is made, does it make sense to commit additional resources to further measurements involving the oscillator (purchasing another soon-to-be obsolete oscillator is a further waste of funds).

So regardless of which route is taken, it is believed the following redefinition of the SADT-DISH interface is the most sensible way forward:

- SADT is responsible for the provision of a stabilised frequency reference signal to the receptor (either pedestal or indexer) with a coherence budget of 1.9% (or TBD) for timescales slower than the clean-up OCXO cross-over frequency (perhaps 10 Hz or 100 Hz, TBD).

- DISH is responsible for the clean-up OCXO (as it should really be closely tied to the ADC) and therefore the coherence budget (1.9%, TBD) at timescales faster than the clean-up OCXO cross-over frequency.

This proposal was first raised with Andre in Perth at the end of March 2017, and then with others at the SKAO during a meeting at JBO at the end of April 2017.

*§5.1.2, p.8* *"The clarification questions had sought to determine whether the oscillator would still be effective at 60s".*

This is explicitly stated in the STFR.FRQ DDD; the optimal cross-over frequency likely to be between 1 Hz and 1 kHz (depending on performance of OCXO and additive phase noise on the particular fibre link in question).

### 5.3.2 Normal Operating Conditions

#### 5.3.2.1 Rate of Change of Temperature/Temperature Testing/Thermal Equilibrium in Pedestals

*§4.3.1.d, p.4* *"it isn't clear what effects the thermal equilibrium of the equipment may have."*

Equilibrium is defined as a state of no change or balance; hence, it is clear that no change would be expected. The largest magnitude of any effect will occur during greatest rate of change of temperature.

*§5.1.1, p.7* *"the steps in temperature are too short and the resolution of the counter is too low. It is not possible to see anything except noise".*

The resolution of the counter is actually extremely high; and as outlined above, only noise is visibly because any effects of temperature change are removed by the stabilisation system. This is exactly what the stabilisation system is designed to do.

*§5.1.1, p.7* *"The current test doesn't mean anything as the time between measurements is so short that the equipment has not had time to reach thermal equilibrium".*

This is incorrect. If the system is in thermal equilibrium, there can be no thermal effects. Rather the opposite is true, to induce maximum stress, the system should be cycled through thermal states as quickly as possible. The rate of change used was the maximum achievable with available equipment (and exceeds the 3°C/10 min requirement).

*§5.1.1, p.8* *"EDFAs can certainly be affected by the range of temperatures in the RPFs."*

The EDFAs are in-loop within the stabilisation servo, so even if temperature changes were to cause an impact, any such changes are suppressed by the stabilisation system. This comments raises concerns that the relevant panel member made little effort to understand the basic concepts of how a phase synchronisation works. "If they have EDFAs they would need to do 3 degree C steps throughout the 18-26C range, and allow enough time (about 3 hours) to reach thermal equilibrium in between". The active components in the EDFA are the size of a mobile phone, and would reach thermal equilibrium in minutes. Also, as outlined above, if the system is in thermal equilibrium, there can be no thermal effects. Rather the opposite is true, to induce maximum stress, the system should be cycled through thermal states as quickly as possible. The rate of change used was the maximum achievable with available equipment (and exceeds the 3°C/10 min requirement).

### 5.3.2.2 Temperature and Wind on Fibre Links

*§5.1.1, p.6* *"UWA showed a measurement result but they used an old design without the MZI."*

This is an invalid criticism for the 'rate of change of temperature' measurements relevant to REQ-2268; all temperature test of equipment housed in CPF was conducted using the 'final equipment' that included the DPM (which I assume is referred to here as the MZI). The test referred to in this comment was only used to support the 'non-weather protected location' (that is, the fibre) requirement.

*§5.1.1, p.6* *"the test was performed using the overhead fibre link at 154 km with the temperature gradient being +/- 1.2 degrees in 10 minutes which the panel felt was inadequate".*

There is strong disagreement with this comment, unless it were possible to test 150 km+ of fibre *cable* within an environmental chamber. (Even testing 150 km of bare fibre is inconceivable; the spools would occupy a much greater volume that any commercially-available environmental chamber I am aware of; and more importantly the tight packing of glass, plastic, and air on a spool would provide high levels of insulation to the inner layers making fast temperature changes across all 150 km unlikely to be practicably achievable.) As only 25 km of spooled fibre is available, it is therefore much better to show a ±1.2°C/10min test on 154 km (185°C·km) of the exact same fibre cable that will be used on SKA1-MID, than ±3°C/10min over 25 km (75°C·km) of bare fibre spool; the magnitude of any effects would be 2.5 times greater with the reported test.

*§5.1.1, p.6* *"It was felt there was not enough evidence for the full range of temperature from -5 to 50 degrees C … the tests should be done in a lab / temperature chamber."*

As above, there is disagreement with this comment. As only 25 km of spooled fibre is available, it is still better to show a range of 0.9°C to 14.8°C on 154 km (2141°C·km) of the exact same fibre cable that will be used on SKA1-MID, than −5°C to 50°C on 25 km (1375°C·km) of bare fibre spool; the magnitude of any effects would be 1.5 times greater with the reported test.

### 5.3.2.3 Seismic Test

*§4.3.1.c, p.5* *"was the entire equipment shaken or just the receiver? It should be the entire equipment".*

In §8.3, p. 41 of the 'Pre-CDR Laboratory Verification' report, it clearly states that the whole "SKA-MID STFR.FRQ" system was exposed to the shaking. Note also, that the comment in Alistair McPherson's email from 21 September 2017 "Seismic in all three directions (not just one)" is plainly not relevant to the UWA system as Figure 37 p. 42 of the 'Pre-CDR Laboratory Verification' report shows data for all three directions (not just one).

### 5.3.2.4 Humidity Tests

*§4.3.1.c, p.5* *"There was a short humidity test which also needed to take all equipment into account, and cover the full required range".*

As stated in §6.4 of the 'Pre-CDR Laboratory Verification' report, the humidity test were conducted on all SKA-MID FRQ equipment, with data presented over a continuous period of three days, and with relative humidity ranging between 9% and 99% (the requirement specifies a range of 40% to 60%). Perhaps this comment was mistakenly applied to the UWA system.

*§5.1.1, p.7* *"One panel member had experienced similar equipment that is sensitive to humidity changes."*

This comments raises concerns that the relevant panel member made little effort to understand the basic concepts the UWA system, and clearly his experience relates to equipment which is not at all similar to the UWA system. Non-condensing humidity cannot affect the system as the photons are kept confined inside the glass medium of the optical fibre at all times. Glass is impermeable to water, for example, windows.

### 5.3.3 Key Additional Requirements

#### 5.3.3.1 RFI Requirements

***§4.3.1.c, p.4*** *"The EMC measurements were not thought to have been adequately performed"*

There is strong disagreement with this comment; the measurements were conducted independently by Franz Schlagenhaufer, one of the SKA community's leading EMC specialists.

*"The UWA solution had only calibrated the antenna up to 18 GHz and then there was a jump…The site requirement goes to 20GHz".*

Actually, the requirement is for test up to 25.5 GHz which was conducted. The measurement was calibrated, it is just that the geometric 'antenna factor' part of the calibration was not available for frequencies above 18 GHz. At worst, this would result in the reporting slightly inaccurate magnitudes of any observed signals. No signals however, were detected above 8 GHz, as the highest frequency present in the UWA system is 8 GHz. The second and third order harmonic at 16 GHz and 24 GHz were measured (via direct sampling of the coaxial cable) to be greater than 40 dB and 60 dB down respectively, so this is a moot point.

*"efforts should be made to carry out measurements under the prescribed conditions, and not rely on calculating / deriving the EMC measurement".*

The EMC measurements were certainly not derived; the data presented are the raw values recorded by the calibrated measurement system. Rather, the threshold masks were adjusted for distance and bandwidth in accordance with standard formulae (and also presented in SKA-TEL-SKO-0000202-AG-RFI-ST-01). The 3m distance could not be achieved in the anechoic chamber used, so measurements were conducted at 1m (as both distances are well in the near-field, if anything the adjustment is actually overly aggressive); and the resolution bandwidth had to be adjusted as it cannot be set as a percentage of the measurement frequency (a higher resolution value was chosen to ensure narrow-band emission was not missed),

*"COTS components can be shielded to try to meet severe specs. UWA had not tried to do this and the panel wondered why".*

The equipment did not need additional shielding as the system was designed at the outset with low RFI emission in mind, and the result indicate that the COTS equipment produces signals at least five orders of magnitude below the requirement.

*"the tests should involve the whole equipment chain with the real equipment, including the power supply".*

The power supply is not part of the FRQ equipment and should therefore not be included in the testing.

*"immunity test should also be done… there are requirements for immunity to radio emissions".*

There may be requirements, but these were not part of the down-select criteria. While no formal immunity from external radiation was not explicitly tested as this was not part of the down-select criteria. While no formal immunity test have been conducted, every time the equipment was used, informal immunity tests were effectively being conducted. The laboratory and neighbouring rooms are filed with numerous pieces of low frequency and high frequency RFI emitting equipment and no unwanted effects were ever experienced (confirmed with antennas set-up in the laboratory). In addition, the equipment inside correlator rooms at ATCA and ASKAP have also been successfully deployed.

#### 5.3.3.2 Availability Requirements

***§4.3, p.3*** *"There was concern that it is hard to get the system to be redundant or robust in the long term with 3 x AOMS".*

The AOMs actually make the system extremely robust as they are intrinsically extremely simple devices.

The point about redundancy is more valid. The detailed design for mass manufacture is critically tied to the physical shape of the particular AOMs recommended for use. In the unlikely event that these (or similarly shaped) AOMs become unavailable in the future, a redesign of the mass manufacture PCBs would be required.

***§4.3.1.c, p.4*** *"there was no MTBF data for the AOMs"*.

This is a valid point, as the AOMs are the dominant cost item in the system. Considerable effort was spent to attempt to obtain these data, however, none of the three leading manufactures were able to provide this information. It is recommended to start a new round of information requests to try to ascertain the MTBF.

### 5.3.4 Miscellaneous Test

#### 5.3.4.1 Synthesiser Testing

***§5.1.3, p.9*** *"panel noted that they had not tested the synthesiser under all the environmental conditions"*.

This is incorrect. The full specified range of all normal operating conditions for all equipment was experimentally tested (the one exception where the full range was not achievable for testing was for the temperature of the fibre link, which has nothing to do with synthesiser). "… or run through a range of the possible frequencies, to demonstrate performance": the UWA Synthesiser is only used to produce one frequency. Perhaps both of these comments were relevant to the THU system and incorrectly attributed to UWA.

#### 5.3.4.2 4GHz vs. 8GHz transfer

***§4.3.1.b, p.4*** *"4GHz is required at the antenna the correction should be at 4 GHz if possible"*.

It is unsure that this comment is understood. Are the panel advocating 4 GHz transfer over 8 GHz transfer as per the 'Recommendation' in §5.2 of the FRQ DDD? Or, is the comment relevant to the THU system, which delivers a static 1 GHz signal, and is incorrectly attributed to UWA? If the former, the argument for the original selection of 8 GHz transfer (lower risk due to improved performance) is clearly outlined in §3.3.2.1 of the FRQ DDD. However, numerous tests have shown that 4 GHz transfer will still comfortably meet the coherence requirements (although the change will require additional detailed design engineering work).

# 6 CONCLUSIONS/RECOMMENDATIONS

## 6.1 Conclusions

To enable the SKA to meet its goal of being a transformational telescope in the 50 MHz to 14 GHz frequency range, active stabilisation of the phase-coherent reference signals transmitted to the antenna sites is required to overcome the environmental disturbances acting to degrade the transmissions. In this report, the detailed design of the SKA phase synchronisation system for the SKA1-MID telescope developed by UWA has been described.

The solution is based on the successful stabilised frequency reference distribution system employed by the ALMA telescope, incorporating recent advances made by the international frequency metrology community, and innovations developed by researchers at UWA, optimised for the needs of the SKA1-MID.

The system receives an electronic reference signal from the SAT clocks at the SKA1-MID CPF, and transfers the full stability of the reference signal across the SAT network to each receptor pedestal. At the receptor pedestal, an electronic copy of the reference signal is provided to DISH. The system is controlled and monitored using the LMC with the required local infrastructure provided by LINFRA.

A complete solution description has been presented, detailing the design justification and all aspects of the detailed design from hardware selection to integration. These details have been used to develop a comprehensive cost model.

An evaluation of the design has been carried out, including an overview of the testing of the design and verification of its performance and compliance with the SKA design requirements. This evaluation is supported by independent assessments from domain experts at the ASTRON Netherlands Institute for Radio Astronomy, the Jet Propulsion Laboratory, and SKA South Africa. The design has been shown to be fully complaint with all SKA design requirements.

Remaining tests and work required to take this system forward into the SKA1 construction phase are detailed in §6.2.

## 6.2 Planned Further Work

UWA has funding and resources for post-CDR tests and intra-consortium design work to help retire two main areas identified as outstanding risk. Namely these are confirming the phase coherence performance of 'Microwave-frequency transfer on overhead fibre' (SKA.SADT.RSK.178); and informing intra-consortium architecture design optimisations for the 'Receiver Module clean-up OCXO and end-to-end STFR.FRQ testing' (SKA.SADT.RSK.179) by demonstrating the operation of the RM on the indexer on MeerKAT. The details of these tests are outlined in §6.2.1 and §6.2.2 respectively.

The schedule and scope of these tests are dictated by available funding and personnel resources. As outlined in SKA.SADT.RSK.180 "While Schediwy can commit 0.6 FTE commit to SADT activities until end-2018; it is still unclear as to the status of Gozzard (International Centre for Radio Astronomy Research (ICRAR) has offered a position only until end-Q1-2018)". Furthermore, Schediwy is presently fully committed with current SADT and other UWA activities until at least mid-2018 (the earliest possible window for MeerKAT testing will be April 2018).

### 6.2.1 MW Frequency Transfer on Long-Haul Overhead Fibre

The plan is to deploy the SKA phase synchronisation system for SKA1-MID on the long-haul overhead fibre at the South African SKA site. This will demonstrate MW frequency transfer on matching infrastructure and similar lengths as is expected for the SKA-mid spiral arms. A map showing the relevant locations and fibre route is given in Figure 85.

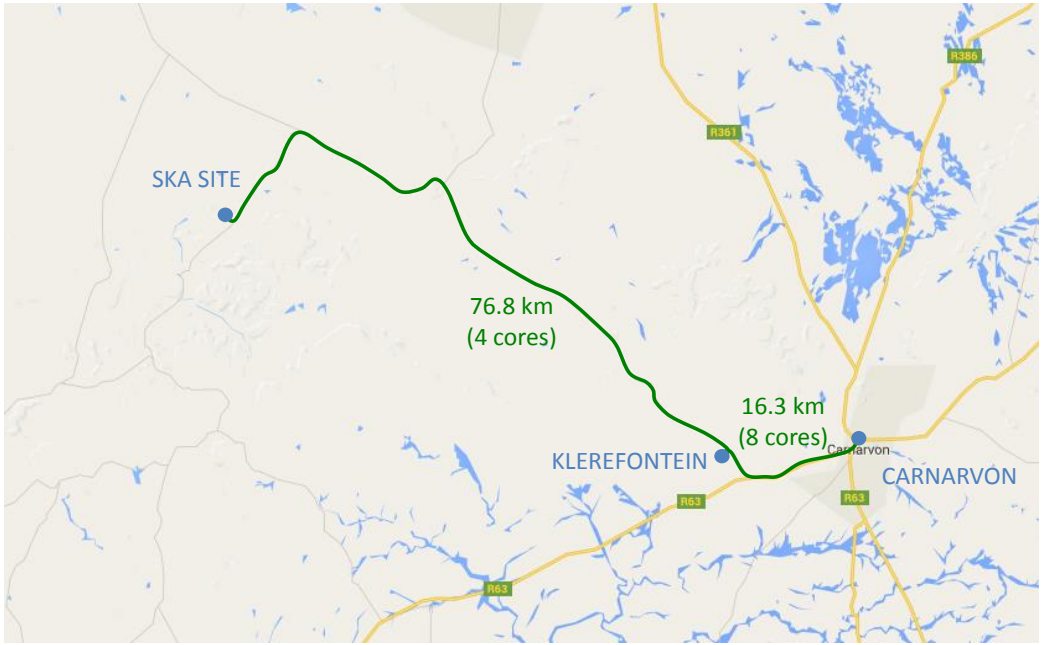

**Figure 85 Map showing the relevant locations and fibre route for the planned South Africa tests**

Field deployable versions of the SKA-MID FRQ TM (see Figure 86) and other required ancillary equipment will be located in the KAPB at the SKA site. A loop-back patch lead and OA will be located at the support base at Klerefontein to create a round-trip fibre distance of 154 km. The photonic MW signal will be transmitted from the KAPB to Klerefontein and then back to the KAPB, where it will be converted to an electronic signal by the RM.

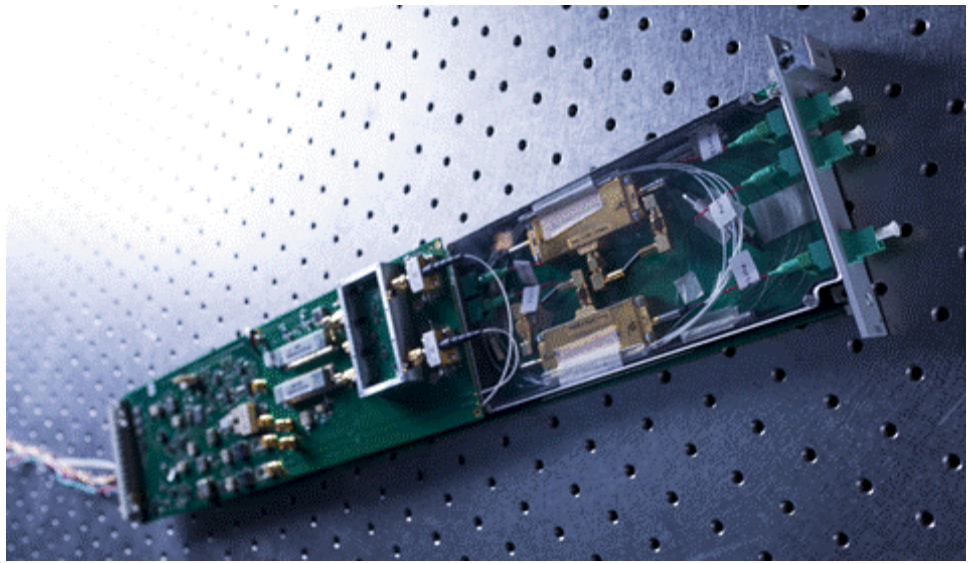

**Figure 86 Prototype SKA-mid FRQ TM PCB**

Standard metrology techniques will be used to determine the transfer performance of the system. Unlike the previous South African field trial in 2015, measurements of all functional performance requirements (coherence, phase drift, and jitter) will look to be conducted. To accommodate this, it is aimed to utilise the optimum test equipment for each measurement, and not just a frequency counter as per the 2015 test. A Microsemi 30201A phase noise test probe has been purchased to aid these tests.

As the aim is to measure the performance of the system operating on the overhead fibre, the RM will not be fitted with a clean-up OCXO. Including a clean-up OCXO would prevent the measurement of fast phase noise originating from the fibre link. This information is important for then developing an optimal clean-up

PLL for minimising the total jitter budget (that is, the optimal cross-over frequency given the link phase noise and clean-up OCXO phase noise). Part of €14.4k in funding to conduct this test in 2018 has been secured.

### 6.2.2 Operation of the RM on the Indexer of the SKA Precursor Telescope

The RM for the SKA phase synchronisation system is capable of being mounted directly on the SKA1-MID antenna indexer alongside the receiver. Just as is done with most telescopes, including MeerKAT, the reference signals are delivered to the location of the receiver ADCs. Currently, the SADT interface with DISH is in antenna pedestal, and the DISH Consortium are required to build a second frequency transfer system to transmit the reference signals up the cable wraps to the indexer where the receiver ADCs are located. After CDR, the DISH consortium and SKAO will enact an ECP to address this costly inefficiency. In 2018 it is planned to deploy the SKA phase synchronisation system on the SKA Precursor MeerKAT to rove the feasibility of the system to be mounted directly on the SKA receptor feed indexer.

To achieve this, the optics and some electronics of the prototype RM will need to be retrofitted into the MeerKAT L-band 'time and frequency' enclosure (see Figure 87).

The MeerKAT L-band receivers are designed to operate between 900 and 1,670 MHz, with the ADC clock at 1,712 MHz. The phase synchronisation system for transmission of 6,848 MHz (and divide by 4) would therefore be modified, rather than the default SKA-mid transmission of 8,000 (and divide by 2). This involves removing the existing passive frequency transfer system components (the largest PCB in the bottom-left) from the MeerKAT L-band 'time and frequency' enclosure, and installing the SKA phase synchronisation system components (everything except the PCB) into the vacated space.

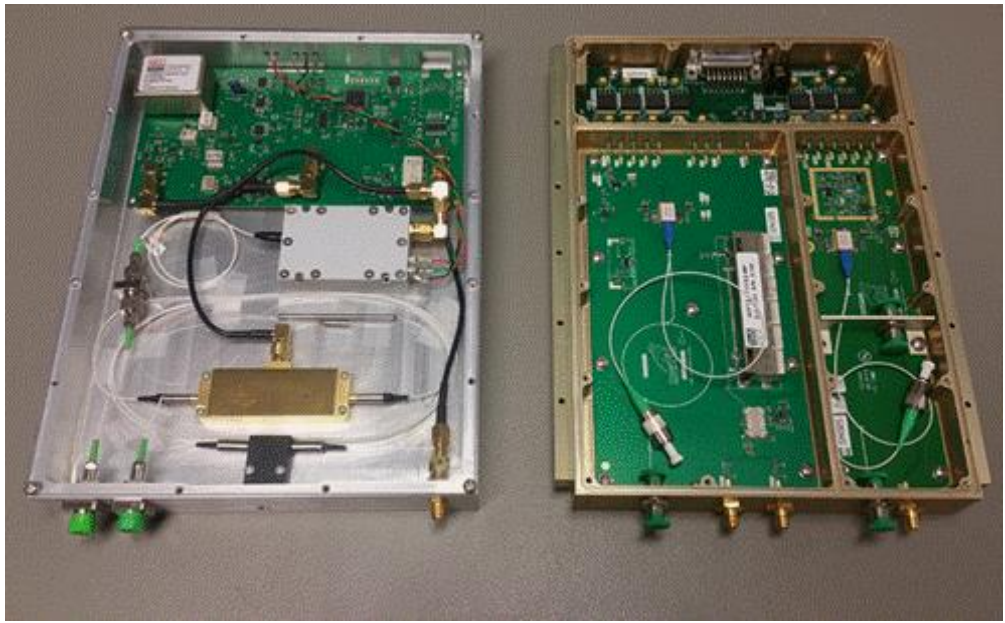

**Figure 87 Prototype of the RM of the SKA phase synchronisation system (left) next to the current MeerKAT L-band 'time and frequency' enclosure**

This retrofit will create an actively-stabilised phase synchronisation system that is fully compatible with the existing MeerKAT L-band receiver. It is planned to use the same 1,712 MHz passive bandpass filter to clean up fast phase noise, just as is done with the current MeerKAT design. This not only keeps the two systems as similar as possible (expect it to be actively stabilised), but it has been determined that a clean-up OCXO could not be physically retrofitted into the MeerKAT L-band 'time and frequency' enclosure. The modified MeerKAT L-band 'time and frequency' enclosure would then need to be qualified for electromagnetic compatibility (EMC) for operation at the Karoo site. This work will provide an opportunity to repeat some of the RM EMC measurements using a professionally-designed, EMC enclosure, rather than the prototype enclosure (in-kind access to the anechoic chamber run ICRAR-Curtin is available). It is therefore expected that the EMC performance will be even better than that reported in §4.3.4.3.

The modified MeerKAT L-band 'time and frequency' enclosure would then replace an existing passive 'time and frequency' enclosure inside the L-band receiver (shown in Figure 88) on one MeerKAT receptor. The L-band receiver is mounted to the feed indexer between the primary and secondary reflectors on the receptor as shown in Figure 88.

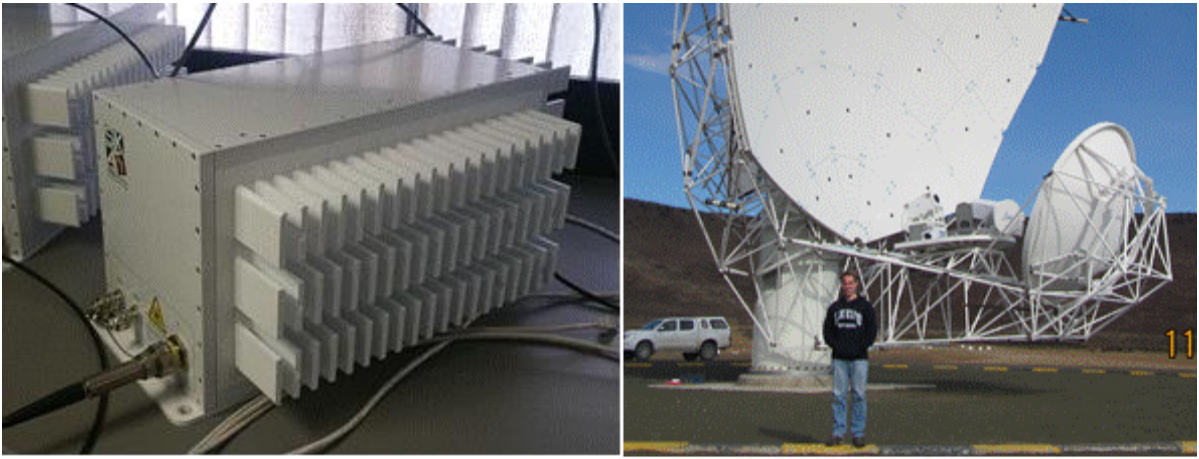

Figure 88 Spare MeerKAT L-band receiver enclosure (left), and a MeerKAT receptor showing a mounted L-band receiver enclosure (along with other receivers) mounted on the receptor's feed indexer

The SKA phase synchronisation system TM would be located at the Karoo site correlator room with the stabilised MW signal sent the ~10 km directly to the RM on the modified MeerKAT receptor. The astronomical signals detected by the modified receptor would then be correlated against the astronomical signals detected by all other MeerKAT receptors, and any additional noise present on interferometric baselines that include the test receptor would be attributed to the additional noise of the STFR system allowing the amount of added phase noise to be quantified. If time and experimental parameters permit, a 150km+ loop-back on overhead fibre between the Karroo site and the Karoo Support Base at Klerefontein could also be included. A part of €14.4k in funding to conduct this test has been secured.

### 6.2.3 Continue work on Design for Mass Manufacture

UWA researchers in partnership with MeerKAT and UoM electronic engineers, progressed the detailed designs described in this document, into a set of mass manufacture prototypes, effectively getting a head-start at addressing manufacturing issues that may be encounter by contractors during SKA construction. The first set of mass manufacture prototypes of the TMs and RMs for SKA1-LOW were completed in Q2, 2016 [RD21]; and for SKA1-MID in Q1, 2017 [RD22]. These systems, however, still comprise multiple PCBs that were developed independently to aid with rapid development and debugging. The final archetypes for not just the TMs and RMs, but the other bespoke elements of the SKA phase synchronisation system (notably the MS) however, are now ready to be built. €33.6k in funding (not including salaries and on-costs) to contribute towards this task has been secured.

## 6.3 Recommendation for Potential Changes/Improvements to the Design

### 6.3.1 Simplify the SKA1-MID RM PLL

The clean-up OCXO currently in the detailed design exceeds the SKA1-MID jitter requirement by a factor of 2.6 (see §4.3.2.3). The output of this oscillator however, is 1 GHz, and so a ×4 multiplier, and some ancillary electronics, are is required to produce the required 4 GHz RM output (see Figure 45). A 4 GHz oscillator at similar prices that matches the jitter performance of the current model has not been able to be sourced, however, were the Jitter requirement be relaxed from the values quoted in the SADT-DISH IICD, to the values indicated by underlying SKA Level 1 requirement (as evaluated by both by the SADT Consortium and the SKA Office), then more cost effective models become relevant. Using these would also simply the RM.

### 6.3.2 Employ a Hybrid Passive/Active Frequency Transfer System

As the SKA phase synchronisation system exceeds the SKA1-MID functional performance requirement by at least two orders of magnitude across most normal operating conditions (see §4.3.2), one could consider a hybrid system that incorporates passive frequency transfer for the shorter links, and only uses the active stabilised system for the longest links. The obvious down-side is the additional complexity in procurement, operation and maintenance. UWA initially considered proposing such a system, however, preliminary calculations showed that while there is some cost gains to be made in equipment capital expenditure, this is rapidly offset in labour capital expenditure, and then also across both equipment and labour operational expenditure.

### 6.3.3 Lowering the Transmission Frequency to 4 GHz

Extensive test campaigns have revealed that the SKA phase synchronisation system exceeds the SKA1-MID functional performance requirement by at least two orders of magnitude across most normal operating conditions (see §4.3.2). All indications are that the system can be redesigned for direct transmission of 4 GHz, and still easily meet the performance requirements. This would moderately simplify the RM electronics (without lowering the cost), however, it would require a complete redesign of detailed design of all MW-frequency elements in both the TM and RM.

### 6.3.4 Using the SKA1-MID System for both MID and LOW

As described in §4.2.2, a variant of the SKA phase synchronisation system specifically optimised for SKA1-LOW has been designed, as well the variant for SKA1-MID described in this report. §4.3.2 shows that the SKA1-MID variant exceed the SKA1-MID functional performance requirement by at least two orders of magnitude across most normal operating conditions. If the SKA1-MID variant were used as the phase synchronisation system for SKA1-LOW, it would exceed functional performance requirement by another order of magnitude. This seems like extreme overkill, however, having one system for both telescopes would undoubtedly generate significant cost savings capital and operation expenditure (with the down-side of having to add the relatively complex MS element to the SKA1-LOW system).

### 6.3.5 Reducing Cost by Ignoring the Possibility of Reflections

One of the key design parameters of the SKA phase synchronisation system is designing for maximum robustness. To this end significant cost and effort has gone into ensuring the system will be immune from unwanted optical reflections that are inevitably present on real-world optical fibre links (see §4.3.7.1). Inexplicably, the need to be able to operate on anything less-than-ideal links was not formulated as a requirement, but achieving this capability requires the addition of two (of the total) thee AOMs per fibre link. Modifying the design to remove the two anti-reflections AOMs would cut 25% from the capital expenditure cost of the system.

## 6.4 Acknowledgements

A large amount of work has gone into the development of the detailed design of the SKA phase synchronisation system presented in this report and the authors are indebted to the many people who have supported this work.

The authors would like to thank our UoM/SADT Consortium colleagues; Keith Grainge, Althea Wilkinson, Bassem Alachkar, Rob Gabrielczyk, Mike Pearson, Jill Hammond, Michelle Hussey, Samantha Lloyd, Richard

Oberland, Paul Boven, and Richard Whitaker for their efforts and support. The authors would especially like to acknowledge the aid of Richard Whitaker in designing the PCB electronics, and Paul Boven for helping to understand many of the SAT systems, concepts, and requirements.

The authors have received a great deal of support from ICRAR and in particular would like to thank Peter Quinn, Lister Staveley-Smith, Andreas Wicenec, and Tom Booler. The authors are very grateful to Franz Schlagenhaufer for conducting the EMC test and compiling the EMC reports. The authors are also very grateful for the input provided by Richard Dodson and Maria Rioja.

The authors have also received a great deal of support from Ian McArthur and Jay Jay Jegathesan at the UWA School of Physics and Astrophysics. The authors are indebted to Michael Tobar, Eugene Ivanov, and the other members of the Frequency and Quantum Metrology group for lending equipment, allowing the raid of their supplies, and discussing important metrology concepts. The authors also wish to thank Gesine Grosche for the use of the dual-drive Mach-Zehnder modulator.

Through ICRAR and the School of Physics and Astrophysics the authors have had the assistance of a number of undergraduate students who have made significant contributions to the project. A huge thank you to Julian Bocking, Gavin Siow, Benjamin Courtney-Barrer, Ben Stone, Maddy Sheard, Charles Gravestock, and Thea Pulbrook for the efforts put into this project. Another very big thank you goes to Simon Stobie who has advanced this project both as an undergraduate research intern, and later by working on the MW electronics for his honours project. Simon's efforts and contribution are evidenced by the fact he was awarded first-class honours for the work he completed in this project.

The authors wish to thank *AARNet* for the provision of light-level access to their fibre network infrastructure. This infrastructure has proven invaluable to the development and verification of UWA's design.

For the astronomical field trials with ATCA, the authors are very grateful for the help received from staff at the CSIRO Paul Wild Observatory and CSIRO Astronomy and Space Science, especially Tasso Tzioumis for helping to arrange the visit. The authors would like to thank Mike Hill, Brett Lennon, Jock McFee, Peter Mirtschin, and Jamie Stevens for their efforts, without which these field trials would not have been a success. At the other end of the continent, the authors would like to thank Aidan Hotan, Maxim Voronkov, Suzy Jackson, and Mark Leach for their help and support during the ASKAP field trials at the Murchison Radio Observatory.

For the South African SKA site overhead fibre field trials, the authors are very grateful for all the local help received from Bruce Wallace, Jaco Müller, Romeo Gamatham, Tim Gibbon, Roufurd Julie, and Johan Burger. Without their contribution, these outcomes would not have been possible. Thanks also to Charles Copley for providing the C-BASS weather station data. Also with SKA South Africa, the authors would like to thank Sias Malan for the mass manufacture design help. The authors are very grateful to Larry D'Addario from JPL, Gijs Schooderbeek from ASTRON, and Johan Burger from SKA South Africa for taking the time to review this solution and an early draft of this report. The authors are also very grateful to William Shillue from the National Radio Astronomy Observatory, Miho Fujieda from the National Institute of Information and Communications Technology, and Sven-Christian Ebenhag from RISE for their valuable comments as part of the technology down-select review.

Apologies to anyone who should be on this list, but have been accidently neglected to mention.

This report describes work being carried out for the SKA SADT Consortium as part of the SKA project. The SKA project is an international effort to build the world's largest radio telescope, led by the SKA organisation with the support of 10 member countries.

# 7 STATEMENT OF COMPLIANCE

The authors confirm that the SKA phase synchronisation system is compliant with all the relevant SKA Level 1 (rev 11) Requirements and associated ECPs. Detailed compliance is shown in Table 49 for all functional performance requirements. Compliance to other requirements may be found in the SADT Compliance Matrix.

Furthermore, how the UWA solution complies with the requirements related to manufacturability and extensibility to SKA2 (SKA1-SYS_REQ-2433, SKA1-SYS_REQ-2462, SKA1-SYS_REQ-2559, SKA1-SYS_REQ-2562, SKA1-SYS_REQ-2594, SKA1-SYS_REQ 2599) are discussed in [RD22] (Appendix 8.3.7).

## 7.1 Functional Performance Requirements

The SKA1-MID functional performance requirements, as defined in [AD3], are listed in Table 49.

| Requirement # | Requirement Description | Compliance | Reference |
|---|---|---|---|
| Coherence requirements | | | |
| SADT.SAT.STFR.FRQ_REQ-2268 | SAT.STFR.FRQ shall distribute a frequency reference with no more than 1.9% maximum coherence loss, within a maximum integration period of 1 second, over an operating frequency range of between 350MHz and 13.8GHz. | Compliant | SADT Rep. **390** (2017) §4.3.2.1<br>Opt. Lett. **42** (2017) 1648 [RD10]<br>SADT Rep. **620** (2017) §4.4.1 [RD13]<br>AJ **154** (2017) 1 [RD20]<br>SADT Rep. **524** (2017) §3.2 [RD18] |
| SADT.SAT.STFR.FRQ_REQ-2692 | SAT.STFR.FRQ shall distribute a reference frequency with no more than 1.9% coherence loss for intervals of 1 minute, over an operating frequency range of between 350MHz and 13.8GHz. | Compliant | SADT Rep. **390** (2017) §4.3.2.1<br>Opt. Lett. **42** (2017) 1648 [RD10]<br>SADT Rep. **620** (2017) §4.4.1 [RD13]<br>AJ **154** (2017) 1 [RD20]<br>SADT Rep. **524** (2017) §3.2 [RD18] |
| Phase drift requirement | | | |
| SADT.SAT.STFR.FRQ_REQ-2693 | SAT.STFR.FRQ shall distribute a reference frequency to a performance allowing a maximum of 1 radian phase drift for intervals up to 10 minutes, over an operating frequency range of between 350MHz and 13.8GHz. | Compliant | SADT Rep. **390** (2017) §4.3.2.2<br>SADT Rep. **620** §4.4.2 [RD13]<br>AJ **154** (2017) 1 [RD20]<br>SADT Rep. **524** (2017) §3.2 [RD18] |
| Jitter requirement | | | |
| EICD *SADT-DSH_ICD* | Jitter shall be equal to or less than 74 femtoseconds for SKA1-MID as defined by EICD *300-0000000-026_02_SADT-DSH_ICD*. | Compliant | SADT Rep. **390** (2017) §4.3.2.3<br>SADT Rep. **620** (2017) §4.4.3 [RD13] |

**Table 49 SKA1-MID functional performance requirements**

## 7.2 Normal Operating Conditions

The SKA1-MID normal operating conditions, as defined in [AD3], are divided into two classes; with the environmental conditions listed in Table 50 and design conditions in Table 51.

| Requirement # | Requirement Description | Compliance | Reference |
|---|---|---|---|
| Ambient Temperature and Humidity Requirements | | | |
| SADT.SAT.STFR.FRQ_REQ-305-075 | SAT.STFR.FRQ components sited within the "EMI Shielded Cabinet" (316-010000) shall withstand, and under normal operating conditions operate within STFR.FRQ Functional Performance requirements in an ambient temperature between -5°C and +50°C, where the rate of change of temperature is a maximum of ±3°C every 10 minutes. | Compliant | SADT Rep. **390** (2017) §4.3.3.1.1<br>SADT Rep. **620** (2017) §6.4.2 [RD13] |
| SADT.SAT.STFR.FRQ_REQ-305-076 | SAT.STFR.FRQ components sited within the "EMI Shielded Cabinet" (316-010000) shall withstand, and under normal operating conditions operate within STFR.FRQ Functional Performance requirements in a non-condensing relative humidity environment between 40% and 60%. | Compliant | SADT Rep. **390** (2017) §4.3.3.1.1<br>SADT Rep. **620** (2017) §6.4.4 [RD13] |
| SADT.SAT.STFR.FRQ_REQ-305-077 | SAT.STFR.FRQ components sited within the "Inner_Repeater_Shelter" (340-052000) or "Outer_Repeater_Shelter" (340-53000) shall withstand, and under normal operating conditions operate within STFR.FRQ Functional Performance requirements in an ambient temperature between +18°C and +26°C. | Compliant | SADT Rep. **390** (2017) §4.3.3.1.2<br>SADT Rep. **620** (2017) §6.4.3 [RD13] |
| SADT.SAT.STFR.FRQ_REQ-305-078 | SAT.STFR.FRQ components sited within the "Inner_Repeater_Shelter" (340-052000) or "Outer_Repeater_Shelter" (340-53000) shall withstand, and under normal operating conditions operate within STFR.FRQ Functional Performance requirements in a non-condensing, relative humidity environment between 40% and 60%. | Compliant | SADT Rep. **390** (2017) §4.3.3.1.2<br>SADT Rep. **620** (2017) §6.4.4 [RD13] |
| SADT.SAT.STFR.FRQ_REQ-305-079 | SAT.STFR.FRQ components sited within the CPF shall withstand, and under normal operating conditions operate within specification, a fluctuating thermal environment between +18°C and +26°C. | Compliant | SADT Rep. **390** (2017) §4.3.3.1.3<br>SADT Rep. **620** (2017) §6.4.1 [RD13] |
| SADT.SAT.STFR.FRQ_REQ-305-080 | SAT.STFR.FRQ components sited within the CPF shall withstand, and operate within, a fluctuating non-condensing, relative humidity environment between 40% and 60%. | Compliant | SADT Rep. **390** (2017) §4.3.3.1.3<br>SADT Rep. **620** (2017) §6.4.4 [RD13] |
| SADT.SAT.STFR.FRQ_REQ-2798 | SAT.STFR.FRQ equipment and fibre located in non-weather protected locations shall be sufficiently environmentally protected to survive, and perform to specification for all ambient temperatures of between -5°C and +50°C. | Compliant | SADT Rep. **390** (2017) §4.3.3.1.4<br>Submitted TUFFC (2017) [RD16]<br>SADT Rep. **109** (2015) §2.5 [RD14] |

| SADT.SAT.STFR.FRQ_REQ-2798 | SAT.STFR.FRQ equipment and fibre located in non-weather protected locations shall be sufficiently environmentally protected to survive, and perform to specification for rates of change of ambient temperature of up to ±3°C every 10 minutes. | Compliant | SADT Rep. **390** (2017) §4.3.3.1.4<br>Submitted to TUFFC (2017) [RD16]<br>SADT Rep. **109** (2015) §2.5 [RD14] |
|---|---|---|---|
| Wind Speed Requirement | | | |
| SADT.SAT.STFR.FRQ_REQ-2798 | SAT.STFR.FRQ equipment and fibre located in non-weather protected locations shall be sufficiently environmentally protected to survive and perform to specification under normal SKA telescope operating wind conditions up to wind speeds of 40km/hr. | Compliant | SADT Rep. **390** (2017) §4.3.3.2<br>Submitted TUFFC (2017) [RD16]<br>SADT Rep. **109** (2015) §2.5 [RD14] |
| Seismic Resilience Requirement | | | |
| SADT.SAT.STFR.FRQ_REQ-2650 | SAT.STFR.FRQ components shall be fully operational subsequent to seismic events resulting in a maximum instantaneous peak ground acceleration of 1 m/s$^2$. Note: Seismic events include underground collapses in addition to earthquakes. | Compliant | SADT Rep. **390** (2017) §4.3.3.3<br>SADT Rep. **620** (2017) §8.4 [RD13] |

**Table 50 SKA1-MID environmental conditions**

| Requirement # | Requirement Description | Compliance | Reference |
|---|---|---|---|
| Telescope Configuration Requirement | | | |
| SADT.SAT.STRF.FRQ_REQ-2712 | For SKA1_MID, SAT.STFR.FRQ shall provide the MID Reference Frequency "Disseminated Reference Frequency Signal" to 133 SKA receptors and 64 MeerKAT receptors as defined by *SKA-TEL-INSA-0000537 SKA1_Mid Configuration Coordinates*. | Compliant | SADT Rep. **390** (2017) §4.3.3.4<br>In Prep. PASA (2017) [RD5] |
| Overhead Fibre Requirement | | | |
| ECP 160013 | ECP 160013 recommending overhead fibre be used between the CPF and a location close to each receptor for SKA1 Mid has been approved. This means that the above coherence, phase drift and jitter requirements must be met over at least the longest overhead fibre span as defined by *SKA-TEL-INSA-0000537 SKA1_Mid Configuration Coordinates*. | Compliant | SADT Rep. **390** (2017) §4.3.3.5<br>Submitted to TUFFC (2017) [RD16]<br>SADT Rep. **109** (2015) §2.5 [RD14] |

**Table 51 SKA1-MID design conditions**

## 7.3 Key Additional Requirements

The key additional requirements for SKA1-MID are given in Table 52.

| Requirement # | Requirement Description | Compliance | Reference |
|---|---|---|---|
| Offset Frequency Requirements | | | |
| SKA1-SYS_REQ-3385 | It shall be possible to set the sample-clock frequency for each receptor for each band to the nominal sample rate plus or minus N times a frequency offset, where N is an integer, from zero up to half the number of receptors. | Compliant | SADT Rep. **390** (2017) §4.3.4.1 |
| SKA1-SYS_REQ-3386 | There shall be an identical integer number of samples between the time-stamped 1 second marks from each receptor. | Compliant | SADT Rep. **390** (2017) §4.3.4.1 |
| SKA1-SYS_REQ-3387 | The frequency offset shall be such that there is at least 10 KHz of frequency difference to prevent unwanted cross-correlator output. | Compliant | SADT Rep. **390** (2017) §4.3.4.1 |
| SKA1-SYS_REQ-3388 | The maximum of the frequency offset shall be 1% (TBC) of the science bandwidth. | Compliant | SADT Rep. **390** (2017) §4.3.4.1 |
| Monitoring Requirement | | | |
| SADT.SAT.STFR.FRQ_REQ-2280 | At least the following STFR component parameters shall be monitored: the Lock signal (indicating that the STFR system is functioning correctly); the Control voltage (giving an indication of how much control is still available to keep the STFR locked); and the Phase measurement (showing the corrections which have been applied to the frequency to compensate for the effects of changes in the fibre connecting the transmit and receive units of the STFR.FRQ system). | Compliant | SADT Rep. **390** (2017) §4.3.4.2<br>SADT Rep. **390** (2017) §4.5.2 |
| RFI Requirements | | | |
| SADT.SAT.STFR.FRQ_REQ-2462 | SAT.STFR.FRQ components emitting electromagnetic radiation within frequency intervals for broad and narrow band cases shall be within the SKA RFI/EMI Threshold Levels as defined in *SKA-TEL-SKO-0000202-AG-RFI-ST-01*. | Compliant | SADT Rep. **390** (2017) §4.3.4.3<br>SADT Rep. **620** (2017) §10.4 [RD13] |
| SADT.SAT.STFR.FRQ_REQ-3383 | SAT.STFR.FRQ shall provide a "Disseminated Reference Frequency Signal" to each Receptor (MID), which shall be independently and uniquely settable for each Receptor. RFI emissions due to the dissemination of common frequency signals into the array shall not interact detrimentally with the Telescope's Receivers. | Compliant | SADT Rep. **390** (2017) §4.3.4.3<br>SADT Rep. **620** (2017) §10.4 [RD13] |

| Space Requirements | | | |
|---|---|---|---|
| SADT NWA Model SKA1-mid Rev. 03 | The Candidate's solution shall meet the following maximum space requirements as defined by the space allocated to SAT.STFR.FRQ by the SADT NWA model *SKA-TEL-SADT-0000523-MOD_NWAModelMid Revision 3.0*. | Compliant | SADT Rep. **390** (2017) §4.3.4.4<br>SADT Rep. **390** (2017) §4.4.1 |
| Availability Requirement | | | |
| SADT.SAT.STFR.FRQ_REQ-3245 | SAT.STFR.FRQ (end-to-end system excluding fibre) shall have 99.9% "Inherent Availability." | Compliant | SADT Rep. **390** (2017) §4.3.5 |
| Power Requirements | | | |
| SADT.SAT.STFR.FRQ_REQ-305-113 | SAT.STFR.FRQ components, sited within the receptor shielded cabinet, shall not exceed nominal power consumption of 50 Watts. | Compliant | SADT Rep. **390** (2017) §4.3.6<br>Appendix 8.9.1<br>SADT Rep. **523** (2017) 'Pedestal(Inner)' and 'Pedestal(Outer)' tabs [RD2] |
| SADT.SAT.STFR.FRQ_REQ-305-163 | The total power consumption of combined SAT.STFR.FRQ components located in CPF (MID) shall be no more than 2.3 kWatts. | Compliant | SADT Rep. **390** (2017) §4.3.6<br>Appendix 8.9.1<br>SADT Rep. **523** (2017) 'CPF (KAPB) inc MeerKAT' tab [RD2] |
| SADT.SAT.STFR.FRQ_REQ-305-164 | The total power consumption of combined SAT.STFR.FRQ components located in any single instance of the "Inner_Repeater_Shelter" (340-052000) or "Outer_Repeater_Shelter" (340-053000) shall be no more than 160 Watts. | Compliant | SADT Rep. **390** (2017) §4.3.6<br>Appendix 8.9.1<br>SADT Rep. **523** (2017) 'Shelter(Outer)' tab [RD2] |

**Table 52 SKA1-MID key additional requirements**

# 8 APPENDICES

The appendices include information relevant to the SKA phase synchronisation system for SKA1-MID.

## 8.1 Concept Documents

### 8.1.1 Concept document – Time and Frequency Dissemination for the Square Kilometre Array

Appendices\Concept document - Time and Frequency Dissemination for the Square Kilometre Array.pdf

### 8.1.2 Concept document – Transfer of microwave-frequency reference signals over optical fibre links

Appendices\Concept_document_-_Transfer_of_microwave-frequency_reference_signals_over_optical_fibre_links.pdf

## 8.2 Journal Papers

### 8.2.1 Journal paper – Square Kilometre Array: The Radio Telescope of the XXI Century

Appendices\Journal paper - Square Kilometre Array The Radio Telescope of the XXI Century.pdf [RD4].

### 8.2.2 Journal paper – A Clock for the Square Kilometre Array

Appendices\Journal paper - A Clock for the Square Kilometre Array.pdf [RD23].

### 8.2.3 Journal paper – A Phase Synchronization System for the Square Kilometre Array

Appendices\Journal paper - A Phase Synchronization System for the Square Kilometre Array.pdf [RD5].

### 8.2.4 Journal paper – Simultaneous transfer of stabilized optical and microwave frequencies over fiber

Appendices\Journal_paper_-_Simultaneous_transfer_of_stabilized_optical_and_microwave_frequencies_over_fiber.pdf [RD24].

### 8.2.5 Journal paper – Simple Stabilized RF Transfer with Optical Phase Actuation

Appendices\Journal_paper_-_Simple_Stabilized_Radio-Frequency_Transfer_with_Optical_Phase_Actuation.pdf [RD11].

### 8.2.6 Journal paper – Stabilized microwave-frequency transfer using optical phase sensing and actuation

Appendices\Journal_paper_-_Stabilized_microwave-frequency_transfer_using_optical_phase_sensing_and_actuation.pdf [RD10].

### 8.2.7 Journal paper – Characterization of Optical Frequency Transfer over 154 km of Aerial Fiber

Appendices\Journal paper - Characterization of optical frequency transfer over 154 km of aerial fiber.pdf [RD15].

### 8.2.8 Journal paper – Stabilized Modulated Photonic Signal Transfer Over 186 km of Aerial Fiber

Appendices\Journal paper - Stabilized Modulated Photonic Signal Transfer Over 186 km of Aerial Fiber.pdf [RD16].

### 8.2.9 Journal paper – Astronomical verification of a stabilized frequency reference transfer system for the Square Kilometre Array

Appendices\Journal paper - Astronomical verification of a stabilized frequency reference transfer system for the SKA.pdf [RD17].

## 8.3 SADT Reports

### 8.3.1 SADT report – Pre-PDR Laboratory Verification of the SKA Phase Synchronisation System

Appendices\SKA-TEL-SADT-0000616 - Pre-PDR Laboratory Verification of UWAs SKA Synchronisation System.pdf [RD12].

### 8.3.2 SADT report – Pre-CDR Laboratory Verification of the SKA Phase Synchronisation System

Appendices\SKA-TEL-SADT-0000620 - Pre-CDR Laboratory Verification of UWAs SKA Synchronisation System.pdf [RD13].

### 8.3.3 SADT report – UWA South African SKA Site Long-Haul Overhead Fibre Field Trial Report

Appendices\SKA-TEL-SADT-0000109 - UWA South African SKA Site Long-Haul Overhead Fibre Field Trial Report.pdf [RD14].

### 8.3.4 SADT report – SKA-LOW Astronomical Verification

Appendices\SKA-TEL-SADT-0000617 - SKA-low Astronomical Verification.pdf [RD19].

### 8.3.5 SADT report – SKA-mid Astronomical Verification

Appendices\SKA-TEL-SADT-0000524 - SKA-mid Astronomical Verification.pdf [RD18].

### 8.3.6 SADT report – Notes on Calculating the Relationship between Coherence Loss and Allan Deviation

Appendices\SKA-TEL-SADT-0000619 - Notes on Calculating the Relationship between Coherence Loss and Allan Deviation.pdf [RD26].

### 8.3.7 SADT report – Design-for-Manufacture of the SKA1-Mid Frequency Synchronisation System

Appendices\SKA-TEL-SADT-0000618 - Design-for-Manufacture of the SKA1-Mid Frequency Synchronisation System.pdf [RD22].

## 8.4 Detailed Design Overview Files

### 8.4.1 SKA-mid Detailed Design – Overview

Appendices\SKA-mid Detailed Design - Overview.pptx

## 8.5 Mechanical Detailed Design Files

### 8.5.1 SKA-mid Detailed Design – Solid Edge computer aided design files

Appendices\SKA-mid Detailed Design - Solid Edge computer aided design files.zip

## 8.6 Optical Detailed Design Files

### 8.6.1 SKA-mid Detailed Design – Optical Schematics

Appendices\SKA-mid Detailed Design - Optical Schematics.pptx

## 8.7 Electronic Detailed Design Files

### 8.7.1 SKA-mid Detailed Design – Electronic Schematics

Appendices\SKA-mid Detailed Design - Electronic Schematics.pptx

### 8.7.2 SKA-mid Detailed Design – Design Spark circuit schematic and PCB layout files

Appendices\SKA-mid Detailed Design - Design Spark circuit schematic and PCB layout files.zip

## 8.8 Modelling Software

### 8.8.1 Modelling software – Java interactive tool for modelling frequencies in optical fibre networks

Appendices\Modelling software - Java interactive tool for modelling frequencies in optical fibre networks.zip

### 8.8.2 Modelling software – SKA Excel Frequency Calculator

Appendices\Modelling software - SKA Excel Frequency Calculator.xlsx

## 8.9 Bill of Materials and Cost Model Database

### 8.9.1 SKA-mid Detailed Cost Model – Capex Equipment

Appendices\SKA-mid Detailed Cost Model - Capex Equipment.xlsx

### 8.9.2 SKA-mid Detailed Cost Model – Capex Labour

Appendices\SKA-mid Detailed Cost Model - Capex Labour.xlsx

### 8.9.3 Light Touch Solutions – Capex Labour Costing Analysis report

Appendices\SKA-mid and SKA-low Labour cost review.pdf

## 8.10 Independent Assessment of Solution

### 8.10.1 ASTRON Netherlands Institute for Radio Astronomy – Gijs Schoonderbeek

[Appendices\Independent Assessment - ASTRON Netherlands Institute for Radio Astronomy.pdf](Appendices/Independent%20Assessment%20-%20ASTRON%20Netherlands%20Institute%20for%20Radio%20Astronomy.pdf)

### 8.10.2 Jet Propulsion Laboratory – Larry D'Addario

[Appendices\Independent Assessment - Jet Propulsion Laboratory.pdf](Appendices/Independent%20Assessment%20-%20Jet%20Propulsion%20Laboratory.pdf)

### 8.10.3 Square Kilometre Array South Africa – Johan Burger

[Appendices\Independent Assessment - Square Kilometre Array South Africa.pdf](Appendices/Independent%20Assessment%20-%20Square%20Kilometre%20Array%20South%20Africa.pdf)

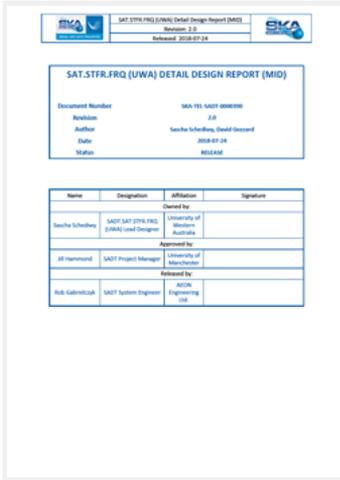

# SKA-TEL-SADT-0000390_DDD_SAT.STFR.FRQ(UWA)DetailDesignReport(MID)

Adobe Sign Document History          24/07/2018

| | |
|---|---|
| Created: | 24/07/2018 |
| By: | Michelle Hussey (michelle.hussey@manchester.ac.uk) |
| Status: | Signed |
| Transaction ID: | CBJCHBCAABAAJRumPV2_hRrlYMpfVPmQm5V6MMYjWcWI |

# "SKA-TEL-SADT-0000390_DDD_SAT.STFR.FRQ(UWA)Detail DesignReport(MID)" History

- Document created by Michelle Hussey (michelle.hussey@manchester.ac.uk)
  24/07/2018 - 02:14:39 PDT- IP address: 2.122.139.205

- Document emailed to Sascha Schediwy (sascha.schediwy@uwa.edu.au) for signature
  24/07/2018 - 02:16:17 PDT

- Document viewed by Sascha Schediwy (sascha.schediwy@uwa.edu.au)
  24/07/2018 - 02:43:17 PDT- IP address: 59.101.134.36

- Document e-signed by Sascha Schediwy (sascha.schediwy@uwa.edu.au)
  Signature Date: 24/07/2018 - 03:58:04 PDT - Time Source: server- IP address: 59.101.134.36

- Document emailed to Jill Hammond (jill.hammond@manchester.ac.uk) for signature
  24/07/2018 - 03:58:06 PDT

- Document viewed by Jill Hammond (jill.hammond@manchester.ac.uk)
  24/07/2018 - 03:58:27 PDT- IP address: 130.88.9.95

- Document e-signed by Jill Hammond (jill.hammond@manchester.ac.uk)
  Signature Date: 24/07/2018 - 03:58:51 PDT - Time Source: server- IP address: 130.88.9.95

- Document emailed to Robert Gabrielczyk (Rob@aeon-eng.com) for signature
  24/07/2018 - 03:58:53 PDT

- Document viewed by Robert Gabrielczyk (Rob@aeon-eng.com)
  24/07/2018 - 04:15:58 PDT- IP address: 185.230.101.8

- Document e-signed by Robert Gabrielczyk (Rob@aeon-eng.com)
  Signature Date: 24/07/2018 - 04:16:33 PDT - Time Source: server- IP address: 185.230.101.8

Adobe Sign

- Signed document emailed to Robert Gabrielczyk (Rob@aeon-eng.com), Jill Hammond (jill.hammond@manchester.ac.uk), Sascha Schediwy (sascha.schediwy@uwa.edu.au) and Michelle Hussey (michelle.hussey@manchester.ac.uk)

  24/07/2018 - 04:16:33 PDT

Adobe Sign